\documentclass[phd,cornellheadings]{cornell}
\usepackage{amsmath}
\usepackage{euscript}
\usepackage{epsfig}
\usepackage{hanging,hangcaption}
\usepackage{color,ifthen}

\title{Soft-Collinear Factorization 
and Sudakov Resummation 
of Heavy Meson Decay Amplitudes 
with Effective Field Theories}
\author{Bj\"orn Olaf Lange}
\conferraldate{January}{2005}

\def\pslash{\rlap{\hspace{0.02cm}/}{p}}
\def\delslash{\rlap{\hspace{0.02cm}/}{\partial}}
\def\lslash{\rlap{/}{l}}

\def\nslash{\rlap{\hspace{0.02cm}/}{n}}
\def\nbslash{\rlap{\hspace{0.02cm}/}{\bar n}}
\def\vslash{\rlap{\hspace{0.02cm}/}{v}}

\def\Dslash{\rlap{\hspace{0.07cm}/}{D}}
\def\Aslash{\rlap{\hspace{0.07cm}/}{A}}
\def\calAslash{\rlap{\hspace{0.08cm}/}{{\EuScript A}}}

\def\calDslash{\rlap{\hspace{0.1cm}/}{{\EuScript D}}}
\def\delslash{\rlap{\hspace{0.02cm}/}{\partial}}
\def\epsslash{\rlap{\hspace{0.02cm}/}{\varepsilon}}

\def\A{{\EuScript A}}

\def\D{{\EuScript D}}
\def\H{{\EuScript H}}
\def\Q{{\EuScript Q}}
\def\X{{\EuScript X}}
\def\F{{\EuScript F}}
\def\J{{\EuScript J}}

\def\bm#1{\mbox{\boldmath$#1$\unboldmath}}

\def\capfont{\normalsize}

\renewcommand{\caption}[2][]
{\singlespacing
  \ifthenelse{\equal{}{#1}}
  {\hangcaption{\capfont #2}}
  {\hangcaption[#1]{\capfont #2}}
  \normalspacing
}
\newcommand{\captionstar}[2][]
{\singlespacing
  \ifthenelse{\equal{}{#1}}
  {\caption*{\capfont #2}}
  {\caption*[#1]{\capfont #2}}
  \normalspacing
}

\begin{document}
\maketitle

\iffinal
\makecopyright

\begin{abstract}
Decays of the $B$ meson into light and energetic particles are
discussed. The calculation of the corresponding decay amplitudes is
non-trivial in many respects. Strong interaction effects are always
present and cannot be computed reliably using analytic
techniques. However, besides the intrinsic energy scale $\Lambda_{\rm
QCD}$ of Quantum Chromo Dynamics, there also exists a much larger
scale, the $b$-quark mass $m_b$, at which perturbation theory can be
applied. QCD-factorization is the idea of separating the contributions
that arise at these different scales and performing a systematic
expansion in the ratio $\Lambda_{\rm QCD}/m_b$. Hard processes at the
large scale can be computed using perturbation theory, while soft
processes are encoded in non-perturbative structure functions. The
observation that for many decay modes the same structure functions are
needed make this a useful approach. However, factorization theorems
need to be proved for every single decay, since some amplitudes do not
factorize. The intent of this thesis is to study examples of
factorizable and non-factorizable amplitudes in a systematic
framework, by using effective field theory techniques.

The advancing precision of experimental measurements of $B$-decays
make it necessary to improve the accuracy of theoretical
predictions. To achieve this, it is necessary to perform a resummation
of Sudakov logarithms, which enter at every non-trivial order in
perturbation theory.

In this thesis we present an introduction to Soft-Collinear Effective
Theory, which can be used to prove (or disprove) factorization
theorems to all orders in the strong coupling constant for some $B$
decays into light and energetic particles. Specifically, the
factorizable amplitudes for inclusive $B$ $\to$ $X_u\,l^-\bar\nu$ and
exclusive $B^-$ $\to$ $\gamma\,l^-\bar\nu$ are calculated in
renormalization-group improved perturbation theory to first
non-trivial order. Form factors encoding the exclusive decay
amplitudes for $\bar B$ $\to$ $P\,l^-\bar\nu$ and $B$ $\to$ $V\,l^-\bar\nu$
($P=$ light pseudoscalar meson, $V=$ light vector meson)
are studied and proved to be dominated by the non-factorizable Feynman
mechanism.
\end{abstract}

\begin{biosketch}
Bjorn O.~Lange was born in Minden, Germany on the 24$^{\rm th}$ of
September, 1974, as the second son of the primary school teacher
Christa Lange and the patent attorney Gerd Lange. He attended the
primary school ``Bierpohlschule'' for four years, starting in 1981,
and continued his education at the ``Besselgymnasium'' in Minden. In
1994 he graduated with the ``Abitur'' at the top of his class and was
awarded the ``Besselpreis'' for extra-curriculum activities. After
spending twelve months as a navigator in the German navy in
fulfillment of his military service duties, he began his academic
education at the university of Heidelberg, Germany, where he studied
Physics. In 1998 he was awarded one year of full scholarship at
Cornell University, where he enrolled in the PhD physics program on a
non-degree status in the academic year 1998/99. After his return to
Germany he finished the program in Heidelberg and received the degree
``Diplom-Physiker'' in 2001. In the fall of that year he came to
Cornell as a regular graduate student. This document is the outcome of
that course of action.
\end{biosketch}






\begin{acknowledgements}
Many thanks to my collaborators of the past three years: Thomas
Becher, Stefan Bosch, Richard Hill, Matthias Neubert, and Gil Paz.
Much of the content of this thesis has been found in teamwork, and I
am deeply indebted to them. I am especially grateful to my advisor
Matthias Neubert for his encouragement, support, patience, and, most
of all, for being an outstanding teacher. 

The research has been supported by the National Science Foundation
under Grant PHY-0098631. 

\end{acknowledgements}
\fi

\contentspage
\tablelistpage
\figurelistpage

\normalspacing
\setcounter{page}{1}
\pagenumbering{arabic}
\pagestyle{cornell}

\chapter{Introduction}

\section{Preface}

The conscious and active interplay between the observation of Nature
and its mathematical description is at the core of the physical
sciences since the 16$^{\rm th}$ century. In the quest of finding the
ever-more fundamental processes, this interplay has led to emergence
of the ``Standard Model'' in the second half of the 20$^{\rm th}$
century. Our current theoretical understanding of elementary processes
at energies testable through terrestrial experiments is based on the
description within quantum field theories in combination with the
gauge principle.  The effects of electromagnetic, weak and strong
nuclear forces through which the matter, quarks and leptons,
interacts, is captured in a quantum field theory which is invariant
under the gauge group $SU(3)_C \times SU(2)_L \times U(1)_Y$. The
electromagnetic and weak forces are unified in the gauge group
$SU(2)_L \times U(1)_Y$ around the energy scale of roughly 100
GeV. Below the electroweak scale this gauge group is spontaneously
broken down to $U(1)_{\rm em}$, which is the group of Quantum Electro
Dynamics, and the $W^\pm$ and $Z$ bosons mediating the weak
interactions become massive. In the most simple mechanism that breaks
the electroweak symmetry, the Higgs mechanism or
Glashow-Weinberg-Salam Model, masses are also given to quarks and
leptons. Fermion masses are generated through Yukawa interactions with
the Higgs field. Using global unitary transformations in flavor space,
the Yukawa interactions are diagonalized to obtain the physical mass
eigenstates. It can be chosen such that all up-type quarks (up, charm,
top) are not rotated in flavor space, whereas the weak eigenstates of
the down-type quarks (down, strange, bottom) are related to the mass
eigenstates through the unitary matrix $V_{\rm CKM}$. Numerically, the
Cabbibo-Kobayashi-Maskawa matrix $V_{\rm CKM}$ is found to be almost
diagonal, as the non-diagonal entries, that allow for transitions
between quarks of different generations, are small. Such transitions
are mediated by the weak currents. Since the $W^\pm$ and $Z$ bosons
have masses very close to the electroweak scale, it is possible to
construct an effective theory for weak processes far below this
scale. This theory is called the Fermi theory of weak interactions, or
the weak effective Hamiltonian.

The theory of strong interactions is Quantum Chromo Dynamics (QCD),
the non-abelian gauge theory of $SU(3)_C$. The condition to be
renormalizable determines the Lagrangian (almost entirely) to be 
\begin{equation}
\mathcal L_{\rm QCD} = \sum\limits_{\psi = u, d, \ldots}
\bar \psi ( i\Dslash - m_\psi) \psi \;
-\; \frac14\,G^a_{\mu\nu} G^{a,\mu\nu}\;,
\end{equation}
where $i D^\mu = i\partial^\mu + g_s A^\mu$ is the covariant
derivative. We have suppressed any spinor indices and the color index
$a$ on the gluon field $A_\mu = A^a_\mu T^a$, which reflects the fact
that it is a color-octet. The gluon field strength is $G_{\mu\nu}^a =
\partial_\mu A^a_\nu - \partial_\nu A^a_\mu + g_s f^{abc} A^b_\mu
A^c_\nu$. Finally, the generators of $SU(N)$ are denoted by $t^a$ with
$a = 1,\ldots,N^2-1$, $f^{abc}$ are the structure constants, and $g_s$
is the strong coupling constant. Every procedure to renormalize the
theory necessitates the introduction of a dimensionful parameter
$\mu$. It is generally thought of as the energy scale at which the
theory is defined. The dependence of various quantities in the theory
on this scale is governed by renormalization-group equations (RGEs),
which can be calculated in perturbatively as long as the coupling is
reasonably small compared to unity. Qualitatively, the strong coupling
``constant'' $g_s(\mu)$ is large for values of $\mu$ around the
intrinsic QCD scale $\Lambda_{\rm QCD} \sim 0.5$ GeV, and tends
asymptotically to zero for increasing $\mu$. This effect is in
contrast to the other sectors of the Standard Model and reflects the
fact that quarks are confined in hadrons for low scales, but are
``asymptotically free'' for highly energetic processes such as deep
inelastic scattering at scales much larger than $\Lambda_{\rm QCD}$.

All of the above is standard class-work and text-book material, which
is why we do not feel the need to cite \cite{classes} and discuss the
original works in great detail here.

Corrections to the Standard Model, generally referred to as ``New
Physics'', can be searched for in two complementary ways: in direct
searches where new particles are produced and observed, or in indirect
searches in which new particles enter as quantum corrections. The
former method is obviously a very effective way to find New Physics,
albeit a formidable financial endeavor. Indirect searches, on the
other hand, require both precision measurements on the experimental
side, as well as a level of theoretical Standard Model predictions
that compares well with the experimental uncertainties. Only if both
sides of the scientific medallion, the observation of Nature through
High-Energy Experiments and its mathematical description within the
Standard Model, are met with equal precision can we conclude on the
validity of the underlying theory. 

Needless to say, the knowledge of the parameter values entering the
Standard Model is crucial for theoretical predictions to be
reliable. The flavor sector is in this respect very challenging. In
particular, the smallest of the CKM matrix elements, $V_{ub}$ and
$V_{td}$, are least well known, but are responsible for a wealth of
phenomena, such as $\mathcal C \mathcal P$ violation. In general, the
flavor sector of the Standard Model displays a broad spectrum of
phenomena and mechanisms that need to be understood in order to engage
in indirect searches for New Physics. Primarily one is interested in
the study of flavor-changing heavy quark decays, which are governed by
the weak interactions. However, it is the strong force that is
responsible for the formation of the hadrons that can be observed in
experiments. The $B$ meson is a unique system in that it is the
simplest hadron containing a heavy quark. Simple, because it is a
bound state of only one heavy and one light (anti-) quark. A heavy
quark is favorable for New Physics searches, because flavor-changing
processes from heavy particles of the third family to light particles
of the first two families are very rare. If the Standard Model
predicts a quantity to be small or even vanishing, the chances of
observing New Physics are more pronounced. Even though top quarks are
by far the heaviest, they are less suitable because they decay before
hadronization can occur.

The broader aim of the ``$B$-physics community'' is the study of weak
decays of $B$ mesons and the interpretation of the data collected at
dedicated $B$ factories, such as the BaBar, Belle and CLEO
experiments. The major goals are to test the Standard Model, extract
its fundamental parameters, explore the phenomenon of $\mathcal C
\mathcal P$ violation, and to search for New Physics at higher
energies. This quest will continue in the future with the help of the
Large Hadron Collider and the LHCb detector in particular, as well as
BTeV and possible High-Luminosity $B$ factories. The theoretical
challenge is the development and application of tools that enable us
to control the effects from strong interactions. These theoretical
tools include effective field theories, factorization theorems,
symmetries, and heavy-quark expansions. It is our belief that the
present work contributes a significant step toward meeting this challenge.

\section{Methodology}

The calculation of hadronic processes is extremely difficult because
of our lack of understanding of the connection between quark and
hadron properties at low-energy scales. While it is possible to apply
the perturbative technique to ``hard'' processes (for example
$\alpha_s(m_b) \approx 0.22$), soft processes make it difficult for
analytical methods to make reliable predictions, since the expansion
in the coupling constant $\alpha_s(\mu) = g_s^2(\mu)/4\pi$ no longer
converges. This puts some grave limitations on the amount of
information about the quark level, that can be extracted from the
hadronic level, and vice versa. Often times it is only possible to
access such information if certain symmetries are present, such as the
chiral symmetry for massless quarks, or the heavy-quark symmetry in
the limit of infinite quark masses \cite{Neubert:1993mb}.

In the present work, we will concentrate on the subgroup of $B$ decays
into light particles only. We are thus facing the problem of having
neither of the two symmetries mentioned above, but rather a system
that involves both heavy and light quarks. As we shall see below, in
some cases a certain ``universality'' emerges, that all soft processes
are captured in a set of structure functions entering the calculation
of factorizable decay amplitudes, regardless of the specific decay
channel. To achieve this it is necessary to disentangle the
short-distance effects associated with the large scale $m_b$, from the
long-distance effects at the low scale $\Lambda_{\rm QCD}$. This is
the idea of QCD-factorization, as put forward by Beneke, Buchalla,
Neubert, and Sachrajda
\cite{Beneke:1999br,Beneke:2000ry,Beneke:2000fw,Beneke:2001ev}. It
states that factorizable amplitudes can be expressed in the
heavy-quark limit and to all orders in perturbation theory as
convolution integrals of a perturbatively calculable ``hard-scattering
kernel'' with some of the universal structure functions, which are
treated as input parameters, i.e. quantities that need to be extracted
by other non-perturbative techniques or directly from experiment. For
practical purposes they may simply be replaced by a model. We shall
stress, however, that this would obviously introduce a model
dependence to the prediction, whereas QCD-factorization itself is a
model independent property of QCD. We were careful to add the pronoun
``factorizable'' to the amplitudes in question, because not all
processes share this feature. Factorization needs to be proved for
every individual decay channel.

QCD-factorization combines the disentangling of physics effects from
different energy scales with the idea of ``naive factorization'' of
matrix elements. For a simple example of naive factorization consider
the weak semileptonic decay amplitude of a neutral kaon into a charged
pion and an electron-neutrino pair. The parton-level tree diagram
mediating this decay is as follows: the strange quark couples to an up
quark and an highly off-shell $W$ boson, which in turn decays into a
lepton-neutrino pair. After integrating out the $W$ boson, the decay
is mediated by $G_F\, V_{us}$ times a four-fermion operator. The
matrix element of this operator can be factorized to all orders in the
strong coupling constant into
\begin{equation}\label{naiveFact}
  \langle e^-\bar\nu_e\, \pi^+|\;
  (\bar l\nu)_{\rm V-A}(\bar us)_{\rm V-A}\;|\bar K^0\rangle
= \langle e^-\bar\nu_e|(\bar l\nu)_{\rm V-A}|0\rangle\;
  \langle \pi^+|(\bar us)_{\rm V-A}|\bar K^0\rangle\;,
\end{equation}
since gluons cannot couple to the lepton-neutrino pair. Note that this
argument holds for any lepton-neutrino pair that is kinematically
allowed, e.g. also for a muon-neutrino pair. The purely hadronic
matrix element on the right-hand side is identified with a structure
function called the transition form factor. Therefore equation
(\ref{naiveFact}) serves as a simple example of the ``universality''
mentioned above. Note also that an attempt to further factorize this
hadronic matrix element must fail because of unsuppressed soft gluon
exchange. The question of whether this is nevertheless possible for a
$B\to\pi$ transition form factor (exchange $s\leftrightarrow b$ quarks
and $K\leftrightarrow B$ states) in the limit of $m_b\to\infty$ is a
very non-trivial one and will be addressed in this thesis. (The answer
is that the form factor does {\em not} factorize. However, the study
reveals many surprising properties of this quantity.) To see how this
might possibly work consider a different example, the non-leptonic
$B\to D \pi$ decays in which the $D$ meson picks up the spectator
quark of the $B$ meson, i.~e. $\bar B\to D^+\pi^-$ and $B^-\to
D^0\pi^-$ \cite{Bauer:2001cu}. The spectator quark and the other light
degrees of freedom inside the $B$ meson can easily form a $D$ meson
after the weak $b\to c$ transition. The remaining two light quarks are
very energetic. To form a pion they must carry momenta collinear to
the pion momentum and must form a color-singlet state. Such an
energetic ``color-transparent'' compact object can leave the decay
region without interfering with the formation of the $D$ meson. The
decoupling of soft degrees of freedom from collinear ones therefore
lies at the core of the color-transparency argument, and likewise of
soft-collinear factorization.

Let us now turn to the question of energy scale separation.
Perturbative effects are included in the hard-scattering kernel. In
many processes one has to deal with multiple scales that enter the
calculation. Typically, aside from a large scale $E$ and the soft
scale $\Lambda_{\rm QCD}$ there exists also an intermediate scale of
order $\sqrt{E\Lambda_{\rm QCD}}$. For the sake of the argument, let
us simply state that two perturbative scales $M_1$ and $M_2$ enter the
calculation of the hard-scattering kernel. Whereas one would naturally
expand the kernel in the strong coupling constant $\alpha_s(\mu)$ if
$\mu$ can be chosen close to either of the two scales, it could also
happen that $\alpha_s(\mu)$ is multiplied by powers of the large
logarithm $\ln M_1/M_2$, thus upsetting the expansion procedure. The
effects of a single power of such a logarithm at one-loop order is
well understood. However, double (Sudakov) logarithms appearing in the
series $\alpha_s \ln^2 + \alpha_s^2 \ln^4 +\ldots$ are more
troublesome. A clean separation of scales requires the resummation of
large logarithms to all orders in perturbation theory.

Both issues, the soft-collinear factorization and the separation of
energy scales including Sudakov resummation, can be most elegantly
addressed using effective field theory technology. By definition the
use of effective field theories achieves the separation of scales in
that infra-red physics are reproduced in the matrix elements of
effective operators, while ultra-violet physics effects give rise to
Wilson coefficients. A resummation of large Sudakov logarithms is
performed by solving the RGEs of effective operators and running
Wilson coefficients from a high scale $\mu_h \sim M_1$ down to a low
scale $\mu_i \sim M_2$. 

The theories we have in mind are the familiar heavy-quark effective
theory (HQET) and two versions of soft-collinear effective theory
(SCET), called SCET$_I$ and SCET$_{II}$. These theories have been
proposed and pioneered in the year 2000 by Bauer, Pirjol, Stewart
et.~al. \cite{Bauer:2000ew,Bauer:2000yr,Bauer:2001ct,Bauer:2001yt},
with further development by Beneke, Feldman, et.~al.
\cite{Beneke:2002ph,Beneke:2002ni} and Becher, Hill, Neubert, et.~al.
\cite{Hill:2002vw,Becher:2003qh,Becher:2003kh} as of today. (Here we
refer to the major developments only. Many more researchers
contributed as well and were not mentioned, for which we apologize.)
These theories can be applied to any process in which soft particles
interact with light but highly energetic particles such as
heavy-to-light decays, deep inelastic scattering and jet physics. In
the present thesis SCET will be used to prove (or disprove)
QCD-factorization for some exemplary inclusive and exclusive $B$ meson
decays.

\section{Structure of Thesis}

The thesis is structured as follows: 

We begin with a pedagogical introduction to SCET in
Chapter~\ref{chap:scet}. Since the theory is a relatively recent
development within the Standard Model (its proposal dates only four
years back), a comprehensive introduction on a basic level seems
worthwhile. It should be understood that the presentation does not
follow the chronology of development, but rather follows the logical
steps in the construction as understood in retrospect\footnote{In
particular we will not formulate SCET in the ``label formalism'' which
was used in early papers by Bauer et.~al. The formulation in position
space, first used by Beneke et.~al. and then by Neubert et.~al., is
equally intuitive but easier on the technical level. We will use the
latter framework in this thesis.}. As mentioned above, two different
theories, called SCET$_I$ and SCET$_{II}$, are needed in later
Chapters. They differ in the field content and will be discussed
separately. Conceptually, SCET$_I$ is less complicated in that it
contains only two distinct sectors of momentum modes, whereas
SCET$_{II}$ contains three and therefore displays a richer gauge
structure. Before we address the two theories separately and in
detail, a brief motivation and an outline of the general strategy is
given. As a first application, we consider SCET currents and general
four-quark operators, including their renormalization. These
quantities are of special interest to any heavy-to-light decay
process, in which a local operator product expansion (OPE) fails due
to the appearance of soft and collinear singularities. We close this
Chapter with a discussion on reparameterization invariance, which
provides useful information on the structure of SCET operators and the
large-scale dependence of their Wilson coefficients.

Hadronic matrix elements of low-energy effective operators cannot be
calculated perturbatively. They define long-distance structure
functions, whose precise functional form must be extracted by other
means, for example Lattice QCD or directly from experiment. However,
much can be learned apart from the functional form using perturbative
techniques. For two structure functions, the $B$-meson light-cone
distribution amplitude (LCDA) and the so-called shape function, such
calculations are performed in Chapter~\ref{chap:strucfunc}. While
LCDAs enter the factorization theorems of exclusive $B$ decays, the
shape function encodes the ``Fermi motion'' of the $b$ quarks inside
the $B$ meson and is needed to factorize inclusive $B$ decay
amplitudes. Besides a study of renormalization properties of these
structure functions, several novel constraints emerging from a moment
analysis of the shape function are derived. 

As a first application we factorize the decay amplitude for inclusive
$\bar B\to X_u\,l^-\bar\nu$ in Chapter~\ref{chap:inclusive}, using a
two-step matching procedure QCD $\to$ SCET$_I$ $\to$ HQET at
next-to-leading order in perturbation theory and leading power. Such a
calculation is appropriate in the phase-space region of large energy
and moderate invariant mass of the $X_u$ system. (It would be possible
to avoid this two-step matching procedure and apply a local operator
product expansion instead, only if the $X_u$ invariant mass is also
large. Such a kinematic setup is, however, not suitable due to large
backgrounds, see below.) The inclusive decay is, in a sense, simpler
than exclusive ones because the more complicated theory SCET$_{II}$
does not enter the computation.  However, the matching calculation and
subsequent resummation of Sudakov logarithms by analytically solving
renormalization-group equations is far from trivial. To discriminate
the background $\bar B\to X_c\,l^-\bar\nu$ decays it is necessary to
restrict the kinematics to a subspace in phase-space by means of
certain experimental cuts. We present results for event fractions that
pass such cuts. To give a final numerical answer the functional form
of the shape function is required. Here we adopt a model that is
consistent with all constraints derived in the previous
Chapter~\ref{chap:strucfunc}. One of the discussed methods of
kinematic cuts, the $P_+$ cut, is theoretically favored and deserves a
closer study. This is done at the end of this Section, including a
thorough estimate of all theoretical uncertainties entering a
$|V_{ub}|$ determination using this method.

In the remainder of this thesis we turn our attention to exclusive
decays. In Chapter~\ref{chap:radiative}, the outlined methodology is
applied to $B\to \gamma\,l^-\bar\nu$, which is one of the
simplest exclusive decay modes in that there are no hadrons in the
final state. Still there is sensitivity to the light-cone structure of
the $B$ meson, because of the coupling of the high-energy photon to
the soft spectator quark inside the heavy meson. Many techniques can
be learned in this environment. With minor modifications they can then
be applied to the hard-scattering term of such important decay
amplitudes as the ones for $B\to K^*\gamma$ or $B\to
\pi\pi$, etc. One of the important remaining contributions to the
corresponding amplitudes involves $B\to M$ form factors, where
$M$ is some light meson. While it is not the goal of this thesis to
study these difficult decay modes in all completeness, we wish to
contribute by analyzing some of their ingredients in a clean
environment.

For that reason we devote the Chapter~\ref{chap:formfactor} to analyse
$B\to M$ form factors in the high recoil region, meaning that the
light meson $M$ is highly energetic. The underlying processes are
$B\to \pi\,l^-\bar\nu$ and $B\to \rho\,l^-\bar\nu$. The calculation of
these form factors using a two-step matching procedure QCD $\to$
SCET$_I$ $\to$ SCET$_{II}$ reveals some almost expected outcomes (such
as the fact that form factors do {\em not} factorize, and that form
factors at large recoil are power suppressed quantities) and also some
surprises. The most important one might be that, although going
through SCET$_I$ enables us to resum large {\em perturbative}
logarithms, there exists a purely long-distance contribution for which
the intermediate theory is without any physical significance. In other
words, the non-perturbative ``soft overlap'' or Feynman mechanism (of
lots of small momentum kicks to bring the spectator quark up to speed)
exists and dominates over the hard-scattering picture. Furthermore we
find that there are unsuppressed contributions from higher Fock states
of both the $B$ and $M$ meson. More precisely, three-particle
configurations (two quarks plus one gluon) contribute at the same
power as two-particle ones. This is in contrast to the previous
understanding as found by e.~g.~light-cone sum rules. It is tempting
to try to estimate the contribution of this effect; however, this
would be accompanied by large uncertainties. Note also that we only
perform the matching at tree-level, due to the complexity of the
problem. A numerical analysis is therefore left for future work and
will not be given here.  Finally we conclude in
Chapter~\ref{chap:conclusion}.

\addtocontents{toc}{\protect\contentsline{part}{Theory}{}}
\chapter{Soft-Collinear Effective Theory}\label{chap:scet}

\section{General Considerations}
Many processes in $B$-decay physics involve hadrons or jets of hadrons
with energies much larger than their masses. The assumption, that
underlies the construction of an effective field theory describing the
dynamics of the process, is that the constituents of these hadrons or
jets carry momenta collinear to the hadronic momenta. Consider a light
but energetic hadron with momentum $P^\mu$ in the $z$-direction and
invariant mass $P^2=m_H^2$ much smaller than its energy $E$. We may
decompose $P^\mu = (E-m_H^2/4E+\ldots)n^\mu + (m_H^2/4E+ \ldots)\bar
n^\mu$ with the light-like vectors $n^\mu=(1,0,0,1)$ and $\bar
n^\mu=(1,0,0,-1)$ satisfying $n^2=\bar n^2 = 0$ and $n\cdot \bar n =
2$. Up to small corrections of order $m_H^2/E^2$ the coefficient of
$n^\mu$ is given by the large energy $E$ of the hadron. A natural
description of these kind of momenta is given in terms of the
light-cone decomposition
\begin{equation}\label{lightconeCoords}
  P^\mu = (n\cdot P)\,\frac{\bar n^\mu}{2} 
        + (\bar n\cdot P)\,\frac{n^\mu}{2} 
        + P_{\perp}^\mu \;.
\end{equation}
Four-vectors are characterized by the scaling of the individual
light-cone components. The three terms in (\ref{lightconeCoords}) are
referred to as $P_+$, $P_-$ and $P_\perp$. In other words, $P_+$ is a
four-vector $P_+^\mu = (n\cdot P)\,\bar n^\mu/2$, etc. In the example
above, we have $(P_+^\mu,P_-^\mu,P_\perp^\mu) =
(\frac{m_H^2}{2E}\,\frac{\bar n^\mu}{2},2E\,\frac{n^\mu}{2},0)+\ldots$ 
as the leading contributions. We assume that a parton inside this
hadron carries momentum $p^\mu$, which also points essentially in the
$n^\mu$ direction, but it may also have non-vanishing momentum in the
$\bar n$ and perpendicular direction. Those are, however, dynamically
generated and typically much smaller than the large minus
component. By introducing a dimensionless scaling parameter $\lambda$
we may characterize the order of the light-cone components $p^\mu \sim
(\lambda^2, 1, \lambda)$. The scaling of the ``plus component'' $p_+
\sim \lambda^2$ follows from the assumptions that the invariant mass
of the constituent $p^2=2p_+\cdot p_-+p_\perp^2$ is not larger than
$O(\lambda^2)$. All the above assumptions are well justified for the
processes described below. For example, consider a decay $B \to \pi$
in the kinematic situation where the $B$ meson is at rest and the pion
takes away almost half of the $B$-meson mass as kinetic energy $E
\approx M_B/2$. The quarks and gluons inside the pion are massless,
but off-shell by parametrically $\Lambda_{\rm QCD}^2$. Throughout this
work, $\Lambda_{\rm QCD}$ denotes the soft QCD scale of
non-perturbative physics. We would define $\lambda = \Lambda_{\rm
QCD}/E$ and assume $\bar n\cdot p = 2xE$ where $x$ is a positive dimensionless
variable less than (but of order) 1. Generally, the precise definition
of the scaling parameter $\lambda$ depends on the particular process
and should be thought of as the ratio of two scales whenever we are
facing a multi-scale problem. The effective theory is constructed
around the systematic expansion in inverse powers of the larger of the
two scales. We shall distinguish between labels and dynamical
components. While large momentum components appear as labels and are
not altered by soft interactions (only by collinear interactions), the
remaining ones are dynamical. This is similar to the notion of the
velocity label $v$ and ``residual'' momentum $k$ in the HQET expansion
$p_b = m_b v+k$ \cite{Neubert:1993mb}. For the sake of simplicity, we
will not write out explicitly the label $v$ on heavy quark fields or
the label $E n$ on collinear fields. In fact, we will not use the
label formalism of earlier papers, and keep the large momentum
dependence explicit.

The relevant degrees of freedom of SCET are such that for any given
perturbative process in QCD involving the kind of momenta described
above, each infra-red contribution is reproduced by the effective
fields. In this spirit, SCET is founded on the method of regions
\cite{Beneke:1997zp}. However, the approach of going through the
effective theory allows for a systematic power counting already at the
beginning of a calculation.

There is a certain spin-symmetry realized in the large energy
limit. This can be understood when decomposing the collinear
QCD spinor $\psi_c$ into two separate fields $\xi$ and $\eta$, which
are both two-component spinors and subject to the constraints $\nslash
\xi = 0$ and $\nbslash \eta = 0$. One defines
\begin{equation}
  \xi = \frac{\nslash \nbslash}{4} \psi_c \; ,\quad 
  \eta = \frac{\nbslash \nslash}{4} \psi_c \; .
\end{equation}
The power counting associated with these two-component spinor fields
is chosen such that propagation of the fields do not enter the power
counting of any given diagram, i.e. count as unity. Denoting the
time-ordered product by $T$, the propagator reads
\begin{equation}
  \langle 0 | T \left\{ \psi_i(x), \bar \psi_j(y) \right\} |0 \rangle
  = \int \frac{d^4 p}{(2\pi)^4}\, \frac{i\pslash_{ij}}{p^2+i \epsilon}
  \, e^{-i p\cdot(x-y)} \; .
\end{equation}
When assuming collinear momentum scaling $\sim (\lambda^2, 1,
\lambda)$, the integration measure $d^4 p = \frac{1}{2} dp_+ dp_- d^2
p_\perp$ scales like $O(\lambda^4)$, and $p^2 \sim O(\lambda^2)$. It
follows that the individual fields count as $\xi \sim O(\lambda)$ and
$\eta \sim O(\lambda^2)$.  In the limit of large energy, the
4-component QCD spinors $\psi_c$ reduce to a two-component spinor
$\xi$, thus giving rise to a reduced Dirac basis. This is the
manifestation of large-energy spin-symmetry
\cite{Charles:1998dr}. Similarly, analyzing the gluon propagator in
covariant gauge
\begin{equation}
    \langle 0 | T \left\{ A_c^\mu(x), A_c^\nu(y) \right\} |0 \rangle =
  \int \frac{d^4 p}{(2\pi)^4}\, \frac{i}{p^2+i\epsilon} \left[
  -g^{\mu\nu} +(1+\alpha)\, \frac{p^\mu p^\nu}{p^2} \right] e^{-i
  p\cdot(x-y)}
\end{equation}
reveals that $A_c^\mu$ counts in its light-cone components like the
momentum $p^\mu$ itself.

After the identification of the small component $\eta$, the
SCET-Lagrangian, which contains only the leading collinear field
$\xi$, is constructed by integrating out $\eta$. Let $\psi_c$ be the
full QCD quark field which carries collinear momentum.
\begin{equation}
  \mathcal L_q^{\rm QCD} 
  = \bar \psi_c i\Dslash \psi_c 
  = \left( \bar\xi + \bar\eta\right)
    \left(in\cdot D \frac{\nbslash}{2} + i\bar n\cdot D \frac{\nslash}{2} 
      + i \Dslash_\perp \right) 
    \left( \xi + \eta\right)
\end{equation}
The small component $\eta$ is integrated out by solving its equation
of motion \linebreak $\delta \mathcal L_q^{\rm QCD} / \delta \bar\eta=0$. The
solution is
\begin{equation} 
  \eta=-\frac{\nbslash}{2}\,\frac{1}{i\bar n\cdot D+i\epsilon}\,
       i\Dslash_\perp\,\xi\;.
\end{equation}
The SCET Lagrangian is therefore identical to the QCD Lagrangian, but
boosted to a reference frame in which the fields carry collinear
momenta. As a result, the collinear Lagrangian is exact to all orders
in $\lambda$ and not renormalized \cite{Beneke:2002ph}. At this stage
we have
\begin{equation}
  \mathcal L_c=\bar \xi \,\frac{\nbslash}{2} \left(
  in\cdot D + i\Dslash_\perp \frac{1}{i\bar n\cdot D+i\epsilon}\,
  i\Dslash_\perp \right) \xi\;.
\end{equation}
Note that the Lagrangian itself scales like $\lambda^4$, which is
consistent with the action $S = \int d^4x \;\mathcal L_c \sim O(1)$
since the integration measure $d^4x \sim O(\lambda^{-4})$ for
collinear momenta.  The factor $\bar n\cdot D$ in the denominator is
symbolic. After the introduction of Wilson lines, it leads to an
integration of the fields along the $n$ direction. This point will be
explained later on in more detail. We have also inserted a
$+i\epsilon$ prescription, which is an arbitrary choice and not
dictated by the QCD Lagrangian \cite{Beneke:2002ph}. This regulator is
necessary for practical purposes but bears no physical
implications. From now on we will omit this prescription in the
notation and only refer to it explicitly when needed.

The next step toward a consistent power counting is to achieve a
``homogeneous'' scaling in $\lambda$, i.e. an expansion. In the above
formulation, $D$ denotes the full covariant derivative, which means
that the gluon fields can carry any momentum. As mentioned before, the
strategy for the construction of SCET is to assign an independent
effective field for each momentum configuration that contributes to a
given process at leading power. Depending on the process, there can be
many. Typically, we work in a reference frame in which there are
``soft'' momenta present, e.g. fields that carry momenta with small
light-cone components. The largest soft mode that is kinematically
allowed to couple to a collinear field without changing its momentum
scaling is an ``ultrasoft'' field with momentum $\sim(\lambda^2,
\lambda^2, \lambda^2)$. Let us assume for the moment, and for the sake
of simplicity, that only these two momentum configurations
contribute. (Such a theory is called SCET$_I$.) It is then necessary
to split up $iD^\mu = i\partial^\mu + A_c^\mu + A_{us}^\mu$. The
subscripts $c$ and $us$ label the collinear and ultrasoft fields,
respectively. Ultrasoft gluons scale like ultrasoft momenta, and it is
therefore necessary to expand the Lagrangian in powers of $\lambda$
for a homogeneous scaling. This, and also the so-called ``multipole
expansion'' below, provide a rigorous definition of leading and
subleading Lagrangians in terms of inverse powers of the large scale
$E$.

Many ingredients of SCET, including the definition of the expansion
parameter $\lambda$ in terms of physical quantities, and the number
and nature of degrees of freedom in the effective theory, depend on
the particular process that is analyzed. In the next sections we will
construct two different theories called SCET$_{I}$ and SCET$_{II}$,
suitable for inclusive and exclusive heavy-to-light decays. The heavy
system provides for a natural reference frame in which to carry out
the calculations, namely its rest frame. The constituents of the heavy
system contain both heavy and light soft fields. In the effective
theory, the mass of the heavy system is a large scale and should not
enter the low-energy description. This is provided by the heavy-quark
effective theory, which will be part of SCET. In the decay
applications considered here, the $b$-quark mass $m_b$ and the large
energy $E$ of the collinear fields will be of the same order.

\begin{table}[t!]
  \begin{center}
    \caption[Nomenclature for various momentum modes in
      SCET]{\label{tab:field.scaling} Nomenclature for various
      momentum modes. Quark fields carrying these momenta scale
      according to the last column, whereas gluon fields scale in
      their light-cone components like the corresponding momenta. The
      effective field theories will contain a subset of these fields,
      depending on the application.  }
    \begin{tabular}{cc|ccc}
      Name & Abbreviation & Momentum & mom. squared & Fermion field \\ 
           &              & scaling $[E]$& $[E^2]$        & scaling \\
      \hline \hline
      hard           & $h$  & $(1,1,1)$ & $1$ & $$ \\
      hard-collinear & $hc$ & $(\lambda,1,\lambda^{1/2})$ 
                     & $\lambda$ & $\lambda^{1/2}$ \\
      collinear      & $c$  & $(\lambda^2,1,\lambda)$ 
                     & $\lambda^2$ & $\lambda$ \\
      soft           & $s$  & $(\lambda,\lambda,\lambda)$ 
                     & $\lambda^2$ & $\lambda^{3/2}$ \\
      soft-collinear & $sc$ & $(\lambda^2,\lambda,\lambda^{3/2})$ 
                     & $\lambda^3$ & $\lambda^2$ \\
      ultrasoft      & $us$ & $(\lambda^2,\lambda^2,\lambda^2)$ 
                     & $\lambda^4$ & $\lambda^3$ \\ \hline
    \end{tabular}
  \end{center}
\end{table}

\begin{figure}[t!]
\begin{center}
\epsfig{file=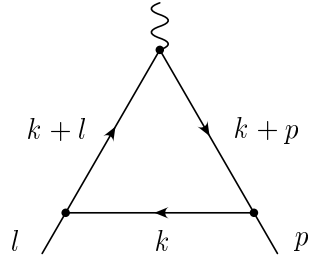, width=4cm}
\caption[An example for the method of regions.]{
\label{fig:exampleTriangleFull} An Example: Scalar triangle graph with
external momenta $l$ (soft) and $p$ (collinear). The loop momentum is
denoted $k$.  }
\end{center}
\end{figure}

Before going into detail, let us set the nomenclature for the various
fields and their scaling. The names are tied to the scaling in
$\lambda$, regardless of the physical definition of $\lambda$. They
are listed in Table \ref{tab:field.scaling}. Not all fields listed
here will be included in the final low-energy theory. Nevertheless, it
is convenient to define a set of momentum modes which allow to address
contributions to processes as they are found by the method of
regions. This can be demonstrated in the following example
\cite{Becher:2003qh}: Consider a scalar triangle graph in the
kinematic setup where the external momenta are
$l^\mu\sim(\lambda,\lambda,\lambda)$ soft,
$p^\mu\sim(\lambda^2,1,\lambda)$ collinear, and
$q^\mu=(l-p)^\mu\sim(\lambda,1,\lambda)$ hard-collinear, as shown in
Fig.~\ref{fig:exampleTriangleFull}. ($q^\mu$ is called a
hard-collinear momentum even though the perpendicular component is
smaller than $\sqrt{\lambda}$. Only the first two components, ``plus''
and ``minus'' are of importance.) One defines the loop integral
\begin{equation}
   I = i\pi^{-d/2} \mu^{4-d} \int d^dk\,
   \frac{1}{(k^2+i0)\,[(k+l)^2+i0]\,[(k+p)^2+i0]}
\end{equation}
in $d=4-2\epsilon$ space-time dimensions and analyzes it for arbitrary
external momenta obeying the above scaling relations. It will be
convenient to define the invariants
\begin{equation}
   L^2\equiv -l^2 - i0 \,, \qquad
   P^2\equiv -p^2 - i0 \,, \qquad
   Q^2\equiv -(l-p)^2-i0 = 2l_+\cdot p_- - i 0 + \dots \,,
\end{equation}
which scale like $L^2\sim\lambda^2$, $P^2\sim\lambda^2$, and
$Q^2\sim\lambda$.  (This $P$ has nothing to do the momentum defined in
(\ref{lightconeCoords}), where $P$ served as an example for the
light-cone decomposition of a 4-vector.) The exact result is
\begin{equation}
   I = \frac{1}{Q^2} \left[ \ln\frac{Q^2}{L^2} \ln\frac{Q^2}{P^2}
   + \frac{\pi^2}{3} + O(\lambda) \right] .
\end{equation}

The next step is to reproduce the result using the method of regions
\cite{Beneke:1997zp}, which means that one assumes a certain scaling
for the loop momentum $k$ and expands the integrand in powers of
$\lambda$ {\em before} performing the integration. Going through the
list of modes in Table \ref{tab:field.scaling} we find leading power
contributions for $k$ being hard-collinear, collinear, soft, and
soft-collinear. The individual regions give the following
contributions \cite{Becher:2003qh}:
\begin{eqnarray} \label{funIntegral}
   I_{\rm HC}
   &=& i\pi^{-d/2} \mu^{4-d} \int d^dk\,
    \frac{1}{(k^2+i0)\,(k^2 + 2k_-\cdot l_+ + i0)\,
             (k^2 + 2k_+\cdot p_- + i0)} \nonumber\\   
   &=& \frac{\Gamma(1+\epsilon)}{Q^2} \left( \frac{1}{\epsilon^2}
    + \frac{1}{\epsilon} \ln\frac{\mu^2}{Q^2}
    + \frac12 \ln^2\frac{\mu^2}{Q^2} - \frac{\pi^2}{6} \right)
    + O(\epsilon) \,,
\end{eqnarray}
\begin{eqnarray}
   I_{\rm C}
   &=& i\pi^{-d/2} \mu^{4-d} \int d^dk\,
    \frac{1}{(k^2+i0)\,(2k_-\cdot l_+ + i0)\,[(k+p)^2+i0]} \nonumber\\
   &=& \frac{\Gamma(1+\epsilon)}{Q^2} \left( - \frac{1}{\epsilon^2}
    - \frac{1}{\epsilon} \ln\frac{\mu^2}{P^2}
    - \frac12 \ln^2\frac{\mu^2}{P^2} + \frac{\pi^2}{6} \right)
    + O(\epsilon) \,.
\end{eqnarray}
\begin{eqnarray}
   I_{\rm S}
   &=& i\pi^{-d/2} \mu^{4-d} \int d^dk\,
    \frac{1}{(k^2+i0)\,[(k+l)^2+i0]\,(2k_+\cdot p_- + i0)} \nonumber\\
   &=& \frac{\Gamma(1+\epsilon)}{Q^2} \left( - \frac{1}{\epsilon^2}
    - \frac{1}{\epsilon} \ln\frac{\mu^2}{L^2}
    - \frac12 \ln^2\frac{\mu^2}{L^2} + \frac{\pi^2}{6} \right)
    + O(\epsilon) \,.
\end{eqnarray}
\begin{eqnarray}
   I_{\rm SC}
   &=& i\pi^{-d/2} \mu^{4-d} \int d^dk\,
    \frac{1}{(k^2+i0)\,(2k_-\cdot l_+ + l^2 + i0)\,
             (2k_+\cdot p_- + p^2 + i0)} \nonumber\\
   &=& \frac{\Gamma(1+\epsilon)}{Q^2} \left( \frac{1}{\epsilon^2}
    + \frac{1}{\epsilon} \ln\frac{\mu^2 Q^2}{L^2 P^2}
    + \frac12 \ln^2\frac{\mu^2 Q^2}{L^2 P^2} + \frac{\pi^2}{6} \right)
    + O(\epsilon) \,.
\end{eqnarray}
A couple of comments are in order.

(i) It is a simple check to add all the above contributions and find
that they indeed reproduce the full result $I = I_{\rm HC} + I_{\rm C}
+ I_{\rm S} + I_{\rm SC}$.

(ii) Each individual integral is, of course, Lorentz invariant. In
this sense Lorentz invariance was never broken by picking a particular
frame. Had we picked a different reference frame, for example the
Breit frame, the results would not change. However, we would have
found different momentum modes that contribute. Clearly, we could
perform a boost to this frame before evaluating the integrals, thus
finding a different set of momentum modes. In the example of the Breit
frame, the formerly soft leg $l$ becomes collinear in the $\bar n$
direction (anti-collinear), and the collinear momentum $p$ stays
collinear in the $n$ direction. Leading contributions would come from
hard loop momenta (formerly hard-collinear), anticollinear (formerly
soft), collinear (formerly collinear) and ultrasoft (formerly
soft-collinear). It is therefore not only the physical process that
defines which modes must be included in the effective theory, but also
the frame in which we choose to work. One further point can be made by
boosting the frame even further, such that the formerly collinear
momenta become soft. Clearly, the formerly soft momenta then become
collinear in the $\bar n$ direction and we find the following
``symmetry'' (this is, of course, nothing else but Lorentz
invariance): The contribution from soft and collinear loop momenta are
related to each other under the simultaneous exchange of soft and
collinear external momenta and the interchange of $n$ and $\bar
n$. However, if a heavy quark is present there is a preferred Lorentz
frame, namely the rest frame of the heavy quark, and the above
symmetry is broken.

\begin{figure}[t!]
\begin{center}
\epsfig{file=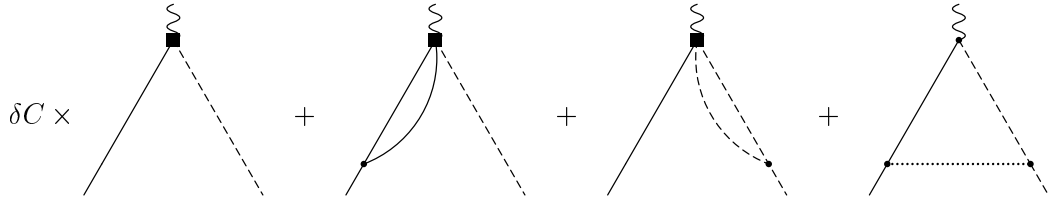, width=14cm}
\caption[Corresponding effective theory diagrams]{
\label{fig:exampleTriangleSCET} Corresponding contributions in the
effective low-energy theory. $\delta C$ denotes a Wilson coefficient,
the solid lines are soft fields, the dashed lines collinear, and the
dotted line soft-collinear.  }
\end{center}
\end{figure}

(iii) We went through a lengthy discussion of the various
contributions to the full theory diagram. The benefit is the
separation of short- and long-distance physics.  The strategy for
constructing a low-energy effective theory is to let the collinear,
soft, and soft-collinear contributions be reproduced by exchange of
collinear, soft, and soft-collinear fields, while the hard-collinear
contribution in this case will be absorbed into a Wilson
coefficient. In fact, the sum of the collinear, soft, and
soft-collinear integrals is free of infra-red singularities, which are
regularized by keeping the external legs off-shell (e.g. the sum is
free of $\ln(\mu^2/L^2)$ and $\ln(\mu^2/P^2)$ ). The view of
associating the infra-red contributions with diagrams in the effective
theory is depicted in Fig.~\ref{fig:exampleTriangleSCET}. Note that
when a soft gluon connects to a collinear line it will lead to a
(hard-collinear) off-shell mode, which will be integrated
out. Therefore the soft gluon emerges out of the vertex in the
figure. A similar argument applies to the collinear gluon. A deeper
investigation of this point will follow in later sections and lead
to Wilson lines. Here we wish to stress the close connection between
the method of regions and the idea of constructing SCET as a
low-energy effective theory.

Before beginning the discussion of SCET$_{I}$ in greater detail, note
that the same (physical) momentum modes may have different names,
depending on the (physical) definition of the expansion parameter
$\lambda$. Consider, for example, a momentum $p\sim (\Lambda_{\rm
QCD}, E, \sqrt{E\Lambda_{\rm QCD}})$ with some scales $E \gg
\Lambda_{\rm QCD}$. This mode can be called collinear when defining
$\lambda = \sqrt{\Lambda_{\rm QCD}/E}$. It can, however, also be
called hard-collinear when defining $\lambda = \Lambda_{\rm
QCD}/E$. Similarly, a momentum $l\sim (\Lambda_{\rm QCD}, \Lambda_{\rm
QCD}, \Lambda_{\rm QCD})$ might be called ultrasoft or soft for
$\lambda = \sqrt{\Lambda_{\rm QCD}/E}$ or $\lambda = \Lambda_{\rm
QCD}/E$, respectively.

\section{Renormalization}

For consistency we review the main strategies of renormalization using
dimensional regularization \cite{'tHooft:1972fi} in $4-2\epsilon$
space-time dimensions. This scheme has already been adopted in the
previous section and will be used throughout this work. As apparent in
(\ref{funIntegral}), for example, loop integrals which are
logarithmically divergent in four dimensions become finite for
$\epsilon \ne 0$. The general strategy is to absorb terms that diverge
in the limit $\epsilon \to 0$ into factors that relate bare quantities
to renormalized ones. In the end, renormalized quantities are finite
for $\epsilon \to 0$ and have meaningful relations to physical
observables.

Since the action $\int d^{4-2\epsilon}x\, \mathcal L(x)$ is
dimensionless, it follows that the bare QCD coupling constant
$g_s^{\rm bare}$ has a mass-dimension of $\epsilon$. Defining
$(g_s^{\rm bare})^2 = 4\pi \alpha_s^{\rm bare}$, the relation to the
dimensionless renormalized coupling $\alpha_s(\mu)$ can be written as
\cite{Gross:1975vu}
\begin{equation}
\alpha_s^{\rm bare} = \mu^{2\epsilon}\,Z_\alpha(\alpha_s(\mu))\;
\alpha_s(\mu)\;,
\end{equation}
where $\mu$ is an arbitrary mass parameter, called the renormalization
scale.  In a minimal subtraction scheme, the renormalization factor
$Z_\alpha = 1+\sum\limits_{k=1}^{\infty} \epsilon^{-k} Z_\alpha^{(k)}$
is expanded in inverse powers of $\epsilon$, and the evolution of
$\alpha_s(\mu)$ is encoded in the ``beta function''
\begin{equation} \label{betafunc1}
\frac{d}{d\ln\mu}\,\alpha_s(\mu) = \beta(\alpha_s(\mu),\epsilon)\;.
\end{equation}
We will consistently work in the modified minimal subtraction scheme
$\overline{\rm MS}$, in which all terms multiplying powers of
$1/\epsilon - \gamma_E + \ln 4\pi$ are absorbed into the
renormalization factors. Since the beta function
$\beta(\alpha_s,\epsilon)$ is finite in the limit $\epsilon \to 0$ (in
a renormalizable theory), it can be Taylor expanded around $\epsilon =
0$. It follows \cite{Gross:1975vu} that only the first term in this
expansion is non-zero and the exact relation reads
\begin{equation} \label{betafunc2}
\begin{aligned}
\beta(\alpha_s, \epsilon)&=\beta(\alpha_s) - 2\epsilon\,\alpha_s\;,\\
\beta(\alpha_s) &= 2\alpha_s^2 \,\frac{d}{d\alpha_s}\,
Z_\alpha^{(1)}(\alpha_s) \;.
\end{aligned}
\end{equation}
Throughout this work we will use the following expansion of the beta
function in terms of $\alpha_s$:
\begin{equation}\label{TheBetaFunc}
\beta(\alpha_s) = -2\alpha_s \sum\limits_{k=0}^\infty \beta^{(k)}
\left( \frac{\alpha_s}{4\pi} \right)^{k+1} \;,
\end{equation}
with \cite{Hagiwara:2002fs}

\begin{equation}
\begin{aligned}
\beta^{(0)} &= 11-\frac23\, n_f \;,\\ 
\end{aligned}
\end{equation}
\begin{equation}
\begin{aligned}
\beta^{(1)} &= 102 - \frac{38}{3}\, n_f \;,\\ 
\beta^{(2)} &= \frac{2857}{2}-\frac{5033}{18}\,n_f +
\frac{325}{54}\, n_f^2 \;,
\end{aligned}
\end{equation}
where $n_f$ is the number of flavors that contribute in the loop
calculation.

We will also need the wave-function renormalization of the quark
fields. In QCD, the (massless or massive) bare and renormalized quark
fields are related by $\psi^{\rm bare} = Z_\psi^{1/2}\, \psi$ with
$Z_\psi = 1 - \alpha_s/(3\pi \epsilon)$. The only quark field in SCET
that renormalizes differently is the heavy quark field. Its
wave-function renormalization is obviously identical to the heavy
quark field in HQET and reads \cite{Neubert:1993mb} \linebreak 
$Z_h = 1 + 2 \alpha_s/(3\pi \epsilon)$.

Composite operators $Q_i$ require renormalization beyond that of its
field and coupling components, and are renormalized in the same
manner. Under renormalization they can mix among each other if they
share the same quantum numbers and dimensions. We will only consider
operators that are gauge invariant and do not vanish under the
application of equation of motions. Therefore one relates
\begin{equation}
Q_i^{\rm bare} = Z_{i j}(\mu)\, Q_j(\mu)
\end{equation}
with $Z_{i j} = \delta_{i j} + \sum\limits_{k=1}^\infty \epsilon^{-k}
Z^{(k)}_{i j}$ in the $\overline{\rm MS}$ scheme.  The $\mu$
dependence of an operator $Q_i$ is encoded in the renormalization
group equation (RGE). Since the set of all allowed operators
$\{Q_1,Q_2,\ldots\}$ is a complete basis, one can express the operator
$\partial Q_i/\partial\ln\mu$ as a linear combination of
$Q_1,Q_2,\ldots$.  Defining the anomalous dimension matrix $\gamma$
through the RGE
\begin{equation}
\frac{d}{d\ln \mu}\,Q_j = -\gamma_{j k} Q_k \;,
\end{equation}
it follows from the fact that the bare operator $Q_i^{\rm bare}$ is
independent of $\mu$, that
\begin{equation}
\gamma_{j k} = (Z^{-1})_{j i}\,\frac{d}{d\ln \mu}\, Z_{i k} 
             = - 2\alpha_s \,\frac{d}{d\alpha_s}\,Z_{j k}^{(1)}\;.
\end{equation}
Here we have used (\ref{betafunc1}) and (\ref{betafunc2}). In words,
{\em the anomalous dimension is, apart from a minus sign, twice the
coefficient of the $1/\epsilon$ term}, when all ultra-violet
divergences are regularized dimensionally and all infra-red effects
are regularized by other means. For any anomalous dimension
$\gamma(\alpha_s)$ we will adopt the expansion
\begin{equation}\label{anomDimExpansion}
\gamma(\alpha) = \sum\limits_{k=0}^\infty \gamma^{(k)} \left(
\frac{\alpha_s}{4\pi} \right)^{k+1} \;,
\end{equation}
unless otherwise stated. 

Physical amplitudes are given as products of operator matrix elements
and Wilson coefficients $C_i(\mu)\, \langle Q_i(\mu) \rangle$ and are
independent of the renormalization scale $\mu$. This implies that the
Wilson coefficients must obey the RGE
\begin{equation}
\frac{d}{d\ln\mu} C_j = \left( \gamma^T \right)_{j k} C_k\;,
\end{equation}
where $T$ denotes the transpose of the matrix $\gamma$. The above
equation will be used below to resum large logarithms to all orders.

\section{SCET$_{I}$}
\label{sec:scet1}

This theory is applicable to processes in which there are only two
different momentum modes present, namely hard-collinear and soft
ones. An example for such a process is the inclusive decay of a heavy
meson such as the $B$ meson. In the rest frame of the $B$ meson the
partons carry, with the exception of the massive $b$ quark, momenta
that scale like $(\Lambda_{\rm QCD}, \Lambda_{\rm QCD}, \Lambda_{\rm
QCD})$, with $\Lambda_{\rm QCD}$ being the soft QCD scale. The $b$
quark momentum can be split up into the static component $m_b v$ and
the dynamic residual momentum $k$, which also scales like
$(\Lambda_{\rm QCD}, \Lambda_{\rm QCD}, \Lambda_{\rm QCD})$. A
flavor-changing current turns the $b$ quark into some light but
energetic quark. This quark could be a parton of a particular hadron
or, more generally, of a hadronic jet of invariant mass much larger
than $\Lambda_{\rm QCD}$. We are then faced with the kinematic
situation in which a hierarchy of scales emerged: $\Lambda_{\rm QCD}
\ll M \ll E$, where $E$ is the energy of the jet (close to half of the
$b$-quark mass) and $M$ is the jet invariant mass. Specifically, we
assume that $M$ is of the ``intermediate''\footnote{For instance, in
inclusive $B\to X_u$ decays we need to restrict the jet invariant mass
to be less than the charm mass for discrimination
purposes. Numerically, $m_c$ compares very well with $\sqrt{m_b
\Lambda_{\rm QCD}}$.} order $M\sim \sqrt{E\Lambda_{\rm QCD}}$. In this
kinematic setup the two momentum modes are therefore $(\Lambda_{\rm
QCD}, \Lambda_{\rm QCD}, \Lambda_{\rm QCD})$ and $(\Lambda_{\rm QCD},
E, \sqrt{E\Lambda_{\rm QCD}})$. When choosing the expansion parameter
$\lambda = \Lambda_{\rm QCD}/E$ we can refer to them as
soft and hard-collinear, respectively. (Alternatively, they are also
referred to as ultrasoft and collinear when choosing $\lambda =
\sqrt{\Lambda_{\rm QCD}/E}$.) 

\subsection{Power Counting}
\label{sec:scetPowCou}

The full massless QCD fields are separated into hard-collinear and
soft momentum modes, e.g. the quark field splits up into the soft
quark field $q$ and the hard-collinear quark field $\psi_{hc} =
\xi_{hc}+\eta_{hc}$. Analogously, the QCD gluon field $A$ is
decomposed, yielding
\begin{equation}
\psi = \xi_{hc}+\eta_{hc}+q\;, \qquad 
iD^\mu=i\partial^\mu+g A_{hc}^\mu+g A_s^\mu\;.
\end{equation}
Let us study the effect on the kinetic term in the Lagrangian
$\mathcal L =\bar\psi\,i\Dslash\,\psi+\ldots$. Since the spinors
$\xi_{hc}$ and $\eta_{hc}$ are constrained by $\nslash
\xi_{hc}=\nbslash \eta_{hc} = 0$, we decompose the Dirac matrix
$\gamma^\mu$ in light-cone coordinates to $n^\mu \nbslash/2 + \bar
n^\mu \nslash/2 + \gamma_\perp^\mu$ and expand the Lagrangian.
\begin{equation} \label{masslessL}
  \begin{aligned}
\bar\psi\,i\Dslash\,\psi\;= &\;
\bar\xi_{hc}\,\frac{\nbslash}{2}\,in\cdot D \,\xi_{hc}
  + \bar\eta_{hc}\,\frac{\nslash}{2}\,i\bar n\cdot D\,\eta_{hc}
  + \bar\eta_{hc} \,i\Dslash_\perp \xi_{hc}
  + \bar\xi_{hc}\, i\Dslash_\perp \eta_{hc} \\
+ & \bar\xi_{hc}\, g\Aslash_{hc} \,q + \bar\eta\, g\Aslash_{hc} \,q 
  + \bar q\, g\Aslash_{hc}\,\xi_{hc} + \bar q\, g\Aslash_{hc}\,\eta_{hc}
  + \bar q\, (i\delslash + g\Aslash_s)\,q
  \end{aligned}
\end{equation}
Note that we do not allow for interaction terms that are forbidden by
kinematics, for example two soft fields cannot couple to a single
hard-collinear field. The small component $\eta_{hc}$ is then
integrated out by solving its equation of motion. This yields
\begin{equation}
  \eta_{hc} = -\frac{\nbslash}{2}\,\frac{1}{i\bar n\cdot D}
  \left( i\Dslash_\perp\,\xi_{hc}+g\Aslash_{hc}\,q \right)\;,
\end{equation}
which must be plugged back into (\ref{masslessL}). So far we have
merely rewritten the Lagrangian into a more complicated but exact
form. In a next step one needs to perform a systematic expansion in
$\lambda$ of this Lagrangian. To do this, it is important to take the
scaling of the integration measure in $S=\int d^4x\,\mathcal L$ into
account. (So far, we omitted the explicit dependence of the various
fields on the position $x$. We will continue to do so, unless
otherwise stated.) For terms involving only soft fields, $d^4x$ scales
like $\lambda^{-4}$. To see this consider a soft field at position
$y$. Translational invariance relates this field to one at position
$x+y$ through an exponential $\exp(i\, l\cdot x)$, where $l$ is the
soft momentum, and $l\cdot x = l_+\cdot x_- + l_- \cdot x_+ + l_\perp
\cdot x_\perp$. Because of the scaling properties of the soft
momentum, leading contributions to the action are obtained when $x$
scales like $(\lambda^{-1},\lambda^{-1},\lambda^{-1})$. Similarly, for
terms involving hard-collinear fields, $x$ scales like
$(1,\lambda^{-1},\lambda^{-1/2})$ and the integration measure $d^4x
\sim \lambda^{-2}$.

Inverse covariant derivative operators are also expanded. For brevity,
let us introduce the intuitive notation $D_{hc} \equiv \partial -i g A_{hc}$
(and $D_{s} \equiv \partial -i g A_{s}$ will also be used below).
When $(i\bar n\cdot D)^{-1}$ acts on hard-collinear fields, we must expand
\begin{equation}
  \frac{1}{i\bar n\cdot D} = \frac{1}{i\bar n\cdot D_{hc}}
  -\frac{1}{i\bar n\cdot D_{hc}}\,g\,\bar n\cdot A_{s}\,
  \frac{1}{i\bar n\cdot D_{hc}} + O(\lambda^4) \;.
\end{equation}
Every term in the effective theory should have a single and
homogeneous scaling behaviour. This requirement leads to a subtlety in
interaction terms involving soft and hard-collinear fields. In the
above notation we have omitted the dependence on the position and
treated all terms as local in that all fields are evaluated at the
same position $x$. According to the above argument, the light-cone
components of $x$ in interactions of hard-collinear and soft fields
scale like $(1,\lambda^{-1},\lambda^{-1/2})$, because the vertex
carries hard-collinear momentum (since two hard-collinear momenta can
not add up to a soft momentum). One must therefore perform a
``light-front multipole expansion'' \cite{Beneke:2002ni} of the
translation operator with $-i\partial \sim (\lambda,\lambda,\lambda)$
\begin{eqnarray}
\phi_{s}(x) &=& \exp(x\cdot \partial)\; \phi_{s}(0) \nonumber \\
&=& \exp(x_+ \cdot\partial_- + x_\perp\cdot\partial_\perp)\;
    \exp(x_-\cdot\partial_+)\;\phi_{s}(0) \nonumber \\
&=& \phi_{s}(x_-) + \left[ x_\perp \cdot \partial_\perp 
    \phi_{s}\right](x_-) \nonumber \\
&& + \frac12 x_+ \cdot\left[ \partial_- \phi_{s}\right](x_-)
   + \frac12\left[ x_\perp^\mu x_\perp^\nu \partial_\mu \partial_\nu 
     \phi_{s}\right](x_-) + \ldots
\end{eqnarray}
for any soft field $\phi_{s}$ in such interaction terms. This
means that all soft fields must be evaluated at a position
$x_-^\mu = (\bar n\cdot x)\, n^\mu/2$ on the light cone, whereas the
hard-collinear fields remain at position $x$. This non-locality is a
reflection of the fact that at leading power in $\lambda$ the total
momentum in such interactions is not conserved. However, to put this
shocking observation into the right context, we emphasize that
translational invariance will be restored order by order in the
expansion in $\lambda$. At leading power only the ``plus component''
of the soft momentum is kept when adding it to a hard-collinear
momentum: $(\lambda,1,\lambda^{1/2}) + (\lambda,\lambda,\lambda) =
(\lambda,1,\lambda^{1/2})$. The expanded Lagrangian is then homogeneous in
the power counting term by term.  This concludes the systematic
expansion in powers of $\lambda$. We will address the important
question of gauge invariance in a separate section below.

Let us turn our attention to the treatment of the $b$ quark in
SCET. Heavy quarks are described by the familiar HQET Lagrangian
\cite{Neubert:1993mb}. In the low energy effective theory, the heavy
quarks do not couple to hard-collinear fields because such an interaction
would put the quark off-shell. The HQET Lagrangian reads
\begin{equation} \label{HQETLagrangian}
\mathcal L_h = \bar h\,iv\cdot D_{s}\,h 
   + \frac{1}{2m_b} \left[ \bar h\,(i D_{s})^2\,h 
   + C_{mag}(\mu)\;\bar h\,\frac{g}{2}\sigma_{\mu\nu}G^{\mu\nu}_{s}\,h 
   \right] + O(1/m_b^2)\;.
\end{equation}

Finally the pure glue (Yang-Mills) Lagrangian $\mathcal L_{\rm glue}$
can be split into soft plus hard-collinear plus interaction terms. The
soft and hard-collinear sectors take the same form as in full QCD,
including gauge-fixing and ghost terms. The interaction terms are
restricted to those that are kinematically allowed, using the same
arguments as above.

\subsection{Wilson lines}

The introduction of Wilson lines is a well known concept in
HQET. Consider the effect of coupling a gluon with momentum of order
$O(\Lambda_{\rm QCD})$ (soft) to a heavy quark field with
momentum $m_b v$. This will lead to a propagator that is off-shell by
an amount $(m_b v+k)^2-m_b^2 \approx 2 m_b v\cdot k \sim O(m_b
\Lambda_{\rm QCD})$. If the effective theory describes physics of
energies below this scale, such propagators should be integrated
out. We can repeat this argument consecutively and find that attaching
$n$ gluons with momenta $k_i$, $i=1\ldots n$, to a heavy quark line
leads to
\begin{equation}
\frac{-g\,v^{\mu_n} t^{a_n}}{v\cdot(k_1 + \ldots + k_n)} 
    \;\cdot \ldots\cdot\;
\frac{-g\,v^{\mu_2} t^{a_2}}{v\cdot(k_1 + k_2)} \;\cdot \;
\frac{-g\,v^{\mu_1} t^{a_1}}{v\cdot k_1} \;.
\end{equation}
These are precisely the Feynman rules for the object
\begin{equation}
S_v(x)\equiv P\exp\left(i g\int_{-\infty}^0\!d s\;
   v\cdot A_{s}(x+s v)\right)\;,
\end{equation}
which is called a Wilson (or Schwinger) line extending from minus
infinity to the interaction point $x$ along the $v$ direction. The
path-ordering symbol $P$ denotes that the gluon fields are ordered
from left to right in order of decreasing $s$. In the spirit of
effective theories, we integrate the off-shell propagators out by
performing the field redefinition $h(x) \to S_v(x)\, h^{(0)}(x)$. The
Wilson line $S_v$ has some very useful properties, $S_v^\dagger S_v =
S_v S_v^\dagger = 1$ and, most importantly, $S_v^\dagger\,iv\cdot
D_{s}\,S_v = iv\cdot \partial$. Performing the field redefinition
leads to a free-particle Lagrangian at leading power, $\mathcal L_h =
\bar h^{(0)}\,iv\cdot \partial\,h^{(0)}+\ldots$, so that the
``sterile'' field $h^{(0)}$ no longer couples to soft gluons. In
practice we will use the Wilson lines $S_v$ only when discussing the
renormalization properties of operators that include heavy fields. For
explicit loop calculations one can choose to work with either $h$ or
$S_v h^{(0)}$, since they lead to the same Feynman rules.

The purpose of this recapitulation is to draw a close analogy to the
treatment of hard-collinear gluon attachments, which can be dealt with in
the same manner. In full QCD the attachment of a gluon with hard-collinear
momentum $E\,n+\ldots$ to a massive quark line carrying momentum 
$m_b v$ will provide for a Dirac structure
\begin{equation}
  \frac{m_b(1+\vslash)+E\nslash}{2m_b E}\,g\Aslash_{hc} \,
  \frac{1+\vslash}{2}\,h = \frac{1}{2E}\,g\,\bar n\cdot A_{hc}
  \,\frac{1+\vslash}{2}\,h \; + \ldots\;,
\end{equation}
where the dots represent power corrections. An infinite succession of
such attachments results in the Wilson line $W_{hc}$ \cite{Bauer:2001ct},
which can be written in position space as
\begin{equation}
W_{hc}(x) = P\exp \left( i g\int_{-\infty}^0\!d s\;
   \bar n\cdot A_{hc}(x+s \bar n)\right)\;.
\end{equation}
We should point out that the above (somewhat historic) discussion does
{\em not} suggest that the massive $b$ field matches onto the
product $(W_{hc}\,h)$. In fact, it is the combination with a 
hard-collinear field, $(\bar \xi_{hc}\,W_{hc})$, that appears in explicit
matching calculations, e.~g.~$(\bar \psi\,\Gamma\,b)\to (\bar
\xi_{hc}\,W_{hc}\,\Gamma\,h)$.  As discussed below, this combination is also
dictated by hard-collinear gauge invariance.

We also need the conjugate operator $W_{hc}^\dagger$, which has the
opposite path ordering and the factor $i g$ replaced by $-i g$. The
most important operator properties are that $W_{hc}^\dagger W_{hc} =
W_{hc} W_{hc}^\dagger=1$ and $W_{hc}^\dagger \, i\bar n\cdot D_{hc}\,
W_{hc} = i\bar n\cdot \partial$, following from the fact that $(i\bar
n\cdot D_{hc}\,W_{hc})=0$ by definition. As a consequence, one obtains
the very useful identity \cite{Bauer:2001yt,Beneke:2002ph,Hill:2002vw}
\begin{equation}
  \frac{1}{i\bar n\cdot D_{hc}}\,\phi(x) = W_{hc}(x)\, \frac{1}{i\bar n\cdot
  \partial}\,W_{hc}^\dagger(x)\,\phi(x) = -i\int_{-\infty}^0\!d s\; W_{hc}(x)
  \left[ W_{hc}^\dagger\, \phi \right](x+s\bar n)\;,
\end{equation}
for any field (or product of fields) $\phi(x)$. This enables us to
write the purely hard-collinear quark Lagrangian $\mathcal L_{hc}$ 
(the interaction terms with soft fields will be denoted by 
$\mathcal L_{int}$) in the following form:
\begin{eqnarray}
  \mathcal L_{hc}(x) &=& 
  \bar \xi_{hc}(x)\,\frac{\nbslash}{2}\,in\cdot D_{hc}\;\xi_{hc}(x) \\
&& - i\int_{-\infty}^0\!d s\; \left[ \bar
  \xi_{hc}\,i\overleftarrow{\Dslash}_{hc\perp} W_{hc} \right](x)\,
  \frac{\nbslash}{2}\, \left[
  W_{hc}^\dagger\,i\overrightarrow{\Dslash}_{hc\perp}\, \xi_{hc} \right]
  (x+s\bar n)\;. \nonumber
\end{eqnarray}
This Lagrangian sums up an infinite number of leading-order couplings
between hard-collinear quarks and hard-collinear gluons. For later purposes,
note that $W_{hc}=1$ in ``collinear light-cone gauge'' 
$\bar n\cdot A_{hc} = 0$.

One can also introduce the soft Wilson line $S_s$, analogous to the
hard-collinear Wilson line, with $\bar n \leftrightarrow n$ interchanged
\cite{Bauer:2001yt,Beneke:2002ph}. Again, $S_s=1$ if we choose the
$n\cdot A_{s}=0$ gauge. Although $S_s$ is not necessary to construct
the Lagrangian for soft fields, it turns out to be very useful in
understanding how soft--collinear factorization arises as a
property of the effective theory at lowest order. This is similar to
the decoupling of HQET fields $h$ from soft fields, which leaves
the heavy field $h^{(0)}$ sterile to gluon interactions. It has been
shown \cite{Bauer:2001yt} that the field redefinitions
\begin{equation}\label{scet1-decoupling}
  \xi_{hc}=S_s\,\xi_{hc}^{(0)}\;, \quad 
  A_{hc}^\mu=S_s\,A_{hc}^{(0)\mu}\,S_s^\dagger\;,\quad
  c = S_s\,c^{(0)}\,S_s^\dagger\;,
\end{equation}
{\em remove all hard-collinear -- soft interactions} from the
leading-order Lagrangian. (Above, $c$ denotes the ghost fields.) We
will find (and explain in some more detail) an analogous mechanism for
the decoupling transformations in SCET$_{II}$ below.

\subsection{Gauge Transformations}

Gluon fields have been split up into two components, hard-collinear
and soft. The gauge symmetry of QCD also decomposes into
hard-collinear and soft gauge symmetries. It is clear that soft fields
do not transform under hard-collinear gauge transformations $U_{hc}$,
since this would change their momentum scaling. Only hard-collinear
fields can transform. On the other hand, soft gauge transformations
$U_{s}$ do not alter the kinematic scalings, and therefore all fields
transform, including hard-collinear ones. For that matter, soft gluons
can be viewed as slowly varying background fields. Hard-collinear
gluons transform then covariantly under soft gauge transformations,
while they transform inhomogeneously in the hard-collinear sector.
\begin{equation}
\begin{aligned}
& \hbox{hard-coll.:}\quad & A_{hc} &\to U_{hc}\,A_{hc}\,U_{hc}^\dagger+
  \frac{i}{g}\, U_{hc}\left[ D_s\,, U_{hc}^\dagger\right]\;,\qquad &
  \xi_{hc} & \to U_{hc}\,\xi_{hc}\;, \\
& & A_s & \to A_s\;, & q &\to q \;, \\[5pt]
& \hbox{soft:} & A_{hc} &\to U_s\, A_{hc}\, U_s^\dagger\;, & \xi_{hc}
  & \to U_s\,\xi_{hc}\;, \\
& & A_s & \to U_s\, A_s\, U_s^\dagger + \frac{i}{g}\,
  U_s\left[ \partial\,, U_s^\dagger \right]\;, 
  & q &\to U_s q \;.
\end{aligned}
\end{equation}

Note that the sum $A_{hc} + A_s$ transforms in the usual way under
both hard-collinear and soft transformations. We have also introduced
the hard-collinear Wilson line $W_{hc}(x)$, which extends from ($-
\infty$) to $x$, and therefore transforms like $W_{hc}(x) \to
U_{hc}(x)\,W_{hc}(x)\,U_{hc}^\dagger(-\infty)$. Let us agree that the
gauge transformations $U_{hc}(x)$ become irrelevant for $x\to -\infty$
and set $U_{hc}(-\infty)=1$. This will obviously not alter the
transformation properties of finite Wilson lines
$W_{hc}(0)\,W_{hc}^\dagger(t \bar n)$ along the light cone $\bar n$,
because the transformations $U_{hc}(-\infty)$ cancel. It will,
however, simplify the discussion since now the Wilson lines transform
like $W_{hc}(x) \to U_{hc}(x)\, W_{hc}(x)$ under hard-collinear gauge
transformations. Note that, as a simple but important example, the
combinations
\begin{equation}\label{scet1-buildblocks}
  \X_{hc} = W_{hc}^\dagger \,\xi_{hc}\;, \qquad 
  \A_{hc}^\mu = \left[ W_{hc}^\dagger \,i D_{hc}^\mu \,W_{hc} \right]\;,
\end{equation}
are left invariant. In light-cone gauge $\bar n\cdot A_{hc} = 0$, the
calligraphic gluon field $\A_{hc}$ is identical to the QCD gluon field,
which will make matching calculations particularly simple. We will
sometimes refer to the calligraphic fields as ``gauge-invariant
building blocks'', because it is easy to construct gauge-invariant
operators using these fields \cite{Hill:2002vw}.

The effective Lagrangian before multipole expanding is invariant under
the above gauge transformations. However, such a transformation
reintroduces terms of different power counting in the Lagrangian. In
the end, we wish to construct the theory in such a way that every term
in the Lagrangian has a simple (homogeneous) scaling behaviour. This
leads to the notion of ``homogeneous gauge transformations'', and is
discussed in detail in \cite{Beneke:2002ni}. We have also explored
that in interactions with hard-collinear fields, soft fields need to
be positioned at $x_-$ rather than $x$. A new Wilson line extending
from $x$ to $x_-$ is then necessary to restore gauge invariance. These
Wilson lines need to be expanded in $\lambda$, too, which leads to new
interaction terms in the subleading Lagrangians. The above arguments
will be expanded and discussed in more detail in the case of
SCET$_{II}$.

\subsection{Heavy-to-Collinear Currents}
\label{sec:SCET1:heavy-to-coll}

An important application for SCET is the decay of the heavy $b$ quark
to a highly energetic light quark $q$. Such flavor-changing currents
are the result of the weak interactions and appear in the weak
effective Hamiltonian description of the Standard Model at energy
scales well below the weak scale $\sim 100$ GeV. The QCD current is
$\bar q\,\Gamma\,b$, where we treat $\Gamma$ as an arbitrary Dirac
structure. At leading power in $\lambda$, this current matches onto
the SCET current $\bar \X_{hc}(x)\,\Gamma\,h(x_-)$. As noted before,
the dynamical heavy field is treated as a soft particle and is
therefore positioned at $x_-$. The Wilson line in the definition of
the gauge-invariant field $\X_{hc} = W_{hc}^\dagger\,\xi_{hc}$
correctly accounts for an arbitrary power of $\bar n\cdot A_{hc}$
gluon insertions. Only the large components $\bar n\cdot A_{hc}$ need
to be exponentiated since all other components are parametrically
smaller and can be expanded.  Soft gluon insertions enter at
subleading power. Although these subleading currents are known in the
literature \cite{Beneke:2002ph,Chay:2002vy,Hill:2004if}, they are not
needed for this work at present (although very important for future
continuations of it). There are two categories (labeled as ``A'' and
``B'') which are separately invariant under reparameterization
transformations (see section \ref{sec:RPI}). The leading current $\bar
\X_{hc}(x)\,\Gamma\,h(x_-)$ belongs to the ``A'' category and at order
$\lambda^{1/2}$ one finds, for example,
\begin{equation}
  \begin{aligned}
    J^{(A1)} &= \bar \X_{hc}\,\Gamma\,x_\perp\cdot D_{s}\,h 
    - \bar \X_{hc}\,i\overleftarrow \calDslash_{hc\perp}\,
      \frac{1}{i\bar n\cdot \overleftarrow \partial}\,
      \frac{\nbslash}{2}\,\Gamma\,h \\
    J^{(B1)} &= \bar \X_{hc}\,\Gamma\,\calAslash_{hc\perp}\,
    \frac{\nslash}{2m_b}\,h \;.
  \end{aligned}
\end{equation}
The calligraphic covariant derivative is defined as $i\D_{hc} = i\partial
+ \A_{hc}$. All hard-collinear fields are located at position $x$, and all
soft fields at position $x_-$.  ``A''- and ``B''-type currents
are also known at order $\lambda$. For the remainder of this section
we will consider the most simple case and discuss the matching at
leading power only.

The effective fields $\bar \X_{hc}$ and $h$ are two-component spinors, in
contrast to the four-component spinors $\bar q$ and $b$. Thus,
the Dirac basis $\{ 1, \gamma_5, \gamma^\mu, \gamma^\mu \gamma_5,
i\sigma^{\mu \nu}, i\sigma^{\mu \nu} \gamma_5 \}$ will be matched onto
a smaller basis, which is commonly chosen as $\{1, \gamma_5,
\gamma^\mu_\perp \}$. Let us define the currents
\begin{equation}\label{HtCcurrents}
  J_1 = \bar \X_{hc} \,h\;, \qquad J_5 = \bar \X_{hc}\,\gamma_5\, h\;,\qquad 
  J_\perp^\mu = \bar \X_{hc} \,\gamma_\perp^\mu\, h\;,
\end{equation}
and the following symmetric and anti-symmetric tensors in the
transverse plane
\begin{equation}\label{transtensors}
g_\perp^{\mu\nu} = 
  g^{\mu\nu} - \frac{n^\mu \bar n^\nu + n^\nu \bar n^\mu}{2} \;,\qquad 
\epsilon_\perp^{\mu\nu} = \epsilon^{\mu\nu\alpha\beta} v_\alpha n_\beta\;,
\end{equation}
for convenience. We use $\epsilon_{0123} = 1$ and $\gamma_5 =
i\gamma^0 \gamma^1 \gamma^2 \gamma^3$.

Omitting the dependence on the renormalization scale and denoting the
Wilson coefficients by $C_i$, the matching of heavy-to-light currents
onto the above SCET currents reads to all orders in the strong
coupling constant and at leading power
\begin{eqnarray} \label{DiracReduction}
\bar q\,b &\to& C_1\,J_1\;, \\
\bar q\,\gamma_5\,b &\to& C_2 \,J_5\;, \nonumber \\
\bar q\,\gamma^\mu\,b &\to& C_3\,J_\perp^\mu 
  + (C_4 \,n^\mu+C_5\, v^\mu)\,J_1\;, \nonumber \\
\bar q\,\gamma^\mu\gamma_5\,b &\to& 
  - C_6\, i\epsilon_{\mu \nu}^\perp\, J_\perp^\nu 
  - (C_7 \,n^\mu+C_8\, v^\mu)\,J_5\;, \nonumber \\
\bar q\,i\sigma^{\mu\nu}\,b &\to&  C_9\,(n^\mu g_\perp^{\nu \lambda} 
  - n^\nu g_\perp^{\mu \lambda})\,J^\perp_\lambda 
  - C_{10} \, i\epsilon_\perp^{\mu\nu}\, J_5 \nonumber \\
&& + C_{11}\,(v^\mu n^\nu - v^\nu n^\mu) \, J_1 + 
  C_{12}\, (v^\mu g_\perp^{\nu \lambda} - v^\nu g_\perp^{\mu \lambda})\,
  J^\perp_\lambda \;, \nonumber \\
\bar q\,i\sigma^{\mu\nu}\gamma_5\,b &\to&  
  (C_9 + C_{12})\, (n^\mu i\epsilon_\perp^{\nu \lambda} 
  - n^\nu i\epsilon_\perp^{\mu \lambda}) \,J^\perp_\lambda 
  - C_{11}\,i\epsilon^{\mu\nu}_\perp\,J_1 \nonumber \\
&& + C_{10} \,(v^\mu n^\nu - v^\nu n^\mu) \, J_5 
  - C_{12}\, (v^\mu i\epsilon_\perp^{\nu \lambda} 
  - v^\nu i\epsilon_\perp^{\mu \lambda}) \,J^\perp_\lambda\;. \nonumber
\end{eqnarray}
This follows from the reduction of the Dirac basis. Furthermore, it
has been shown \cite{Bauer:2000yr} that in the naive dimensional
regularization scheme of anticommuting $\gamma_5$ in $4-2\epsilon$
dimensions, there are relations $C_{1} = C_{2}$, $C_{3} = C_{6}$,
$C_{4} = C_{7}$, $C_{5} = C_{8}$, $C_{10} = C_{11}$, $C_{12} = 0$,
among the coefficient functions that hold true to all orders for
massless light quarks.  At tree-level, the coefficient functions are
\begin{equation}
C_{1,2,3,4,6,7,9,10,11} = 1\;, \qquad C_{5,8,12}=0 \;.
\end{equation}
One-loop corrections to these Wilson coefficients have been computed
and can be found in \cite{Bauer:2000yr}. Apart from the renormalization
scale $\mu$ they depend on the large scales $\bar n\cdot p$ and
$m_b$. This matching calculation achieves the first step in scale
separation, and we will typically need some combination of these
Wilson coefficients in later applications (e.g. factorization of the
hadronic tensor in inclusive $B\to X_u\, l^-\bar \nu$ decays).

In this short section we do not present the radiative corrections, and
therefore we will also delay the discussion of renormalization-group
flow of the coefficient functions to later chapters, i.e. when we need
it. Let us just mention without proof that all Wilson coefficient
functions $C_i$ share the same anomalous dimension
\cite{Bauer:2000yr}, which will be the same (up to a minus sign) as
the anomalous dimension of the SCET\footnote{We will see later that
the anomalous dimension is independent of the hard-collinear momentum
square $p^2$. This must be true, because $p^2$ serves as an infra-red
regulator, which does not enter the ultra-violet behaviour encoded in
the anomalous dimension. Therefore the anomalous dimension will be the
same in both theories SCET$_I$ ($p^2\sim E\Lambda_{\rm QCD}$) and
SCET$_{II}$ ($p^2\sim \Lambda_{\rm QCD}^2$), and we will discuss it in
the latter theory.} current $\bar \X_{hc}\,\Gamma\,h$.

\section{SCET$_{II}$}

The goal is to construct an effective theory in which momentum
fluctuations are at most of order the soft QCD scale $\Lambda_{\rm
QCD}$. The main application for such a theory is the calculation of
exclusive Heavy-to-Light decay amplitudes. Introducing the expansion
parameter as $\lambda = \Lambda_{\rm QCD}/E$, the constituents of the
$B$ meson have a soft momentum scaling $(\lambda,\lambda,\lambda)$. If
the decay product is a light energetic hadron, we assign collinear
scaling $(\lambda^2,1,\lambda)$ to the partons inside this
hadron. These are the only momenta that appear on external legs.
However, as we have seen in the introductory example in
Fig.~\ref{fig:exampleTriangleFull}, there is an additional
long-distance mode (soft-collinear) which scales like
$(\lambda^2,\lambda,\lambda^{3/2})$. Such a field communicates with
either the soft or the collinear sector but does not appear on
external legs, and is therefore referred to as a ``messenger''.
Whether or not this mode will contribute in the end depends on choices
such as infra-red regulators and quark masses. We shall work
consistently in the limit of vanishing quark masses for the first
family, and therefore include this mode in the effective theory.  We
will see later that the soft-collinear sector of SCET$_{II}$ provides
us with very elegant arguments to discuss important physics such as
factorizability and endpoint singularities of decay amplitudes.

Let us deal with the various sectors of SCET$_{II}$ one by one. The
effective Lagrangian contains interactions between soft and collinear
fields by either the exchange of messenger modes or induced
interactions resulting from the exchange of hard-collinear off-shell
propagators. The latter are, however, kinematically forbidden in the
applications considered in this work \cite{Becher:2003qh}, because
they require exceptional momentum configurations which are typically
not generated at leading power in loop graphs. We may therefore
neglect such terms. The Lagrangian can be split up as
\begin{equation}\label{Lscet}
  \mathcal L = \mathcal L_c + \mathcal L_s + \mathcal L_h + 
  \mathcal L_{sc} + \mathcal L_{\rm int} \;.
\end{equation}
The construction of the collinear, soft, and heavy Lagrangians proceed
analogously to our previous discussions \cite{Beneke:2002ph,Bauer:2000yr}:
\begin{equation}
   {\mathcal L}_{c} = \bar\xi\,\frac{\nbslash}{2}\,in\cdot D_{c}\,\xi
   - \bar\xi\,i\Dslash_{c\perp}\,\frac{\nbslash}{2}\,
   \frac{1}{i\bar n\cdot D_{c}}\,i\Dslash_{c\perp}\,\xi + 
   \mathcal L_c^{\rm glue}\;, \qquad
{\mathcal L}_{s} = \bar q\,i\Dslash_s\, q + \mathcal L_s^{\rm glue}\;,
\end{equation}
and the HQET Lagrangian as given in (\ref{HQETLagrangian}). Note that,
in contrast to the construction of SCET$_I$, the above Lagrangians are
exact to all orders in $\lambda$. The gluon Lagrangians in the three
sectors retain the same form as in full QCD, but with the gluon fields
restricted to the corresponding subspaces of their soft, collinear, or
soft-collinear Fourier modes. The soft-collinear Lagrangian resembles
in its form the collinear Lagrangian. This is not surprising, since
the light-cone components of a soft-collinear momentum displays the
same hierarchy of scales. For completeness (and at the same time
reviewing the strategy for $\mathcal L_c$), let us repeat the
derivation briefly.  Using the projectors $\nslash \nbslash /4$ and
$\nbslash \nslash /4$, the quark field $q_{sc} = \theta + \sigma$
splits up into small and large components. Analyzing the propagator
reveals that $\theta \sim \lambda^2$ is the large component, and
$\sigma \sim \lambda^{5/2}$ can be eliminated using the equation of
motion
\begin{equation}
   \sigma = - \frac{\nbslash}{2}\,\frac{1}{i\bar n\cdot D_{sc}}\,
   i\Dslash_{sc\perp}\,\theta \,.
\end{equation}
Inserting this result back into the Lagrangian yields the exact result
\begin{equation}\label{Lsc}
   {\cal L}_{sc} = \bar\theta\,\frac{\nbslash}{2}\,in\cdot D_{sc}\,\theta
   - \bar\theta\,i\Dslash_{sc\perp}\,\frac{\nbslash}{2}\,
   \frac{1}{i\bar n\cdot D_{sc}}\,i\Dslash_{sc\perp}\,\theta \,.
\end{equation}
The most important new ingredient of SCET$_{II}$ is the interaction
sector between soft-collinear fields with soft or collinear degrees of
freedom. The formalism follows along the same lines as the discussion
of hard-collinear -- soft interactions in SCET$_{I}$, but is richer (and
more interesting). It has been developed and discussed in detail in
\cite{Becher:2003qh} and will be outlined in the next section. 

In interactions with other fields, the soft-collinear fields (but not the 
soft and collinear fields) are multipole expanded as
\begin{equation}\label{multipoleInts}
\begin{aligned}
   \phi_{sc}(x) &= \phi_{sc}(x_-) + x_\perp\cdot\partial_\perp\,
    \phi_{sc}(x_-) + \dots \quad && \mbox{in collinear interactions}
    \,, \\
   \phi_{sc}(x) &= \phi_{sc}(x_+) + x_\perp\cdot\partial_\perp\,
   \phi_{sc}(x_+) + \dots \quad && \mbox{in soft interactions} \,.
\end{aligned}
\end{equation}
As before, the four-vectors $x_- = (\bar n\cdot x)\, n/2$ and $x_+ =
(n\cdot x)\,\bar n/2$ are light-cone positions. The first correction
terms in (\ref{multipoleInts}) are of $O(\lambda^{1/2})$, and the
omitted terms are of $O(\lambda)$ and higher.

\subsection{Gauge Transformations and Interaction Terms}

Soft-collinear fields can couple to soft or collinear fields without
altering their scaling properties. This motivates the treatment of the
soft-collinear gluon field as a background field. However, in order to
preserve the scaling properties of the fields under gauge
transformations one must expand the transformation laws in
$\lambda$. \linebreak In close analogy to the discussion in SCET$_I$,
this leads to the following set of ``homogeneous'' gauge
transformations for the quark fields:
\begin{equation}
\begin{aligned}
   \mbox{soft:} \quad &
    q_s(x)\to U_s(x)\,q_s(x) \,, &&
    \mbox{coll. and soft-coll. fields invariant} \\
   \mbox{collinear:} \quad &
    \xi(x)\to U_c(x)\,\xi(x) \,, &&
    \mbox{soft and soft-coll. fields invariant} \\
   \mbox{soft-collinear:} \quad & 
    q_s(x)\to U_{sc}(x_+)\,q_s(x) \,, ~\,&&
    \xi(x)\to U_{sc}(x_-)\,\xi(x) \,, \\
    &&& q_{sc}(x)\to U_{sc}(x)\,q_{sc}(x) \;.
\end{aligned}
\end{equation}
The transformation laws for gluons are more complicated, but
straightforward \cite{Becher:2003qh}. Treating the slowly varying
soft-collinear gluons as background fields, we have the following
transformation laws under soft and collinear gauge transformations
before multipole expansion:
\begin{equation}\label{gauge1}
   A_s^\mu \to U_s\,A_s^\mu\,U_s^\dagger + U_s\,[iD_{sc}^\mu,U_s^\dagger]
    \,, \qquad
   A_c^\mu \to U_c\,A_c^\mu\,U_c^\dagger + U_c\,[iD_{sc}^\mu,U_c^\dagger]
    \,,
\end{equation}
while the collinear and soft-collinear fields remain invariant under
$U_s$, and soft and soft-collinear fields remain invariant under
$U_c$. Finally, under soft-collinear gauge transformations $U_{sc}$
the fields transform as
\begin{equation}\label{Usc}
\begin{aligned}
   A_c^\mu &\to U_{sc}\,A_c^\mu\,U_{sc}^\dagger \,,
    &\xi &\to U_{sc}\,\xi \,, \\
   A_s^\mu &\to U_{sc}\,A_s^\mu\,U_{sc}^\dagger \,,
    &q_s &\to U_{sc}\,q_s \,, \\
   A_{sc}^\mu &\to U_{sc}\,A_{sc}^\mu\,U_{sc}^\dagger
    + U_{sc}\,[i\partial^\mu,U_{sc}^\dagger] \,, \qquad
    &q_{sc} &\to U_{sc}\,q_{sc} \,.
\end{aligned}
\end{equation}
It can be seen from these relations that the combination
$(A_s^\mu+A_{sc}^\mu)$ transforms in the usual way under both soft and
soft-collinear gauge transformations, while $(A_c^\mu+A_{sc}^\mu)$
transforms in the usual way under both collinear and soft-collinear
gauge transformations. In order to preserve a homogeneous power
counting, we expand the above rules and keep only the leading-order
terms for consistency. This means that the transformation laws are
replaced by the homogeneous transformations
\begin{equation}
\begin{aligned}
   \mbox{soft:}
    \quad &
    n\cdot A_s\to U_s\,n\cdot A_s\,U_s^\dagger 
    + U_s\,[in\cdot\partial,U_s^\dagger] \,, 
    & & q_s\to U_s\,q_s \,, & \\
    \quad &
    A_{s\perp}^\mu\to U_s\,A_{s\perp}^\mu\,U_s^\dagger 
    + U_s\,[i\partial_\perp^\mu,U_s^\dagger] \,, & \\
    \quad & 
    \bar n\cdot A_s\to U_s\,\bar n\cdot A_s\,U_s^\dagger 
    + U_s\,[i\bar n\cdot D_{sc}(x_+),U_s^\dagger] \,, & \\
   \mbox{soft-collinear:}
    \quad & A_s^\mu\to U_{sc}(x_+)\,A_s^\mu\,
    U_{sc}^\dagger(x_+) \,,
    & & q_s\to U_{sc}(x_+)\,q_s \,,
\end{aligned} 
\end{equation}
for soft gluon fields and
\begin{equation}
\begin{aligned}
   \mbox{collinear:}
    \quad &
    \bar n\cdot A_c\to U_c\,\bar n\cdot A_c\,U_c^\dagger 
    + U_c\,[i\bar n\cdot\partial,U_c^\dagger] \,, 
    & & \xi\to U_c\,\xi \,, & \\
    \quad &
    A_{c\perp}^\mu\to U_c\,A_{c\perp}^\mu\,U_c^\dagger 
    + U_c\,[i\partial_\perp^\mu,U_c^\dagger] \,, & \\
    \quad & 
    n\cdot A_c\to U_c\,n\cdot A_c\,U_c^\dagger 
    + U_c\,[i n\cdot D_{sc}(x_-),U_c^\dagger] \,, & \\
   \mbox{soft-collinear:}
    \quad & A_c^\mu\to U_{sc}(x_-)\,A_c^\mu\,
    U_{sc}^\dagger(x_-) \,,
    & & \xi\to U_{sc}(x_-)\,\xi \,,
\end{aligned} 
\end{equation}
for collinear gluon fields. The soft-collinear gauge sector transforms
in the usual form. Let us now return to the interaction
Lagrangian. There are no interactions between soft and collinear
fields, since such interactions would lead to off-shell momenta of
$O(E\Lambda_{\rm QCD})$. Soft and collinear fields can, however, couple
separately to soft-collinear fields. At leading power we have
\begin{eqnarray}\label{SCints}
   {\cal L}_{\rm int}^{(0)}(x)
   &=& \bar q_s(x)\,\frac{\nslash}{2}\,g\bar n\cdot A_{sc}(x_+)\,q_s(x)
    + \bar h(x)\,\frac{n\cdot v}{2}\,g\bar n\cdot A_{sc}(x_+)\,h(x)
    \nonumber\\
   &+& \bar\xi(x)\,\frac{\nbslash}{2}\,g n\cdot A_{sc}(x_-)\,
    \xi(x) + \mbox{pure glue terms} \,.
\end{eqnarray}
Subleading terms have been calculated and can be found in
\cite{Becher:2003qh}. 

Momentum conservation implies that soft-collinear fields can only
couple to either soft or collinear modes, but not both. More than one
soft or collinear particle must be involved in such interactions. The
gluon self-couplings can be derived by substituting $A_s^\mu\to
A_s^\mu+\frac12 n^\mu\,\bar n\cdot A_{sc}(x_+)$ for the gluon field in
the soft Yang--Mills Lagrangian and $A_c^\mu\to A_c^\mu+\frac12 \bar
n^\mu\,n\cdot A_{sc}(x_-)$ for the gluon field in the collinear
Yang--Mills Lagrangian, and isolating terms containing the
soft-collinear field. The precise form of these interactions will not
be relevant to our discussion. Finally, let us note that none of the
terms in the SCET Lagrangian (\ref{Lscet}) is renormalized beyond the
usual renormalization of the strong coupling and the fields
\cite{Beneke:2002ph,Becher:2003qh}.

The multipole expansion of the soft-collinear fields implies that
momentum is {\em not\/} conserved at these vertices. When a soft
(light or heavy) quark with momentum $p_s$ absorbs a soft-collinear
gluon with momentum $k$, the outgoing soft quark carries momentum
$p_s+k_-$.  Likewise, when a collinear quark with momentum $p_c$
absorbs a soft-collinear gluon with momentum $k$, the outgoing
collinear quark carries momentum $p_c+k_+$.

In order to match the quark and gluon fields of the full theory onto
SCET fields obeying the homogeneous gauge transformations one first
adopts specific gauges in the soft and collinear sectors, namely soft
light-cone gauge $n\cdot A_s=0$ (SLCG) and collinear light-cone gauge
$\bar n\cdot A_c=0$ (CLCG). At leading order in $\lambda$, one then
introduces the corresponding SCET fields via the substitutions
\cite{Beneke:2002ni,Becher:2003qh}
\begin{equation}\label{newfields}
   \psi_s \big|_{\rm SLCG} \to R_s\,S_s^\dagger\,q_s \,, \qquad
   b \big|_{\rm SLCG} \to R_s\,S_s^\dagger\,h \,, \qquad
   \psi_c \big|_{\rm CLCG} \to R_c\,W_c^\dagger\,\xi \,.
\end{equation}
The corresponding replacements for gluon fields can be found in
\cite{Becher:2003qh}. They are a little more complicated, because the
homogeneous gauge transformation mixes the soft and the collinear
gluon fields with soft-collinear ones. For example, the two non-zero
components of the soft gluon in SLCG are given by
\begin{equation}
\begin{aligned}
     A_{s\perp}^\mu \big|_{\rm SLCG} 
&\to R_s\,S_s^\dagger(iD_{s\perp}^\mu\,S_s)\,R_s^\dagger \;, \\
     \bar n\cdot A_s \big|_{\rm SLCG} 
&\to R_s\,\big[ S_s^\dagger(i\bar n\cdot D_{s\perp}^\mu\,S_s)
     + S_s^\dagger [\bar n\cdot A_{sc}(x_+), S_s] \big]R_s^\dagger \;.
\end{aligned}
\end{equation}
Similarly for collinear gluons. The quantities
\begin{equation}\label{WSdef}
\begin{aligned}
   S_s(x) &= \mbox{P}\exp\left( ig\int_{-\infty}^0 dt\,
    n\cdot A_s(x+tn) \right) , \\
   W_c(x) &= \mbox{P}\exp\left( ig\int_{-\infty}^0 dt\,
    \bar n\cdot A_c(x+t\bar n) \right) 
\end{aligned} 
\end{equation}
are SCET Wilson lines in the soft and collinear sectors
\cite{Beneke:2002ph,Bauer:2001yt}, which effectively put the SCET
fields into light-cone gauge.  Small subscripts ``s'' and ``c'' are
provided to distinguish these Wilson lines from corresponding ones
with the soft or collinear gluon fields replaced by soft-collinear
ones. The objects $R_s$ and $R_c$ are short gauge strings of
soft-collinear fields from $x_+$ to $x$ (for $R_s$) and $x_-$ to $x$
(for $R_c$). They differ from 1 by terms of order $\lambda^{1/2}$ and
so must be Taylor expanded. Note that $S_s$ transforms as $S_s(x)\to
U_s(x)\,S_s(x)$ and $S_s(x)\to
U_{sc}(x_+)\,S_s(x)\,U_{sc}^\dagger(x_+)$ under soft and
soft-collinear gauge transformations and is invariant under collinear
gauge transformations. Likewise, $W_c$ transforms as $W_c(x)\to
U_c(x)\,W_c(x)$ and $W_c(x)\to
U_{sc}(x_-)\,W_c(x)\,U_{sc}^\dagger(x_-)$ under collinear and
soft-collinear gauge transformations and is invariant under soft gauge
transformations. The short strings only transform under soft-collinear
gauge transformations, in such a way that $R_s(x)\to
U_{sc}(x)\,R_s(x)\,U_{sc}^\dagger(x_+)$ and $R_c(x)\to
U_{sc}(x)\,R_c(x)\,U_{sc}^\dagger(x_-)$. It follows that the
expressions on the right-hand side of (\ref{newfields}) are invariant
under soft and collinear gauge transformations and transform as
ordinary QCD quark fields under soft-collinear gauge transformations.

Again it is sometimes useful to introduce the calligraphic fields that
are gauge-invariant building blocks
\begin{equation}\label{blocks}
\begin{aligned}
   W_c^\dagger\,\xi &= S_{sc}(x_-)\,\X \,, \qquad
    S_s^\dagger\,q_s = W_{sc}(x_+)\,\Q_s \,, \qquad
    S_s^\dagger\,h = W_{sc}(x_+)\,\H \,, \\
   W_c^\dagger\,(iD_{c\perp}^\mu W_c)
   &= S_{sc}(x_-)\,\A_{c\perp}^\mu\,S_{sc}^\dagger(x_-) \,, \qquad
    S_s^\dagger\,(iD_{s\perp}^\mu S_s)
    = W_{sc}(x_+)\,\A_{s\perp}^\mu\,W_{sc}^\dagger(x_+) \,,
\end{aligned}
\end{equation}
where fields without argument live at position $x$. The quantities
$W_{sc}$ and $S_{sc}$ are a new set of Wilson lines defined in analogy
with $W_c$ and $S_s$ in (\ref{WSdef}), but, as mentioned above, with
the gluon fields replaced by soft-collinear gluon fields in both
cases. The calligraphic fields in (\ref{blocks}) are invariant under
soft, collinear, and soft-collinear gauge transformations.  This
follows from the fact that the new Wilson lines transform as
\begin{equation}
   W_{sc}(x_+)\to U_{sc}(x_+)\,W_{sc}(x_+) \,, \qquad
   S_{sc}(x_-)\to U_{sc}(x_-)\,S_{sc}(x_-) \,
\end{equation}
under soft-collinear gauge transformations.

\subsection{Currents}\label{sec:currents}

\subsubsection{Heavy-to-Collinear Currents}

Let us recall from the discussion of Wilson lines that $W_{hc}$ emerged in
heavy-to-collinear currents in SCET$_I$ by keeping leading-order
attachments of collinear gluons to the heavy quark line. In
SCET$_{II}$ we denote this Wilson line by $W_c$. Note that attachments
of soft gluons to the collinear quark line will also lead to off-shell
modes, and thus to the appearance of $S_s$. Naively, one might
therefore expect to build up the current $\bar \xi\,S_s^\dagger\,
\Gamma\,W_c\,h$, and in a non-abelian theory $S_s^\dagger$ and $W_c$
do not commute. However, adding diagrams involving non-abelian gluon
couplings reverses the order of the two Wilson lines
\cite{Bauer:2001yt}, as dictated by soft and collinear gauge
invariance.

The matching procedure (\ref{newfields}) reproduces this explicit
finding in the following simple way. At tree-level and leading power
the QCD current is matched onto the gauge-invariant object (omitting
the large HQET phase $e^{-im_b v\cdot x}$)
\begin{eqnarray}
   \bar\psi_c(x)\,\Gamma\,b(x)
   &\to& \big[ \bar\xi\,W_c\,R_c^\dagger \big](x)\,
    \Gamma\,\big[ R_s\,S_s^\dagger\,h \big](x) \nonumber\\
   &=& \big[ \bar\xi\,W_c \big](x_+ + x_\perp)\,
    \Gamma\,\big[ S_s^\dagger\,h \big](x_- + x_\perp) + O(\lambda) \,.
\end{eqnarray}
Note that the expression in the first line is not homogeneous in
$\lambda$. In interactions of soft and collinear fields, the soft
fields must be multipole expanded about $x_+=0$, while the collinear
fields must be multipole expanded about $x_-=0$. Also, as mentioned
above, the quantities $R_s$ and $R_c$ must be expanded and equal 1 to
first order. This leads to the result shown in the second line. The
terms of $O(\lambda^{1/2})$ in the expansions of $R_s$ and $R_c$
cancel each other \cite{Becher:2003qh}. The leading-order SCET current
in the final expression is gauge invariant even without the $R_s$ and
$R_c$ factors, since soft fields at $x_+=0$ and collinear fields at
$x_-=0$ both transform with $U_{sc}(0)$ under soft-collinear gauge
transformations.

When radiative corrections are taken into account, the current mixes
with analogous operators at different positions on the light cone, and
for the case of the heavy-to-collinear currents different Dirac
structures can be induced by hard gluon exchange.  The correct
matching relation reads (setting $x=0$ for simplicity)
\begin{equation}\label{hc}
\begin{aligned}
   \bar\psi_c(0)\,\Gamma\,b(0)
   &\to \sum_i \int ds\,\widetilde C_i(s,\mu)\,
    \big[ \bar\xi\,W_c \big](s\bar n)\,\Gamma_i\,
    \big[ S_s^\dagger\,h \big](0) + O(\lambda)  \\
    &= \sum_i C_i(\bar n\cdot P^c,\mu)\,
    \big[ \bar\xi\,W_c\,\Gamma_i\,S_s^\dagger\,h \big](0)
    + O(\lambda) \,,
\end{aligned}
\end{equation}
where translation invariance is used to rewrite the expression in a
local form. As in the case of SCET$_I$, there are only three
independent Dirac structures $\Gamma_i$ allowed. The coefficient
functions $C_i(\bar n\cdot P^c) = \int ds\, e^{is \bar n\cdot
P^c}\widetilde C_i(s)$ are the Fourier transforms of the
position-space Wilson coefficients $\widetilde C_i$ and depend on the
total collinear momentum operator $P^c = P^c_{\rm out} - P^c_{\rm
in}$. In fact, by reparameterization invariance (see
Section~\ref{sec:RPI} below) the dependence of the Wilson coefficient
functions on $P^c$ must be through the combination $2v\cdot P^c_- =
(v\cdot n)(\bar n\cdot P^c)$. Unless otherwise stated, we set $v\cdot
n=1$ for convenience, and work with the above expressions.

\subsubsection{Soft-to-Collinear Currents}

Arguments along the same line as in the previous section would suggest
that the QCD current $\bar \psi_c(0)\,\Gamma\,\psi_s(0)$ matches
simply onto terms of the form $\left[\bar \xi\, W_c\right]
\Gamma\left[S_s^\dagger\,q\right]$ with the fields evaluated at
different positions.  This is, however, not quite correct. Note that
the soft massless quark field is a full four-component QCD spinor, so
that we cannot use the reduced Dirac basis $\Gamma_i$.  A careful
perturbative matching analysis shows that there is a second structure
that appears at leading power, involving a perpendicularly polarized
collinear gluon \cite{Hill:2002vw}. At tree-level, the expression
\begin{equation} \label{currentops}
  \left[\bar \xi\,W_c\right]\!(0)\,\Gamma\left[S_s^\dagger\,q\right]\!(0)
  \;-\; \int_{-\infty}^0 d t\; 
  \bar \xi(0)\,\Gamma\left[\frac{\nslash}{2}\,
    \Dslash_{c\perp}\,W_c\right]\!(0)\left[S_s^\dagger\,q \right]\!(t n)
\end{equation}
sums up an infinite set of leading-order interactions. Despite of the
extra perpendicular derivative (which scales like $\lambda$), this
operator is not subleading because of the large non-localities
associated with soft fields (integration in $t$ over a large domain of
order $\lambda^{-1}$). Radiative corrections mix these currents with
similar operators at different positions on the light-cone. Note that
because of the factor of $\nslash$, the operator vanishes if $\Gamma$
commutes with $\nslash$. The existence of this second operator
therefore depends on the particular Dirac structure. For the present
work we will only need Dirac structures that are of the form $\Gamma
\nslash$, so that the second term in (\ref{currentops}) vanishes since
$n^2=0$. The matching relation of interest therefore reads
\pagebreak
\begin{equation}\label{sc}
\begin{aligned}
   \bar\psi_c(0)\,\Gamma \frac{\nslash}{2}\,\psi_s(0)
   &\to \int ds\,dt\,\widetilde D(s,t,\mu)\,
    \big[ \bar\xi\,W_c \big](s\bar n)\,\Gamma \frac{\nslash}{2}\,
    \big[ S_s^\dagger\,q_s \big](t n) + O(\lambda)  \\
    &= D(P^s_+ \cdot P^c_-,\mu)\,
    \big[ \bar\xi\,W_c\,\Gamma \frac{\nslash}{2}\,S_s^\dagger\,q_s \big](0)
    + O(\lambda) \,.
\end{aligned}
\end{equation}
Again, reparameterization invariance dictates that the Wilson
coefficient can only depend on the scalar product $2P^s_+ \cdot P^c_- =
(n\cdot P^s)(\bar n \cdot P^c)$.

\subsubsection{Radiative Corrections and Anomalous Dimensions}

The momentum-space coefficient functions in (\ref{hc}) and (\ref{sc})
are renormalized multiplicatively and obey renormalization-group (RG)
equations of the Sudakov type \cite{Bauer:2000yr,Bosch:2003fc},
i.e. contain an explicit logarithmic renormalization scale dependence,
\begin{equation}
\begin{aligned}
   \frac{d}{d\ln\mu}\,C_i(v\cdot p_{c-},\mu)
   &= \gamma_{\xi h}(v\cdot p_{c-},\mu)\,C_i(v\cdot p_{c-},\mu) \,, \\
   \frac{d}{d\ln\mu}\,D(p_{s+}\!\cdot p_{c-},\mu)
   &= \gamma_{\xi q}(p_{s+}\!\cdot p_{c-},\mu)\,
    D(p_{s+}\!\cdot p_{c-},\mu) \,, 
\end{aligned}
\end{equation}
where the anomalous dimensions take the form
\begin{equation}\label{andims}
\begin{aligned}
   \gamma_{\xi h}(v\cdot p_{c-},\mu)
   &= -\frac12\,\Gamma_{\rm cusp}[\alpha_s(\mu)]\,
    \ln\frac{\mu^2}{(2v\cdot p_{c-})^2}
    + \Gamma_{\xi h}[\alpha_s(\mu)] \,, \\
   \gamma_{\xi q}(p_{s+}\!\cdot p_{c-},\mu)
   &= -\Gamma_{\rm cusp}[\alpha_s(\mu)]\,
    \ln\frac{\mu^2}{2p_{s+}\!\cdot p_{c-}}
    + \Gamma_{\xi q}[\alpha_s(\mu)] \,.
\end{aligned}
\end{equation}
The coefficients of the logarithmic terms are determined in terms of
the universal cusp anomalous dimension $\Gamma_{\rm
cusp}=C_F\,\alpha_s/\pi+O(\alpha_s^2)$, which plays a central role in
the renormalization of Wilson lines with light-like segments
\cite{Korchemsky:wg}. The one-loop expressions for the non-logarithmic
terms in the anomalous dimensions can be deduced from the explicit
results for the Wilson coefficients derived in
\cite{Bauer:2000yr,Hill:2002vw}. They are
\begin{equation}\label{Gammas}
   \Gamma_{\xi h}(\alpha_s) = -\frac54\,\frac{C_F\alpha_s}{\pi}
    + O(\alpha_s^2) \,, \qquad
   \Gamma_{\xi q}(\alpha_s) = -\frac32\,\frac{C_F\alpha_s}{\pi}
    + O(\alpha_s^2) \,.
\end{equation}

It may seem surprising that after hard and hard-collinear scales have
been integrated out the operators of the low-energy theory still know
about the large scales $v\cdot p_{c-}\sim E$ and $p_{s+}\cdot
p_{c-}\sim E\Lambda_{\rm QCD}$, as is evident from the appearance of
the logarithms in (\ref{andims}). The reason is that in interactions
involving both soft and collinear particles there is a large Lorentz
boost $\gamma\sim p_s\cdot p_c/\sqrt{p_s^2\,p_c^2}\sim E/\Lambda_{\rm
QCD}$ connecting the rest frames of soft and collinear hadrons, which
is fixed by external kinematics and enters the effective theory as a
parameter. This is similar to applications of HQET to $b\to c$
transitions, where the fields depend on the external velocities of the
hadrons containing the heavy quarks, and $\gamma=v_b\cdot v_c=O(1)$ is
an external parameter that appears in matrix elements and anomalous
dimensions of velocity-changing current operators
\cite{Neubert:1993mb,Falk:1990yz}. In the case of SCET, the boost
parameter $\gamma$ can, in some cases, reintroduce a ``long-distance
type'' dependence on the large energy through low-energy operator
matrix elements (see for example Section~\ref{chap:formfactor}).

\begin{figure}
\begin{center}
\epsfig{file=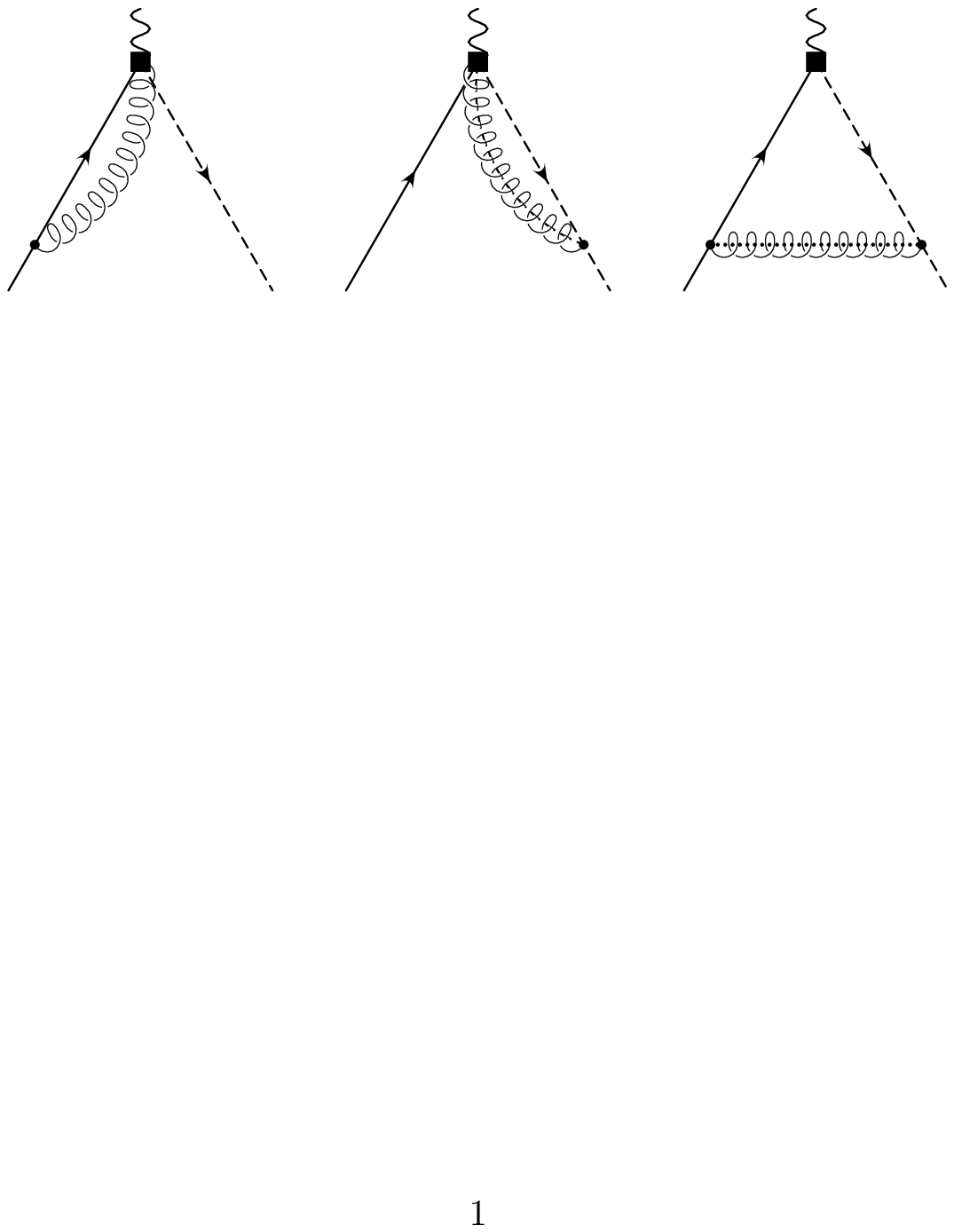, width=9cm}
\end{center}
\vspace{-0.2cm} \centerline{\parbox{14cm}{\caption[One-loop SCET current
diagrams.]{ \label{fig:SCETgraphs} SCET graphs contributing to the
anomalous dimension of a soft-collinear current. Full lines denote
soft fields, dashed lines collinear fields, and dotted lines
soft-collinear fields.}}}
\end{figure}

We will now explain how the results for the anomalous dimensions can
be obtained from a calculation of UV poles of SCET loop diagrams
\cite{Becher:2003kh}. The relevant diagrams needed at one-loop order
are shown in Fig.~\ref{fig:SCETgraphs}. They must be supplemented by
wave-function renormalization of the quark fields. The gluons
connected to the current are part of the Wilson lines $W_c$ and
$S_s$. We regularize IR singularities by keeping the external lines
off-shell. The results for the sum of all UV poles must be independent
of the IR regulators. For the heavy-to-collinear current the pole
terms obtained from the three diagrams are (here and below we omit the
$-i0$ in the arguments of logarithms)
\begin{equation}\label{eq:hc}
\begin{aligned}
   & \left( \frac{1}{\epsilon^2}
   - \frac{2}{\epsilon}\,\ln\frac{-2v\cdot p_s}{\mu}
   + \frac{1}{\epsilon} \right)
   + \left( \frac{2}{\epsilon^2}
   - \frac{2}{\epsilon}\,\ln\frac{-p_c^2}{\mu^2}
   + \frac{3}{2\epsilon} \right) \\
   +& \left( - \frac{2}{\epsilon^2} + \frac{2}{\epsilon}\,
   \ln\frac{(-2v\cdot p_s)(-p_c^2)}{2v\cdot p_{c-}\,\mu^2} \right)
   \;\;=\;\; \frac{1}{\epsilon^2}
   + \frac{2}{\epsilon}\,\ln\frac{\mu}{2v\cdot p_{c-}}
   + \frac{5}{2\epsilon} \,,
\end{aligned}
\end{equation}
while for the soft-to-collinear current we obtain
\begin{equation}
\begin{aligned}
\label{eq:sc}
   &\left( \frac{2}{\epsilon^2}
   - \frac{2}{\epsilon}\,\ln\frac{-p_s^2}{\mu^2}
   + \frac{3}{2\epsilon} \right)
   + \left( \frac{2}{\epsilon^2}
   - \frac{2}{\epsilon}\,\ln\frac{-p_c^2}{\mu^2}
   + \frac{3}{2\epsilon} \right) \\
   +& \left( - \frac{2}{\epsilon^2} + \frac{2}{\epsilon}\,
   \ln\frac{(-p_s^2)(-p_c^2)}{2p_{s+}\!\cdot p_{c-}\,\mu^2} \right)
   = \frac{2}{\epsilon^2}
   + \frac{2}{\epsilon}\,\ln\frac{\mu^2}{2p_{s+}\!\cdot p_{c-}}
   + \frac{3}{\epsilon} \,.
\end{aligned}
\end{equation}
We quote the contributions to the operator renormalization constants
$Z^{-1}$ in units of $C_F\alpha_s/4\pi$ (in the $\overline{\rm MS}$
subtraction scheme in $4-2\epsilon$ dimensions). The three parentheses
in the first line of the above equations correspond to the soft,
collinear, and soft-collinear contributions, where the first two terms
include the corresponding contributions from wave-function
renormalization. The $1/\epsilon$ poles of the soft and collinear
graphs depend on the IR regulators, but this dependence is precisely
canceled by the soft-collinear contribution. By construction, the sum
of the soft, collinear, and soft-collinear contributions is IR finite
and only contains UV poles, whose coefficients depend on the ratios
$v\cdot p_{c-}/\mu$ and $p_{s+}\!\cdot p_{c-}/\mu^2$.  This follows
since IR divergences in both the full and the effective theory (which
are equivalent at low energy) are regularized by the off-shellness of
the external quark lines. The one-loop contributions to the anomalous
dimensions $\gamma_{\xi h}$ and $\gamma_{\xi q}$ are given by
$-C_F\alpha_s/2\pi$ times the coefficients of the $1/\epsilon$ poles
in the above expressions. They are in agreement with the results
(\ref{andims}) and (\ref{Gammas}) obtained from the scale dependence
of Wilson coefficients.

The calculations presented above make it evident that there is an
intricate interplay between the soft, collinear, and soft-collinear
diagrams. In dimensional regularization the dependence of the
anomalous dimensions on the hard or hard-collinear scale enters
through the loop integral involving the soft-collinear exchange and
thus seems to be related to very small momentum scales. However, care
must be taken when assigning physical significance to the scales
associated with individual diagrams in SCET, because in the soft and
collinear diagrams a cancellation of IR and UV divergences takes
place. The logarithms appearing in their divergent parts should be
interpreted as [cf.~(\ref{eq:sc})]
\begin{equation}
   \ln\frac{-p_s^2}{\mu^2} + \ln\frac{-p_c^2}{\mu^2}
   = \ln\frac{Q^2}{\mu^2} + \ln\frac{m_{sc}^2}{\mu^2} \,, 
   \qquad \mbox{with} \quad
   m_{sc}^2 = \frac{(-p_s^2)\,(-p_c^2)}{2 p_{s+}\!\cdot p_{c-}} \,,
\end{equation}
and thus arise from a cancellation of physics at the hard scale 
$Q^2=2p_{s+}\!\cdot p_{c-}$ and at the soft-collinear scale $m_{sc}^2$.
In the sum of all graphs, the soft-collinear contribution precisely 
cancels the IR piece of the soft and collinear parts, see (\ref{eq:sc}).
This interpretation is consistent with the fact that the anomalous
dimensions measure the change of operator matrix elements under
infinitesimal variations of the UV cutoff $\mu$. They are therefore
insensitive to the physics at low scales by construction, and
the large logarithms in (\ref{andims}) are really of short-distance
nature. 

We would like to close this Section with a (somewhat historic) remark
that dates back to the time before soft-collinear fields were
introduced to SCET$_{II}$. After all, for exclusive $B$ decays one
identifies the expansion parameter $\lambda$ with the ratio
$\Lambda_{\rm QCD}/E$, and therefore the soft-collinear scale $m_{sc}$
is parametrically smaller than $\Lambda_{\rm QCD}$. This raises some
concerns as to whether the soft-collinear fields have a right to
exist. So consider for a few seconds (and not much longer) the
soft-collinear sector to be absent. Then the following puzzle arises:
We know from the explicit expressions for the Wilson coefficients of
the currents that their anomalous dimensions must depend on a scalar
product of the collinear momentum with a momentum characterizing the
soft quark (i.e., $v\cdot p_{c-}$ or $p_{s+}\!\cdot p_{c-}$). However,
the SCET Feynman rules imply that the first graph in
Fig.~\ref{fig:SCETgraphs} can only be a function of the soft momentum
(i.e., $v\cdot p_s$ or $p_s^2$), while the second one can only depend
on the collinear momentum (i.~e.~, $p_c^2$), as is in fact confirmed
by the explicit calculation. The apparent ``factorization'' of soft
and collinear degrees of freedom in SCET$_{II}$ (in the hypothetical
absence of soft-collinear fields) would thus lead to the conclusion
that the anomalous dimensions of the currents are independent of the
products $v\cdot p_{c-}$ or $p_{s+}\cdot p_{c-}$, in contradiction
with the results for the Wilson coefficients. In fact, factorization
would be a generic property of SCET$_{II}$, since the interaction
Lagrangian in (\ref{SCints}) would no longer exist. This is obviously
incorrect; for example it is well known that $B\to\pi$ transition form
factors do not factorize, although they can be described using
SCET$_{II}$.

Although the above argument strongly motivates the use of
soft-collinear fields in SCET$_{II}$, it does not necessitate
it. Alternative prescriptions include the introduction of ``analytic
regulators'' \cite{Beneke:2003pa} where propagators are raised to a
non-integer power, and explicit infra-red regulators in the Lagrangian
\cite{Bauer:2003td}. However, these schemes are not favorable in our
opinion because analytic regulators are very complicated and cannot be
introduced on the Lagrangian level in a gauge-invariant fashion, and
the proposed infra-red regulators in \cite{Bauer:2003td} break many
important symmetries such as gauge invariance and Lorentz invariance.
The soft-collinear sector solves the problem in a much
more simple and elegant way.

\subsection{Four-quark Operators}\label{sec:fourquark}

In many cases in $B$-decay physics, amplitudes calculated in SCET
receive contributions from hadronic matrix elements of four-quark
operators, which can be expressed in terms of the leading-order
light-cone distribution amplitudes of a light final-state meson and of
the initial state $B$ meson. The relevant SCET operators can be taken
as \cite{Hill:2002vw,Becher:2003kh}
\begin{eqnarray}\label{QRdef}
   Q_{(C)}(s,t)
   &=& \big[ \bar\xi\,W_c \big](s\bar n)\,\frac{\nbslash}{2}\,
    \Gamma_1\,T_1\,\big[ W_c^\dagger\,\xi \big](0)\,\,
    \big[ \bar q_s\,S_s \big](tn)\,\frac{\nslash}{2}\,\Gamma_2\,
    T_2\,\big[ S_s^\dagger\,h \big](0) \nonumber\\
   &\equiv& \int_0^\infty\!d\omega\,e^{-i\omega t}
    \int_0^{\,\bar n\cdot P}\!d\sigma\,e^{i\sigma s}\,
    Q_{(C)}(\omega,\sigma) \,,
\end{eqnarray}
where the color label $C=S$ or $O$ refers to the color singlet-singlet
and color octet-octet structures $T_1\otimes T_2={\bf 1}\otimes{\bf
1}$ or $T_A\otimes T_A$, respectively. Again, $\bar n\cdot P$ is the
total momentum carried by all collinear particles, which is fixed by
kinematics. (Strictly speaking, this is a momentum operator.) In
light-cone gauge, $\omega=n\cdot p_s$ corresponds to the plus
component of the momentum of the spectator anti-quark in the $B$
meson, while $\sigma=\bar n\cdot p_\xi$ denotes the minus component of
the momentum of the quark inside a light final-state meson. It is
conventional to introduce a dimensionless variable $u=\sigma/\bar
n\cdot P\in[0,1]$ corresponding to the longitudinal momentum fraction
carried by the quark.

\begin{figure}[t]
\begin{center}
\epsfig{file=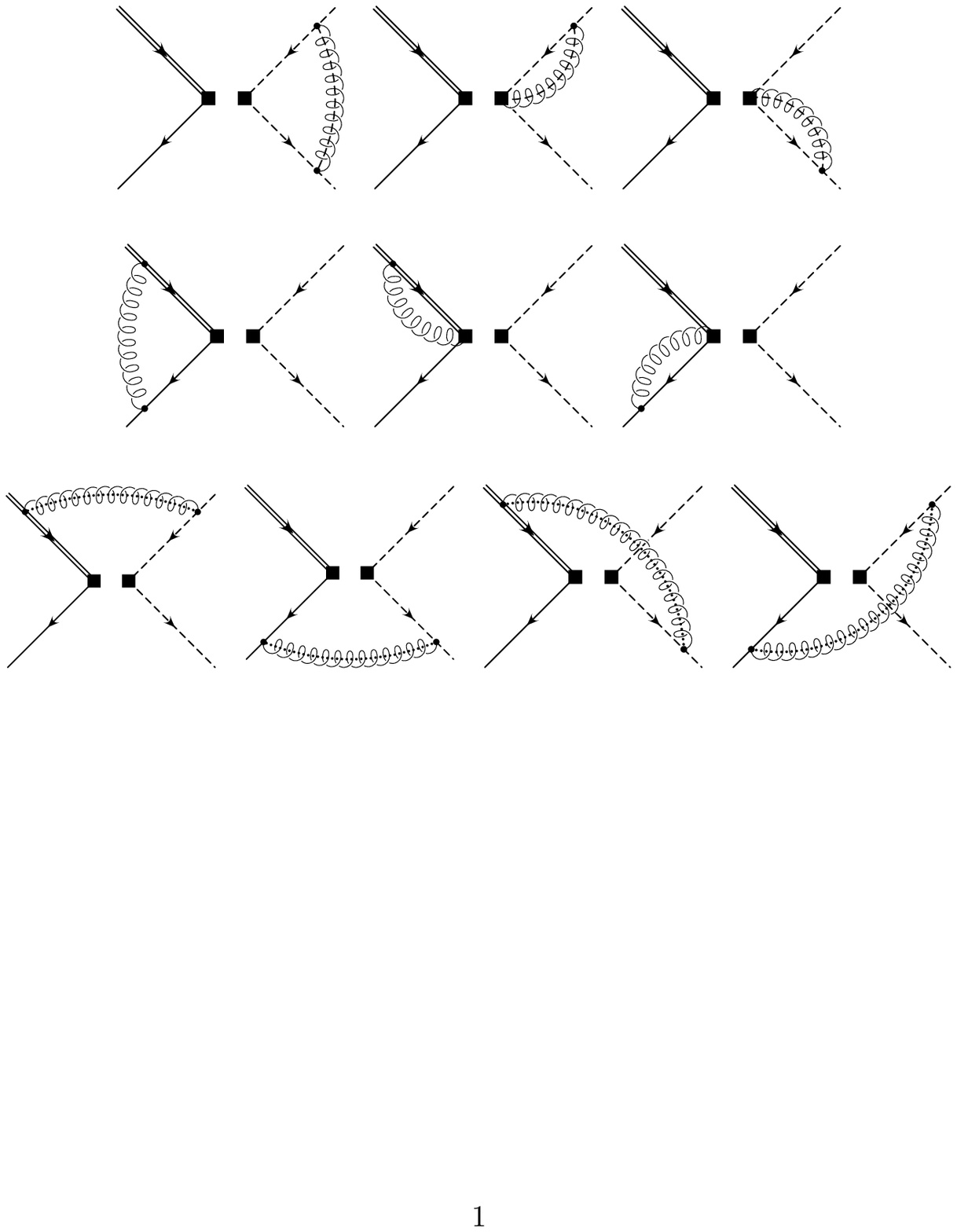, width=12cm}
\end{center}
\vspace{-0.2cm} \centerline{\parbox{14cm}{\caption[One-loop SCET
four-quark operator diagrams]{ \label{fig:fourquark} SCET graphs
contributing to the anomalous dimension of the four-quark operators
$Q_{(C)}(\omega,\sigma)$. Full lines denote soft fields, dashed lines
collinear fields, and dotted lines soft-collinear fields.}}}
\end{figure}

The momentum-space operators $Q_{(C)}(\omega,\sigma)$ obey the rather
complicated RG integro-differential equation
\begin{equation}
   \frac{d}{d\ln\mu}\,Q_{(C)}(\omega,u\,\bar n\cdot P)
   = - \int_0^{\infty}\!\!d\omega^\prime \int_0^1\!du^\prime\,
   \gamma_{(C)}(\omega,\omega^\prime,u,u^\prime,\bar n\cdot P,\mu)\,
   Q_{(C)}(\omega^\prime,u^\prime\,\bar n\cdot P) \,.
\end{equation}
To obtain the anomalous dimensions at leading order we compute the 
$1/\epsilon$ poles of the diagrams shown in Fig.~\ref{fig:fourquark} in
dimensional regularization and add the contributions from wave-function 
renormalization. Note that only soft-collinear gluons can be exchanged 
between the soft and collinear currents. For the color-singlet case 
$T_1\otimes T_2={\bf 1}\otimes{\bf 1}$ we find that the sum of the four 
diagrams with soft-collinear exchanges (but not each diagram separately) 
is  UV finite. The anomalous dimension is then a combination of the 
anomalous dimensions for the two non-local currents in (\ref{QRdef}). At 
one-loop order we obtain
\begin{equation}
   \gamma_{(S)}(\omega,\omega^\prime,u,u^\prime,\bar n\cdot P,\mu)
   = \frac{C_F\alpha_s}{\pi}\,\Big[ \delta(\omega-\omega^\prime)\,V(u,u^\prime)
   + \delta(u-u^\prime)\,H(\omega,\omega^\prime,\mu) \Big] \,,
\end{equation}
where (with $\bar u\equiv 1-u$) 
\begin{equation}
\begin{aligned}
   V(u,u^\prime) 
   =& - \left[
    \frac{u}{u^\prime} \left( \frac{1}{u^\prime-u} + c(\Gamma_1) \right) 
    \theta(u^\prime-u) 
    + \frac{\bar u}{\bar u^\prime} \left( \frac{1}{u-u^\prime} + c(\Gamma_1) 
    \right) 
    \theta(u-u^\prime) \right]_+ \\
   &\mbox{}+ \frac{1-c(\Gamma_1)}{2}\,\delta(u-u^\prime)
\end{aligned}
\end{equation}
with $c(1)=c(\gamma_5)=1$, $c(\gamma_\perp^\mu)=0$ is the
Brodsky--Lepage kernel \cite{Lepage:1980fj} for the evolution of the
leading-twist light-cone distribution amplitudes of a light meson,
which we have reproduced here using the Feynman rules of SCET. The
plus distribution is defined as
\begin{equation}
   [f(u,u^\prime)]_+ = f(u,u^\prime) - \delta(u-u^\prime) \int_0^1\!dw\,
   f(w,u^\prime) \,,
\end{equation}
which coincides with the conventional definition if the distribution acts 
on functions $g(u)$ but not if it acts on functions $g(u^\prime)$. The function
\begin{equation}
   H(\omega,\omega^\prime,\mu)
   = \left( \ln\frac{\mu\,v\cdot n}{\omega} - \frac54 \right)
   \delta(\omega-\omega^\prime)
   - \omega \left[ \frac{\theta(\omega-\omega^\prime)}{\omega(\omega-\omega^\prime)}
   +  \frac{\theta(\omega^\prime-\omega)}{\omega^\prime(\omega^\prime-\omega)} \right]_+
\end{equation}
is the analogous kernel governing the evolution of the leading-order
$B$-meson light-cone distribution amplitude \cite{Lange:2003ff}. Here
the plus distribution is symmetric in the two arguments and defined as
\begin{equation}
   \int_0^\infty\!d\omega^\prime\,[f(\omega,\omega^\prime)]_+\,g(\omega^\prime)
   =  \int_0^\infty\!d\omega^\prime\,f(\omega,\omega^\prime)\,
   \big[ g(\omega^\prime) - g(\omega) \big] \,.
\end{equation}

For the color-octet case $T_1\otimes T_2=T_A\otimes T_A$ things are
more complicated. In that case the sum of the soft-collinear exchange
graphs shown in the last line in Fig.~\ref{fig:fourquark} does not
vanish. However, it provides contributions such that, in the end, the
dependence on the IR regulators drops out. Our final result for the
anomalous dimension in the octet-octet case is
\begin{eqnarray}
   \gamma_{(O)}(\omega,\omega^\prime,u,u^\prime,\bar n\cdot P,\mu)
   &=& - \frac{1}{2N}\,\frac{\alpha_s}{\pi}\,\Big[
    \delta(\omega-\omega^\prime)\,V(u,u^\prime) 
    + \delta(u-u^\prime)\,H(\omega,\omega^\prime,\mu) \Big] \nonumber \\
   && \hspace{-40mm} - \frac{N}{2}\,\frac{\alpha_s}{\pi}\,
    \delta(\omega-\omega^\prime)\,\delta(u-u^\prime) \left(
    \ln\frac{\mu^3}{n\cdot v\,\omega\,(\bar n\cdot P)^2}
    - \ln u\bar u + \frac{11}{4} \right) \;. 
\end{eqnarray}

\subsection{Decoupling transformation\label{sec:decoupling}}

The leading-order interactions between soft-collinear fields and soft or 
collinear fields in the SCET Lagrangian can be removed by 
a redefinition of the soft and collinear fields \cite{Becher:2003qh}. In 
analogy with the decoupling of soft gluons in SCET$_{\rm I}$ 
\cite{Bauer:2001yt}, we define new fields
\begin{equation}\label{another}
\begin{aligned}
   q_s(x) = W_{sc}(x_+)\,q_s^{(0)}(x) \,, \qquad
   h(x) = W_{sc}(x_+)\,h^{(0)}(x) \,, \qquad
   \xi(x) = S_{sc}(x_-)\,\xi^{(0)}(x) \,, \\
   A_s^\mu(x) = W_{sc}(x_+)\,A_s^{(0)\mu}(x)\,W_{sc}^\dagger(x_+)
    \,, \qquad 
   A_c^\mu(x) = S_{sc}(x_-)\,A_c^{(0)\mu}(x)\,S_{sc}^\dagger(x_-) \,.
    \quad~
\end{aligned}
\end{equation}
The quantities $W_{sc}$ and $S_{sc}$ have already been introduced in
(\ref{blocks}), where we studied their gauge transformation
properties. The new fields with ``(0)'' superscripts are invariant
under soft-collinear gauge transformations, but there is no longer a
soft-collinear background field in their transformation
law. Essentially, the leading power effect of the above field
redefinition is that the substitutions $A_s^\mu\to A_s^\mu+\frac12
n^\mu\,\bar n\cdot A_{sc}(x_+)$ and $A_c^\mu\to A_c^\mu+\frac12 \bar
n^\mu\,n\cdot A_{sc}(x_-)$ from the multipole expansion are
reversed. As a result, when the new fields are introduced in the SCET
Lagrangian the terms ${\cal L}_s$, ${\cal L}_c$, and ${\cal L}_{sc}$
retain their original form, while the {\em leading-order interaction
Lagrangian ${\cal L}_{\rm int}^{(0)}$ vanishes!} Residual interactions
between soft-collinear and soft or collinear fields start at
$O(\lambda^{1/2})$ \cite{Becher:2003qh}.  After the field redefinition
it is convenient to introduce the gauge-invariant building blocks
\cite{Hill:2002vw,Becher:2003qh}
\begin{equation}
\begin{aligned}
   S_s^{(0)\dagger}(x)\,q_s^{(0)}(x)
    &= W_{sc}^\dagger(x_+)\,S_s^\dagger(x)\,q_s(x) = \Q_s(x) \,, \\
   S_s^{(0)\dagger}(x)\,h^{(0)}(x)
    &= W_{sc}^\dagger(x_+)\,S_s^\dagger(x)\,h(x) = \H(x) \,, \\
   W_c^{(0)\dagger}(x)\,\xi^{(0)}(x)
    &= S_{sc}^\dagger(x_-)\,W_c^\dagger(x)\,\xi(x) = \X(x) \,,
\end{aligned}
\end{equation}
which are obviously identical to (\ref{blocks}) and invariant under
all three types of gauge transformations. 

The fact that interactions of soft-collinear fields with other fields
can be decoupled from the strong-interaction Lagrangian does not
necessarily imply that these fields can be ignored at leading order in
power counting. The question is whether the decoupling transformation
(\ref{another}) leaves external operators such as weak-interaction
currents invariant. The analysis of the previous sections indicates
that in some cases the soft-collinear exchange graphs contribute to
the calculation of the anomalous dimensions. Let us then study what
happens when the decoupling transformation is applied to the various
types of operators.

Under the transformation (\ref{another}), the heavy-to-collinear and
soft-to-collinear currents in (\ref{hc}) and (\ref{sc}) transform into
(setting $x_\perp=0$ for simplicity)
\begin{equation}
\begin{aligned}
   \big[ \bar\xi\,W_c \big](x_+)\,\Gamma\,\big[ S_s^\dagger\,h \big](x_-)
   &\to \bar\X(x_+)\,S_{sc}^\dagger(0)\,\Gamma\,W_{sc}(0)\,\H(x_-) \,, \\
   \big[ \bar\xi\,W_c \big](x_+)\,\Gamma\,
   \big[ S_s^\dagger\,q_s\big](x_-)
   &\to \bar\X(x_+)\,S_{sc}^\dagger(0)\,\Gamma\,W_{sc}(0)\,\Q_s(x_-) \,.
\end{aligned}
\end{equation}
We observe that the soft-collinear fields {\em do not} decouple from
these currents but rather form a light-like Wilson loop with a cusp at
$x=0$ \cite{Becher:2003kh}. The anomalous dimension of the combination
$(S_{sc}^\dagger\,W_{sc})$ is the universal cusp anomalous dimension
times a logarithm of the soft-collinear scale, see the first
expressions in the second lines in (\ref{eq:hc}) and
(\ref{eq:sc}). After adding the contributions from the soft and
collinear sectors, the dependence on the IR regulators drops
out. However, the coefficient of the logarithm of $v\cdot p_{c-}$ in
the heavy-collinear current and $p_{s+}\!\cdot p_{c-}$ in the
soft-collinear current is unchanged, since both the soft and the
collinear part are independent of these large scales. This
cancellation also explains why the anomalous dimensions of the
heavy-to-collinear and soft-to-collinear currents involve
$-\Gamma_{\rm cusp}$ and $-\frac12\Gamma_{\rm cusp}$, respectively:
\begin{equation}
\begin{aligned}
   \Gamma_{\rm cusp} \left[
   \ln\frac{2p_{s+}\!\cdot p_{c-}\,\mu^2}{(-p_s^2)(-p_c^2)}
   + \ln\frac{-p_s^2}{\mu^2} + \ln\frac{-p_c^2}{\mu^2} \right]
   &= - \Gamma_{\rm cusp}\,\ln\frac{\mu^2}{2p_{s+}\!\cdot p_{c-}}
    \,, \\
   \Gamma_{\rm cusp} \left[
   \ln\frac{2v\cdot p_{c-}\,\mu^2}{(-2v\cdot p_s)\,(-p_c^2)}
   + \ln\frac{-2v\cdot p_s}{\mu} + \ln\frac{-p_c^2}{\mu^2} \right]
   &= -\frac12\,\Gamma_{\rm cusp}\,\ln\frac{\mu^2}{(2v\cdot p_{c-})^2}
    \,.
\end{aligned}
\end{equation}
Similar arguments were used by Korchemsky in his analysis of the 
off-shell Sudakov form factor \cite{Korchemsky:1988hd}.

The effect of the field redefinition (\ref{another}) on the four-quark
operators is different. The color singlet-singlet operator {\em is
invariant}, namely (setting $x=0$ for simplicity)
\begin{equation}\label{singlet}
   Q_{(S)}(s,t)\to \bar\X(s\bar n)\,\Gamma_1\,\X(0)\,\,
   \bar\Q_s(tn)\,\Gamma_2\,\H(0) \,,
\end{equation}
since the additional soft-collinear Wilson lines come in pairs
$W_{sc}^\dagger\,W_{sc}=1$ and $S_{sc}^\dagger\,S_{sc}=1$. The color
octet-octet operator is however not invariant, since in that case the
cancellation cannot happen. Instead, one obtains the soft-collinear
object \cite{Becher:2003kh}
\begin{equation}
  \mbox{Tr}\big[ S_{sc}\,T_A\,S_{sc}^\dagger\,
  W_{sc}\,T_B\,W_{sc}^\dagger \big](0)\;.
\end{equation}
The presence of this functional in the octet-octet case explains why
soft-collinear modes give a non-zero contribution to the anomalous
dimension of the operator $Q_{(O)}$. However, since this operator does
not mix into the singlet-singlet operator $Q_{(S)}$, this effect does
not propagate into physical decay amplitudes (as hadronic matrix
elements of color-octet currents vanish). The decoupling of
soft-collinear fields from the color singlet-singlet operator implies
that, to all orders in perturbation theory, the anomalous dimension of
the four-quark operator $Q_{(S)}$ is the sum of the anomalous
dimensions of the two currents $\bar\X(s\bar n)\,\Gamma_1\,\X(0)$ and
$\bar\Q_s(tn)\,\Gamma_2\,\H(0)$, in accordance with the one-loop
result obtained in the previous section. This observation has
important implications for applications of SCET to proofs of QCD
factorization theorems.

\section{Reparameterization Invariance}
\label{sec:RPI}

In section \ref{sec:currents} we have briefly touched on
reparameterization invariance, on which we expand here. Operators in
SCET must be invariant under redefinitions of the light-cone basis
vectors $n$ and $\bar n$ that leave the scaling properties of fields
and momenta unchanged \cite{Hill:2002vw,Chay:2002vy,Manohar:2002fd}.
This property is referred to as reparameterization invariance, and has
proved to be a quite powerful tool. For example, it can be used to
derive constraints on the Wilson coefficients of SCET operators, often
relating the coefficients of some operators to those of
others. Reparameterization invariance is a consequence of the
invariance of QCD under Lorentz transformations, which is not explicit
(but still present) in SCET because of the introduction of the
light-cone vectors $n$ and $\bar n$.

Commonly one distinguishes between three classes of infinitesimal
transformations, corresponding to two different transverse boosts and
a longitudinal boost:
\begin{equation}
\begin{aligned}
   &\mbox{Type~I:} & \qquad &n^\mu\to n^\mu + \epsilon_\perp^\mu \,,
   & \qquad &\bar n^\mu \mbox{~invariant} & \qquad
    &\mbox{(with $\epsilon_\perp^\mu\sim\lambda$)} \\
   &\mbox{Type~II:} & &\bar n^\mu\to \bar n^\mu + e_\perp^\mu \,, &
    &n^\mu \mbox{~invariant} &
    &\mbox{(with $e_\perp^\mu\sim 1$)} \\
   &\mbox{Type~III:} & &n^\mu\to n^\mu/\alpha \,, &
    &\bar n^\mu\to \alpha\bar n^\mu &
    &\mbox{(with $\alpha\sim 1$)}
\end{aligned}
\end{equation}
The scaling properties for the parameters are given in parenthesis.
As an example, note that the collinear Wilson line $W_c$ is invariant
under type~I and type~III transformations. To derive the
transformation property under type~II, note that by definition
$\left[\bar n\cdot D_c\,W_c \right]=0$. The variation implies that
with $\delta\,\bar n\cdot D_c = e_\perp\cdot D_{c\perp}$ one can
derive the change $\delta W_c$ from $\delta\left[\bar n\cdot D_c\,W_c
\right]=0$ \cite{Manohar:2002fd}. It follows that the correct type~II
transformation behaviour is $W_c \to W_c - (\bar n\cdot
D_c)^{-1}\,e_\perp\cdot D_{c\perp}W_c$. The transformation laws for
soft Wilson lines are found using arguments along the same line.

It is then straightforward to show that, for example, the current
operators in (\ref{currentops}) are separately invariant under type~I
and type~II transformations to leading order in $\lambda$. In other
words, reparameterization invariance links these operators with
operators that appear at subleading order in $\lambda$. The only
non-trivial point in this analysis concerns the transverse collinear
derivative $D_{c\perp}^\mu$, which has non-vanishing variations at
leading order in $\lambda$ under both type~I and type~II
transformations,
\begin{equation}\label{eq:RPIgivesNslash}
   D_{c\perp}^\mu \stackrel{{\rm type~I}}{\to} D_{c\perp}^\mu
   - \frac{\epsilon_\perp^\mu}{2}\,\bar n\cdot D_c + O(\lambda^2) \,,
   \qquad
   D_{c\perp}^\mu \stackrel{{\rm type~II}}{\to} D_{c\perp}^\mu
   - \frac{n^\mu}{2}\,e_\perp\cdot D_{c\perp} + O(\lambda^2) \,.
\end{equation} 
Note that in both cases the object
$\nslash\,W^\dagger(i\Dslash_{c\perp} W)=\nslash\,\calAslash_{c\perp}$
is left invariant. For type~I transformations this follows from $(\bar
n\cdot D_c\,W)=0$, whereas for type~II transformations it follows
since $\nslash^2=0$. This is an important result, because without the
extra factor of $\nslash$, the second operator in (\ref{currentops})
would not be invariant under a type~II reparameterization.

The type~III transformations are the most powerful ones and have
non-trivial consequences. For example, consider the current operator
in (\ref{sc}). Type~III transformations dictates the dependence of the
Wilson coefficient on the plus-component of the soft momentum $p^s_+$
to all orders in perturbation theory. This provides valuable
information about the convergence of convolution integrals of
hard-scattering kernels (Wilson coefficients) with the $B$-meson
light-cone distribution amplitudes, which will be an important
ingredient to factorization proofs.  With a slight abuse of notation,
the soft-to-collinear current can be written in the following form, in
which we used translation invariance to extract the dependence on the
large energy $E$:
\begin{equation}
   \bar\psi_c(0)\,\Gamma\,\psi_s(0)
   \to \int d t\,\widetilde D(t,E,\mu)\,
    \big[ \bar\xi\,W_c \big](0)\,\Gamma\,
    \big[ S_s^\dagger\,q_s \big](t n) + O(\lambda)\;.
\end{equation}
Performing a variable substitution $t \to \alpha t$ and a subsequent
Type~III transformation (which also rescales the energy) one finds
that the current operator is invariant only if its Wilson coefficient
obeys the homogeneity relation $\widetilde D_i(t,E,\mu)=
\alpha\,\widetilde D_i(\alpha t,\alpha E,\mu)$.  Taking into account
the canonical dimension of the coefficient, it follows that
\begin{equation}
   \widetilde D(t,E,\mu)
   = \delta(t)\,d^{(1)}[\alpha_s(\mu)] + \frac{1}{t}\,
    d^{(2)}\big[\ln \left(\frac{\mu^2\,t}{E}\right),\alpha_s(\mu)\big] \,.
\end{equation}
where the coefficient functions $d_i^{(j)}$ are dimensionless. Since
the dependence of the Wilson coefficients on the renormalization scale
$\mu$ is logarithmic, we conclude that to all orders of perturbation
theory $\widetilde D(t,E,\mu)\sim 1/t$ modulo logarithms. Therefore
the above argument determines the behavior of the momentum-space
coefficients on the soft momentum $p^s_+$ to all orders in
perturbation theory.

\chapter{Structure Functions}\label{chap:strucfunc}

\section{The $B$-meson Light-Cone Distribution Amplitudes}
\label{sec:LCDA}

In section \ref{sec:fourquark} on four-quark operators we learned a
valuable lesson that touches the core of QCD-Factorization: Recall
that color singlet-singlet operators decouple from the soft-collinear
messenger sector, and therefore the collinear and soft sectors do not
interact with each other. Exclusive decay amplitudes are then
expressible in terms of convolution integrals and matrix elements of
operators of the form (\ref{singlet}). The product $\bar
Q_s(tn)\,\Gamma_2\,\H(0)$ has a non-vanishing overlap with the $B$
meson, while $\bar \X(s \bar n)\,\Gamma_1\,\X(0)$ overlaps with the
final light meson. The use of gauge-invariant building blocks provides
the necessary (finite length) Wilson line to make the bilocal
operators gauge invariant, since e.g.~$\bar \X(s \bar
n)\,\Gamma_1\,\X(0) = [\bar \xi W_c](s\bar n)\,\Gamma_1\,[W_c^\dagger
\xi](0) = \bar\xi(s\bar n)\, [s\bar n,0]\,\Gamma_1\,\xi(0)$. The
light-like separation of the fields explains why the corresponding
non-perturbative wave functions after projection onto meson states are
called "light-cone distribution amplitudes" (LCDAs). For the $B$
meson, two such functions, $\phi_+^B$ and $\phi_-^B$, arise for the
two-particle Fock state configurations. They are defined in the HQET
trace formalism as \cite{Grozin:1996pq}
\begin{eqnarray}\label{BLCDA}
   &&\frac{1}{\sqrt{M_B}}\,
    \langle\,0\,|\,\bar \Q_s(t n)\,\Gamma\, \H(0)\,|\bar B(v)\rangle \\
   &=& - \frac{iF(\mu)}{2}\, \int d\omega \; e^{-i\omega t} \;
    \mbox{tr}\bigg[ \left( \phi_+^B(\omega,\mu)
    - \frac{\nslash}{2}\,
    \Big[ \phi_-^B(\omega,\mu) - \phi_+^B(\omega,\mu)
    \Big] \right) \Gamma\,\frac{1+\vslash}{2}\,\gamma_5 \bigg] \;,
    \nonumber
\end{eqnarray}
where $F(\mu)$ denotes the asymptotic value of $\sqrt{M_B}\,f_B$ in
the heavy-quark limit. ($M_B$ is the $B$-meson mass and $f_B$ its
decay constant.) The variable $\omega$ can be interpreted as the plus
component of the spectator momentum. Often times it suffices to only
consider these two-particle LCDAs, since currents with an extra soft
gluon are power suppressed. However, they become also important if the
quantity of interest is itself power suppressed, which is the case for
e.g. $B \to$ {\em light meson} form factors at large recoil.  It will
then be necessary to also define wave functions containing an
additional soft gluon or a derivative. The latter can, however, be
eliminated using the equations of motion.  Let us define the
three-particle LCDAs of the $B$ meson in analogy with (\ref{BLCDA}) as
\cite{in:prep}
\begin{equation}
\begin{aligned}
{\langle0|\bar{\Q_s}(t\bar n)\,\A_{s\bot}^{\mu}(s \bar n)\,\Gamma\,\H(0)
|B\rangle} = & -\frac{i F(\mu)\sqrt{M_B}}{2} \\ 
& {\rm tr} \left[\Gamma\frac{1+\vslash}{2}
\gamma_{5}\gamma_{\perp}^{\mu}\left(\frac{1}{2}{\tilde A}_1(t,s) 
+\frac{\nslash}{2}{\tilde A}_2(t,s)\right)\right] \; .
\label{B3particle}
\end{aligned}
\end{equation}
Since $n\cdot \A_s=0$, the full light-cone decomposition of the gluon
field $\A_s$ reads $\calAslash_s=\calAslash_{s,\perp} + \nslash
\,v\cdot \A_s$ and there are in principle two more wave functions one
can define, namely
\begin{equation}
\begin{aligned}
{\langle0|\bar{\Q_s}(t\bar n)\,v\cdot\A_{s}(s \bar n)\,
  \Gamma\,\H(0)|B\rangle} =& -\frac{i F(\mu)\sqrt{M_B}}{2} \\
&  {\rm tr} \left[\Gamma\frac{1+\vslash}{2}\gamma_{5}
  \left({\tilde A}_3(t,s)+\frac{\nslash}{2}{\tilde A}_4(t,s)\right)
  \right] \; .
\end{aligned}
\end{equation}
The three-particle LCDAs $A_1(\omega_1,\omega_2), \ldots,
A_4(\omega_1,\omega_2)$ are defined as the Fourier transforms of
$A_1(t,s),\ldots,A_4(t,s)$, in close analogy to (\ref{BLCDA}).

\paragraph{Equation of Motion constraints.} 
The above LCDAs are related through the equation of motions. In
particular, the equation of motion for the light quark field
$i\Dslash\,q_s=0$, which in terms of collinear fields implies
$i\delslash\;\Q_s(x) = - \calAslash_s(x)\;\Q_s(x)$, allows us to express
$\phi_-^B$ entirely in terms of the leading LCDA $\phi_+^B$ and the
three-particle wave function $A_1$. It is straightforward to use
(\ref{BLCDA}) with $\Gamma = \gamma^\mu$ and act with a partial
derivative with respect to the position $z$ of the light quark
field. Since $\delslash = (\nbslash/2)\,n\cdot\partial +
(\nslash/2)\,\bar n\cdot\partial + \delslash_\perp$ is the partial
derivative in all directions, it is crucial to keep $z$ slightly off
the light cone, and set it to the final value $z=t n$ only {\em after}
the derivative has been performed. When $z$ is not light-like, the
Wilson line $[z,0]=P\exp[i\int_0^z\!dz\cdot A_s(z)]$ is no longer
given in terms of the product $S_s(z) S_s^\dagger(0)$, but rather as the
product of path-ordered exponentials
\begin{equation}
\left[z,0\right] = S_s(z) \, P\exp \left[ i \int_0^1 d\alpha \, 
                   z\cdot \A_s(\alpha z) \right] S_s^\dagger(0) \;.
\end{equation}
Expanding the Wilson line and taking a partial derivative will then
lead to the combination\footnote{This combination coincides with a
gluon field in Fock-Schwinger gauge $\Aslash_{\rm FSG}(z) = \int_0^1
d\alpha\, \alpha \, G^{\mu \nu}(\alpha z)\, n_\mu \gamma_\nu$.}
$\calAslash_s(z) - \int_0^1 d\alpha\, \calAslash_s(\alpha z)$ on the
left-hand side of (\ref{BLCDA}), while the wave functions on the
right-hand side carry additional dependence on $z^2 \ne 0$, which
needs to be taken into account when performing the derivative. The
result is \cite{in:prep}
\begin{eqnarray}
&& \omega\, \phi_-^B(\omega) 
   + \int_0^\omega \!d\omega^\prime\,
     \left[\phi_+^B(\omega^\prime)-\phi_-^B(\omega^\prime)\right] \nonumber \\
= && - \int_0^\infty \!d\omega_1\,d\omega_2\, 
     \delta(\omega-\omega_1-\omega_2)\, A_1(\omega_1,\omega_2) \\
&& + \int_0^1 \!d\alpha \int_0^\infty \!d\omega_1\,d\omega_2\, 
     \delta(\omega-\omega_1-\alpha \omega_2)\, A_1(\omega_1,\omega_2) 
     \;. \nonumber 
\end{eqnarray}
Finally we state the explicit expression that eliminates
$\phi_-^B(\omega)$ by solving the above equation. Taking a derivative
with respect to $\omega$ leads to the desired result
\begin{equation}
\begin{aligned}
\phi_-^B(\omega) = - \int_0^\omega\!d\omega^\prime \,
  \frac{\phi_+^B(\omega^\prime)}{\omega^\prime} 
&+ \int\limits_{\omega_1+\omega_2 \le \omega}\!\!d\omega_1\,d\omega_2\;
   \frac{\omega_2}{\omega_1(\omega_1+\omega_2)^2}\,A_1(\omega_1,\omega_2) \\
&+ \int_0^\omega\!d\omega_1\int_{\omega-\omega_1}^\infty \!d\omega_2\,
   \frac{\omega-\omega_1}{\omega\,\omega_1\omega_2}\,A_1(\omega_1,\omega_2)\;.
\end{aligned}
\end{equation}
The remaining LCDAs $A_2,\ldots,A_4$ are not relevant to the present
work.

As mentioned earlier, many applications require only one of the above
LCDAs at leading power, $\phi^B_+(\omega,\mu)$.  Physically, the
sensitivity to this universal, non-perturbative function arises in
processes which involve hard interactions with the soft spectator
quark in the $B$-meson, for example in $B \to \gamma \, l^-\bar\nu$,
$B \to K^*\gamma$, etc. \linebreak A generic factorizable decay
amplitude may be written as
\begin{equation}\label{conv}
  {\mathcal A}=\int_0^\infty\frac{d\omega}{\omega}\,T(E \omega,\mu)\,
   \phi_+^B(\omega,\mu) \,,
\end{equation}
where it is assumed that the $\mu$ dependence cancels between the
hard-scattering kernel $T$ and the LCDA. In writing the amplitude in
this way, a first step of scale separation was achieved, since the
kernel depends on the physics associated with large energy scales,
i.e. does not contain large logarithms for $\mu \sim
\sqrt{E\Lambda_{\rm QCD}}$, whereas the LCDA is a universal,
non-perturbative function which ``lives'' on low scales $\mu \sim
\Lambda_{\rm QCD}$. Since there is no one scale $\mu$ at which neither
of the two quantities contain large logarithms it is crucial to resum
those large (Sudakov) logarithms to gain full control over the
separation of physics at different scales. We thus have to derive and
solve the renormalization-group evolution for the LCDA or,
equivalently, the hard-scattering kernel
\cite{Bosch:2003fc,Lange:2003ff}.

\subsection{Renormalization-Group Evolution}
\label{sec:LCDA:RGEsolution}

\begin{figure}[t!]
\begin{center}
\epsfig{file=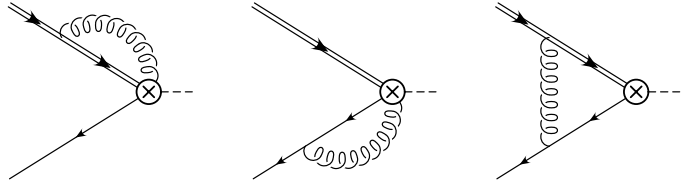, width=9cm}
\caption{One-loop diagrams for the calculation of the anomalous
dimension of the $B$-meson light-cone distribution amplitude.}
\label{fig:oneloop}
\end{center}
\end{figure}

Different operators that share the same quantum numbers can mix under
renormalization. It is therefore appropriate to write the relation
between bare operator $O_+^{bare}(\omega) = \int dt\,
e^{i\omega t}\, \bar \Q_s(t n)\,\Gamma\, \H(0)$ and renormalized one
as
\begin{equation}
O_+^{ren}(\omega,\mu) = 
\int d\omega^\prime\; Z_+(\omega, \omega^\prime,\mu) \; 
O_+^{bare}(\omega^\prime) \; .
\label{eq:op}
\end{equation}
In the case at hand the operator is, up to a Dirac trace, the product
of $F(\mu)$ and the LCDA $\phi_+^B(\omega,\mu)$. The
renormalization-group equation is an integro-differential equation in
which the anomalous dimension is convoluted with the LCDA
\cite{Lange:2003ff}:
\begin{equation}
\frac{d}{d\ln \mu} \phi_+^B(\omega,\mu) = 
- \int\limits_0^\infty d\omega^\prime \; \gamma_+(\omega, \omega^\prime,\mu) \;
  \phi_+^B(\omega^\prime,\mu)\;.
\label{eq:RGE}
\end{equation}
The anomalous dimension $\gamma_+$ is related to the renormalization
factor $Z_+$ through
\begin{equation}
   \gamma_+(\omega,\omega^\prime,\mu)
   = - \int d\tilde\omega\,
    \frac{dZ_+(\omega,\tilde\omega,\mu)}{d\ln\mu}\,
    Z_+^{-1}(\tilde\omega,\omega^\prime,\mu) 
    \;-\;\gamma_F(\alpha_s)\,\delta(\omega-\omega^\prime) \,.
\end{equation}
Here $\gamma_F$ is the universal anomalous dimension of local
heavy-light currents in HQET, which determines the scale dependence of
$F(\mu)$. We may separate the on- and off-diagonal terms of $\gamma_+$
and express it as
\begin{equation}
\gamma_+(\omega, \omega^\prime,\mu) = 
\Big[ \Gamma_{\rm cusp}(\alpha_s) \ln \frac{\mu}{\omega} 
     + \gamma(\alpha_s) \Big] \delta(\omega-\omega^\prime) 
     + \omega \; \Gamma(\omega, \omega^\prime, \alpha_s) 
\label{eq:anomdim}
\end{equation}
to all orders in perturbation theory. The cusp anomalous dimension
$\Gamma_{\rm cusp}(\alpha_s)$ appears as the coefficient of the $\ln
\mu$ term and has a geometric origin \cite{Korchemsky:wg}: Since an
effective heavy-quark field $h(0)$ can be expressed as the product of
a free field $h^{(0)}(0)$ and a Wilson line $S_v(0)$ extending from
$(-\infty)$ to $0$ along the $v$-direction, the matrix element in
(\ref{BLCDA}) contains $S_s(z)\,S_s^\dagger(0)\,S_v(0)$ which can be
combined to form a single Wilson line with a cusp at the origin, as
shown in Fig.~\ref{fig:2}(a). (Recall that soft Wilson lines $S_s$
extend along the $n$-direction.) The appearance of the single $\ln
\mu$ term distinguishes the anomalous dimension of the $B$-meson LCDA
from the familiar Brodsky-Lepage kernel \cite{Lepage:1980fj}. On the
one-loop level the $\ln \mu$ term appears in the calculation of the
first diagram in Fig. \ref{fig:oneloop}, where the gluon from the
Wilson line $S_s(z)\,S_s^\dagger(0)$ connects to the heavy-quark
Wilson line $S_v(0)$.

By evaluating the diagrams in Fig.~\ref{fig:oneloop}, we find the
one-loop expressions \cite{Lange:2003ff} (denoted by the superscript
$^{(1)}$) to be $\Gamma_{\rm cusp}^{(1)}=4$, $\gamma^{(1)}=-2$, and
\begin{equation}
\Gamma^{(1)}(\omega, \omega^\prime) = - \Gamma_{\rm cusp}^{(1)} \left[
  \frac{\theta(\omega^\prime-\omega)}{\omega^\prime (\omega^\prime-\omega)}
+ \frac{\theta(\omega-\omega^\prime)}{\omega (\omega-\omega^\prime)} 
\right]_+ \; ,
\label{eq:1-loop}
\end{equation}
where we have used that $\gamma_F^{(1)}=-3$ is the one-loop
coefficient of the anomalous dimension of heavy-to-light currents. The
subscript $+$ denotes the standard ``plus distribution'' which is defined as
\begin{equation} \label{plusdistribution}
\int d\omega^\prime \Big[ \ldots \Big]_+ f(\omega^\prime) 
= \int d\omega^\prime \Big[ \ldots \Big] \Big( f(\omega^\prime) -  
f(\omega) \Big)
\end{equation}
for any (test-) function $f(\omega^\prime)$. This ensures that $\int
d\omega^\prime \Gamma(\omega, \omega^\prime) = 0$.

The first key toward solving the RG equation (\ref{eq:RGE}) concerns
the off-diagonal term $\omega\, \Gamma(\omega, \omega^\prime,\mu)$ in
the anomalous dimension (\ref{eq:anomdim}). We observe that 
\begin{equation}
\int\limits_0^\infty d\omega^\prime \omega 
\Gamma(\omega, \omega^\prime,\alpha_s) (\omega^\prime)^a
= \omega^a \F(a,\alpha_s)
\label{eq:F}
\end{equation}
on dimensional grounds. The dimensionless function $\F$ can only
depend on the (in general complex) exponent $a$. We can use a
power-law ansatz
\begin{equation}
f(\omega,\mu,\mu_0,a(\mu)) = \left( \frac{\omega}{\mu_0} \right)^{a(\mu)} \; 
                              e^{U(a(\mu),\mu)}
\end{equation}
with an arbitrary mass parameter $\mu_0$. The function $f$ solves the
RG equation (\ref{eq:RGE}), if the exponent $a(\mu)$ and the
normalization $U(a(\mu),\mu)$ obey the differential equations
\begin{eqnarray}
\frac{d}{d\ln \mu} \, a(\mu) &=& \Gamma_{\rm cusp}(\alpha_s) \; , \\
\frac{d\, U(a(\mu),\mu)}{d\ln \mu} \, &=& -\gamma(\alpha_s) 
  - \F(a(\mu),\alpha_s) - \ln \frac{\mu}{\mu_0} \, 
  \Gamma_{\rm cusp}(\alpha_s) \; . \nonumber
\end{eqnarray}
The first equation can be immediately integrated and yields $a(\mu) =
\eta + g(\mu,\mu_0)$ with initial value $\eta = a(\mu_0)$ and
\begin{equation}
g(\mu,\mu_0) = \int\limits_{\alpha_s(\mu_0)}^{\alpha_s(\mu)} 
         \frac{d\alpha}{\beta(\alpha)} \; \Gamma_{\rm cusp}(\alpha) \; .
\label{eq:g}
\end{equation}
With this solution at hand, the second equation integrates to
\begin{equation}
U(a(\mu),\mu) = - \int\limits_{\alpha_s(\mu_0)}^{\alpha_s(\mu)} 
         \frac{d\alpha}{\beta(\alpha)} \; \Big[ 
         \gamma(\alpha) + g_\mu(\alpha) 
	 + \F(\eta+ g_0(\alpha), \alpha) \Big] \; ,
\label{eq:U}
\end{equation}
where $g_\mu(\alpha)=g(\mu,\mu_\alpha)$, $g_0(\alpha)=g(\mu_\alpha,
\mu_0)$, and $\mu_\alpha$ is defined such that $\alpha_s(\mu_\alpha) =
\alpha$. Note that $g(\mu_0,\mu_0)=0$ and $U(\eta,\mu_0)=0$ in this
construction.

The second key to the solution concerns the initial condition
$\phi_+^B(\omega, \mu_0)$. Defining the Fourier transform
$\varphi_0(t)$ of the LCDA at scale $\mu_0$ with respect to $\ln
(\omega/\mu_0)$ allows us to express the $\omega$ dependence in the
desired power-law form
\begin{equation}
\phi_+^B(\omega, \mu_0) = \frac{1}{2\pi} \int\limits_{-\infty}^\infty dt \; 
             \varphi_0(t) \left( \frac{\omega}{\mu_0} \right)^{it} \; .
\end{equation}
We therefore obtain an exact analytic expression for the solution of
the RG equation (\ref{eq:RGE}) as the single integral \cite{Lange:2003ff}
\begin{equation}
\phi_+^B(\omega, \mu) = \frac{1}{2\pi} \int\limits_{-\infty}^\infty dt \; 
             \varphi_0(t) \; f(\omega, \mu, \mu_0, \eta=it) \; .
\label{eq:sol}
\end{equation}

\subsection{Asymptotic behaviour}
\label{sec:asymptotic}

The solution (\ref{eq:sol}) enables us to extract the asymptotic
behaviour of the LCDA $\phi_+^B(\omega,\mu)$ as $\omega \to 0$ and
$\omega \to \infty$ by deforming the integration contour in the
complex $t$ plane. We hence need to study the analytic structure of
the integrand $\varphi_0(t) \; f(\omega, \mu, \mu_0, it) \sim
\omega^{it + g(\mu,\mu_0)}$. If $\omega$ is very small we can deform
the contour into the lower half plane, and then the position of the
nearest pole to the real axis determines the $\omega$ dependence of
$\phi_+^B(\omega,\mu)$. Similarly the nearest pole in the upper half
plane dominates for very large $\omega$.

\begin{figure}[b!]
\begin{center}
$\left.\right.$\\[10mm]
{\bf (a)} \epsfig{file=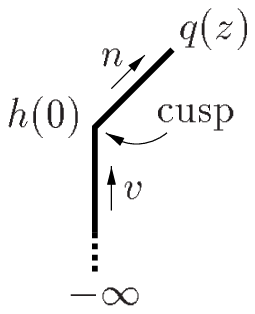, height=30mm} \hspace{30mm} 
{\bf (b)} \epsfig{file=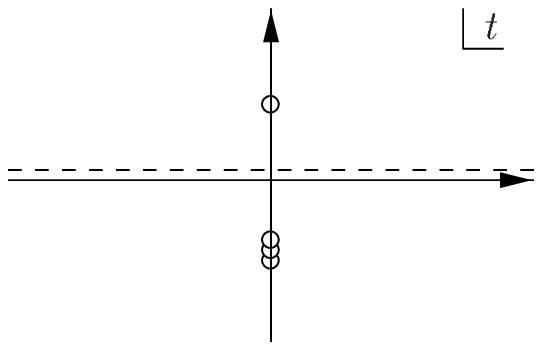, height=30mm}
\caption[Left: The cusp in the Wilson line. Right: Poles in the
complex plane]{{\label{fig:2}
\bf (a)} Left: The cusp in the Wilson line 
$S_s(z)S_s^\dagger(0)S_v(0)$.  {\bf (b)} Right: Poles of the function
$f(\ldots,it)$ in the complex $t$ plane.  The upper pole remains
stationary under renormalization flow, whereas the position of the
lower pole moves toward the real axis for increasing $\mu$. }
\end{center}
\end{figure}

Let us study the analytic structure of $f(\omega, \mu, \mu_0, it)$ at
leading order in RG-improved perturbation theory. Using the one-loop
expressions (\ref{eq:1-loop}) and the definition (\ref{eq:F}) we find
\pagebreak
\begin{equation}\label{eq:1-loop-F}
\F^{(1)}(a) = \Gamma_{\rm cusp}^{(1)} \left[ \Psi(1+a) + \Psi(1-a) + 
                          2 \gamma_E \right] \; .
\end{equation}
$\Psi$ and $\gamma_E$ denote the logarithmic derivative of the
Euler-Gamma function and the Euler-constant, respectively. Plugging
this result into eq. (\ref{eq:U}) with $\eta = it$ one obtains (using
the short-hand notation $g \equiv g(\mu,\mu_0)$)
\begin{equation}
e^{U(it+g,\mu)} \propto 
\frac{\Gamma(1+it)\; \Gamma(1-it-g)}{\Gamma(1-it)\; \Gamma(1+it+g)} \; .
\end{equation}
The function $f(\omega, \mu, \mu_0, it)$ has poles along the imaginary
axis, and the closest to the real axis are located at $t=i$ and
$t=-i(1-g)$. Using the one-loop expression (\ref{eq:g}) we observe
that the function $g$ vanishes at $\mu=\mu_0$ by definition and grows
monotonously as $\mu$ increases.  Therefore the position of the pole
in the lower complex plane approaches the real axis under
renormalization evolution, as illustrated in Fig. \ref{fig:2}(b).
These poles ``compete'' with the singularities arising from
$\varphi_0(t)$ for the nearest position to the real axis. Let us
assume that, for a given model of $\phi_+^B(\omega, \mu_0)$, the LCDA
grows like $\omega^\delta$ for small $\omega$ and falls off like
$\omega^{-\xi}$ for large $\omega$. The corresponding poles of the
function $\varphi_0(t)$ are then located at $t=-i\delta$ and
$t=i\xi$. We therefore obtain the asymptotic behaviour of the
renormalized LCDA as
\begin{equation}\label{eq:asympt}
   \phi_+^B(\omega,\mu)\sim \left\{
   \begin{array}{ll}
     \omega^{\min(1,\delta+g)} \,; & \hbox{for}\;\; \omega\to 0 \;, \\
     \omega^{-\min(1,\xi)+g} \,; & \hbox{for}\;\; \omega\to\infty \;.
   \end{array} \right.
\end{equation}
The two immediate observations are that, regardless of how small the
value of $\delta$ is, evolution effects will drive the small $\omega$
behaviour toward linear growth, and that the renormalized LCDA at a
scale $\mu>\mu_0$ will fall off slower than $1/\omega$ irrespective of
how fast it vanishes at $\mu=\mu_0$.

The emergence of a radiative tail after (even infinitesimally small)
evolution seems, at first sight, a very strange property of the LCDA,
because it implies that the normalization integral of
$\phi_+^B(\omega,\mu)$ is UV divergent. This can be understood as the
corresponding {\em local} operator $\bar \Q(0)\,\Gamma\,\H(0)$
requires an additional subtraction when renormalized
\cite{Grozin:1996pq}. However, this is not an obstacle for practical
applications, since only $\phi_+^B(\omega,\mu)/\omega$ modulo
logarithms appears. An integral over this function remains UV finite as
long as $g(\mu,\mu_0)<1$, at which point the pole at $t=-i(1-g)$
reaches the real axis and the formalism presented above breaks down.

It is evident from (\ref{eq:asympt}) that evolution effects mix
different moments of the LCDA. For example, the first inverse moment
of $\phi_+^B(\omega, \mu)$ defines a parameter 
\begin{equation}
\frac{1}{\lambda_B(\mu)} = \int\limits_0^\infty \frac{d\omega}{\omega} \; 
                 \phi_+^B(\omega,\mu) \; ,
\end{equation}
which is connected to a fractional inverse moment of order
$1-g(\mu,\mu^\prime)$ at a different scale $\mu^\prime$. This makes it
impossible to calculate the scale dependence of $\lambda_B(\mu)$ in
perturbation theory without knowledge of functional form of the LCDA.

\section{The Shape Function}
\label{sec:SF}

The shape function is a non-perturbative structure function that
encodes the Fermi motion of the heavy quark inside the $B$
meson. While the LCDA discussed in the last section has a rough
interpretation as a probability distribution for the plus component of
the spectator-quark momentum (the evolution effects do not quite fit
into this interpretation), the shape function describes the plus
component of the residual momentum of the heavy quark. It enters the
calculation of inclusive $B \to$ {\em light particles} decays such as
$\bar B\to X_u\,l^-\bar\nu$ and $\bar B\to X_s\,\gamma$, and is defined as the
forward matrix element of the bilocal heavy-to-heavy current
\begin{equation}\label{SFdefinition}
   \frac{\langle\bar B(v)|\,\bar h\,\Gamma\,\delta(\omega-in\cdot D)\,
   h\,|\bar B(v)\rangle}{2M_B}
   = S(\omega)\,\frac12\,\mbox{tr}\left( \Gamma\,\frac{1+\vslash}{2}
   \right) \;.
\end{equation}
The soft function $S(\omega)$ coincides with the shape function
$f(k_+)$ introduced in \cite{Neubert:1993ch}. Only one function arises
in the HQET trace formalism, because $v$ is the only vector available
by external kinematics. The shape function $S(\omega)$ has support for
$\omega \in\; ]\!-\infty, \bar\Lambda ]$, which can be understood as
follows:

The $B$-meson momentum can be decomposed into $P_B = m_b v+k$, where
$m_b$ is the $b$-quark mass, and $k$ is the dynamical ``residual''
momentum. For the sake of the argument, the variable $\omega$ can be
thought of as the plus component $n\cdot k$. Let us work in the rest
frame and choose the coordinate system such that the three-momentum
$\vec k$ points in the $+z$ direction. One may then parameterize
$k^\mu=\omega\,v^\mu+|\vec k| n^\mu$, so that $\omega = n\cdot k$
holds. The calculation of
\begin{equation}
  M_B^2 = P_B^2 = (m_b+\omega)^2 + 2|\vec k|(m_b+\omega)
\end{equation}
leads to the endpoints of the allowed $\omega$ interval for the cases
$|\vec k|=0$ and $|\vec k|\to \infty$, which can be read off as
$\omega = (M_B - m_b)$ and $\omega \to - m_b$, respectively. In the
heavy-quark limit $m_b \to \infty$, one therefore finds the support
interval mentioned above, where $(M_B - m_b)
\stackrel{m_b\to\infty}{\longrightarrow} \bar\Lambda$ is used. The
quantity $\bar\Lambda \sim O(\Lambda_{\rm QCD})$ is an HQET parameter
and depends on the particular definition of the heavy-quark mass.

\subsection{Renormalization-Group Evolution}

\subsubsection{Renormalized shape function}
\label{RenShaFunc}

\begin{figure}
\begin{center}
\epsfig{file=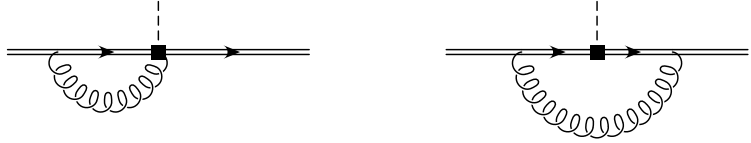,width=10cm}
\end{center}
\centerline{\parbox{14cm}{\caption[One-loop diagrams for the
shape-function operator.]{\label{fig:hqet} Radiative corrections to
the shape function. The bilocal HQET operator is denoted by the black
square. A mirror copy of the first graph is not shown.}}}
\end{figure}

According to (\ref{SFdefinition}), the shape function is defined in
terms of a hadronic matrix element in HQET and thus cannot be
calculated perturbatively. However, the renormalization properties of
this function can be studied using perturbation theory. To this end,
we evaluate the matrix element (\ref{SFdefinition}) in HQET using
external heavy-quark states with residual momentum $k$. For the time
being, $v\cdot k$ is kept non-zero to regularize infra-red
singularities. The relevant one-loop graphs are depicted in
Fig.~\ref{fig:hqet}. Adding the tree contribution, we obtain for the
matrix element of the bare shape-function operator $O(\omega)=\bar
h\,\Gamma\,\delta(\omega-in\cdot D)\,h$ (expressed in terms of
renormalized fields) \cite{Bosch:2004th}
\begin{eqnarray}\label{Sbare}
   S_{\rm bare}(\omega)
   &=& Z_h\,\delta(\omega-n\cdot k)
    - \frac{4C_F g_s^2}{(4\pi)^{2-\epsilon}}\,\Gamma(1+\epsilon)
    \nonumber\\
   &\times& \Bigg\{ \frac{1}{\epsilon}
    \int_0^\infty\!dl\,l^{-1-2\epsilon}\,\Big[ \delta(\omega-n\cdot k)
    - \delta(\omega-n\cdot k+l) \Big] \left( 1 + \frac{\delta}{l} 
    \right)^{-\epsilon} \nonumber\\
   &&\quad\mbox{}+ \theta(n\cdot k-\omega)\,(n\cdot k-\omega)^{-\epsilon}
    (n\cdot k-\omega+\delta)^{-1-\epsilon} \Bigg\} \,,
\end{eqnarray}
where $\delta=-2v\cdot k$, and
\begin{equation}
   Z_h = 1 + \frac{4C_F g_s^2}{(4\pi)^{2-\epsilon}}\,\Gamma(2\epsilon)\,
   \Gamma(1-\epsilon)\,\delta^{-2\epsilon}
\end{equation}
is the off-shell wave function renormalization constant of a heavy
quark in HQET. The next step is to extract the ultra-violet poles from
this result, which determine the anomalous dimension of the shape
function. The renormalized shape function is related to the bare
shape function through
\begin{equation}\label{Sren}
\begin{aligned}
   S_{\rm ren}(\omega) &= \int_{-\infty}^{\bar\Lambda}\!d\omega^\prime\,
    Z_S(\omega,\omega^\prime)\,S_{\rm bare}(\omega^\prime) \,, 
    \qquad \hbox{with}\\
   Z_S(\omega,\omega^\prime) &= \delta(\omega-\omega^\prime)
    + \frac{C_F\alpha_s}{4\pi}\,z_S(\omega,\omega^\prime) + \dots \,.
\end{aligned}
\end{equation}
The result for $Z_S$ following from (\ref{Sbare}) must be interpreted
as a distribution on test functions $F(\omega^\prime)$ with support on the
interval $-\infty<\omega^\prime\le\bar\Lambda$. We find
\begin{eqnarray}\label{zres}
   z_S(\omega,\omega^\prime)
   &=& \left( \frac{2}{\epsilon^2}
    + \frac{4}{\epsilon}\,\ln\frac{\mu}{\bar\Lambda-\omega}
    - \frac{2}{\epsilon} \right) \delta(\omega-\omega^\prime)
    - \frac{4}{\epsilon}\,\left( 
    \frac{\theta(\omega^\prime-\omega)}{\omega^\prime-\omega} \right)_{\!+}
    \nonumber\\
   &=& \left( \frac{2}{\epsilon^2} - \frac{2}{\epsilon} \right)
    \delta(\omega-\omega^\prime)
    - \frac{4}{\epsilon}\,\left( \frac{1}{\omega^\prime-\omega}
    \right)_{\!*}^{\![\mu]} .
\end{eqnarray}
Evidently, the renormalization factor $Z_S$ depends on the parameter
$\bar\Lambda$ setting the upper limit on the integration over
$\omega^\prime$ in (\ref{Sren}), which combines with the plus
distribution (see (\ref{plusdistribution})) to form a star
distribution in the variable $(\omega^\prime-\omega)$. The star
distributions are generalized plus distributions defined as
\cite{Bosch:2004th,DeFazio:1999sv}
\begin{equation}\label{star}
\begin{aligned}
   \int_{\le 0}^z\!dx\,F(x) 
   \left( \frac{1}{x} \right)_{\!*}^{\![u]}
   &= \int_0^z\!dx\,\frac{F(x)-F(0)}{x} + F(0)\,\ln\frac{z}{u} \,, \\
   \int_{\le 0}^z\!dx\,F(x)
    \left( \frac{\ln(x/u)}{x} \right)_{\!*}^{\![u]}
   &= \int_0^z\!dx\,\frac{F(x)-F(0)}{x}\,\ln\frac{x}{u} 
    + \frac{F(0)}{2}\,\ln^2\frac{z}{u} \,,
\end{aligned}
\end{equation}
where $F(x)$ is a smooth test function. For later purposes, we note
the useful rescaling identities
\begin{equation}\label{ids}
\begin{aligned}
   \lambda \left( \frac{1}{\lambda x} \right)_{\!*}^{\![u]}
   &= \left( \frac{1}{x} \right)_{\!*}^{\![u/\lambda]}
    = \left( \frac{1}{x} \right)_{\!*}^{\![u]} + \delta(x)\,\ln\lambda 
    \,, \\
   \lambda \left( \frac{\ln(\lambda x/u)}{\lambda x}
   \right)_{\!*}^{\![u]}
   &= \left( \frac{\ln(\lambda x/u)}{x} \right)_{\!*}^{\![u/\lambda]}
    = \left( \frac{\ln(x/u)}{x} \right)_{\!*}^{\![u]}
    + \left( \frac{1}{x} \right)_{\!*}^{\![u]} \ln\lambda 
    + \frac{\delta(x)}{2}\,\ln^2\lambda \,.
\end{aligned}
\end{equation}

We can now determine the renormalized shape function from
(\ref{Sren}).  The result must once again be interpreted as a
distribution, this time on test functions $F(\omega)$ integrated over
a {\em finite\/} interval $-\Lambda_{\rm
had}\le\omega\le\bar\Lambda$. (In practice, the value of $\Lambda_{\rm
had}$ is set by kinematics or by virtue of some experimental cut.) The
result is
\begin{eqnarray}
   S_{\rm parton}(\omega) &\hspace{-2mm} =& \hspace{-2mm} 
     \delta(\omega-n\cdot k)\,\left\{ 1
    - \frac{C_F\alpha_s}{\pi} \left[ \frac{\pi^2}{24}
    + L_2\!\left( \frac{-\delta}{\Lambda_{\rm had}+n\cdot k} \right)
    \right] \right\} \\
   &&\hspace{-2mm} - \frac{C_F\alpha_s}{\pi}\,\Bigg\{
    \left[ \frac{\theta(n\cdot k-\omega)}{n\cdot k-\omega} \left(
    \ln\frac{n\cdot k-\omega}{\mu} + \ln\frac{n\cdot k-\omega+\delta}{\mu}
    \right) \right]_+ \nonumber\\
   &&\hspace{-2mm} + \delta(n\cdot k-\omega)\,
    \ln^2\frac{\Lambda_{\rm had}+n\cdot k}{\mu}
    + \frac{\theta(n\cdot k-\omega)}{n\cdot k-\omega+\delta}
    + \delta(n\cdot k-\omega)\,\ln\frac{\delta}{\mu} \Bigg\} \,.\nonumber
\end{eqnarray}
While it was useful to keep the heavy quark off-shell in the
calculation of the ultra-violet renormalization factor, the limit
$\delta=-2v\cdot k\to 0$ can be taken in the result for the
renormalized shape functions without leading to infra-red
singularities. This gives
\begin{eqnarray}\label{Sonshell}
   S_{\rm parton}(\omega) &=& \delta(\omega-n\cdot k)\,\left( 1
    - \frac{C_F\alpha_s}{\pi}\,\frac{\pi^2}{24} \right) \\
   &&\mbox{}- \frac{C_F\alpha_s}{\pi} \left[
    2 \left( \frac{1}{n\cdot k-\omega} \ln\frac{n\cdot k-\omega}{\mu}
    \right)_{\!*}^{\![\mu]} 
    + \left( \frac{1}{n\cdot k-\omega} \right)_{\!*}^{\![\mu]}
    \right] ,\nonumber
\end{eqnarray}
where the star distributions must now be understood as distributions
in the variable $(n\cdot k-\omega)$. We stress that these results for
the renormalized shape function are obtained in the parton model and
can in no way provide a realistic prediction for the functional form
of $S(\omega)$. Only the dependence on the ultra-violet
renormalization scale $\mu$ can be trusted.

\subsubsection{Evolution of the shape function}

The next goal is to solve the integro-differential evolution equation
\begin{equation}\label{Sevol}
   \frac{d}{d\ln\mu}\,S(\omega,\mu)
   = - \int d\omega^\prime\,\gamma_S(\omega,\omega^\prime,\mu)\,
   S(\omega^\prime,\mu) \;.
\end{equation}
In the upcoming calculation of the differential decay rate in
inclusive $\bar B\to X_u\,l^-\bar\nu$ decays we will require the shape
function at an intermediate scale $\mu_i \sim \sqrt{m_b \Lambda_{\rm
QCD}}$. If the shape function was known at some lower scale $\mu_0$,
it would be necessary to evolve it according to (\ref{Sevol}).

At one-loop order, the anomalous dimension for the shape function is
twice the coefficient of the $1/\epsilon$ pole in the renormalization
factor $Z_S$. From (\ref{zres}), we obtain
\begin{equation}\label{ourgamma}
   \gamma_S(\omega,\omega^\prime,\mu)
   = \frac{C_F\alpha_s}{\pi} \left[
   \left( 2\ln\frac{\mu}{\bar\Lambda-\omega} - 1 \right)
   \delta(\omega-\omega^\prime)
   - 2 \left( \frac{\theta(\omega^\prime-\omega)}{\omega^\prime-\omega}
   \right)_{\!+} \right] .
\end{equation}

The evolution equation (\ref{Sevol}) can be solved analytically using
the aforementioned general method presented in the previous section
for the case of the $B$-meson LCDA \cite{Lange:2003ff}. It is
convenient to change variables from $\omega$ to
$\hat\omega=\bar\Lambda-\omega\in[0,\infty[$ and define $\hat
S(\hat\omega)\equiv S(\bar\Lambda-\hat\omega)$. The
renormalization-group equation then reads
\begin{equation}\label{oureqn}
   \frac{d}{d\ln\mu}\,\hat S(\hat\omega,\mu)
   = - \int_0^\infty\!d\hat\omega^\prime\,
   \hat\gamma_S(\hat\omega,\hat\omega^\prime,\mu)\,
   \hat S(\hat\omega^\prime,\mu) \,,
\end{equation}
where the anomalous dimension can be written in the general form
\begin{equation}\label{hatGs}
   \hat\gamma_S(\hat\omega,\hat\omega^\prime,\mu)
   = 2 \left[ \,\Gamma_{\rm cusp}(\alpha_s)\,\ln\frac{\mu}{\hat\omega}
   + \gamma(\alpha_s) \right] \delta(\hat\omega-\hat\omega^\prime)
   + 2 {\cal G}(\hat\omega,\hat\omega^\prime,\alpha_s) \,.
\end{equation}
This is obviously analogous to the RGE (\ref{eq:RGE}) governing the
LCDA.  The logarithmic term containing the cusp anomalous dimension
can again be interpreted geometrically. Since the heavy-quark field
$h(x)$ in HQET can be represented as a Wilson line along the $v$
direction, the field $\H(x)$ entering the SCET formalism contains the
product of a light-like Wilson line (along $n$) and a time-like Wilson
line (along $v$), which form a cusp at point $x$. The shape function
contains two such cusps, each of which produces a contribution to the
anomalous dimension proportional to $\Gamma_{\rm cusp}\,\ln\mu$
\cite{Korchemsky:wg}. The one-loop coefficients of the remaining terms
in (\ref{hatGs}) are
\begin{equation}\label{G1l}
   \gamma_0 = - 2C_F \,, \qquad
   {\cal G}_0(\hat\omega,\hat\omega^\prime) = - \Gamma_0 \left(
    \frac{\theta(\hat\omega-\hat\omega^\prime)}
         {\hat\omega-\hat\omega^\prime} \right)_{\!+} .
\end{equation}
The general solution of (\ref{oureqn}) can be obtained (in close
analogy to our previous discussion) using the fact that on dimensional
grounds
\begin{equation}\label{Fdef}
   \int_0^\infty\!d\hat\omega^\prime\,
   {\cal G}(\hat\omega,\hat\omega^\prime,\alpha_s)\,(\hat\omega^\prime)^a
   \equiv \hat\omega^a\,\F(a,\alpha_s) \,,
\end{equation}
where the function $\F$ only depends on the exponent $a$ and the
coupling constant. We set $\F(0,\alpha_s)=0$ by definition, thereby
determining the split between the terms with $\gamma$ and ${\cal G}$
in (\ref{hatGs}). The integral on the left-hand side is convergent as
long as $\mbox{Re}\,a>-1$. At one-loop order we find from (\ref{G1l})
\begin{equation}\label{F1loop}
   \F(a,\alpha_s) = \Gamma_0\,\frac{\alpha_s}{4\pi}\,
   \Big[ \psi(1+a) + \gamma_E \Big] + \dots \,,
\end{equation}
where $\psi(z)$ is the logarithmic derivative of the Euler $\Gamma$
function. Relation (\ref{Fdef}) implies that the ansatz
\begin{equation}
   f(\hat\omega,\mu,\mu_0,\tau)
   = \left( \frac{\hat\omega}{\mu_0} \right)^{\tau+2g(\mu,\mu_0)}
   \exp U_S(\tau,\mu,\mu_0)
\end{equation}
with
\begin{equation}\label{myeqs}
\begin{aligned}
   g(\mu,\mu_0) &= \int\limits_{\alpha_s(\mu_0)}^{\alpha_s(\mu)}\!
    d\alpha\,\frac{\Gamma_{\rm cusp}(\alpha)}{\beta(\alpha)} \,, \\
   U_S(\tau,\mu,\mu_0)
   &= - 2\!\int\limits_{\alpha_s(\mu_0)}^{\alpha_s(\mu)}\!
    \frac{d\alpha}{\beta(\alpha)}\,\Big[
    g(\mu,\mu_\alpha) + \gamma(\alpha) + 
    \F\big(\tau + 2g(\mu_\alpha,\mu_0),\alpha \big) \Big] \,,
\end{aligned}
\end{equation}
provides a solution to the evolution equation (\ref{oureqn}) with
initial condition \\
$f(\hat\omega,\mu_0,\mu_0,\tau)=(\hat\omega/\mu_0)^\tau$ at some scale
$\mu_0$. Here $\mu_\alpha$ is defined such that
$\alpha_s(\mu_\alpha)=\alpha$, and $\tau$ can be an arbitrary complex
parameter. Note that $g(\mu,\mu_0)>0$ if $\mu>\mu_0$. We now assume
that the shape function $\hat S(\hat\omega,\mu_0)$ is given at the low
scale $\mu_0$ and define its Fourier transform with respect to
$\ln(\hat\omega/\mu_0)$ through
\begin{equation}\label{S0def}
   \hat S(\hat\omega,\mu_0)
   = \frac{1}{2\pi} \int_{-\infty}^\infty\!dt\,{\cal S}_0(t)
   \left( \frac{\hat\omega}{\mu_0} \right)^{it} \,.
\end{equation}
The exact result for the shape function at a different scale $\mu$ is
then given by
\begin{equation}\label{Sexact}
   \hat S(\hat\omega,\mu)
   = \frac{1}{2\pi} \int_{-\infty}^\infty\!dt\,{\cal S}_0(t)\,
   f(\hat\omega,\mu,\mu_0,it) \,. 
\end{equation}
With the help of this formula, it is straightforward to derive
explicit expressions for the evolution of the shape function from the
hadronic scale $\mu_0$ up to the intermediate scale $\mu_i$ at any
order in renormalization-group improved perturbation theory. Setting
$r_2=\alpha_s(\mu_0)/\alpha_s(\mu_i)>1$, we obtain for the evolution
function at leading order
\begin{equation}\label{NLOapprox}
   f(\hat\omega,\mu_i,\mu_0,it)
   = e^{V_S(\mu_i,\mu_0)} \left( \frac{\hat\omega}{\mu_0}
   \right)^{it+\frac{\Gamma_0}{\beta_0}\ln r_2}
   \frac{\Gamma(1+it)}{\Gamma(1+it+\frac{\Gamma_0}{\beta_0}\,\ln r_2)} \,,
\end{equation}
where
\begin{eqnarray}\label{VS}
   V_S(\mu_i,\mu_0) 
   &=& \frac{\Gamma_0}{2\beta_0^2} \Bigg[
    - \frac{4\pi}{\alpha_s(\mu_0)}\,(r_2 -1 - \ln r_2)
    + \frac{\beta_1}{2\beta_0}\,\ln^2 r_2 \nonumber \\
   && \mbox{} \qquad \;+ \left( \frac{\Gamma_1}{\Gamma_0} 
    - \frac{\beta_1}{\beta_0} \right) 
    \left( 1 - \frac{1}{r_2} - \ln r_2 \right) \Bigg] \\
   && - \frac{\Gamma_0}{\beta_0}\,\gamma_E\,\ln r_2
    - \frac{\gamma_0}{\beta_0}\,\ln r_2
    + O\Big[(r_2-1)\,\alpha_s(\mu)\Big] \,. \nonumber
\end{eqnarray}

This result is valid as long as $(\Gamma_0/\beta_0)\,\ln r_2<1$, which
is the case for all reasonable parameter values. Missing for a
resummation at next-to-leading order are the $O(\alpha_s)$
contributions to $V_S$, which vanish for $\mu\to\mu_0$. For all
practical purposes, given the intrinsic uncertainties in our knowledge
of the shape function, it will be sufficient to use the equations
given above. As mentioned earlier, we typically have $\mu_i\sim m_c$,
and so the running between $\mu_i$ and $\mu_0$ should be performed in
a theory with $n_f=3$ light quark flavors.  The relevant expansion
coefficients are then $\Gamma_0=\frac{16}{3}$,
$\Gamma_1=\frac{304}{3}-\frac{16}{3}\pi^2$, $\gamma_0=-\frac{8}{3}$,
and $\beta_0=9$, $\beta_1=64$.

The leading-order result presented above can be simplified
further. When (\ref{NLOapprox}) is inserted into (\ref{Sexact}), the
integration over $t$ can be performed analytically. Setting
$\eta=(\Gamma_0/\beta_0)\,\ln r_2>0$, the relevant integral is
\begin{equation}
   I = \frac{1}{2\pi} \int_{-\infty}^\infty\!dt\,{\cal S}_0(t)
    \left( \frac{\hat\omega}{\mu_0} \right)^{it}
    \frac{\Gamma(1+it)}{\Gamma(1+it+\eta)} \,,
\end{equation}
where
\begin{equation}
   {\cal S}_0(t) = \int_0^\infty\!\frac{d\hat\omega^\prime}{\hat\omega^\prime}
   \,    \hat S(\hat\omega^\prime,\mu_0)
    \left( \frac{\hat\omega^\prime}{\mu_0} \right)^{-it}
\end{equation}
is the Fourier transform of the shape function as defined in
(\ref{S0def}). The integrand of the $t$-integral has poles on the
positive imaginary axis located at $t=in$ with $n\ge 1$ an
integer. For $\hat\omega<\hat\omega^\prime$ the integration contour can be
closed in the lower half-plane avoiding all poles, hence yielding
zero. For $\hat\omega>\hat\omega^\prime$ we use the theorem of residues to
obtain
\begin{equation}
   I = \int_0^{\hat\omega}\!d\hat\omega^\prime\,R(\hat\omega,\hat\omega^\prime)
   \,   \hat S(\hat\omega^\prime,\mu_0) \,,
\end{equation}
where
\begin{equation}
   R(\hat\omega,\hat\omega^\prime)
   = \frac{1}{\hat\omega} \sum_{j=0}^\infty
    \left( - \frac{\hat\omega^\prime}{\hat\omega} \right)^j
    \frac{1}{\Gamma(j+1)\,\Gamma(\eta-j)}
   = \frac{1}{\Gamma(\eta)}\,
    \frac{1}{\hat\omega^{\eta}\,(\hat\omega-\hat\omega^\prime)^{1-\eta}} \,.
\end{equation}
Note that
$R(\hat\omega,\hat\omega^\prime)\to\delta(\hat\omega-\hat\omega^\prime)$
in the limit $\eta\to 0$, corresponding to $\mu_i\to\mu_0$, as it
should be.  

Our final result for the shape function at the
intermediate hard-collinear scale, valid at leading order in
renormalization-group improved perturbation theory, can now be written
in the simple form \cite{Bosch:2004th,Balzereit:1998yf,Grozin:1994ni}
(valid for $\mu_i>\mu_0$, so that $\eta>0$)
\begin{equation}\label{wow}
   \hat S(\hat\omega,\mu_i) = e^{V_S(\mu_i,\mu_0)}\,
   \frac{1}{\Gamma(\eta)} \int_0^{\hat\omega}\!d\hat\omega^\prime\,
   \frac{\hat S(\hat\omega^\prime,\mu_0)}
        {\mu_0^{\eta}\,(\hat\omega-\hat\omega^\prime)^{1-\eta}} \,,
\end{equation}
with $V_S$ as given in (\ref{VS}). 

From the above equation one can derive scaling relations for the
asymptotic behavior of the shape function for $\hat\omega\to 0$ and
$\hat\omega\to\infty$ (corresponding to $\omega\to\bar\Lambda$ and
$\omega\to-\infty$). If the function $\hat S(\hat\omega,\mu_0)$ at the
low scale $\mu_0$ vanishes proportional to $\hat\omega^\zeta$ near the
endpoint, the shape function at a higher scale $\mu_i>\mu$ vanishes
faster, proportional to $\hat\omega^{\zeta+\eta}$. Similarly, if $\hat
S(\hat\omega,\mu_0)$ falls off like $\hat\omega^{-\xi}$ for
$\hat\omega\to\infty$, the shape function renormalized at a higher
scale vanishes like $\hat\omega^{-\min(1,\xi)+\eta}$. Irrespective of
the initial behavior of the shape function, evolution effects generate
a radiative tail that falls off slower than $1/\hat\omega$. This fact
implies that the normalization integral of $\hat S(\hat\omega,\mu)$ as
well as all positive moments are ultra-violet divergent. The
field-theoretic reason is that the bilocal shape-function operator
contains ultra-violet singularities as $z_-\to 0$, which are not
subtracted in the renormalization of the shape function. The situation
is analogous to the case of the $B$-meson light-cone distribution
amplitude discussed in \cite{Lange:2003ff,Grozin:1996pq}, see
Section~\ref{sec:asymptotic}.  These divergences are never an obstacle
in practice. Convolution integrals with the shape function are always
cut off at some finite value of $\hat\omega$ by virtue of phase-space
or some experimental cut.

\subsection{Properties of the shape function}

Information about the shape function can be extracted from a study of
its moments using a local operator product expansion. Naively, one
would define the moments $M_N=\int_{-\infty}^{\bar\Lambda}
d\omega\,\omega^N S(\omega)$. Prior to our work in
\cite{Bosch:2004th}, the general understanding was that, at tree
level, $M_0=1$ is fixed by the normalization of the shape function,
\linebreak $M_1=0$ by virtue of the HQET equation of motion $iv\cdot
D\,h=0$ (which implicitly uses the pole-mass definition of $m_b$), and
$M_2 = -\lambda_1/3$ is given by the matrix element of the
kinetic-energy operator. However, it was not clear how to
systematically include radiative corrections to these relations. Apart
from the obvious dependence on HQET parameters such as $\bar\Lambda$
and $\lambda_1$, the moments $M_N$ must also depend on the
renormalization scale $\mu$. However, as was studied in the last
Section, quantum corrections result in the appearance of a radiative
tail, rendering all moments with $N\ge 0$ ultra-violet
divergent. Since in any physical process the integration over the
shape function is restricted to finite intervals, it suffices to
define the moments with an ultra-violet cutoff, such that
\begin{equation}\label{MNdef}
   M_N(\Lambda_{\rm UV},\mu) = \int_{-\Lambda_{\rm UV}}^{\bar\Lambda}\!
   d\omega\,\omega^N S(\omega,\mu) \,.
\end{equation}
In the following Subsections we will expand the finite quantities
$M_N(\Lambda_{\rm UV},\mu)$ in a local HQET operator product
expansion, assuming that $\Lambda_{\rm UV}$ is large compared to
$\Lambda_{\rm QCD}$. The only relevant operators up to dimension 5 are
of the form $\bar h\,(in\cdot D)^m\,h$ with $m=0,1,2$, whose matrix
elements are given as $1,0,-\lambda_1/3$, respectively
\cite{Neubert:1993ch,Mannel:1994pm,Kagan:1998ym,Bigi:2002qq}.

\subsubsection{Moments in the pole scheme}

We can write an expansion of the form
\begin{equation}\label{beauty}
   M_N(\Lambda_{\rm UV},\mu) = \Lambda_{\rm UV}^N
   \left\{ K_0^{(N)}(\Lambda_{\rm UV},\mu)
   + K_2^{(N)}(\Lambda_{\rm UV},\mu)\,\cdot
    \frac{(-\lambda_1)}{3\Lambda_{\rm UV}^2}
   + O\bigg[ \left( \frac{\Lambda_{\rm QCD}}{\Lambda_{\rm UV}} \right)^3
   \bigg] \right\} ,
\end{equation}
where the coefficients $K_i^{(N)}$ can be determined from a
perturbative matching calculation using on-shell external $b$-quark
states with residual momentum $k$. By evaluating the moments of the
renormalized shape function in (\ref{Sonshell}), one finds at one-loop
order \cite{Bosch:2004th}
\begin{eqnarray}\label{mom1}
   && \hspace{-6mm} M_N^{\rm parton}(\Lambda_{\rm UV},\mu)
   = (n\cdot k)^N\,\Bigg\{ 1 - \frac{C_F\alpha_s}{\pi} \left( 
    \ln^2\frac{\Lambda_{\rm UV}+n\cdot k}{\mu}
    + \ln\frac{\Lambda_{\rm UV}+n\cdot k}{\mu} + \frac{\pi^2}{24}
    \right) \nonumber\\
   && - \frac{C_F\alpha_s}{\pi}
    \sum_{j=1}^N \frac{1}{j} \left( 1
    + 2\ln\frac{\Lambda_{\rm UV}+n\cdot k}{\mu}
    - \sum_{l=j}^N \frac{2}{l} \right)
    \left[ \left( - \frac{\Lambda_{\rm UV}}{n\cdot k} \right)^j - 1
    \right] \Bigg\} \,. \qquad
\end{eqnarray}
We then expand this result in powers of $n\cdot k/\Lambda_{\rm UV}$.
Keeping the first three terms in the expansion yields
\pagebreak
\begin{equation}\label{mom2}
\begin{aligned}
   M_0^{\rm parton}(\Lambda_{\rm UV},\mu) &= 1 - \frac{C_F\alpha_s}{\pi}
    \left( \ln^2\frac{\Lambda_{\rm UV}}{\mu}
    + \ln\frac{\Lambda_{\rm UV}}{\mu} + \frac{\pi^2}{24} \right) \\
   &\hspace{-9mm} - \frac{C_F\alpha_s}{\pi} \left[
    \frac{n\cdot k}{\Lambda_{\rm UV}}
    \left( 2\ln\frac{\Lambda_{\rm UV}}{\mu} + 1 \right)
    + \frac{(n\cdot k)^2}{\Lambda_{\rm UV}^2}
    \left( - \ln\frac{\Lambda_{\rm UV}}{\mu} + \frac12 \right) + \dots
    \right] , \\
   M_1^{\rm parton}(\Lambda_{\rm UV},\mu) &= n\cdot k \left[ 1
    - \frac{C_F\alpha_s}{\pi}
    \left( \ln^2\frac{\Lambda_{\rm UV}}{\mu}
    - \ln\frac{\Lambda_{\rm UV}}{\mu} + \frac{\pi^2}{24} - 1 \right) 
    \right] \\
   &\quad\mbox{}- \frac{C_F\alpha_s}{\pi} \left[
    \Lambda_{\rm UV}
    \left( - 2\ln\frac{\Lambda_{\rm UV}}{\mu} + 1 \right)
    + \frac{(n\cdot k)^2}{\Lambda_{\rm UV}}\,
    2\ln\frac{\Lambda_{\rm UV}}{\mu} + \dots \right] , \\
   M_2^{\rm parton}(\Lambda_{\rm UV},\mu) &= (n\cdot k)^2 \left[ 1
    - \frac{C_F\alpha_s}{\pi}
    \left( \ln^2\frac{\Lambda_{\rm UV}}{\mu}
    - 2\ln\frac{\Lambda_{\rm UV}}{\mu} + \frac{\pi^2}{24} - \frac12
    \right) \right] \\
   &\quad\mbox{}- \frac{C_F\alpha_s}{\pi} \left[
    \Lambda_{\rm UV}^2\,\ln\frac{\Lambda_{\rm UV}}{\mu}
    + n\cdot k\,\Lambda_{\rm UV}
    \left( - 2\ln\frac{\Lambda_{\rm UV}}{\mu} + 3\right) + \dots
    \right] .
\end{aligned}
\end{equation}
On the other hand, we need to calculate the one-loop expressions for
the local operators $\bar h\,(in\cdot D)^m\,h$. The relevant diagrams
are the same as in Fig.~\ref{fig:hqet}, where now the black square
represents the local operators. The result is non-trivial when keeping
$v\cdot k$ non-zero to regularize infra-red singularities. However,
for the matching calculation we need the limit $v\cdot k \to 0$, which
can be taken without problems in the sum of all diagrams (but not for
each individual diagram). In that case, the one-loop corrections
vanish and the matrix elements reduce simply to their tree-level
values. It follows that in (\ref{mom2}) we must identify $(n\cdot
k)^n\to\langle\bar h\,(in\cdot D)^n h\rangle$. Substituting the
results for the HQET matrix elements given earlier, we obtain for the
Wilson coefficients of the first three moments
\begin{equation}\label{Dresults}
\begin{aligned}
   K_0^{(0)} &= 1 - \frac{C_F\alpha_s}{\pi}
    \left( \ln^2\frac{\Lambda_{\rm UV}}{\mu}
    + \ln\frac{\Lambda_{\rm UV}}{\mu} + \frac{\pi^2}{24} \right) ,
    \qquad
   K_2^{(0)} = \frac{C_F\alpha_s}{\pi} 
    \left( \ln\frac{\Lambda_{\rm UV}}{\mu} - \frac12 \right) , \\
   K_0^{(1)} &= \frac{C_F\alpha_s}{\pi} 
    \left( 2\ln\frac{\Lambda_{\rm UV}}{\mu} - 1 \right) , \hspace{3.5cm}
   K_2^{(1)} = -2\,\frac{C_F\alpha_s}{\pi}\, 
    \ln\frac{\Lambda_{\rm UV}}{\mu} \,, \\
   K_0^{(2)} &= - \frac{C_F\alpha_s}{\pi}\,
    \ln\frac{\Lambda_{\rm UV}}{\mu} \,, \hspace{6mm}
   K_2^{(2)} = 1 - \frac{C_F\alpha_s}{\pi}
    \left( \ln^2\frac{\Lambda_{\rm UV}}{\mu}
    - 2\ln\frac{\Lambda_{\rm UV}}{\mu} + \frac{\pi^2}{24} - \frac12 
    \right) .
\end{aligned}
\end{equation}
We observe that this result reduces to the naive moment relations
mentioned above in the tree-level approximation. The perturbative
quantum corrections can be trusted as long as the ratio $\Lambda_{\rm
UV}/\mu$ is of $O(1)$.

\subsubsection{The shape-function scheme}

We have stressed before that the value of the moments depend on the
definition of the heavy-quark mass. So far, our calculations have
assumed the definition of the heavy-quark mass as a pole mass,
$m_b^{\rm pole}$, which is implied by the HQET equation of motion
$iv\cdot D\,h=0$. Results such as (\ref{Dresults}) are valid in this
particular scheme. A more general choice is to allow for a residual
mass term $\delta m$ in HQET, such that $iv\cdot D\,h=\delta m\,h$
with $\delta m=O(\Lambda_{\rm QCD})$ \cite{Falk:1992fm}. It is well
known that the pole mass is an ill-defined concept, which suffers from
infra-red renormalon ambiguities \cite{Bigi:1994em,Beneke:1994sw}. The
parameter $\bar\Lambda_{\rm pole}=M_B-m_b^{\rm pole}$, which
determines the support of the shape function in the pole-mass scheme,
inherits the same ambiguities. It is therefore advantageous to
eliminate the pole mass in favor of some short-distance mass. For the
analysis of inclusive $B$-meson decays, a proper choice is to use a
so-called low-scale subtracted heavy-quark mass $m_b(\mu_f)$
\cite{Bigi:1996si}, which is obtained from the pole mass by removing a
long-distance contribution proportional to a subtraction scale
$\mu_f$, writing $m_b^{\rm pole} = m_b(\mu_f) + \mu_f\,\,g\Big(
\alpha_s(\mu), \frac{\mu_f}{\mu} \Big) \equiv m_b(\mu_f) + \delta m$.
As long as $m_b(\mu_f)$ is defined in a physical way, the resulting
perturbative expressions after elimination of the pole mass are
well-behaved and not plagued by renormalon ambiguities. Replacing the
pole mass by the physical mass shifts the values of $n\cdot k$ and
$\omega$ by an amount $\delta m$, since $n\cdot(m_b^{\rm pole}
v+k)=m_b(\mu_f)+(n\cdot k+\delta m)$, and because the covariant
derivative in the definition of the shape function in
(\ref{SFdefinition}) must be replaced by $in\cdot D-\delta m$
\cite{Falk:1992fm}.  At the same time, $\bar\Lambda_{\rm
pole}=\bar\Lambda(\mu_f)-\delta m$, where
$\bar\Lambda(\mu_f)=M_B-m_b(\mu_f)$ is a physical parameter. Note that
this leaves the parameter $\hat\omega=\bar\Lambda-\omega$ and hence
the shape function $\hat S(\hat\omega,\mu)$ invariant! This follows
since $\bar\Lambda_{\rm pole}-\omega_{\rm pole}=\bar\Lambda(\mu_f)%
-(\omega_{\rm pole}+\delta m)$, where $\omega_{\rm pole}$ denotes the
value in the pole-mass scheme used so far. We now {\em choose\/}
$\delta m$ such that the first moment $M_1$ vanishes, thereby defining
a low-scale subtracted heavy-quark mass to all orders in perturbation
theory (with $\mu_f=\Lambda_{\rm UV}$), called the ``shape-function
mass'' $m_b^{\rm SF}$ \cite{Bosch:2004th}. The shape-function mass can
be related to any other short-distance mass using perturbation theory.
To this end, one uses the fact that (\ref{mom2}) implies a relation to
the pole mass
\begin{equation}\label{mSFmpole}
   m_b^{\rm pole} = m_b^{\rm SF}(\mu_f,\mu) 
   + \mu_f\,\frac{C_F\alpha_s(\mu)}{\pi} \left[
   \left( 1 - 2\ln\frac{\mu_f}{\mu} \right) 
   + \frac23\,\frac{(-\lambda_1)}{\mu_f^2}\,\ln\frac{\mu_f}{\mu}
   + \dots \right] .
\end{equation}
Given the above expression, it is easy to relate also to the
potential-subtracted mass introduced in \cite{Beneke:1998rk} and to
the kinetic mass defined in \cite{Bigi:1997fj,Benson:2003kp}.
\begin{equation}\label{mSFmPS}
   m_b^{\rm SF}(\mu_f,\mu_f) = m_b^{\rm PS}(\mu_f)
   = m_b^{\rm kin}(\mu_f) + \mu_f\,\frac{C_F\alpha_s(\mu_f)}{3\pi}
\end{equation}

After the introduction of the shape-function mass, the coefficients
$K_n^{(1)}$ in (\ref{Dresults}) vanish by definition. At one-loop
order, the remaining coefficients of the zeroth and second moment
remain unchanged since $\delta m=O(\alpha_s)$. Proceeding in an
analogous way, we can use the second moment to define a physical
kinetic-energy parameter, commonly called $\mu_\pi^2$. This quantity
can be used to replace the HQET parameter $\lambda_1$, which like the
pole mass suffers from infra-red renormalon ambiguities
\cite{Martinelli:1995zw}. At one-loop order, we obtain
\cite{Bosch:2004th}
\begin{equation}\label{mupi2pole}
\begin{aligned}
   & \hspace{25mm} \frac{\mu_\pi^2(\Lambda_{\rm UV},\mu)}{3}
   \equiv \frac{M_2^{\rm phys}(\Lambda_{\rm UV},\mu)}
                 {M_0^{\rm phys}(\Lambda_{\rm UV},\mu)} \\
   = & - \frac{C_F\alpha_s(\mu)}{\pi}\,\Lambda_{\rm UV}^2\,
    \ln\frac{\Lambda_{\rm UV}}{\mu} 
    + \frac{(-\lambda_1)}{3} \left[ 1 + \frac{C_F\alpha_s(\mu)}{\pi}
    \left( 3\ln\frac{\Lambda_{\rm UV}}{\mu} + \frac12 \right) \right]
    + \dots \,. 
\end{aligned}
\end{equation}
This definition is similar to the running parameter $\mu_\pi^2$
defined in the kinetic scheme \cite{Bigi:1997fj,Benson:2003kp}. At
one-loop order, the two parameters are related by
\begin{equation}\label{mupi2kin}
   \mu_{\pi}^2(\mu_f,\mu_f) = - \mu_f^2\,\frac{C_F\alpha_s(\mu_f)}{\pi}
   + [\mu_{\pi}^2(\mu_f)]_{\rm kin}
   \left[ 1 + \frac{C_F\alpha_s(\mu_f)}{2\pi} \right] .
\end{equation}
Given a value for the kinetic energy in the shape-function scheme for
some choice of scales, we can solve (\ref{mSFmpole}) and
(\ref{mupi2pole}) to obtain values for $m_b^{\rm SF}$ and
$\mu_{\pi}^2$ at any scale, using the fact that $m_b^{\rm pole}$ and
$\lambda_1$ are scale independent.

\subsection{Moments of the scheme-independent function 
$\hat S(\hat\omega,\mu)$}

The variable $\hat\omega=\bar\Lambda-\omega$ and with it the function
$\hat S(\hat\omega,\mu) = S(\omega)$ are independent under
redefinition of the heavy-quark mass. It will be useful to rewrite the
moment relations derived above in terms of these quantities, defining
a new set of moments
\begin{equation}\label{hatMNdef}
   \hat M_N(\mu_f,\mu) = \int\limits_0^{\mu_f+\bar\Lambda(\mu_f,\mu)}\!\!
   d\hat\omega\,\hat\omega^N\,\hat S(\hat\omega,\mu) \,.
\end{equation}
This yields
\begin{eqnarray}\label{M0toM2}
   \hat M_0(\mu_f,\mu)
   &=& 1 - \frac{C_F\alpha_s(\mu)}{\pi}
    \left( \ln^2\frac{\mu_f}{\mu} + \ln\frac{\mu_f}{\mu}
   + \frac{\pi^2}{24} \right) \\
   && \mbox{} \hspace{7pt} + \frac{C_F\alpha_s(\mu)}{\pi}
    \left( \ln\frac{\mu_f}{\mu} - \frac12 \right)
    \frac{\mu_\pi^2(\mu_f,\mu)}{3\mu_f^2} + \dots , \nonumber\\
   \frac{\hat M_1(\mu_f,\mu)}{\hat M_0(\mu_f,\mu)}
   &=& \bar\Lambda(\mu_f,\mu) \,, \qquad \qquad
   \frac{\hat M_2(\mu_f,\mu)}{\hat M_0(\mu_f,\mu)}
    = \frac{\mu_\pi^2(\mu_f,\mu)}{3} + \bar\Lambda(\mu_f,\mu)^2 \,,
    \nonumber
\end{eqnarray}
where the parameters $\bar\Lambda(\mu_f,\mu)=M_B-m_b^{\rm
SF}(\mu_f,\mu)$ and $\mu_\pi^2(\mu_f,\mu)$ should be considered as
known physical quantities. The moments with $N\ge 1$ give simply a
restatement of the shape-function scheme definition. Another
interesting aspect is that the expression for $\hat M_0(\mu_f,\mu)$
can be used to extract the precise form of the asymptotic tail of the
shape function, since $\mu_f$ is considered to be much larger than
$\Lambda_{\rm QCD}$ in our calculation.

\subsubsection{Asymptotic behavior of the shape function}

Taking the derivative of the zeroth moment $\hat M_0$ in
(\ref{hatMNdef}) with respect to $\mu_f$, one obtains
\begin{equation}
   \hat S(\hat\omega,\mu) \Big|_{\hat\omega=\mu_f+\bar\Lambda(\mu_f,\mu)}
   = \left( 1 - \frac{dm_b^{\rm SF}(\mu_f,\mu)}{d\mu_f} \right)^{-1}
   \frac{d}{d\mu_f}\,\hat M_0(\mu_f,\mu) \,.
\end{equation}
From (\ref{M0toM2}) we find at one-loop order
\begin{equation}\label{Sasymp}
\begin{aligned}
   \hat S(\hat\omega,\mu) = - \frac{C_F\alpha_s(\mu)}{\pi}\,
   \frac{1}{\hat\omega-\bar\Lambda}
   \Bigg[ & \left( 2\ln\frac{\hat\omega-\bar\Lambda}{\mu} + 1 \right) \\
   & + \frac23\,\frac{\mu_\pi^2}{(\hat\omega-\bar\Lambda)^2}
   \left( \ln\frac{\hat\omega-\bar\Lambda}{\mu} - 1 \right)
   + \dots \Bigg] .
\end{aligned}
\end{equation}
The precise definitions of $\bar\Lambda$ and $\mu_\pi^2$ are not
specified at this order. (Note that the shape function cannot depend
on the value of the cutoff $\mu_f$.) Relation (\ref{Sasymp}) is a
model-independent result as long as $\hat\omega\gg\Lambda_{\rm
QCD}$. We stress the remarkable fact that this radiative tail of the
shape function is {\em negative}, in contrast with the naive
expectation based on a probabilistic interpretation of the shape
function as a momentum distribution function. The point is that the
definition of the renormalized shape function requires
scheme-dependent ultra-violet subtractions. From (\ref{Sasymp}) it
follows that the shape function must have a zero, which for
sufficiently large $\mu$ is located at a value
$\hat\omega_0\approx\bar\Lambda+\mu/\sqrt{e}$.

\addtocontents{toc}{\protect\contentsline{part}{Phenomenological Applications}{}}
\chapter{Inclusive Semileptonic Decays}\label{chap:inclusive}

Due to experimental cuts in the measurement of the inclusive $B\to
X_u\,l^-\bar\nu$ decays one is generally faced with a situation in
which the hadronic final state $X_u$ is constraint to have large energy
$E_H \sim M_B$, but only moderate invariant mass $s_H = m_X^2 \sim
\Lambda_{\rm QCD}\,M_B$. This kinematic region, called the ``shape-function
region'', is the dominant phase space in which the final state cannot
contain charmed hadrons. The calculation of the inclusive differential
decay rate uses the optical theorem and assumes quark-hadron
duality. This assumption is justified when integrating over large
portions of the phase space, corresponding to the summation over many
final hadronic states. To calculate the total decay rate,
i.e. including all of phase space, all kinematic quantities are
integrated over a domain of order $M_B$, and one can perform an
operator product expansion (OPE) to systematically compute power
corrections. However, as mentioned above, when it is necessary to
calculate the differential decay rate in the shape-function region,
the calculation becomes more difficult because of the presence of
three separated mass scales: the hard scale $M_B$, the hard-collinear
scale $\sqrt{\Lambda_{\rm QCD}\,M_B}$, and the soft scale
$\Lambda_{\rm QCD}$. To properly disentangle the physics associated
with these scales requires a sophisticated effective field-theory
machinery.

A systematic treatment consists of matching QCD onto SCET$_I$ in a
first step, in which hard quantum fluctuations are integrated out. The
degrees of freedom in SCET$_I$ are referred to as soft and
hard-collinear to indicate that the invariant final state momenta have
fluctuations of order the intermediate scale $\sqrt{\Lambda_{\rm
QCD}\,M_B}$, much larger than $\Lambda_{\rm QCD}$. It is therefore
possible to treat them perturbatively. The expansion parameter of
SCET$_I$ is hence defined as $\lambda = \Lambda_{\rm QCD}/E_H$.

In a second step, SCET$_I$ is matched onto HQET, and hard-collinear
modes are integrated out. The resulting expressions for inclusive
differential decay rates have the factorized form $d\Gamma\sim
H\,J\otimes S$ \cite{Korchemsky:1994jb}. The function $H$ contains the
hard corrections, the jet function $J$, which describes the properties
of the final-state hadronic jet, contains the hard-collinear effects,
and the shape function $S$ accounts for the internal soft dynamics in
the $B$ meson \cite{Neubert:1993ch,Bigi:1993ex}. The $\otimes$ symbol
implies a convolution over a light-cone momentum variable $\omega$
associated with the residual momentum of the $b$ quark inside the $B$
meson.

\section{Factorization theorem}
\label{sec:fact}

Using the optical theorem, the hadronic physics relevant to the
inclusive semileptonic decay $\bar B\to X_u\,l^-\bar\nu$ can be
related to a hadronic tensor $W^{\mu\nu}$ defined via the
discontinuity of the forward $B$-meson matrix element of a correlator
of two flavor-changing weak currents $J^\mu=\bar
u\gamma^\mu(1-\gamma_5) b$
\cite{Chay:1990da,Bigi:1992su,Manohar:1993qn}. We define
\begin{equation}\label{WandTdef}
   W^{\mu\nu} = \frac{1}{\pi}\,\mbox{Im}\,
    \frac{\langle\bar B(v)|\,T^{\mu\nu}\,|\bar B(v)\rangle}{2 M_B} \,,
    \qquad
   T^{\mu\nu} = i \int d^4x\,e^{iq\cdot x}\,\,
    \mbox{T}\,\{ J^{\dagger\mu}(0), J^\nu(x) \} \,.
\end{equation}
Here $v$ is the $B$-meson velocity and $q$ the momentum carried by the
lepton pair.  After the field redefinition $b(x)=e^{-im_b v\cdot
x}\,b^\prime(x)$, which is always the first step in the construction
of an effective heavy field, the phase factor in (\ref{WandTdef})
becomes $e^{i(q-m_b v)\cdot x}\equiv e^{-ip\cdot x}$, where $p=m_b
v-q$ corresponds to the momentum of the jet of light partons into
which the $b$-quark decays. We will assume that these partons can be
described by hard-collinear fields. This is justified in the
shape-function region, in which case the current correlator can be
expanded in non-local light-cone operators
\cite{Neubert:1993ch,Mannel:1994pm,Bigi:1993ex}. 

The hadronic tensor in (\ref{WandTdef}) factorizes at leading power
\cite{Bosch:2004th}, as can be shown in the following way:
The jet momentum (and likewise the momentum of the hadronic final
state) scales like $p^\mu\sim E(\lambda,1,\sqrt{\lambda})$. (For the
jet momentum, $p_\perp=0$ by choice of the coordinate system.) The jet
invariant mass, $p^2\sim E\Lambda_{\rm QCD}$, defines a hybrid,
intermediate short-distance scale. The appropriate effective field
theory for integrating out the short-distance fluctuations associated
with the hard scale $p_-$ is SCET$_{I}$, see Section~\ref{sec:scet1}
but with hard-collinear and soft degrees of freedom instead of
collinear and ultrasoft ones. Below a matching scale $\mu_h\sim m_b$,
the semileptonic current can be expanded as
\begin{equation}\label{current}
   \bar u(x)\gamma^\mu(1-\gamma_5) b^\prime(x)
   = \sum_{i=1}^3 \int ds\,\widetilde C_i(s)\,
   \bar\X_{hc}(x+s\bar n)\,\Gamma_i^\mu\,\H(x_-) + \dots \,,
\end{equation}
where the dots denote higher-order terms in the SCET expansion, which
can be neglected at leading power in $\Lambda_{\rm QCD}/m_b$. The
hard-collinear light-quark field
$\X_{hc}(x)=S_s^\dagger(x_-)\,W_{hc}^\dagger(x)\,\xi_{hc}(x)$ and the
soft heavy-quark field $\H(x_-)=S_s^\dagger(x_-)\,h(x_-)$ are SCET
building blocks that are invariant under a set of homogeneous soft and
hard-collinear gauge transformations
\cite{Beneke:2002ni,Hill:2002vw,Becher:2003qh}. These are the
ingredients of SCET$_I$ after the field redefinitions
(\ref{scet1-decoupling}) and (\ref{scet1-buildblocks}).  Because of
the reduced Dirac basis between collinear spinors in (\ref{current}),
we may choose any three independent Dirac structures that do not
vanish between the spinors. A convenient choice is
\begin{equation}
   \Gamma_1^\mu = \gamma^\mu(1-\gamma_5) \,, \qquad
   \Gamma_2^\mu = v^\mu(1+\gamma_5) \,, \qquad
   \Gamma_3^\mu = \frac{n^\mu}{n\cdot v}\,(1+\gamma_5) \,.
\end{equation}
The current correlator in (\ref{WandTdef}) then becomes
\begin{equation}\label{T0}
\begin{aligned}
   T^{\mu\nu} = i & \int\! d^4x\,e^{-ip\cdot x} 
   \sum_{i,j=1}^3 \int\! ds\,dt\,\widetilde C_j^*(t)\,\widetilde C_i(s)
   \,\,\times \\
    & \mbox{T} \left\{ \bar\H(0)\,\bar\Gamma_j^\mu\,\X_{hc}(t\bar n),
    \bar\X_{hc}(x+s\bar n)\,\Gamma_i^\nu\,\H(x_-) \right\} 
    + \dots \,.
\end{aligned}
\end{equation}
In a second step, the hard-collinear fluctuations associated with the
light-quark jet can be integrated out by matching SCET onto HQET at an
intermediate scale $\mu_i\sim\sqrt{m_b\Lambda_{\rm QCD}}$. At leading
order the SCET Lagrangian (when written in terms of the
gauge-invariant fields such as $\X_{hc}$) does not contain interactions
between hard-collinear and soft fields, due to the decoupling
transformation (\ref{scet1-decoupling}). Since the external $B$-meson
states only contain soft constituents, we can take the vacuum matrix
element over the hard-collinear fields, defining a jet function
\begin{equation}\label{Jdef}
   \langle\,\Omega|\,\mbox{T} \left\{ \X_{hc,k}(t\bar n),
     \bar\X_{hc,l}(x+s\bar n)
   \right\} |\Omega\rangle\equiv \delta_{kl}\,
   \widetilde{\cal J}(x+(s-t)\bar n) + \dots \,,
\end{equation}
where $k,l$ are color indices, and we have used translational
invariance to determine the dependence on the coordinate
vectors. Shifting the integration variable from $x$ to $z=x+(s-t)\bar
n$, with $z_-=x_-$, and introducing the Fourier-transformed Wilson
coefficient functions $C_i(\bar n\cdot p) = \int ds\,e^{is\bar n\cdot
p}\,\widetilde C_i(s)$, we then obtain
\begin{equation}\label{T1}
   T^{\mu\nu} = i \sum_{i,j=1}^3 C_j^*(\bar n\cdot p)\,C_i(\bar n\cdot p)
   \int d^4z\,e^{-ip\cdot z}\,\bar\H(0)\,\bar\Gamma_j^\mu\,
   \widetilde{\cal J}(z)\,\Gamma_i^\nu\,\H(z_-) + \dots \,.
\end{equation}
In the next step, we rewrite the bilocal heavy-quark operator as 
\cite{Neubert:1993ch}
\begin{eqnarray}\label{SFops}
   \bar\H(0)\,\Gamma\,\H(z_-)
   &=& (\bar h\,S_s)(0)\,\Gamma\,e^{z_-\cdot\partial_+}\,
    (S_s^\dagger\,h)(0) 
    = \bar h(0)\,\Gamma\,e^{z_-\cdot D_+}\,h(0) \nonumber\\
   &=& \int d\omega\,e^{-\frac{i}{2}\omega\bar n\cdot z}\,
    \bar h(0)\,\Gamma\,\delta(\omega-in\cdot D)\,h(0) \,,
\end{eqnarray}
where $\Gamma$ may be an arbitrary Dirac structure, and we have used
the property $in\cdot D\,S_s=S_s\,in\cdot\partial$ of the soft Wilson line
$S_s$. When this expression is used in (\ref{T1}), the resulting formula
for the correlator involves the Fourier transform of the jet function,
\begin{equation}
   \int d^4z\,e^{-ip\cdot z}\,\widetilde{\cal J}(z)
   = \pslash_-\,{\cal J}(p^2) \,,
\end{equation}
however with $p^\mu$ replaced by the combination $p_\omega^\mu\equiv
p^\mu+\frac{1}{2}\omega\bar n^\mu$. Intuitively, this happens as only
the plus component of the soft residual momentum of the $b$ quark adds
to the hard-collinear momentum $p$. Using that $p_{\omega-}=p_-$, we
now obtain
\begin{equation}\label{Tres}
   T^{\mu\nu} = i\sum_{i,j=1}^3 H_{ij}(\bar n\cdot p)
   \int d\omega\,{\cal J}(p_\omega^2)\,
   \bar h\,\bar\Gamma_i^\mu\,\pslash_-\Gamma_j^\nu\,
   \delta(\omega-in\cdot D)\,h + \dots \,,
\end{equation}
where $H_{ij}(\bar n\cdot p)=C_j^*(\bar n\cdot p)\,C_i(\bar n\cdot p)$ 
are called the hard functions.

We need the discontinuity of the jet function, $J(p^2) =
\frac{1}{\pi}\,\mbox{Im}\,[i{\cal J}(p^2)]$, in order to compute the
hadronic tensor.  The $B$-meson matrix element of the soft operator is
evaluated using the HQET trace formalism, which allows us to write
\cite{Neubert:1993mb}
\begin{equation}\label{Sdef}
   \frac{\langle\bar B(v)|\,\bar h\,\Gamma\,\delta(\omega-in\cdot D)\,
   h\,|\bar B(v)\rangle}{2M_B}
   = S(\omega)\,\frac12\,\mbox{tr}\left( \Gamma\,\frac{1+\vslash}{2}
   \right) + \dots 
\end{equation}
at leading power in the heavy-quark expansion. Clearly, the soft
function $S(\omega)$ is the shape function, which was studied in
Section~\ref{sec:SF}. This gives the factorization formula
\begin{equation}
\label{Wres}
   W^{\mu\nu} = \sum_{i,j=1}^3 H_{ij}(\bar n\cdot p)\,
   \mbox{tr}\left( \bar\Gamma_i^\mu\,\frac{\pslash_-}{2}\,\Gamma_j^\nu\,
   \frac{1+\vslash}{2} \right)
   \int d\omega\,J(p_\omega^2)\,S(\omega) + \dots \,.
\end{equation}

In the final expressions (\ref{Tres}) and (\ref{Wres}) the dependence
on the three scales $\bar n\cdot p\sim m_b$, $p_\omega^2\sim
m_b\Lambda_{\rm QCD}$ and $\omega\sim\Lambda_{\rm QCD}$ has been
factorized into the hard, jet, and shape functions, respectively. The
factorization formula (\ref{Wres}) was derived at tree level in
\cite{Neubert:1993ch,Bigi:1993ex}, and was conjectured to hold to all
orders in perturbation theory in \cite{Korchemsky:1994jb}. The
derivation presented above \cite{Bosch:2004th} is equivalent to an
all-order proof of this formula first presented in \cite{Bauer:2001yt}
(see also \cite{Bauer:2000ew}). The limits of integration in the
convolution integral are determined by the facts that the jet function
has support for $p_\omega^2\ge 0$, and the shape function has support
for $-\infty<\omega\le\bar\Lambda$. The argument $p_\omega^2$ of the
jet function can be rewritten in the ``partonic variable'' $\bar
n\cdot p$ and the ``hadronic variable'' $n\cdot P_H \sim
O(\Lambda_{\rm QCD})$:
\begin{equation}\label{pw2}
   p_\omega^2 = p^2 + \bar n\cdot p\,\omega
   = \bar n\cdot p\,(n\cdot P_H - (\bar\Lambda-\omega))
   \equiv \bar n\cdot p\,(n\cdot P_H - \hat\omega) \,,
\end{equation}
where $P_H=M_B\,v-q=p+\bar\Lambda v$ is the 4-momentum of the hadronic
final state, and the variable $\hat\omega=\bar\Lambda-\omega\ge 0$.
Finally, $n\cdot P_H=E_H-|\vec{P}_H|=s_H/2E_H+O(\Lambda_{\rm
QCD}^2/m_b)$ is related to the hadronic invariant mass and energy of
the final state. The usefulness of this variable has also been
emphasized in \cite{Balzereit:1998yf,Aglietti:2002md}. We shall see
below that expressing the convolution integral in terms of the new
variable $\hat\omega$ eliminates any spurious dependence of the decay
spectra on the $b$-quark pole mass. 

Using the fact that the Wilson coefficients $C_i$ are real and hence
$H_{ij}$ is symmetric in its indices, we find
\begin{eqnarray}
   \sum_{i,j=1}^3 H_{ij}\,\mbox{tr}\left( \bar\Gamma_i^\mu\,
   \frac{\pslash_-}{2}\,\Gamma_j^\nu\,\frac{1+\vslash}{2} \right)
   &=& 2 H_{11} \left( p_-^\mu v^\nu + p_-^\nu v^\mu
    - g^{\mu\nu}\,v\cdot p_-
    - i\epsilon^{\mu\nu\alpha\beta} p_{-\alpha} v_\beta \right) 
    \nonumber \\
   &&\hspace{-6.0cm}\mbox{}+ 2 H_{22}\,v\cdot p_-\,v^\mu v^\nu
    + 2 (H_{12}+H_{23})\,(p_-^\mu v^\nu + p_-^\nu v^\mu)
    + 2 (2H_{13}+H_{33})\,\frac{p_-^\mu p_-^\nu}{v\cdot p_-} \,,
\end{eqnarray}
which may be compared with the general Lorentz decomposition of the
hadronic tensor given in \cite{DeFazio:1999sv}:
\begin{eqnarray}\label{Wdecomp}
   W^{\mu\nu} &=& W_1 \left( p^\mu v^\nu + p^\nu v^\mu
    - g^{\mu\nu}\,v\cdot p
    - i\epsilon^{\mu\nu\alpha\beta} p_\alpha v_\beta \right)
    - W_2\,g^{\mu\nu}
    \nonumber\\
   &&\mbox{}+ W_3\,v^\mu v^\nu + W_4\,(p^\mu v^\nu + p^\nu v^\mu)
    + W_5\,p^\mu p^\nu
\end{eqnarray}
We see that the structure function $W_2$ is not generated at leading
order in the SCET expansion. Since only the Wilson coefficient $C_1$
is non-zero at tree-level, the structure function $W_1$ receives
leading-power contributions at tree level, whereas $W_4$ and $W_5$
receive leading-power contributions at $O(\alpha_s(m_b))$. The
function $W_3$ receives leading-power contributions only at
$O(\alpha_s^2(m_b))$, which is beyond the accuracy of a
next-to-leading order calculation.

\section{Matching calculations}
\label{sec:match}

We derive expressions for the perturbative functions $H_{ij}(\bar
n\cdot p)$ and $J(p_\omega^2)$, that enter the factorization formula
in (\ref{Wres}) at next-to-leading order in $\alpha_s$. To this end,
we match expressions for the hadronic tensor obtained in full QCD,
SCET, and HQET, using for simplicity on-shell external $b$-quark
states.

\subsection{Hard functions}

Perturbative expressions for the hadronic functions $W_i$ in the
decomposition (\ref{Wdecomp}) have been obtained in
\cite{DeFazio:1999sv} by evaluating one-loop Feynman graphs for the
current correlator $T^{\mu\nu}$ using on-shell external quark states
with residual momentum $k$ (satisfying $v\cdot k=0$) in full QCD. The
leading terms in the region of hard-collinear jet momenta are
\begin{equation}\label{Fulvia}
\begin{aligned}
   \frac12\,W_1 &= \delta(p_k^2) \left[ 1 - \frac{C_F\alpha_s}{4\pi}
    \left( 8\ln^2 y - 10\ln y + \frac{2\ln y}{1-y} + 4 L_2(1-y)
    + \frac{4\pi^2}{3} + 5 \right) \right] \\
   &\quad\mbox{}+ \frac{C_F\alpha_s}{4\pi} \left[ -4 
    \left( \frac{\ln(p_k^2/m_b^2)}{p_k^2} \right)_{\!*}^{\![m_b^2]}
    + (8\ln y-7) \left( \frac{1}{p_k^2} \right)_{\!*}^{\![m_b^2]}
    \right] + \dots \,, \\
   \frac12\,W_4 &= \delta(p_k^2)\,\frac{C_F\alpha_s}{4\pi}\,
    \frac{2}{1-y} \left( \frac{y\ln y}{1-y} + 1 \right) + \dots \,, \\
   \frac{m_b}{4}\,W_5 &= \delta(p_k^2)\,\frac{C_F\alpha_s}{4\pi}\,
    \frac{2}{1-y} \left( \frac{1-2y}{1-y}\,\ln y - 1 \right) + \dots \,, 
\end{aligned}
\end{equation}
whereas $W_2$ and $W_3$ do not receive leading-power contributions at
this order. Here $\alpha_s\equiv\alpha_s(\mu)$, $y=\bar n\cdot p/m_b$,
and $p_k^2=p^2+\bar n\cdot p\,n\cdot k$. The star distributions are
generalized plus distributions, which also play an important role in
the renormalization of the shape function. They have been introduced
in (\ref{star}) and obey the rescaling identities (\ref{ids}). In
order to find the hard functions $H_{ij}$, we need the discontinuity
of the current correlator (\ref{T0}) between on-shell heavy-quark
states in SCET, keeping $i,j$ fixed. The corresponding tree diagram
yields
\begin{equation}
   D^{(0)} = K\,\delta(p_k^2) \,, \qquad \mbox{with} \quad
   K = \bar u_b(v)\,\bar\Gamma_j^\mu\,\pslash_-\,\Gamma_i^\nu\,u_b(v) \,,
\end{equation}
where $u_b(v)$ are on-shell HQET spinors normalized to unity, and the
quantity $K$ corresponds to the Dirac trace in (\ref{Wres}). The
interpretation of this result in terms of hard, jet, and soft
functions is that, at tree level, $J^{(0)}(p_\omega^2)=
\delta(p_\omega^2)$ and $S_{\rm parton}^{(0)}(\omega)=
\delta(\omega-n\cdot k)$ in the free-quark decay picture. It follows
that the convolution integral $\int d\omega\,J(p_\omega^2)\,S(\omega)$
in (\ref{Wres}) produces $\delta(p_k^2)$, and comparison with
(\ref{Fulvia}) shows that $H_{11}^{(0)}=1$, while all other hard
functions vanish at tree level.

\begin{figure}
\begin{center}
\epsfig{file=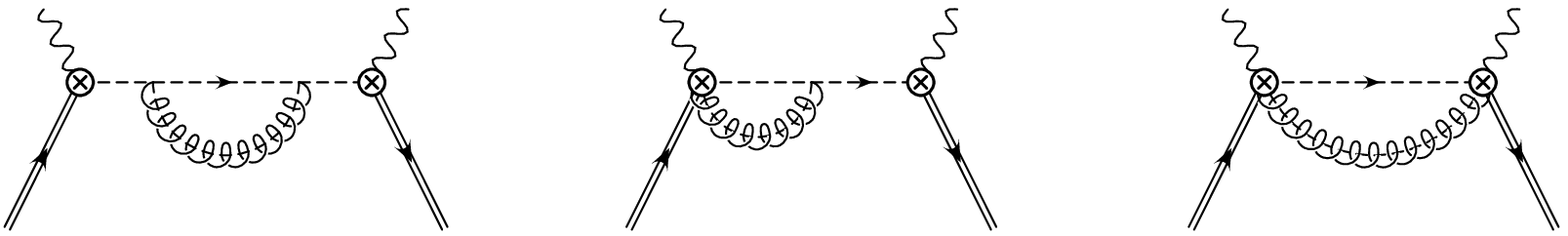,width=14cm}
\epsfig{file=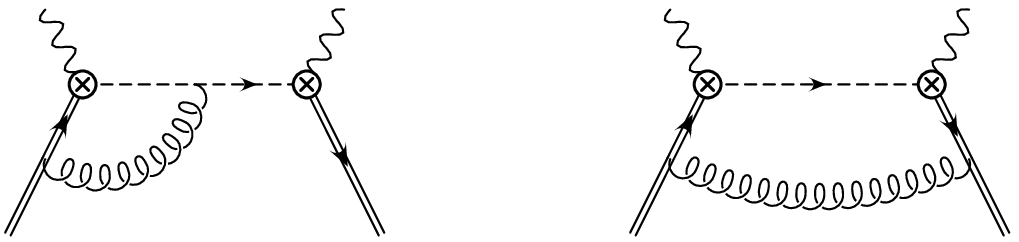,width=9cm}
\end{center}
\centerline{\parbox{14cm}{
\caption[One-loop SCET diagrams of the forward current correlator.]{
\label{fig:scet} One-loop diagrams contributing to the current
correlator in SCET. The effective current operators are denoted by
crossed circles, and hard-collinear propagators are drawn as dashed
lines. Mirror graphs obtained by exchanging the two currents are not
shown.}}}
\end{figure}

At one-loop order we need to evaluate the SCET diagrams shown in
Fig.~\ref{fig:scet}. The first three graphs contain hard-collinear
gluon exchanges, while the last two diagrams contain soft exchanges.
For the sum of all hard-collinear exchange graphs, we find
\cite{Bosch:2004th}
\begin{eqnarray}\label{hcloops}
   D_{hc}^{(1)} &=& K\,\frac{C_F\alpha_s}{4\pi} \Bigg[
   \left( \frac{4}{\epsilon^2} + \frac{3}{\epsilon} + 7 - \pi^2 \right)
   \delta(p_k^2) \\
   &&\quad\mbox{}+ 
   4 \left( \frac{\ln(p_k^2/\mu^2)}{p_k^2} \right)_{\!*}^{\![\mu^2]}
   - \left( \frac{4}{\epsilon} + 3 \right)
   \left( \frac{1}{p_k^2} \right)_{\!*}^{\![\mu^2]} \Bigg] .
\end{eqnarray}
The sum of the soft contributions is given by
\begin{eqnarray}\label{sloops}
   D_{s}^{(1)} 
   &=& K\,\frac{C_F\alpha_s}{4\pi} \Bigg[
    \left( - \frac{2}{\epsilon^2} - \frac{4}{\epsilon}\,L
    + \frac{2}{\epsilon} - 4L^2 + 4L - \frac{\pi^2}{6} \right)
    \delta(p_k^2) \nonumber\\
   &&\quad\mbox{}- 8 \left( \frac{\ln(p_k^2/\mu^2)}{p_k^2} 
    \right)_{\!*}^{\![\mu^2]}
    + \left( \frac{4}{\epsilon} + 8L - 4 \right)
   \left( \frac{1}{p_k^2} \right)_{\!*}^{\![\mu^2]} \Bigg] \,,
\end{eqnarray}
where $L=\ln(\bar n\cdot p/\mu)$. The $1/\epsilon$ poles in the sum of
the hard-collinear and soft contributions are subtracted by a
multiplicative renormalization factor $Z_J^2$ applied to the bare
current correlator in (\ref{T0}), where
\begin{equation}\label{ZJres}
   Z_J = 1 + \frac{C_F\alpha_s}{4\pi} \left( - \frac{1}{\epsilon^2}
    + \frac{2}{\epsilon}\,L - \frac{5}{2\epsilon} \right)
\end{equation}
is the (momentum-space) current renormalization constant in SCET
\cite{Bauer:2000yr}. Since the wave-function renormalization factor of
on-shell heavy quarks is equal to 1, the sum of (\ref{hcloops}) and
(\ref{sloops}) after subtraction of the pole terms is matched with the
results in (\ref{Fulvia}), so that the hard functions equal
\begin{equation}\label{Hres}
\begin{aligned}
   H_{11}(\bar n\cdot p) &= 1 + \frac{C_F\alpha_s}{4\pi}
    \left( -4 L^2 + 10 L - 4\ln y - \frac{2\ln y}{1-y} - 4 L_2(1-y)
    - \frac{\pi^2}{6} -12 \right) , \\
   H_{12}(\bar n\cdot p) &= \frac{C_F\alpha_s}{4\pi}\,
    \frac{2}{1-y} \left( \frac{y\ln y}{1-y} + 1 \right) , \\
   H_{13}(\bar n\cdot p) &= \frac{C_F\alpha_s}{4\pi}\,
    \frac{y}{1-y} \left( \frac{1-2y}{1-y}\,\ln y - 1 \right) . 
\end{aligned}
\end{equation}

\subsection{Jet function}

The SCET loop graphs in Fig.~\ref{fig:scet} determine the one-loop
contributions to the product of the jet function and the shape
function in (\ref{Wres}). We may write this product symbolically as
$J^{(1)}\otimes S^{(0)}+J^{(0)}\otimes S^{(1)}$, where the $\otimes$
symbol means a convolution in $\omega$. Although the realistic shape
function is a hadronic quantity, we may use its partonic version to
determine the perturbative jet function. Its one-loop contribution
$J^{(1)}$ must therefore be extracted from the results (\ref{hcloops})
and (\ref{sloops}). To this end, we need the renormalized shape
function at one-loop order in the parton model, which has been done in
Section~\ref{RenShaFunc}. It follows that the renormalized jet
function is given by the distribution \cite{Bosch:2004th}
\begin{equation}\label{Jres}
   J(p_\omega^2) = \delta(p_\omega^2)
   + \frac{C_F\alpha_s}{4\pi} \left[ (7-\pi^2)\,\delta(p_\omega^2)
   + 4 \left( \frac{\ln(p_\omega^2/\mu^2)}{p_\omega^2} 
   \right)_{\!*}^{\![\mu^2]}
   - 3 \left( \frac{1}{p_\omega^2} \right)_{\!*}^{\![\mu^2]} \right] .
\end{equation}
It will often be useful to separate the dependence on $\bar n\cdot p$
and $n\cdot P_H$ in \pagebreak this result by means of the
substitution $p_\omega^2=y\,\hat p_\omega^2$, where $\hat
p_\omega^2=m_b(n\cdot P_H-\hat\omega)$ according to (\ref{pw2}).
Using the identities (\ref{ids}), we find
\begin{eqnarray}\label{Jrescale}
   y\,J(p_\omega^2)\equiv\hat J(\hat p_\omega^2,y)
   &=& \delta(\hat p_\omega^2) + \frac{C_F\alpha_s}{4\pi} \Bigg[
    \big( 2\ln^2 y - 3\ln y + 7 - \pi^2 \big)\,\delta(\hat p_\omega^2)
    \nonumber\\
   &&+ 4 \left(
    \frac{\ln(\hat p_\omega^2/\mu^2)}{\hat p_\omega^2} 
    \right)_{\!*}^{\![\mu^2]}
    + (4\ln y - 3) 
    \left( \frac{1}{\hat p_\omega^2} \right)_{\!*}^{\![\mu^2]} \Bigg] \,.
\end{eqnarray}

\section{Sudakov resummation}

Equations (\ref{Hres}) and (\ref{Jres}) determine the short-distance
objects $H_{ij}$ and $J$ in the factorization formula (\ref{Wres}) at
one-loop order in perturbation theory. However, there is no common
choice of the renormalization scale $\mu$ that would eliminate all
large logarithms from these results. Likewise, the shape function,
being a hadronic matrix element, is naturally renormalized at some low
scale, whereas the short-distance objects contain physics at higher
scales. The problem of large logarithms arising from the presence of
disparate mass scales can be dealt with using renormalization-group
equations.  Proceeding in three steps, our strategy will be as
follows:

\begin{enumerate}
\item We match QCD onto SCET and extract matching conditions for the 
hard functions $H_{ij}$ at a high scale $\mu_h\sim m_b$. At that
scale, no large logarithms appear and so the hard functions can be
reliably computed using perturbation theory. We then evolve them down
to an intermediate hard-collinear scale
$\mu_i\sim\sqrt{m_b\Lambda_{\rm QCD}}$ by solving the
renormalization-group equation
\begin{equation}\label{Hevol}
   \frac{d}{d\ln\mu}\,H_{ij}(\bar n\cdot p,\mu)
   = 2\gamma_J(\bar n\cdot p,\mu)\,H_{ij}(\bar n\cdot p,\mu) \,,
\end{equation}
where $\gamma_J = \gamma_{\xi h}$ is the anomalous dimension of the
semileptonic heavy-to-collinear current in SCET, see the first line in
(\ref{andims}).

\item Starting from a model for the shape function $S(\omega,\mu_0)$ at 
some low scale $\mu_0=\mbox{few}\times\Lambda_{\rm QCD}$ large enough
to trust perturbation theory. Such a model could be provided by a
QCD-inspired approach such as QCD sum rules or lattice QCD, or it
could be tuned to experimental data. We then solve the
integro-differential evolution equation (\ref{Sevol}) to obtain the
shape function at the intermediate scale $\mu_i$.

\item Lastly we combine the results for the hard functions and for the 
shape function with the jet function $J$ in (\ref{Jres}) at the scale
$\mu_i$, where $J$ is free of large logarithms and so has a reliable
perturbative expansion. The dependence on the matching scales $\mu_h$
and $\mu_i$ cancels in the final result (to the order at which we are
working).
\end{enumerate}

We begin with the evolution of the hard functions. The anomalous
dimension $\gamma_J$ for the SCET current is twice the coefficient of
the $1/\epsilon$ pole in the renormalization factor $Z_J$ in
(\ref{ZJres}). More generally
\cite{Bauer:2000yr,Bosch:2003fc},
\begin{equation}
   \gamma_J(\bar n\cdot p,\mu)
   = - \Gamma_{\rm cusp}(\alpha_s)\,\ln\frac{\mu}{\bar n\cdot p}
    + \gamma^\prime(\alpha_s)
   = \frac{C_F\alpha_s}{\pi}
    \left( - \ln\frac{\mu}{\bar n\cdot p} - \frac54 \right)
    + \dots \,,
\end{equation}
where $\Gamma_{\rm cusp}=C_F\alpha_s/\pi+\dots$ is the universal cusp
anomalous dimension governing the ultra-violet singularities of Wilson
lines with light-like segments \cite{Korchemsky:wg}. The exact
solution to the evolution equation (\ref{Hevol}) can be written as
\cite{Bosch:2004th}
\begin{equation}
   H_{ij}(\bar n\cdot p,\mu_i) =  H_{ij}(\bar n\cdot p,\mu_h)\,
   \exp{U_H(\bar n\cdot p,\mu_h,\mu_i)} \,,
\end{equation}
where
\begin{equation}
   U_H(\bar n\cdot p,\mu_h,\mu_i)
   = 2 \int\limits_{\alpha_s(\mu_h)}^{\alpha_s(\mu_i)}\!\!
   \frac{d\alpha}{\beta(\alpha)}\,\Bigg[ \Gamma_{\rm cusp}(\alpha)\,
   \Bigg( \ln\frac{\bar n\cdot p}{\mu_h}
   - \int\limits_{\alpha_s(\mu_h)}^{\alpha}\!
   \frac{d\alpha^\prime}{\beta(\alpha^\prime)} \Bigg) + \gamma^\prime(\alpha)
   \Bigg] \,.
\end{equation}
Setting $r_1=\alpha_s(\mu_i)/\alpha_s(\mu_h)>1$, and expanding the \pagebreak
evolution function to $O(\alpha_s)$, we obtain
\begin{equation}
\begin{aligned}
   e^{U_H(\bar n\cdot p,\mu_h,\mu_i)}
   = & e^{V_H(\mu_h,\mu_i)} \left( \frac{\bar n\cdot p}{\mu_h}
   \right)^{-\frac{\Gamma_0}{\beta_0} \ln r_1} \\
   & \times
   \left[ 1 - \frac{\alpha_s(\mu_h)}{4\pi}\,\frac{\Gamma_0}{\beta_0}
   \left( \frac{\Gamma_1}{\Gamma_0} - \frac{\beta_1}{\beta_0} \right)
   (r_1-1)\,\ln\frac{\bar n\cdot p}{\mu_h} \right] ,
\end{aligned}
\end{equation}
where
\begin{eqnarray}\label{VHres}
   V_H(\mu_h,\mu_i) 
   &=& \frac{\Gamma_0}{2\beta_0^2} \Bigg[ \frac{4\pi}{\alpha_s(\mu_h)}
    \left( 1 - \frac{1}{r_1} - \ln r_1 \right)
    + \frac{\beta_1}{2\beta_0}\,\ln^2 r_1 \nonumber \\
   &&\qquad-\left( \frac{\Gamma_1}{\Gamma_0} - \frac{\beta_1}{\beta_0} \right) 
    (r_1-1-\ln r_1) \Bigg] \nonumber\\
   &&\mbox{}- \frac{\gamma_0^\prime}{\beta_0}\,\ln r_1
    + O\Big[(r_1-1)\,\alpha_s(\mu_h)\Big] \,.
\end{eqnarray}
(As a reminder, the QCD $\beta$-function is given in
(\ref{TheBetaFunc}), and we expanded all anomalous dimensions as in
(\ref{anomDimExpansion}).) The terms proportional to
$1/\alpha_s(\mu_h)$ resum the leading, double logarithmic terms to all
orders in perturbation theory. The remaining $O(1)$ terms in $V_H$
contribute at leading, single-logarithmic order. At next-to-leading
order, the corrections proportional to the coupling $\alpha_s(\mu_h)$
are included. In our case, the only piece missing for a complete
resummation at next-to-leading order is the $O(\alpha_s)$ contribution
to $V_H$,
\begin{eqnarray}
\frac{\alpha(\mu_{h})}{4\pi} \Bigg\{ 
&& (\ln r_1) \, \left[ \frac{r_1 \beta_1}{2\beta_0^4} \, 
   (\beta_0 \Gamma_1 - \beta_1 \Gamma_0 ) +
   \frac{\Gamma_0}{2\beta_0^4} \, (\beta_1^2 - \beta_0 \beta_2 ) 
   \right] \nonumber \\
&+& \frac{(r_1-1)^2}{4 \beta_0^4} \Big[
    \Gamma_0 (\beta_0\beta_2-\beta_1^2)
    + \beta_0(\beta_1 \Gamma_1 - \beta_0 \Gamma_2) \Big] \\
&+& \frac{(r_1-1)}{2 \beta_0^3} \Big[
    (\beta_2 \Gamma_0 - \beta_1 \Gamma_1)
    + 2\beta_0( \beta_1 \gamma^\prime_0 - \beta_0 \gamma^\prime_1)
    \Big] \qquad \Bigg\} \nonumber \;,
\end{eqnarray}
which is independent of the kinematic variable $\bar n\cdot p$ and
vanishes for $\mu_i\to\mu_h$. To compute these terms would require to
calculate the cusp anomalous dimension to three loops (knowledge of
$\Gamma_2$) and the anomalous dimension $\gamma^\prime$ to two loops
($\gamma^\prime_1$). While the former has recently been computed in
\cite{moch:et:al}, the latter is still missing up to date. This
implies a universal, process-independent small uncertainty in the
normalization of inclusive $B$-decay spectra in the shape-function
region. We stress, however, that this uncertainty cancels in all
ratios of decay distributions, even between $\bar B\to
X_u\,l^-\bar\nu$ and $\bar B\to X_s\gamma$ spectra.

It is appropriate to perform the running between $\mu_h$ and $\mu_i$
in a theory with $n_f=4$ light quark flavors, since the intermediate
scale $\mu_i$ will be of order $m_c$ in our applications below. The
relevant expansion coefficients are, as far as they are known,
\begin{eqnarray}
&& \Gamma_0=\frac{16}{3}\;, \quad 
   \Gamma_1=\frac{2576}{27}-\frac{16}{3}\pi^2\;,\quad 
   \Gamma_2=\frac{96488}{81}-\frac{5152}{27}\pi^2+\frac{176}{15}\pi^4
           -\frac{160}{9}\zeta_3\;, \label{cuspExpand} \nonumber \\
&& \beta_0=\frac{25}{3}\;,\quad 
   \beta_1=\frac{154}{3}\;,\quad 
   \beta_2=\frac{21943}{54}\;, \label{betaExpand} \nonumber \\
&& \gamma_0^\prime=-\frac{20}{3}\;.
\end{eqnarray}

\section{Differential decay rates and spectra}
\label{sec:rates}

As mentioned earlier, the hadronic tensor is most naturally expressed
in terms of the variables $n\cdot P_H$ and $\bar n\cdot p$ in the
shape-function region. It is thus useful to derive expressions for the
decay rates in terms of these variables. Our theoretical results are
valid as long as $n\cdot P_H$ can be considered as being of order a
hadronic scale (say, a $\mbox{few}\times\Lambda_{\rm QCD}$), whereas
$\bar n\cdot p$ is integrated over a domain of order
$m_b\gg\Lambda_{\rm QCD}$. It is this integration which provides a
sampling over sufficiently many hadronic final states needed to ensure
quark--hadron duality \cite{Bigi:2001ys}. The distribution in $\bar
n\cdot p$ will be described in terms of a ``partonic'' scaling
variable $y=\bar n\cdot p/m_b$, while the distribution in the
orthogonal light-cone component is described in terms of the
dimensionfull {\em hadronic\/} variable $P_+\equiv n\cdot
P_H=E_H-|\vec{P}_H|$. At leading order in $\Lambda_{\rm QCD}/m_b$, we
obtain from \cite{DeFazio:1999sv} the triple differential decay rate
\pagebreak
\begin{equation}\label{triple}
   \frac{d^3\Gamma}{d\bar x\,dy\,dP_+}
   = 12m_b\,\Gamma_{\rm tree}\,y(y-\bar x)
   \left[ (1+\bar x-y)\,\frac{W_1}{2}
   + \bar x \left( \frac{W_4}{2} + \frac{m_b W_5}{4} \right) \right]
   + \dots \,,
\end{equation}
where $\bar x=1-x$, and $x=2E_l/m_b$ is a scaling variable
proportional to the energy of the charged lepton measured in the
$B$-meson rest frame. The quantity $\Gamma_{\rm tree}=
G_F^2|V_{ub}|^2(m_b^{\rm pole})^5/(192\pi^3)$ denotes the leading
power, tree-level expression for the total $\bar B\to X_u\,l^-\bar\nu$
decay rate. The hadronic function $W_i$ are written in the factorized
form, which are found to be \cite{Bosch:2004th}
\begin{eqnarray}\label{master}
   \frac{W_1}{2} 
   &=& \Bigg\{ 1 + \frac{C_F\alpha_s(m_b)}{4\pi}
    \Big[ -4\ln^2 y + (6-c)\ln y - \frac{2\ln y}{1-y} - 4 L_2(1-y)
    \nonumber \\
   && \hspace{-8mm} - \frac{\pi^2}{6} - 12 \Big] \Bigg\} \; 
    y^{-1-a}\,e^{V_H(m_b,\mu_i)} 
    \int_0^{n\cdot P_H}\!
    d\hat\omega\,\hat J(\hat p_\omega^2,y,\mu_i)\,
    \hat S(\hat\omega,\mu_i) + \dots \,, \nonumber \\
   \frac{W_4}{2} + \frac{m_b W_5}{4}
   &=& \frac{C_F\alpha_s(m_b)}{4\pi}\,\frac{2\ln y}{1-y} \; y^{-1-a}\,
    e^{V_H(m_b,\mu_i)} \nonumber \\
   && \times \int_0^{n\cdot P_H}\!d\hat\omega\,
    \hat J(\hat p_\omega^2,y,\mu_i)\,\hat S(\hat\omega,\mu_i) + \dots
    \,, 
\end{eqnarray}
where the dots represent power corrections in $\Lambda_{\rm QCD}/m_b$.
The result for the rescaled jet function $\hat J(\hat p_\omega^2,
y,\mu_i)$ is the expression in (\ref{Jrescale}). Furthermore we need
the Sudakov exponent $V_H$ in (\ref{VHres}). This function is
independent of the kinematic variables $y$ and $\hat p_\omega^2$. In
addition, we need
\begin{equation}\label{acdef}
\begin{aligned}
   a &= \frac{\Gamma_0}{\beta_0}\,\ln r_1
    = \frac{16}{25}\,\ln\frac{\alpha_s(\mu_i)}{\alpha_s(m_b)} \,, \\ 
   c &= \frac{4}{\beta_0} \left( \frac{\Gamma_1}{\Gamma_0}
    - \frac{\beta_1}{\beta_0} \right) (r_1-1)
    = \left( \frac{10556}{1875} - \frac{12\pi^2}{25} \right)
    \left( \frac{\alpha_s(\mu_i)}{\alpha_s(m_b)} - 1 \right) .
\end{aligned}
\end{equation}
For simplicity, we have identified the high-energy matching scale
$\mu_h$ with the heavy-quark mass $m_b$. In the variables $\bar x$,
$y$, and $P_+$, the phase space is remarkably simple:
\begin{equation}\label{ps1}
   0\le P_+\le M_B-2E_l = m_b\,\bar x + \bar\Lambda \,, \qquad
   \frac{P_+-\bar\Lambda}{m_b}\le\bar x\le y\le 1 \,.
\end{equation}
If $E_l$ is integrated over a domain of order $m_b\gg\Lambda_{\rm
QCD}$, i.~e.~$\bar x$ is integrated over a domain of order unity, one
can replace the second condition by $0\le\bar x\le y\le 1$ at leading
power in $\Lambda_{\rm QCD}/m_b$. If, on the other hand, the lepton
energy is restricted to be close to its kinematic limit, $E_l\approx
M_B/2$, then $\bar x=O(\Lambda_{\rm QCD}/m_b)$, and at leading order
the rate (\ref{triple}) can be simplified to
\begin{equation}\label{triple2}
   \frac{d^3\Gamma}{dE_l\,dy\,dP_+}
   = 24\Gamma_{\rm tree}\,y^2(1-y)\,\frac{W_1}{2} + \dots \,,
\end{equation}
with $0\le y\le 1$ and $0\le P_+\le M_B-2E_l$.

\begin{figure}[t]
\begin{center}
\epsfig{file=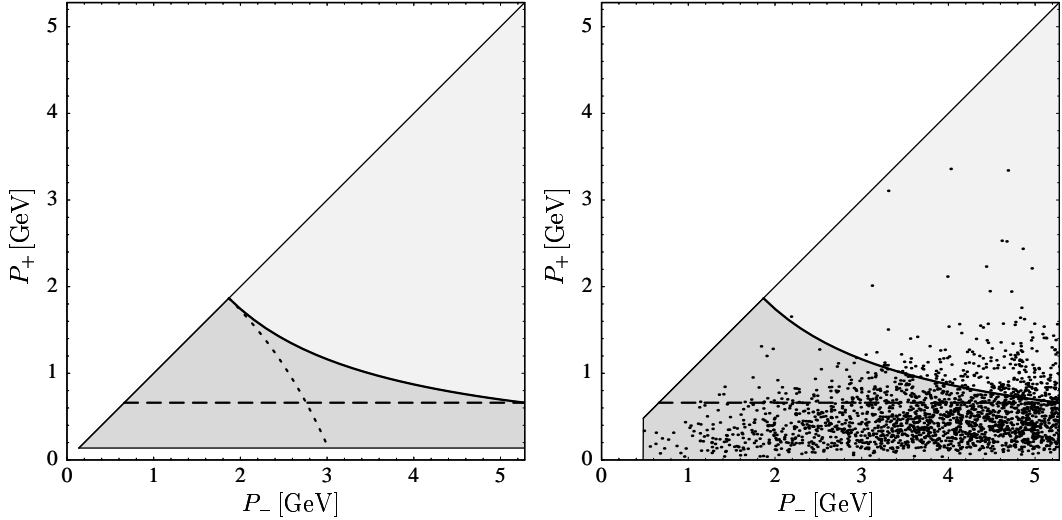,width=14cm}
\caption[Hadronic phase space for the light-cone 
variables $P_-$ and $P_+$.]{\label{fig:phase} Hadronic phase space for
the light-cone variables $P_-$ and $P_+$ (left), and theory phase
space for $m_b=4.8$\,GeV (right). The scatter points indicate the
distribution of events as predicted by the model of
\cite{DeFazio:1999sv}. In each plot the solid line separates the regions 
where $s_H<M_D^2$ (dark gray) and $s_H>M_D^2$ (light gray), whereas the 
dashed line corresponds to $P_+=M_D^2/M_B$. The dotted line in the first 
plot shows the contour where $q^2=(M_B-M_D)^2$.}
\end{center}
\end{figure}

In general, the hadronic tensor can be described in terms of the
quantities $P_\pm = E_H \mp |\vec{P}_H|$, whose true
phase-space is $M_\pi\le P_+\le P_-\le M_B$, corresponding to a
triangular region in the $(P_-,P_+)$ plane. The variable $P_-$ is
related to our parton variables by $P_-=\bar n\cdot
p+\bar\Lambda=m_b\,y+\bar\Lambda$.  In our theoretical description
based on quark--hadron duality $P_+$ starts from 0, while the small
region with $P_-<\bar\Lambda$ is left unpopulated. This is illustrated
in Fig.~\ref{fig:phase}. Contours of constant hadronic or leptonic
invariant mass in the $(P_-,P_+)$ plane are easy to visualize, since
\begin{equation}
   s_H = P_H^2 = P_+ P_- \,, \qquad
   q^2 = (M_B-P_+)(M_B-P_-)
\end{equation}
are given by very simple expressions. The solid and dotted lines in
the left-hand plot in Fig.~\ref{fig:phase} show the contours where
$s_H=M_D^2$ and $q^2=(M_B-M_D)^2$, respectively, which can be used to
separate $\bar B\to X_u\,l^-\bar\nu$ events from semileptonic decays
with charm hadrons in the final state. The dashed horizontal line
shows the maximum allowed value of $P_+$ when a cut
$E_l\ge(M_B^2-M_D^2)/(2M_B)$ is applied to the charged-lepton energy,
which implies $P_+\le M_D^2/M_B$.  We will see later that this cut,
which is obviously another way of eliminating the charm background,
allows for a systematic treatment of the theoretical prediction, and
is therefore of great interest. In the right-hand plot, we indicate
the density of events in theory phase space obtained using the model
of \cite{DeFazio:1999sv}.\footnote{While not rigorously implementing
shape-function effects beyond tree level, the model of
\cite{DeFazio:1999sv} has the advantage that it interpolates between
the shape-function region and the remainder of phase space, where a
local operator product expansion can be employed. On the contrary, our
more rigorous discussion here is limited to the region of
hard-collinear jet momenta. However, the scatter plot shown in
the figure provides a reasonably realistic impression of the
population in phase space.}  It is apparent that the vast majority of
events is located in the shape-function region of small $P_+$ and
large $P_-$.

In the remainder of this section, we present analytic results to the
order of accuracy we have worked so far, for a variety of spectra in
$\bar B\to X_u\,l^-\bar\nu$ decays. They are obtained by integrating
over the scaling variable $y$ before integrating over the hadronic
variable $P_+$, changing variables from $P_+$ to $\hat
p_\omega^2=m_b(P_+-\hat\omega)$. The integral over the shape-function
variable $\hat\omega$ is left until the end, so that our formulae are
model independent. We will always present fractional decay rates
normalized to the total inclusive rate
\begin{equation}\label{totalInclusiveRate}
   \Gamma(\bar B\to X_u\,l^-\bar\nu)\equiv \Gamma_{\rm tot}
   = \Gamma_{\rm tree} \left[ 1 + \frac{C_F\alpha_s(m_b)}{4\pi}
   \left( \frac{25}{2} - 2\pi^2 \right) \right] + \dots \,,
\end{equation}
where the dots represent higher-order perturbative corrections as well
as power corrections of order $(\Lambda_{\rm QCD}/m_b)^2$ and
higher. This procedure offers the advantage of eliminating the strong
sensitivity to the heavy-quark (pole) mass. The integrals over the
parton variable $y$ encountered in our analysis can be reduced to a
set of master integrals defined as
\begin{equation}\label{Idef}
\begin{aligned}
   I_1(b,z) &= \int_0^z\!dy\,y^b
    = \frac{z^{1+b}}{1+b} \,, \\
   I_2(b,z) &= \int_0^z\!dy\,y^b\ln y
    = \frac{z^{1+b}}{1+b} \left( \ln z - \frac{1}{1+b} \right) , \\
   I_3(b,z) &= \int_0^z\!dy\,y^b\ln^2 y
    = \frac{z^{1+b}}{1+b} \left( \ln^2 z - \frac{2\ln z}{1+b}
    + \frac{2}{(1+b)^2} \right) , \\
   I_4(b,z) &= \int_0^z\!dy\,y^b \frac{\ln y}{1-y}
    = \sum_{j=0}^\infty \frac{z^{1+b+j}}{1+b+j}
    \left( \ln z - \frac{1}{1+b+j} \right) , \\
   I_5(b,z) &= \int_0^z\!dy\,y^b L_2(1-y)
    = \frac{z^{1+b}}{1+b}\,L_2(1-z) - \frac{I_4(1+b,z)}{1+b} \,,
\end{aligned}
\end{equation}
where $b>-1$  and $z\le 1$ are a arbitrary real numbers.

\subsection{Charged-lepton energy spectrum}

Near the charged-lepton energy endpoint, where we can assume that
$M_B-2E_l$ is of order a hadronic scale, the underlying event falls
into the shape-function region of large $\bar n\cdot p$ and small
$n\cdot P_H$. Starting from the triple differential rate in
(\ref{triple2}), we obtain for the normalized energy spectrum
\cite{Bosch:2004th}
\begin{eqnarray}\label{Elspectrum}
   \frac{1}{\Gamma_{\rm tot}}\,\frac{d\Gamma}{dE_l}
   &=& \frac{4T(a)}{m_b}\,e^{V_H(m_b,\mu_i)}
    \int_0^{M_B-2E_l}\!d\hat\omega\,\hat S(\hat\omega,\mu_i)\,
    \Bigg\{ 1 + \frac{C_F\alpha_s(m_b)}{4\pi}\,H(a) \nonumber\\
   && + \frac{C_F\alpha_s(\mu_i)}{4\pi} \Bigg[
    2\ln^2\frac{m_b(M_B-2E_l-\hat\omega)}{\mu_i^2} \nonumber\\
   && + \Big( 4f_2(a) - 3 \Big)
    \ln\frac{m_b(M_B-2E_l-\hat\omega)}{\mu_i^2} \nonumber\\
   && + \Big( 7 - \pi^2 - 3f_2(a) + 2f_3(a) \Big) \Bigg] \Bigg\} \,.
\end{eqnarray}
Here
\begin{equation}
\begin{aligned}
   T(a) &= 6 \Big[ I_1(1-a,1) - I_1(2-a,1) \Big] \,, \\
   f_n(a) &= \frac{I_n(1-a,1)-I_n(2-a,1)}{I_1(1-a,1)-I_1(2-a,1)} \,, \\
   H(a) &= \frac{11\pi^2}{6} - \frac{49}{2} + (6-c)f_2(a) - 4f_3(a)
    - 2f_4(a) - 4f_5(a) \,.
\end{aligned}
\end{equation}
At leading power in $\Lambda_{\rm QCD}/m_b$ the heavy-quark mass in the 
denominator of the prefactor on the right-hand side of (\ref{Elspectrum}) 
can be replaced by $m_b+\omega=M_B-\hat\omega$, which removes any 
sensitivity to the definition of $m_b$. This replacement can indeed be 
justified by studying power corrections to the shape function
\cite{Bauer:2002yu,Neubert:2002yx}.

All our results for decay rates will have a similar structure, but the
definitions of the functions $T$, $f_n$, and $H$ will be different in
each case. The tree-level result can be recovered by setting $V_H=0$
and $a=0$, in which case $T(0)=1$, and the spectrum is simply given in
terms of an integral over the shape function \cite{Neubert:1993ch}.
Using the above result, it is straightforward to calculate the fraction 
$F_E=\Gamma(E_l\ge E_0)/\Gamma_{\rm tot}$ of all 
$\bar B\to X_u\,l^-\bar\nu$ events with charged-lepton energy above a 
threshold $E_0$. Defining $\Delta_E=M_B-2E_0$, we find
\begin{eqnarray}\label{FE}
   F_E(\Delta_E) &=& T(a)\,e^{V_H(m_b,\mu_i)}
    \int_0^{\Delta_E}\!d\hat\omega\,
    \frac{2(\Delta_E-\hat\omega)}{M_B-\hat\omega}\,
    \hat S(\hat\omega,\mu_i)\,
    \Bigg\{ 1 + \frac{C_F\alpha_s(m_b)}{4\pi}\,H(a) \nonumber\\
   &&\mbox{}+ \frac{C_F\alpha_s(\mu_i)}{4\pi} \Bigg[
    2\ln^2\frac{m_b(\Delta_E-\hat\omega)}{\mu_i^2}
    + \Big( 4f_2(a) - 7 \Big)
    \ln\frac{m_b(\Delta_E-\hat\omega)}{\mu_i^2} \nonumber\\
   &&\hspace{2.3cm}\mbox{}
    + \Big( 14 - \pi^2 - 7f_2(a) + 2f_3(a) \Big) \Bigg] \Bigg\} \,.
\end{eqnarray}
Note that the fraction $F_E(\Delta_E)$ is given in terms of a weighted
integral over the shape function, with a weight factor of order
$\Lambda_{\rm QCD}/m_b$ that vanishes at the upper end of
integration. As a result, only a small fraction of events is contained
in the lepton endpoint region.

\subsection{Hadronic \boldmath$P_+$ spectrum\unboldmath}

Applying a lower cut on the charged-lepton energy restricts the
variable $P_+$ to be less than $\Delta_E$. However, a cut on $P_+$
does {\em not\/} restrict the lepton energy to be in the endpoint
region. Still, the fraction of events with $P_+\le\Delta_E$ samples
the same hadronic phase space as the lepton-endpoint cut, but it
contains significantly more events. Such a cut therefore offers an
excellent opportunity to determine the CKM matrix element $|V_{ub}|$.

To determine the fraction of events that survive, we integrate over
$\bar x$ and $y$ in the range $0\le\bar x\le y\le 1$ before
integrating over $P_+$. The result is
\begin{eqnarray}\label{FP}
   F_P(\Delta_P) &=& T(a)\,e^{V_H(m_b,\mu_i)}
    \int_0^{\Delta_P}\!d\hat\omega\,\hat S(\hat\omega,\mu_i)\,
    \Bigg\{ 1 + \frac{C_F\alpha_s(m_b)}{4\pi}\,H(a) \nonumber\\
   &&\mbox{}+ \frac{C_F\alpha_s(\mu_i)}{4\pi} \Bigg[
    2\ln^2\frac{m_b(\Delta_P-\hat\omega)}{\mu_i^2}
    + \Big( 4f_2(a) - 3 \Big)
    \ln\frac{m_b(\Delta_P-\hat\omega)}{\mu_i^2} \nonumber\\
   &&\hspace{2.3cm}\mbox{}
    + \Big( 7 - \pi^2 - 3f_2(a) + 2f_3(a) \Big) \Bigg] \Bigg\} \,,
\end{eqnarray}
where now
\begin{equation}\label{TforFP}
\begin{aligned}
   T(a) &= 6 I_1(2-a,1) - 4 I_1(3-a,1) \,, \\
   H(a) &= \frac{11\pi^2}{6} - \frac{49}{2} + (6-c)f_2(a) - 4f_3(a)
    - 2 \Big[ f_4(a) - \Delta f_4(a) \Big] - 4f_5(a) \,,
\end{aligned}
\end{equation}
and
\begin{equation}\label{FforFP}
\begin{aligned}
   f_n(a) &= \frac{3I_n(2-a,1)-2I_n(3-a,1)}{3I_1(2-a,1)-2I_1(3-a,1)} \,,
    \\
   \Delta f_4(a) &= \frac{I_4(3-a,1)}{3I_1(2-a,1)-2I_1(3-a,1)} \,.
\end{aligned}
\end{equation}
The contribution $\Delta f_4$ arises from the terms contained in the
structure functions $W_4$ and $W_5$ in (\ref{master}). A comparison of
the result for $F_P(\Delta_P)$ in (\ref{FP}) with the expression for
$F_E(\Delta_E)$ in (\ref{FE}) yields that the cut on hadronic $P_+$
contains a much larger fraction of all $\bar B\to X_u\,l^-\bar\nu$
events. As a matter of fact, $F_P(\Delta_P)$ is directly given in
terms of an integral over the shape function, without a weight
function of order $\Lambda_{\rm QCD}/m_b$. (At tree level,
$F_P(\Delta_P)=\int_0^{\Delta_P}\!d\hat\omega\,\hat S(\hat\omega)$.)
Since the shape function peaks around $\hat \omega\approx \bar\Lambda
\approx 0.5$ GeV, we expect a high efficiency for values of $\Delta_P$
in the vicinity of the ``optimal cut'' $\Delta_P = M_D^2/M_B$, which
eliminates the charm background.

\subsection{Hadronic invariant mass spectrum}

A cut on the hadronic invariant mass in the final state constitutes
the ideal separator between $\bar B\to X_u\,l^-\bar\nu$ and $\bar B\to
X_c\,l^-\bar\nu$ events, since any final state containing a charm
hadron has invariant mass above $M_D$.  Let us discuss the cut
$\sqrt{s_H}\le M_D$ by examining the phase-space picture of
Fig.~\ref{fig:phase}. The available phase space for a cut on $P_+
\le M_D^2/M_B$ is fully contained. In addition there is a
triangle-shaped region of larger $P_+$, which culminates in a cusp
where $P_+=P_-=M_D$. Near the cusp, both light-cone momentum
components are of the same order, and hence this portion of phase
space should not be treated using our theoretical description based on
the collinear expansion. A priori, it is not evident that we can
compute the fractional rate $F_M(s_0)=\Gamma(s_H\le s_0)/\Gamma_{\rm
tot}$ in a controlled heavy-quark expansion.

To see what happens, it is instructive to first ignore radiative 
corrections. At tree level, it is straightforward to obtain
\begin{equation}\label{FM}
   F_M(s_0) = \int\limits_0^{\Delta_s}\!\!d\hat\omega\,\hat S(\hat\omega)
   + \int\limits_{\Delta_s}^{\sqrt{s_0}}\!\!d\hat\omega\,
   \hat S(\hat\omega) \left( \frac{\Delta_s}{\hat\omega} \right)^3
   \left( 2 - \frac{\Delta_s}{\hat\omega} \right) ,
\end{equation}
where $\Delta_s=s_0/M_B$. The calculation of this event fraction
requires knowledge of the shape function over a wider range in
$\hat\omega$ than in the case of the event fraction with a cut on
$P_+$. The first integral is the same as for $F_P(\Delta_s)$ in
(\ref{FP}) and corresponds to the region in phase space where $P_+\le
s_0/M_B$. The second integral corresponds to the phase space above the
dashed line in Fig.~\ref{fig:phase}. The region near the cusp
corresponds to the upper integration region in the second
integral. Note that, due to the rapid fall-off of the integrand, the
tip of the triangle region only gives a power-suppressed contribution
to the decay rate. When radiative corrections are included, the result
for the integrated hadronic invariant mass spectrum becomes rather
complicated. In general, we may split up the contributions into
\begin{equation}\label{thisIShadrMass}
   F_M(s_0) = F_M^{\rm box}(s_0) + F_M^{\rm triangle}(s_0) \,, \qquad
   \mbox{with} \quad F_M^{\rm box}(s_0) = F_P(\Delta_s) \,,
\end{equation}
The box contribution is given by the expression for the rate 
fraction $F_P(\Delta_P)$ in (\ref{FP}) evaluated with 
$\Delta_P=\Delta_s=s_0/M_B$. For the remaining contribution from the 
triangular region, we obtain \cite{Bosch:2004th}.
\begin{eqnarray}\label{Ftrian}
   F_M^{\rm triangle}(s_0)
   &=& e^{V_H(m_b,\mu_i)} \int_{\Delta_s}^{\sqrt{s_0}}\!d\hat\omega\,
    \hat S(\hat\omega,\mu_i) \left[ G_1(\Delta_s/\hat\omega)
    + \frac{C_F\alpha_s(\mu_i)}{4\pi}\,G_2(\Delta_s,\hat\omega) \right]
    \nonumber\\
   &&\mbox{}+ e^{V_H(m_b,\mu_i)} \int_0^{\Delta_s}\!d\hat\omega\,
    \hat S(\hat\omega,\mu_i)\,\frac{C_F\alpha_s(\mu_i)}{4\pi}\,
    G_3(\Delta_s,\hat\omega) \,,
\end{eqnarray}
where
\begin{equation}
\begin{aligned}
   G_1(z) = T(a,z) \Bigg\{ 1 & + \frac{C_F\alpha_s(m_b)}{4\pi}\,H(a,z) \\
   & + \frac{C_F\alpha_s(\mu_i)}{4\pi} \left[ 7 - \pi^2 - 3f_2(a,z)
   + 2f_3(a,z) \right] \Bigg\}
\end{aligned}
\end{equation}
contains the same functions $T$, $H$ and $f_n$ as defined in 
(\ref{TforFP}) and (\ref{FforFP}), but with all master integrals replaced 
by $I_n(b,1)\to I_n(b,z)$. In addition, we need
\begin{eqnarray}\label{G2G3}
   G_2(\Delta_s,\hat\omega)
   &=& \int_0^{\mu_i^2/m_b}\!\frac{dP}{P} \left\{
    \ln\frac{m_b P}{\mu_i^2} \left[
    k_1\!\left( \frac{\Delta_s}{P+\hat\omega} \right)
    - k_1\!\left( \frac{\Delta_s}{\hat\omega} \right) \right] \right.
    \nonumber \\ 
    && \hspace{33mm} \left.
    + \left[ k_2\!\left( \frac{\Delta_s}{P+\hat\omega} \right)
    - k_2\!\left( \frac{\Delta_s}{\hat\omega} \right) \right] \right\}
    \nonumber \\
    &+& \int_{\mu_i^2/m_b}^{\sqrt{s_0}-\hat\omega}\!\frac{dP}{P}
    \left[ \ln\frac{m_b P}{\mu_i^2}\,
    k_1\!\left( \frac{\Delta_s}{P+\hat\omega} \right)
    + k_2\!\left( \frac{\Delta_s}{P+\hat\omega} \right) \right] , \\
   G_3(\Delta_s,\hat\omega)
   &=& \int_{\Delta_s}^{\sqrt{s_0}}\!\frac{dP}{P-\hat\omega}
    \left[ \ln\frac{m_b(P-\hat\omega)}{\mu_i^2}\,
    k_1\!\left( \frac{\Delta_s}{P} \right)
    + k_2\!\left( \frac{\Delta_s}{P} \right) \right] , \nonumber
\end{eqnarray}
where
\begin{eqnarray}
   k_1(z) &=& 4 \Big[ 6 I_1(2-a,z) - 4 I_1(3-a,z) \Big] = 4 T(a,z) \,, \\
   k_2(z) &=& 4 \Big[ 6 I_2(2-a,z) - 4 I_2(3-a,z) \Big] 
    - 3 \Big[ 6 I_1(2-a,z) - 4 I_1(3-a,z) \Big] \,, \nonumber
\end{eqnarray}
with the constant $a$ given in (\ref{acdef}). For orientation, a
typical numerical value is $a\approx 0.335$ for $\mu_i = 1.5\;\rm
GeV$.

As mentioned above, the phase-space region near the cusp where
$\hat\omega\sim\sqrt{s_0}$ or $P\sim\sqrt{s_0}$ gives a
power-suppressed contribution to the decay rate. More precisely, we
find that the contribution is given by
\begin{equation}\label{Bermuda}
   F_M(s_0)\owns e^{V_H(m_b,\mu_i)}\,
   \frac{C_F\alpha_s(\mu_i)}{\pi}\,\frac{6}{(3-a)^2}
   \left( \frac{\Delta_s}{\sqrt{s_0}} \right)^{3-a}
   \left( \frac74 + \frac{3}{3-a} \right) + \dots \,,
\end{equation}
where the dots represent higher-order power corrections.  For $s_0\sim
m_b\Lambda_{\rm QCD}$, the above result thus scales like
$(\Lambda_{\rm QCD}/m_b)^{(3-a)/2}$, whereas $F_M(s_0)$ is of
$O(1)$. For consistency, we should therefore omit the term in
(\ref{Bermuda}), which can be done by replacing all occurrences of
$\sqrt{s_0}$ in upper integration limits in (\ref{Ftrian}) and
(\ref{G2G3}) with $\infty$, which we will use this prescription in our
numerical analysis in Section~\ref{sec:appls}.

\subsection{Combined cuts on hadronic and leptonic invariant mass}
\label{sec:comb}

\begin{figure}
\begin{center}
\epsfig{file=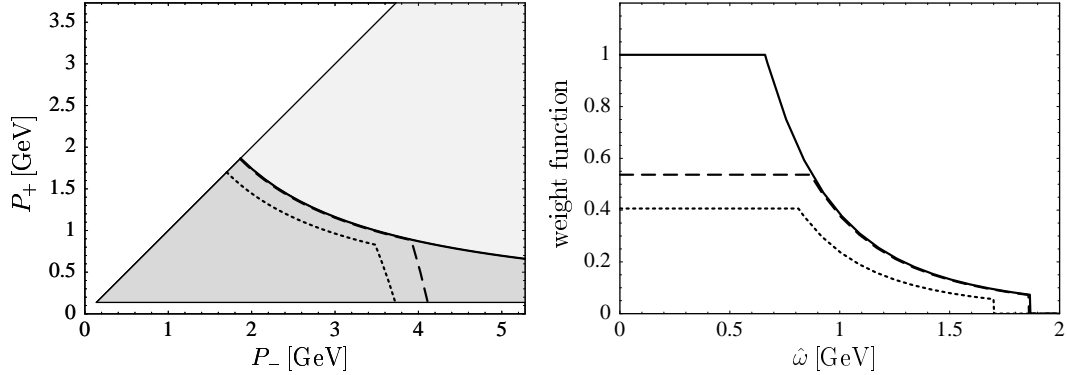,width=14cm}
\end{center}
\centerline{\parbox{14cm}{\caption[Phase-space constraints and weight 
functions for combined cuts on the hadronic and leptonic invariant
mass]{\label{fig:combined} Phase-space constraints (left) and weight
functions (right) for combined cuts on the hadronic and leptonic
invariant mass: $(s_0,\,q_0^2)=(M_D^2,\,0)$ (solid),
$(M_D^2,\,6\,\mbox{GeV}^2)$ (dashed), and
$((1.7\,\mbox{GeV})^2,\,8\,\mbox{GeV}^2)$ (dotted).}}}
\end{figure}

Bauer et al.\ have proposed to reduce the sensitivity to
shape-function effects in the extraction of $|V_{ub}|$ by combining a
cut on hadronic invariant mass with a cut $q^2\ge q_0^2$ on the
invariant mass squared of the lepton pair \cite{Bauer:2001rc}. The
first plot in Fig.~\ref{fig:combined} shows that this eliminates a
large portion of the events with large $P_-$. We indicate the
remaining phase space for several choices of cuts, for example
$(s_0,\,q_0^2)=(M_D^2,\,0)$ (solid line), $(M_D^2,\,6\,\mbox{GeV}^2)$
(dashed line), and $((1.7\,\mbox{GeV})^2,\,8\,\mbox{GeV}^2)$ (dotted
line).  For the corresponding event fraction at tree level, we obtain
\begin{equation}
   F_{\rm comb}(s_H\le s_0, q^2\ge q_0^2)
   = y_0^3\,(2-y_0)\! \int\limits_0^{\Delta_s/y_0}\!d\hat\omega\,
   \hat S(\hat\omega)
   + \!\! \int\limits_{\Delta_s/y_0}^{\sqrt{s_0}}\!\!d\hat\omega\,
   \hat S(\hat\omega) \left( \frac{\Delta_s}{\hat\omega} \right)^3
   \left( 2 - \frac{\Delta_s}{\hat\omega} \right) ,
\end{equation}
where $y_0=1-q_0^2/(m_b M_B)$, and $\Delta_s=s_0/M_B$ as above. For a
fixed hadronic-mass cut $s_0$, the effect of the additional cut on
$q^2$ is to broaden the support of the first integral, while at the
same time reducing its weight due to the prefactor. To illustrate this
point, we show in the second plot in Fig.~\ref{fig:combined} the
weight functions under the integral with the shape function for the
three different choices of $(s_0,\,q_0^2)$ mentioned above. The
sensitivity to the precise form of the shape function is reduced
because the weight functions become progressively more shallow as the
value of $q_0^2$ is raised. However, this reduction comes at the price
of a significant reduction of the rate, raising questions about the
validity of the assumption of quark--hadron duality.  We will see in
Section~\ref{sec:appls} that the {\em relative\/} uncertainty due to
shape-function effects is not strongly (although somewhat) reduced
when imposing an additional cut on $q^2$.

\section{Model-independent relations between spectra}
\label{sec:relations}

As we have seen in the previous section, all spectra and event
fractions are expressed in terms of weighted integrals of perturbative
functions over the (universal, i.e.~process independent) shape
function. Therefore they all require the knowledge of the function
form of the shape function, which cannot be calculated using analytic
techniques. One way would be to adopt a specific model or extract the
shape function from experiment.  Alternatively, it is possible to
derive model-independent relations between different decay
distributions in which the shape function has been eliminated
\cite{Neubert:1993ch}. The most promising strategy is to relate event
fractions in semileptonic $\bar B\to X_u\,l^-\bar\nu$ decays to a
weighted integral over the $\bar B\to X_s\gamma$ photon spectrum,
which at present provides the most direct access to the shape
function.  While it is straightforward to derive such relations at
tree level, radiative corrections introduce non-trivial complications
\cite{Leibovich:1999xfx,Neubert:2001sk,Leibovich:2001ra,Mannel:1999gs}.
Since our formalism \cite{Bosch:2004th} has yet to be applied to the
$\bar B\to X_s\gamma$ photon spectrum, we will instead derive a
relation between the charged-lepton energy spectrum and a weighted
integral over the $P_+$ spectrum, which is in many respects very
similar to the photon spectrum. We wish to construct a {\em
perturbative\/} weight function $w(\Delta,P_+)$ such that at leading
power in $\Lambda_{\rm QCD}/m_b$
\begin{equation}\label{raterel}
   \int_{E_0}^{M_B/2}\!dE_l\,\frac{d\Gamma}{dE_l}
   = \int_0^{\Delta}\!dP_+\,w(\Delta,P_+)\,\frac{d\Gamma}{dP_+} \,,
   \qquad \Delta = M_B - 2E_0 \,.
\end{equation}
This relation is independent of the shape function and hence
insensitive to hadronic physics. The construction of the weight
function is straightforward order by order in perturbation
theory. Using the results of the previous section, we find
\begin{eqnarray}
   w(\Delta,P_+)
   &=& \frac{2(\Delta-P_+)}{M_B-P_+}\,\frac{3(4-a)}{(6-a)(2-a)}\,
    \Bigg\{ 1 + \frac{C_F\alpha_s(m_b)}{4\pi}\,h_1(a) \nonumber
\end{eqnarray}
\begin{eqnarray}\label{wfun}
   &&\mbox{}+ \frac{C_F\alpha_s(\mu_i)}{4\pi} \left[ h_2(a)\,
    \ln\frac{m_b(\Delta-P_+)}{\mu_i^2} + h_3(a) \right] \Bigg\} \,,
\end{eqnarray}
where
\begin{eqnarray}
   h_1(a) &=& 2 - 2\,\frac{3952-5416a+2988a^2-838a^3+120a^4-7a^5}
                      {(6-a)(4-a)^2(3-a)(2-a)^2} \nonumber \\
   && \mbox{}\quad + c\,\frac{20-8a+a^2}{(6-a)(4-a)(2-a)} \,, \nonumber\\
   h_2(a) &=& - 4\,\frac{20-8a+a^2}{(6-a)(4-a)(2-a)} \,, \\
   h_3(a) &=& \frac{5056-6744a+3556a^2-942a^3+127a^4-7a^5}
                   {(6-a)(4-a)^2(3-a)(2-a)^2} \,. \nonumber
\end{eqnarray}
The lesson we learned from this prototype relation (\ref{raterel})
should be applied in the future to the $\bar B\to X_s\,\gamma$ photon
spectrum. Using similar methods, it will be possible to construct a
shape-function independent relation of the form
\begin{equation} \label{reltoBsGamma}
   F_{P}(\Delta) = \frac{1}{\Gamma_{s}} 
   \int_{\frac{M_{B}-\Delta}{2}}^{\frac{M_{B}}{2}}\!dE_{\gamma}\,
   \frac{d\Gamma_{s}}{dE_{\gamma}}\,w_s(\Delta,E_{\gamma}) \,,
\end{equation}
where at tree level the weight function is simply
$w_s(\Delta,E_{\gamma})=1$. It is also possible (although far more
complicated) to relate the $\bar B\to X_{s}\gamma$ photon spectrum to
the hadronic invariant mass spectrum $F_M(s_0)$ in
(\ref{thisIShadrMass}). However, because of the larger integration
domain over the shape function, such a relation would require input of
the photon spectrum beyond the region where it is currently
experimentally accessible.

\begin{figure}
\begin{center}
\epsfig{file=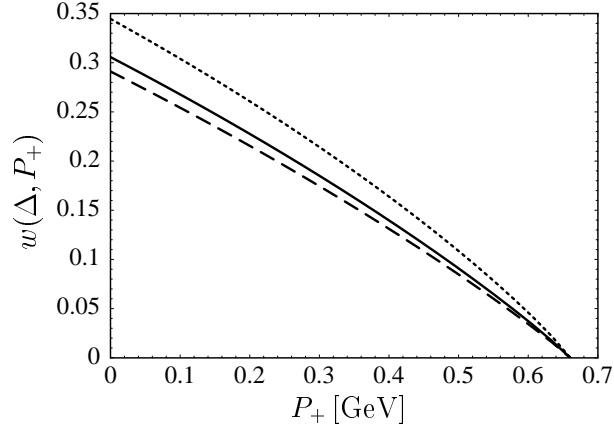,width=8cm}
\end{center}
\centerline{\parbox{14cm}{\caption[Weight function in the prototype 
relation]{\label{fig:wfun} Weight function $w(\Delta,P_+)$ entering
the rate relation (\ref{raterel}) for \linebreak $\Delta=M_D^2/M_B$ and three
different choices of the intermediate scale, namely $\mu_i=1.5$\,GeV
(solid), 2.0\,GeV (dashed), and 1.0\,GeV (dotted). The weight function
is formally independent of $\mu_i$.}}}
\end{figure}

The alert reader might wonder about the appearance of the
renormalization scale $\mu_i$ in the weight function, since $w(\Delta,
P_+)$ is formally independent of the scale $\mu_i$ (because there is
nothing to cancel a potential $\mu_i$ dependence in
(\ref{raterel})). Expanding the resummed result for the weight
function to first order in $\alpha_s$, we obtain the simple expression
\begin{equation}\label{w1loop}
   w(\Delta,P_+)\big|_{\rm 1-loop}
   = \frac{2(\Delta-P_+)}{M_B-P_+} \left[ 1 + \frac{C_F\alpha_s}{4\pi}
   \left( - \frac53\,\ln\frac{\Delta-P_+}{m_b} - \frac{17}{36} \right)
   \right] ,
\end{equation}
in which the dependence on $\mu_i$ has canceled. However, since this
formula contains a large logarithm and the scale to be used in
$\alpha_s$ is undetermined, it should not be used for phenomenological
applications. A visualization of the resummed weight function in
(\ref{wfun}) for different choices of $\mu_i$ is given in
Fig.~\ref{fig:wfun}. As is apparent, the function is pretty
uneventful, since apart from the rational prefactor, the $P_+$
dependence is given through a single logarithm.

\section{Numerical results}
\label{sec:appls}

We are now ready to study the implications of our analysis for
phenomenology. We start by deriving the numerical values for the
shape-function mass and kinetic energy including errors. We then
present a model for the shape function which satisfies all theoretical
constraints, as derived in Chapter~\ref{chap:strucfunc}. Finally, we
present numerical results for the various decay rates and spectra
investigated in Section~\ref{sec:rates}. We use the two-loop running
coupling constant in the $\overline{\rm MS}$ scheme, normalized such
that $\alpha_s(M_Z)=0.119$.  Our standard choice of the intermediate
matching scale is $\mu_i=1.5$\,GeV. This corresponds to setting
$\mu_i^2=m_b\Lambda_{\rm had}$ with a typical hadronic scale
$\Lambda_{\rm had}\approx 0.5$\,GeV. The values of the strong coupling
evaluated at these scales are $\alpha_s(m_b)\simeq 0.222$ and
$\alpha_s(\mu_i)\simeq 0.375$. The corresponding values of the
perturbative parameters $a$ and $c$ defined in (\ref{acdef}) are
$a\simeq 0.335$ and $c\simeq 0.614$. Finally, the leading-order
Sudakov factor in (\ref{master}) takes the value
$e^{V_H(m_b,\mu_i)}\simeq 1.21$.

\subsection{Shape-function mass and kinetic energy}

A value for the shape-function mass can be obtained by combining the
relations (\ref{mSFmpole}) or (\ref{mSFmPS}) with existing predictions
for the $b$-quark mass in the relevant renormalization schemes. The
potential-subtracted mass at the scale $\mu_f=2$\,GeV has been
determined from moments of the $b\bar b$ cross section and the mass of
the $\Upsilon(1S)$ state \cite{Beneke:1999fe}. Using the first
relation in (\ref{mSFmPS}), we find \cite{Bosch:2004th} $m_b^{\rm
SF}(2\,\mbox{GeV},2\,\mbox{GeV})=m_b^{\rm PS}(2\,\mbox{GeV})%
=(4.59\pm 0.08)$\,GeV. From a similar analysis the kinetic mass has
been determined at the scale $\mu_f=2$\,GeV to be $m_b^{\rm
kin}(1\,\mbox{GeV})=(4.57\pm 0.06)$\,GeV \cite{Benson:2003kp}.  From
the second relation in (\ref{mSFmPS}) it then follows that $m_b^{\rm
SF}(1\,\mbox{GeV},1\,\mbox{GeV})=(4.65\pm 0.06)$\,GeV. Computing the
scale dependence of the shape-function mass (using the fact that the
pole mass is RG invariant), we obtain at the intermediate scale the
values $m_b^{\rm SF}(\mu_i,\mu_i)=(4.61\pm 0.08)$\,GeV and $m_b^{\rm
SF}(\mu_i,\mu_i)=(4.65\pm 0.06)$\,GeV, respectively.  Alternatively,
we may use relation (\ref{mSFmpole}) in conjunction with an
experimental determination of the $b$-quark pole mass from moments of
inclusive $\bar B\to X_c\,l^-\bar\nu$ and $\bar B\to X_s\gamma$ decay
spectra. Using the average value $\bar\Lambda_{\rm pole}=(0.375\pm
0.065)$\,GeV obtained from
\cite{Chen:2001fj,Cronin-Hennessy:2001fk,Mahmood:2002tt,Aubert:2003dr}, 
we find $m_b^{\rm SF}(\mu_i,\mu_i)=(4.67\pm 0.07)$\,GeV. It is quite
remarkable that these different determinations of the shape-function mass,
which use rather different physics input, give highly consistent results.
Combining them, we quote our default value for the shape-function mass
at the intermediate scale $\mu_i=1.5$\,GeV as
\begin{equation}\label{mbdefault}
   m_b^{\rm SF}(\mu_i,\mu_i) = (4.65\pm 0.07)\,\mbox{GeV} \,.
\end{equation}
The corresponding $\bar\Lambda$ parameter is 
$\bar\Lambda(\mu_i,\mu_i)=(0.63\pm 0.07)$\,GeV.

A value of the kinetic-energy parameter in the shape-function scheme
can be obtained from (\ref{mupi2pole}) or (\ref{mupi2kin}). Using the
first relation and the experimental value $-\lambda_1=(0.25\pm
0.06)$\,GeV$^2$
\cite{Cronin-Hennessy:2001fk,Mahmood:2002tt,Aubert:2003dr} yields
$\mu_\pi^2(\mu_i,\mu_i)=(0.271\pm 0.064)$\,GeV$^2$. Alternatively, we may 
use the result for the kinetic-energy parameter obtained in the kinetic 
scheme, $[\mu_\pi^2(1\,\mbox{GeV})]_{\rm kin}=(0.45\pm 0.10)$\,GeV$^2$ 
\cite{Benson:2003kp}, to get from (\ref{mupi2kin}) the value
$\mu_\pi^2(\mu_i,\mu_i)=(0.254\pm 0.107)$\,GeV$^2$. Again, the two
determinations are in very good agreement with each other. Combining 
them, we obtain
\begin{equation}\label{mupi2default}
   \mu_\pi^2(\mu_i,\mu_i) = (0.27\pm 0.07)\,\mbox{GeV}^2 \,.
\end{equation}

\subsection{Model shape functions}

In our analysis of decay rates below, we will adopt a model for the
shape function $\hat S(\hat\omega,\mu_i)$ at the intermediate
scale. For the purpose of illustration, we use a two-component ansatz
for the shape function that is a generalization of the model employed
in \cite{DeFazio:1999sv,Kagan:1998ym}. The form we propose is
\cite{Bosch:2004th}
\begin{equation}\label{SFmodel}
   \hat S(\hat\omega,\mu)
   = \frac{N}{\Lambda} \left( \frac{\hat\omega}{\Lambda} \right)^{b-1}
   \!\!\! \exp\left( - b\,\frac{\hat\omega}{\Lambda} \right)
   - \frac{C_F\alpha_s(\mu)}{\pi}\,
   \frac{\theta(\hat\omega-\Lambda-\mu/\sqrt{e})}{\hat\omega-\Lambda}
   \left( 2\ln\frac{\hat\omega-\Lambda}{\mu} + 1 \right) ,
\end{equation}
where $\Lambda$ and $b$ are model parameters, and $\Lambda$ differs
from the pole-scheme parameter $\bar\Lambda_{\rm pole}$ by an amount
of $O(\alpha_s(\mu))$. In the limit $\alpha_s(\mu)\to 0$ this function
reduces to the familiar model used in
\cite{DeFazio:1999sv,Kagan:1998ym}. The radiative tail ensures the
correct leading asymptotic behavior of the shape function as displayed
in (\ref{Sasymp}). This in turn gives the correct power-like
dependence of shape-function moments on the integration cutoff. In our
model, this tail is glued onto a ``primordial'', exponential function
such that the combined result is continuous. The normalization factor
$N$ is given by
\begin{equation}
   N = \left[ 1 - \frac{C_F\alpha_s(\mu)}{\pi}
   \left( \frac{\pi^2}{24} - \frac14 \right) \right]
   \frac{b^b}{\Gamma(b)} ,
\end{equation}
which is determined such that the integral over the shape function
from $\hat\omega=0$ to $\mu_f+\bar\Lambda(\mu_f,\mu)$ coincides with
the first expression in (\ref{M0toM2}) up to second-order
corrections. By evaluating the first moment of the model shape
function, we find that the model parameter $\Lambda$ is related to the
HQET parameter $\bar\Lambda$ in the pole scheme and the shape-function
scheme as
\begin{equation}\label{Lambdavalues}
   \Lambda = \bar\Lambda_{\rm pole}
   + \frac{C_F\alpha_s(\mu)}{\pi}\,\frac{2\mu}{\sqrt e}\;,\quad
   \Lambda = \bar\Lambda(\mu_i,\mu_i)
   + \mu_i \left( \frac{2}{\sqrt e} - 1 \right)
   \frac{C_F\alpha_s(\mu_i)}{\pi}\;.
\end{equation}
Finally, the model parameter $b$ can be adjusted to reproduce a given 
value for the second moment of the shape function. 

\begin{table}[t!]
\centerline{\parbox{14cm}{\caption[Parameters and moments of the model 
shape functions at the intermediate scale. ]{\label{tab:SFproperties}
Parameters and moments of the model shape functions at the
intermediate scale $\mu_i$. The running quantities $m_b^{\rm SF}$,
$\bar\Lambda$, and $\mu_\pi^2$ are defined in the shape-function
scheme and evaluated at $\mu_f=\mu=\mu_i=1.5$\,GeV.}}}
\vspace{0.1cm}
\begin{center}
{\tabcolsep=0.3cm
\begin{tabular}{|cc|ccc|cc|}
\hline\hline
Model & Lines & $m_b^{\rm SF}$\,[GeV] & $\bar\Lambda$\,[GeV]
 & $\mu_\pi^2$\,[GeV$^2$] & $\Lambda$\,[GeV] & $b$ \\
\hline\hline
S1 & Dotted & 4.72 & 0.56 & 0.20 & 0.611 & 2.84 \\
S2 & & & & 0.27 & 0.617 & 2.32 \\
S3 & & & & 0.34 & 0.626 & 1.92 \\
\hline
S4 & Solid & 4.65 & 0.63 & 0.20 & 0.680 & 3.57 \\
S5 & & & & 0.27 & 0.685 & 2.93 \\
S6 & & & & 0.34 & 0.692 & 2.45 \\
\hline
S7 & Dashed & 4.58 & 0.70 & 0.20 & 0.751 & 4.40 \\
S8 & & & & 0.27 & 0.753 & 3.61 \\
S9 & & & & 0.34 & 0.759 & 3.03 \\
\hline\hline
\end{tabular}}
\end{center}
\end{table}

\begin{figure}[!ht]
\begin{center}
\epsfig{file=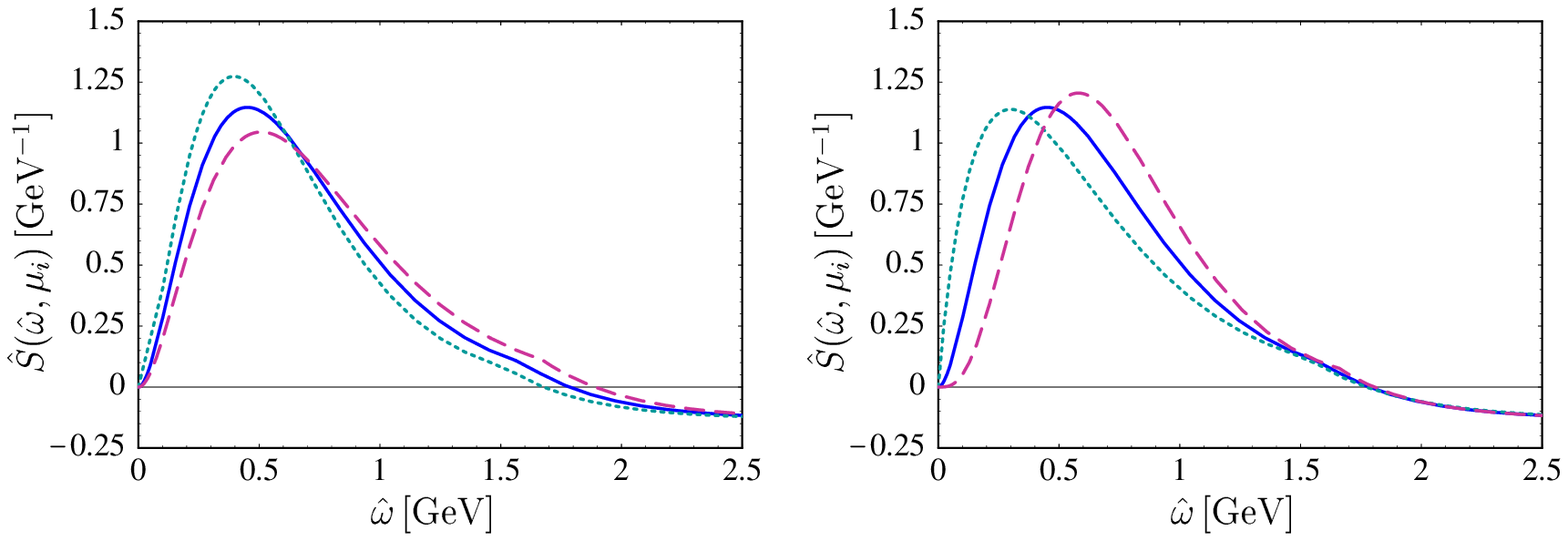,width=14cm}
\end{center}
\centerline{\parbox{14cm}{\caption[Shape function models under variation 
of its parameter settings.]{\label{fig:SFatmui} Various models for the
shape function at the intermediate scale $\mu_i=1.5$\,GeV,
corresponding to different parameter settings in
Table~\ref{tab:SFproperties}. Left: Functions S1, S5, S9 with
``correlated'' parameter variations. Right: Functions S3, S5, S7 with
``anti-correlated'' parameter variations.}}}
\vspace{0.3cm}
\begin{center}
\epsfig{file=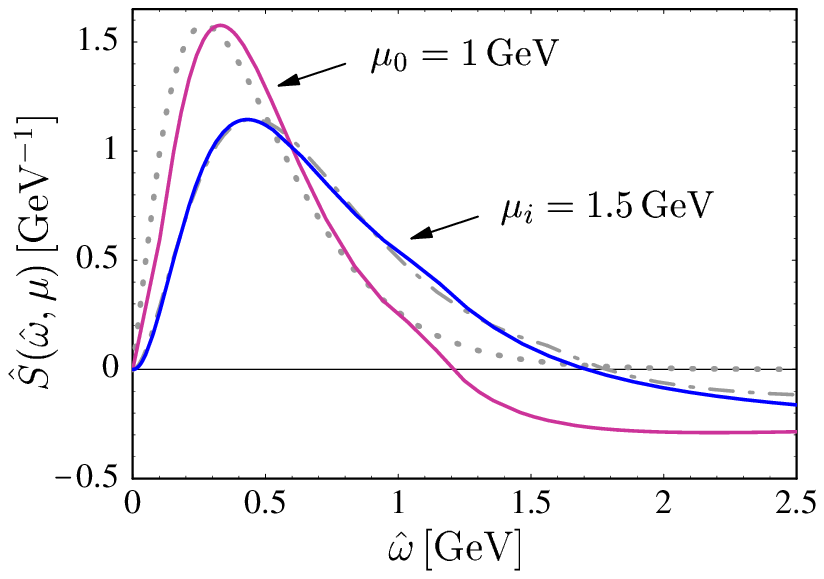,width=7cm}
\end{center}
\centerline{\parbox{14cm}{\caption[Renormalization-group evolution of a 
model shape function.]{\label{fig:SFrge} Renormalization-group
evolution of a model shape function from a low scale $\mu_0$ (sharply
peaked solid curve) to the intermediate scale $\mu_i$ (broad solid
curve). See the text for an explanation of the other curves.}}}
\end{figure}

Table~\ref{tab:SFproperties} collects the parameters of the model
shape functions at the intermediate scale $\mu_i=1.5$\,GeV
corresponding to different values of $\bar\Lambda(\mu_i,\mu_i)$ and
$\mu_\pi^2(\mu_i,\mu_i)$. The left-hand (right-hand) plot in
Fig.~\ref{fig:SFatmui} shows three models for the shape function
obtained by varying the parameters $\bar\Lambda$ and $\mu_\pi^2$ in a
correlated (anti-correlated) way. In both cases, the solid, dashed,
and dotted curves refer to different values of $\bar\Lambda$, as
indicated in the table.
In Fig.~\ref{fig:SFrge} we illustrate the renormalization-group
evolution of the shape function as studied in
Chapter~\ref{chap:strucfunc}. The sharply peaked solid line shows our
model function evaluated with $\Lambda=0.495$\,GeV and $b=3.0$, which
we use as an ansatz for the function $\hat S(\hat\omega,\mu_0)$ at the
low scale $\mu_0=1$\,GeV. (For comparison, the dotted gray curve shows
the default choice for the shape function adopted in
\cite{DeFazio:1999sv,Kagan:1998ym}, which exhibits a very similar shape
except for the missing radiative tail.) The broad solid curve gives
the shape function at the intermediate scale $\mu_i=1.5$\,GeV as
obtained from the evolution equation (\ref{wow}). The barely visible
dashed-dotted curve shows our default model for the function $\hat
S(\hat\omega,\mu_i)$, which coincides with the solid line in the
left-hand plot. The beautiful agreement of the two curves gives us
confidence in the consistency of our models adopted for the shape
function at the intermediate scale.

\subsection{Predictions for decay spectra and event fractions}

We are now ready to present our results for the decay spectra and
partially integrated event fractions in $\bar B\to X_u\,l^-\bar\nu$
decays. In order to illustrate the sensitivity to shape-function
effects we use all nine shape functions S1 through S9 in
Table~\ref{tab:SFproperties}, thus varying the parameters
$\bar\Lambda(\mu_i,\mu_i)$ and $\mu_\pi^2(\mu_i,\mu_i)$
independently. This is quite conservative because we neglect any
possible correlation between them. For each physical quantity we draw
three bands corresponding to the three different values of
$\bar\Lambda$. The width of each band reflects the sensitivity to the
variation of $\mu_\pi^2$.

The following predictions for spectra and rate fractions refer to the
leading term in the heavy-quark expansion. We note that our
calculations would break down if the cuts on kinematic variables were
taken to be too strict, because then the spectra would become
dominated by hadronic resonance effects. Parametrically, this happens
when the quantities $\Delta_P$, $\Delta_s$, or $\Delta_E$ become of
order $\Lambda_{\rm QCD}^2/M_B\sim 50$\,MeV. On the other hand, taking
too large values of $\Delta_P$, $\Delta_s$, or $\Delta_E$ such that
they are not of $O(\Lambda_{\rm QCD})$ anymore invalidates the
collinear expansion.

\begin{figure}[t]
\begin{center}
\epsfig{file=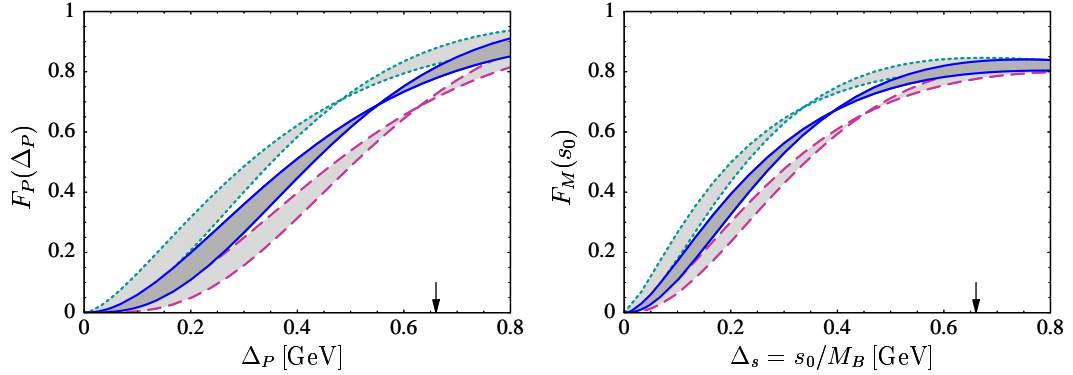,width=14cm}
\end{center}
\centerline{\parbox{14cm}{\caption[Event fractions for a $P_+$ cut and for 
a hadronic invariant mass cut.]{\label{fig:FPandFM} Fraction of $\bar
B\to X_u\,l^-\bar\nu$ events with hadronic light-cone momentum
$P_+\le\Delta_P$ (left), and fraction of events with hadronic
invariant mass $s_H\le s_0$ (right). In each plot, the three bands
correspond to the values $\bar\Lambda=0.63$\,GeV (solid curves),
0.70\,GeV (dashed curves), and 0.56\,GeV (dotted curves). Their width
reflects the sensitivity to the value of $\mu_\pi^2$ varied in the
range between 0.20 and 0.34\,GeV$^2$. The arrow indicates the point at
which the charm background starts.}}}
\end{figure}

In Fig.~\ref{fig:FPandFM} we show predictions for the fractions of all
$\bar B\to X_u\,l^-\bar\nu$ events with hadronic light-cone momentum 
$P_+\le\Delta_P$, and with hadronic invariant mass squared $s_H\le s_0$. 
Recall that, for $\Delta_P=\Delta_s=s_0/M_B$, the hadronic invariant mass 
fraction $F_M$ differs from the fraction $F_P$ by the contribution of the 
events in the triangular region above the dashed line in 
Fig.~\ref{fig:phase}. Comparing the two plots, we observe that this 
additional contribution is predicted to be very small. (Note that for 
large values of $\Delta_s$ we even find a negative contribution to the 
rate from the triangle region for some choices of the shape function. 
This feature is unphysical and should be fixed by the inclusion of 
power corrections to our leading-order predictions.) The arrows on the 
horizontal axes indicate the points $\Delta_{P,s}=M_D^2/M_B$, beyond 
which final states containing charm hadrons are kinematically allowed. 
With this choice of the cut, both rate fractions capture about 80\% of 
all events. While it is well known that a hadronic invariant mass cut 
$\sqrt{s_H}\le M_D$ provides a very efficient discrimination against 
charm background \cite{Bigi:1997dn,Falk:1997gj,Jezabek:2001pg}, here we 
observe that the same is true for a cut on the $P_+$ variable. 

\begin{figure}
\begin{center}
\epsfig{file=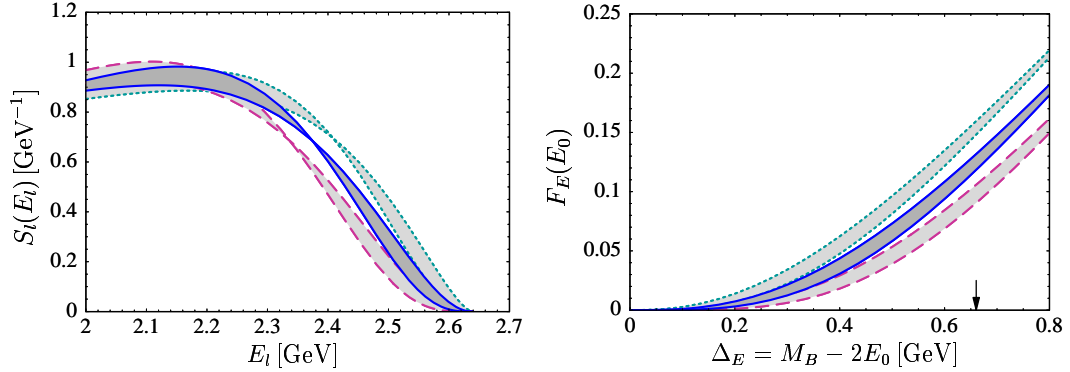,width=14cm}
\end{center}
\centerline{\parbox{14cm}{\caption[Event fraction for a cut on the 
charged-lepton energy.]{\label{fig:FE} Charged-lepton energy spectrum
in the region near the kinematic endpoint (left), and fraction of
events with charged-lepton energy $E_l\ge E_0$ (right). The meaning of
the bands and the arrow is the same as in Fig.~\ref{fig:FPandFM}.}}}
\end{figure}

Our results for the lepton energy spectrum
$S_l(E_l)=(1/\Gamma_{\rm tot})\,(d\Gamma/dE_l)$, and for the event
fraction with a cut $E_l\ge E_0$, are displayed in Fig.~\ref{fig:FE}.
The right-hand plot shows that with $\Delta_E=M_D^2/M_B$ only about
10--15\% of all events are retained, and the theoretical calculation
is very sensitive to shape-function effects. Such a cut is therefore
much less efficient than the cuts on $s_H$ or $P_+$. As a result, an
extraction of $|V_{ub}|$ from the charged-lepton endpoint region is
theoretically disfavored.

The shape-function sensitivity is rather small for values of
$\Delta_P$ and $\Delta_s$ near the charm threshold. This is to some
extent a consequence of our improved knowledge of the shape-function
parameters. On the other hand we observe an interesting ``focus
mechanism'' in that the three bands in the $P_+$ and the hadronic
invariant mass (but {\em not} for the charged lepton energy) event
fractions start to converge near the charm threshold. This is due to a
subtle interplay between the jet function and the shape
function, which is explained in more detail in
\cite{Bosch:2004th}. The important observation is that the leading
logarithm in the jet function has the opposite sign as found in a
straightforward partonic calculation. The convergence is somewhat
surprising since the model shape functions at the intermediate scale
$\mu_i$ are rather different for values around $\hat \omega \sim
\Lambda_{\rm QCD}$ (see Fig.~\ref{fig:SFatmui}), and the argument
that they share the same norm is not applicable yet. One way of
thinking about this mechanism is to notice that the broadening of the
shape function under renormalization-group evolution from a low scale
up to the intermediate scale (Fig.~\ref{fig:SFrge}) is a
perturbative effect, which should not lead to an increased
shape-function sensitivity. Because the convolution of the shape
function with the jet and hard functions is independent of the scale
$\mu_i$, the broadening of the shape function must be compensated by
perturbative logarithms in the jet function.

This focus effect did not take place in earlier studies such as
\cite{DeFazio:1999sv,Bigi:1997dn,Falk:1997gj,Jezabek:2001pg,Dikeman:1997es}, 
where parton-model spectra were convoluted with a primordial shape 
function. As mentioned earlier, in the parton model the leading Sudakov 
logarithm comes with the opposite (negative) sign, hence causing an 
anti-focus effect of the radiative corrections. This also explains 
why our prediction for the hadronic invariant mass fraction $F_M$ 
exhibits a smaller shape-function sensitivity than what has been found 
in most previous analyses.

\begin{figure}
\begin{center}
\epsfig{file=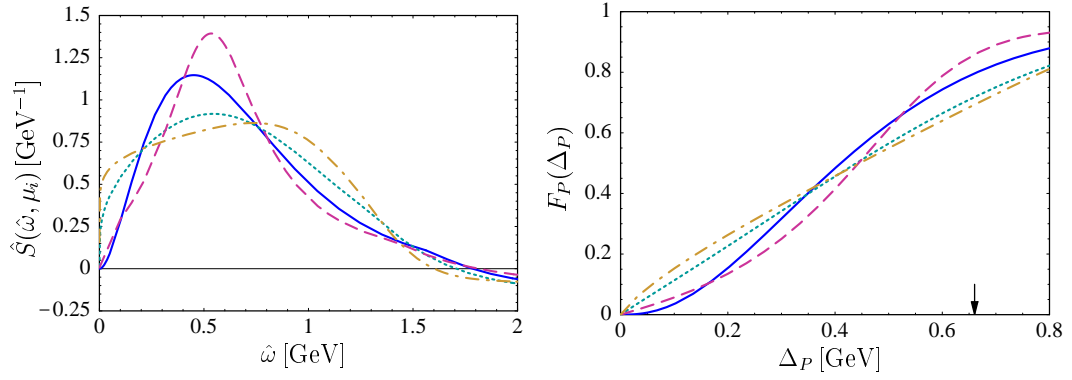,width=14cm}
\end{center}
\centerline{\parbox{14cm}{\caption[Various different functional forms for 
the shape function.]{\label{fig:new} Left: Four examples of shape
functions with identical normalization and first two moments,
corresponding to $\bar\Lambda(\mu_i,\mu_i)=0.63$\,GeV and
$\mu_\pi^2(\mu_i,\mu_i)=0.27$\,GeV$^2$, but different functional
form. Right: Corresponding results for the event fraction
$F_P(\Delta_P)$.}}}
\end{figure}

We shall stress again that the knowledge of the first few moments does
not determine the functional form of the shape function.  In order to
demonstrate this, we have constructed several functions of different
functional forms, but with identical norm, first, and second
moment. To correctly account for the asymptotic behaviour $\hat\omega
\gg \Lambda_{\rm QCD}$, we have added the second term in
(\ref{SFmodel}) to each of the functions. They are shown on the
left-hand side of Fig.~\ref{fig:new}. All of them are thus allowed
by the required moment analysis and could be considered as variations
of the shape function. Some of the curves are admittedly extreme,
given the fact that the $b\to s\gamma$ photon spectrum is identical to
the shape function at tree level. Any information about the shape
function can be used to eliminate some of these curves. At present,
however, we stress that using the model shape functions S1 through S9
sample even the variation of the functional form, as the event
fraction predictions on the right-hand side of Fig.~\ref{fig:new}
fall within the bands studied earlier.

\begin{table}[t!]
\centerline{\parbox{14cm}{\caption[Numerical predictions of event 
fractions for different experimental cuts.]{\label{tab:comp}
Comparison of different theoretical methods using inclusive $B$-decay
rates to extract the CKM matrix element $|V_{ub}|$. The error on the
efficiency represents the sensitivity to the shape function only. All
results refer to the leading term in the heavy-quark expansion.}}}
\vspace{0.1cm}
\begin{center}
{\tabcolsep=0.3cm\renewcommand{\arraystretch}{1.35} 
\begin{tabular}{|c|c|c|}
\hline\hline
Method & Cut & Efficiency \\
\hline\hline
Hadronic invariant mass & $s_H\le M_D^2$ & $(81.4_{\,-3.7}^{\,+3.2})\%$ \\
 & $s_H\le(1.7\,\mbox{GeV})^2$ & $(78.2_{\,-5.2}^{\,+4.9})\%$ \\
 & $s_H\le(1.55\,\mbox{GeV})^2$ & $(72.7_{\,-6.3}^{\,+6.4})\%$ \\
\hline
Hadronic $P_+$ & $P_+\le\frac{M_D^2}{M_B}=0.66$\,GeV
 & $(79.6_{\,-8.2}^{\,+8.2})\%$ \\
 & $P_+\le 0.55$\,GeV & $(69.0_{\,-12.1}^{\,+\phantom{1}9.7})\%$ \\
\hline
Charged-lepton energy & $E_l\ge\frac{M_B^2-M_D^2}{2M_B}=2.31$\,GeV
 & $(12.5_{\,-3.5}^{\,+3.4})\%$ \\
 & $E_l\ge 2.2$\,GeV & $(22.2_{\,-3.6}^{\,+3.2})\%$ \\
\hline
Combined ($s_H$, $q^2$) cuts & $s_H\le M_D^2$, ~$q^2\ge 0$
 & $(74.6_{\,-5.1}^{\,+5.1})\%$ \\
\mbox{[tree level only]} & $s_H\le M_D^2$, ~$q^2\ge 6\,\mbox{GeV}^2$
 & $(45.7_{\,-2.0}^{\,+1.8})\%$ \\
 & $s_H\le(1.7\,\mbox{GeV})^2$, ~$q^2\ge 8\,\mbox{GeV}^2$
 & $(33.4_{\,-1.8}^{\,+1.6})\%$ \\
\hline\hline
\end{tabular}}
\end{center}
\end{table}

We summarize our phenomenological results in Table~\ref{tab:comp}, in
which we compare the shape-function sensitivity (which is the only
uncertainty considered here) of different kinematic cuts. The first
two blocks of entry address the high-efficiency methods of cutting on
the hadronic invariant mass $s_H$ and the variable $P_+$. The
efficiency for the optimal cut $s_H \le M_D^2$ is remarkably precise;
however, due to detector resolution effects it is necessary to relax
that cut somewhat. We state results for upper limits on the invariant
mass of $1.7$ GeV and $1.55$ GeV, for which the relative uncertainty
increases somewhat. Cutting on $P_+$ introduces a relative uncertainty
of order 10\%, which is still acceptable. Again, experimental
constraints might force us to move away from the optimal cut. However,
it might be possible to stay closer to the optimal cut due to the
apparent ``buffer zone''. (We will discuss the charm background in
much more detail below.)

The only low-efficiency method we discuss is a cut on the
charged-lepton energy. Unlike the previous two methods, such a
measurement does not require the reconstruction of the neutrino in the
decay $\bar B \to X_u\,l^-\bar\nu$, and is therefore favored by
experiment. However, due to the low efficiency one has to worry also
about other theoretical uncertainties like weak annihilation effects
\cite{Voloshin:2001xi}, which are expected to contribute less than 3\%
to the total rate and can therefore be safely neglected in
high-efficiency methods like the $s_H$ and $P_+$ cuts. Our prediction for
the lepton energy event fractions are significantly larger than have
been reported in the past, and suggest that the extracted value of
$|V_{ub}|$ needs to be corrected to lower values.

For completeness, we also give results for some combined cuts 
on hadronic and leptonic invariant mass, which have been briefly 
discussed in Section~\ref{sec:comb}. Contrary to the other cases, these 
numbers refer to the tree-level approximation and so should be taken with 
caution. For reference, we quote again the (tree-level) result for the 
pure hadronic invariant mass cut, which differs significantly from the 
corresponding result including radiative corrections. While the 
additional cut on leptonic $q^2$ reduces the shape-function sensitivity, 
it comes along with a strong reduction of the efficiency. For instance, 
the combined cut $\sqrt{s_H}\le 1.7$\,GeV and $q^2\ge 8$\,GeV$^2$ 
employed in a recent analysis of the Belle Collaboration 
\cite{Kakuno:2003fk} has an efficiency of about 33\% (at tree level and 
leading order in $\Lambda_{\rm QCD}/m_b$), which is much smaller than the 
efficiency of the pure hadronic invariant mass cut 
$\sqrt{s_H}\le 1.7$\,GeV. However, the sensitivity to shape-function 
effects is only slightly better in the case of the combined cut.

\section{Charm background}

One of the main advantages of the $P_{+}$ spectrum over the
hadronic-mass spectrum is a better control of the charm background. In
order to study this in more detail, we investigate the OPE prediction
for the normalized $\hat p_+ = n\cdot p/m_b$ spectrum in $\bar B\to
X_c\,l^-\bar\nu$ decays,
\begin{equation}\label{charmOPE}
   \frac{1}{\Gamma_c}\,\frac{d\Gamma_c}{d\hat p_+}
   = \frac{2(\varrho-\hat p_+^2)^2}{f(\varrho)\,\hat p_+^5}
    \Big[ \hat p_+^3 (3 - 2\hat p_+) 
   \mbox{}+ \varrho \,\hat p_+ (3 - 8\hat p_+ + 3\hat p_+^2)
    - \varrho^2 (2 - 3\hat p_+) \Big] \,,
\end{equation}
where $\varrho\le\hat p_+\le\sqrt{\varrho}$ with
$\varrho=(m_c/m_b)^2$, and
\begin{equation}
   f(\varrho) = 1 - 8\varrho + 8\varrho^3 - \varrho^4
   - 12\varrho^2\ln\varrho \,.
\end{equation}
We only include tree-level contributions from dimension-3 operators
\cite{Bosch:2004bt}. 

In this approximation, inclusive charm events are located along a
single line $p_+p_-=m_c^2$, or in terms of the hadronic variables
$(P_+-\bar\Lambda)(P_--\bar\Lambda)=(M_D-\bar\Lambda)^2$, in
Fig.~\ref{fig:phase}. This line starts at the tip of the triangle
$P_+=P_-=M_D$ and extends to the right while always staying above the
solid line $s_H=M_D^2$. Because ${\cal O}(\alpha_{s})$ corrections
redistribute these events into the light-gray segment above that line,
the integral over the spectrum in (\ref{charmOPE}) serves as an upper
bound on the inclusive charm background. In the same approximation,
the hadronic mass distribution is given by
\begin{equation}\label{FMOPE}
   \frac{1}{\Gamma_c}\,\frac{d\Gamma_c}{d\hat s}
   = \frac{2}{f(\varrho)\,\varepsilon} \sqrt{z^2-4\varrho}\,
    \Big[ z(3-2z) - \varrho(4-3z) \Big] \,, \quad 
   \mbox{with~~}
   z = \frac{\hat s-\varepsilon^2-\varrho}{\varepsilon} \,.
\end{equation}
Here, $\hat s=s_H/m_b^2$, $\varepsilon=\bar\Lambda/m_b$, and phase
space is such that $(\sqrt{\varrho}+\varepsilon)^2\le\hat s
\le(\varrho+\varepsilon)(1+\varepsilon)$.

\begin{figure}[t]
\begin{center}
\epsfxsize=15.0cm
\epsfig{file=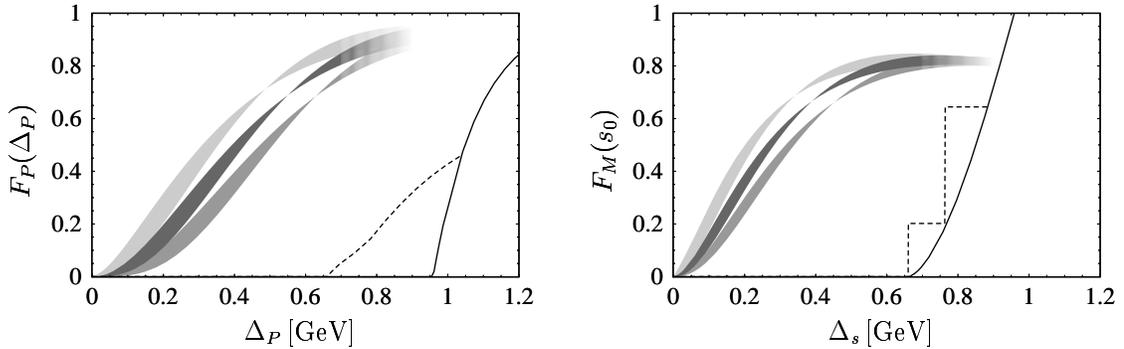, width=14.7cm}
\vspace{-0.3cm}
\parbox{14cm}{\caption[Event fractions and charm background in a 
single plot.]{\label{fig:Fcharm} Fraction of $\bar B\to
X_{u}\,l^{-}\nu$ events with $P_{+}\le\Delta_{P}$ (left) and $s_{H}\le
s_{0}$ (right). The bands are identical to the ones in
Fig.~\ref{fig:FPandFM}. The curves represent the background from
inclusive $\bar B\to X_{c}\,l^-\bar\nu$ (solid) and exclusive $\bar
B\to D^{(*)} l^-\bar\nu$ decays (dashed), normalized to the total
inclusive semileptonic charm rate.}}
\end{center}
\end{figure}

The results for the event fractions $F_{P}$ and $F_{M}$ are summarized
in Fig.~\ref{fig:Fcharm}. The bands show the inclusive $\bar B\to
X_u\,l^-\bar\nu$ event fractions for different shape-function models
as discussed above, while the solid lines give predictions for the
inclusive $\bar B \to X_c\, l^- \bar \nu$ background spectra as
obtained from equations.~(\ref{charmOPE}) and (\ref{FMOPE}). The
latter are normalized to their total decay rate $\Gamma_c$, which is
about 60 times larger than $\Gamma_u$. In the case of the fraction
$F_{M}$, the inclusive charm background starts right at the threshold
$\Delta_{s}=(m_{c}+\bar\Lambda )^{2}/M_{B}\simeq M_{D}^{2}/M_{B}$. In
the case of the $F_{P}$ event fraction, on the contrary, the inclusive
charm background starts at a value
$\Delta_{P}=m_{c}^{2}/m_{b}+\bar\Lambda\approx 0.96\,{\rm GeV}$, which
is significantly larger than the value
$\Delta_{P}=M_{D}^{2}/M_{B}\approx 0.66\,{\rm GeV}$ above which final
states containing charm mesons are kinematically allowed. This is a
consequence of the fact that there is a gap between the charm
threshold $s_H=M_D^2$ and the start of the inclusive $b\to c$
events. In reality this gap is filled by exclusive modes containing
the hadronic final states $D$, $D^*$, or $D \pi$, $D\pi\pi$. The
exclusive contributions to the hadronic $P_+$ spectrum from $\bar B\to
D^{(*)} l^-\bar\nu$ decays are
\begin{eqnarray}
   \frac{d\Gamma_D}{dP_+}
   &=& \frac{G_F^2|V_{cb}|^2 M_B^5}{48\pi^3}\,
    \frac{(1+r)^2\,r^3}{P_+}\,(w^2-1)^2\,|{\cal F_D}(w)|^2 \,, \nonumber\\
   \frac{d\Gamma_{D^*}}{dP_+}
   &=& \frac{G_F^2|V_{cb}|^2 M_B^5}{48\pi^3}\,
    \frac{(1-r_*)^2\,r_*^3}{P_+}\,(w^2-1)\,(w+1)^2 \nonumber\\
   &\times& \left( 1 + \frac{4w}{w+1}\,
    \frac{1 - 2w r_* + r_*^2}{(1-r_*)^2} \right)
    |{\cal F_{D^*}}(w)|^2 \,,
\end{eqnarray}
where $M_{D^{(*)}}^2/M_B\le P_+\le M_{D^{(*)}}$ and
$r_{(*)}=M_{D^{(*)}}/M_B$. The recoil variable $w = v\cdot v^\prime$ is
given by $M_{D^{(*)}}^{2}+P_{+}^{2}=2w\,M_{D^{(*)}}P_{+}$. The
corresponding contributions to the hadronic mass spectrum are given by
\begin{equation}
   \frac{d\Gamma_{D^{(*)}}}{ds_H} = \Gamma_{D^{(*)}} \cdot
   \delta(s_H-M_{D^{(*)}}) \,.
\end{equation}
The exclusive contributions to the event fractions are given by the
dashed lines in Fig.~\ref{fig:Fcharm}. To obtain these curves we
have used an ansatz for the form factors ${\cal F}_{D^{(*)}}(w)$ that
is consistent with experimental data on the recoil spectra and
branching fractions. The fact that these exclusive modes are very well
understood should help to model the background. The smooth onset of
the $D^{(*)}$ background is a direct consequence of the fact that the
ideal $P_+$ cut touches the charm region at only a single point in
phase space (see Fig. \ref{fig:phase}). On the contrary, the region
of phase space when applying the $s_{H}$ cut borders the background
along the curve separating the light- and dark-shaded regions, which
leads to a step increment in the event fraction $F_{M}$. As a
consequence, one needs to move away from the ideal cut $s_H=M_D^2$
because of smearing effects due to experimental resolution. It is our
hope that it will be possible to stay closer to the ideal cut when
performing a $P_+$ analysis. In this case, both the $P_+$ and $s_H$
discrimination methods lead to comparable efficiencies and
shape-function uncertainties, as emphasized earlier in the discussion
after Table~\ref{tab:comp}.

\section{Theoretical accuracy of a $\boldsymbol{|V_{ub}|}$ measurement}

As a final remark to this long Chapter, let us comment on the
applicability of the theoretical framework developed in this
work. According to Fig.~\ref{fig:phase} most of the $\bar B\to
X_u\,l^-\bar\nu$ events are located in the shape-function region of
large $P_-$ and small to moderate $P_+$, in which a systematic
heavy-quark expansion using SCET power counting is valid. It allows us
to calculate inclusive decay rates integrated over domains $\Delta
P_-\sim M_B$ and $\Delta P_+\ll M_B$, where typically $\Delta
P_+\sim\Lambda_{\rm QCD}$. (In the examples above, $\Delta
P_+=\Delta_E$, $\Delta_P$, or $\Delta_s$, respectively.)  While the
corresponding predictions for decay spectra and event fractions are
sufficient to analyze experimental data over most of the phase space
relevant to measurements of the CKM matrix element $|V_{ub}|$, it
would be of interest to extend the validity of the theoretical
description outside the shape-function region. Let us in the following
paragraph comment on such a prospect.

In the case where $\Delta P_-\sim\Delta P_+\sim M_B$ are both large,
the decay spectra can be computed using a local operator product
expansion. The resulting prediction for the normalized $\bar B\to
X_u\,l^-\bar\nu$ spectrum in the variable $\hat
p_+=p_{+}/m_{b}=(P_{+}-\bar\Lambda)/m_b$ reads \cite{Bosch:2004bt}
\begin{eqnarray}
   \frac{1}{\Gamma_u}\,\frac{d\Gamma_u}{d\hat p_+}
   &=& \left( 1 - \frac{463}{36}\,\frac{\alpha_s}{3\pi} \right)
    \delta(\hat p_+) \nonumber\\
   &+& \frac{\alpha_s}{3\pi} \left[
    - 4 \left( \frac{\ln\hat p_+}{\hat p_+} \right)_{\!*}
    - \frac{26}{3} \left( \frac{1}{\hat p_+} \right)_{\!*}
    + h(\hat p_+) \right] \nonumber\\
   &-& \left( \frac{17\lambda_1}{18 m_b^2}
    + \frac{3\lambda_2}{2m_b^2} \right) \delta^\prime(\hat p_+)
    - \frac{\lambda_1}{6m_b^2}\,\delta^{\prime\prime}(\hat p_+) \,,
\end{eqnarray}
where $0\le\hat p_+\le 1$, and
\begin{eqnarray}
   h(p)
   &=& \frac{158}{9} + \frac{407p}{18} - \frac{367p^2}{6}
    + \frac{118p^3}{3} - \frac{100p^4}{9} 
    + \frac{11p^5}{6} - \frac{7p^6}{18} \nonumber\\
   &-& \left( \frac{4}{3} - \frac{46p}{3} - 6p^2
    + \frac{16p^3}{3} \right) \ln p 
    - 4p^2(3-2p) \ln^2 p \,.
\end{eqnarray}
We include the contributions from dimension-3 operators at
$O(\alpha_s)$ and those from dimension-5 operators (whose matrix
elements are proportional to the heavy-quark effective theory
parameters $\lambda_{1,2}$) at tree level, using
\cite{Manohar:1993qn,DeFazio:1999sv}. If the $\hat p_+$ spectrum is
integrated without a weight function, the tree-level power corrections
from dimension-5 operators vanish. This is in accordance with the fact
that subleading shape functions have zero norm at tree level
\cite{Bauer:2002yu}.

An interesting question is whether it will be possible to match the
Factorization approach for the shape-function region $\Delta P_+ \sim
\Lambda_{\rm QCD}$ and the OPE approach for $\Delta P_+ \sim M_B$ in
some intermediate region of $\Delta P_+$ values that are numerically
(but not parametrically) large compared with $\Lambda_{\rm QCD}$. If
the two predictions were to agree in an overlap region, this could be
used to construct a theoretical description of inclusive $\bar B\to
X_u\,l^-\bar\nu$ decay distributions that is valid over the entire
phase space. While this is an exciting prospect, we note that
performing a systematic operator product expansion in the overlap
region is far from trivial. For a hierarchy of scales $\Lambda_{\rm
QCD}\ll\Delta_P\ll m_b$ we obtain with
$\overline{\Delta}=\Delta_P-\bar\Lambda$
\begin{equation}\label{FPOPE}
   F_P(\Delta_P) = 1 - \frac{\alpha_s}{3\pi} \bigg[
    \left( 2\ln^2\frac{m_b}{\overline{\Delta}}
    - \frac{26}{3}\ln\frac{m_b}{\overline{\Delta}} + \frac{463}{36} \right)
    - \frac{\overline{\Delta}}{m_b} \left(
    \frac{4}{3}\,\ln\frac{m_b}{\overline{\Delta}} + \frac{170}{9} \right)
    \bigg] + \dots \,.
\end{equation}
For $\Delta_{P}\sim 1$\,GeV, the power correction in the second term
leads to an enhancement of the fraction $F_{P}$ by 5--10\%. This is of
similar magnitude as tree-level estimates of (zero-norm) subleading
shape-function effects on the $E_{l}$ and $s_{H}$ spectra
\cite{Neubert:2002yx,Burrell:2003cf}.

The leading-order term in Eq.~(\ref{FPOPE}) can also be obtained from
Eq.~(\ref{FP}) by taking the limit $\Delta_{P}\gg\Lambda_{\rm
QCD}$. Interestingly, such an analysis uncovers that in this kinematic
range there is an enhanced class of power corrections of the form
$(\Lambda_{\rm QCD}/\overline{\Delta})^n$ with $n\ge 2$, which arise
first at $O(\alpha_s)$ (see also \cite{Bauer:2003pi}). The leading
corrections to the expression above are given by \pagebreak
\begin{equation}
   \delta F_P(\Delta_P) = - \frac{\alpha_s}{3\pi}
   \left( 2\ln\frac{m_b}{\overline{\Delta}} - \frac{7}{3} \right)
   \frac{\mu_\pi^2}{3\overline{\Delta}^2} \,.
\end{equation}
It follows that for sufficiently large $\overline{\Delta}$ the event
fraction can be expressed as a double expansion in
$\overline\Delta/m_{b}$ and $\Lambda_{\rm
QCD}/\overline{\Delta}$. Because of the hierarchy of scales $\Delta
P_+^2\ll\Delta P_+\,M_B\ll M_B^2$, again a two-step procedure is in
order. (An example of such an approach can be found in
\cite{Neubert:2001ib}.)
Such a multi-scale OPE is non-trivial and left for future work. For
now we merely note the estimate that subleading power contributions to
the event fraction $F_P$ result to about 10\%. 

We are now ready to give an in-depth look at the prospect of
determining $|V_{ub}|$ from a $P_+$ measurement. A list of relevant
theoretical uncertainties to the fraction of events with the optimal
cut $P_{+}\le M_{D}^{2}/M_{B}$ is \cite{Bosch:2004bt}
\begin{equation}\label{FPnum}
   F_{P} = (79.6\pm 10.8\pm 6.2\pm 8.0)\% \,,
\end{equation}
where the errors represent the sensitivity to the shape function (as
stated in Table~\ref{tab:comp}), an estimate of ${\cal
O}(\alpha_{s}^{2})$ contributions, and power corrections,
respectively. The uncertainty due to our ignorance of the shape
function is obtained in part by varying the parameters $\bar\Lambda =
M_B-m_b$ (\ref{mbdefault}) and $\mu_{\pi}^{2}$ (\ref{mupi2default}) in
the shape-function scheme, determining its first two moments ($\delta
F_{P}=\phantom{}^{+8.2}_{-8.1}\%$). Furthermore we vary the functional
form of $S$, as depicted in Fig.~\ref{fig:new} ($\delta
F_{P}=\phantom{}^{+6.3}_{-7.8}\%$).
 
Higher-order perturbative effects are estimated by studying the
dependence on the matching scales $\mu_{i}$ and $\mu_{h}$, varied in
the ranges $1.25\,{\rm GeV}\le\mu_{i}\le 1.75\,{\rm GeV}$ ($\delta
F_{P}=\phantom{}^{+4.7}_{-4.9}\%$) and
$m_{b}/\sqrt{2}\le\mu_{h}\le\sqrt{2}\,m_{b}$ ($\delta
F_{P}=\phantom{}^{+4.1}_{-3.9}\%$). 

Every single error stated here can be reduced in the future within the
systematic framework presented in \cite{Bosch:2004th,Bosch:2004bt}. A
proper treatment of subleading power corrections in the shape-function
region is now feasible. In addition, the leading shape-function
uncertainty can be eliminated using model-independent relations such
as (\ref{reltoBsGamma}). Finally the perturbative uncertainty can be
reduced by computing the ${\cal O}(\alpha_{s}^{2})$ corrections to
equation~(\ref{FP}).

The CKM-matrix element $|V_{ub}|$ can be extracted by comparing a
measurement of the partial rate $\Gamma_{u}(P_{+}\le\Delta_{P})$ with
a theoretical prediction for the product of the event fraction $F_{P}$
and the total inclusive $\bar B\to X_{u} \, l^{-}\bar\nu$ rate. The
resulting theoretical uncertainty on $|V_{ub}|$ up to date is
\begin{equation}
   \frac{\delta |V_{ub}|}{|V_{ub}|} = (\pm 7\pm 4\pm 5\pm 4)\% \,,
\end{equation}
where the last error comes from the uncertainty in the total rate
(\ref{totalInclusiveRate}) as can be found in
\cite{Hoang:1998ng,Uraltsev:1999rr}. Because of the large efficiency
of the $P_{+}$ cut, weak annihilation effects \cite{Voloshin:2001xi}
have an influence on $|V_{ub}|$ of less than 2\% and can be safely
neglected.

QCD-Factorization has provided us with a powerful method to
systematically improve the calculation of event distributions in the
shape-function region. The hadronic $P_+$ spectrum is the preferred
application of this framework, and the possibility of a high-precision
determination of $|V_{ub}|$ through a measurement of the partial decay
rate $\Gamma_{u}(P_{+}\le\Delta_{P})$ is an exciting prospect.

\chapter{Exclusive Radiative Decays}\label{chap:radiative}

The radiative, semileptonic decay $B\to\gamma\,l\nu$ provides a clean
environment for the study of soft-collinear interactions
\cite{Korchemsky:1999qb}. This process is particularly simple in that
no hadrons appear in the final state. Yet, there is sensitivity to the
light-cone structure of the $B$ meson, probed by the coupling of the
high-energy photon to the soft spectator quark inside the heavy meson.
Therefore this decay mode serves as an opportunity to learn QCD
factorization and Sudakov resummation for exclusive decays in a
realistic environment \cite{Bosch:2003fc}. Our main goal is to
establish the QCD factorization formula \cite{Descotes-Genon:2002mw}
\begin{equation}\label{ff}
   {\cal A}(B^-\to\gamma\,l^-\bar\nu_l) 
   \propto M_B f_B\,Q_u \int_0^\infty\!dl_+\,
   \frac{\phi_+^B(l_+,\mu)}{l_+}\,T(l_+,E_\gamma,m_b,\mu)
\end{equation}
to all orders in perturbation theory and at leading power in
$\Lambda_{\rm QCD}/m_b$. Here $Q_u=\frac23$ is the electric charge of
the up-quark (in units of $e$), and $T=1+O(\alpha_s)$ is a
perturbative hard-scattering kernel. The decay constant $f_B$ and the
LCDA $\phi^B_+$ have been introduced in Section~\ref{sec:LCDA}. The
physics underlying the factorization formula is that a high-energy
photon coupling to the soft constituents of the $B$-meson produces
quantum fluctuations far off their mass shell, which can be integrated
out in a low-energy effective theory.  As usual we choose the photon
direction to be the $z$-direction, so that the photon momentum is
$E_\gamma n$. The two transverse polarization states of the photon can
be expressed in terms of the basis vectors
$\varepsilon_\mp^\mu=\frac{1}{\sqrt2}(0,1,\mp i,0)$, which correspond
to left- and right-circular polarization, respectively.

The new complication, as compared to the factorization of the {\em
inclusive} case of the last Chapter, is that the final theory is
SCET$_{II}$. Radiative decays are particularly simple in that the only
external ``collinear'' particle is the electromagnetic photon, which
carries a purely light-like momentum. Therefore the only hadronic
constituents present are the HQET fields describing quarks and gluons
of the $B$ meson.  While it is possible to match QCD directly onto
this theory, a two-step matching procedure through the intermediate
theory SCET$_I$ has the advantage of integrating out off-shell
propagators of order $E_\gamma^2$ and $E_\gamma \Lambda_{\rm QCD}$ in
separate steps. This suggests a second stage of ``perturbative''
factorization \cite{Hill:2002vw,Lunghi:2002ju,Bauer:2002aj}, which we
will establish below. It says that the hard-scattering kernel itself
can be factorized as
\begin{equation}\label{fact2}
   T(l_+,E_\gamma,m_b,\mu)
   = H\left(\frac{2E_\gamma}{\mu},\frac{2E_\gamma}{m_b}\right)
   \cdot J\left(\frac{2E_\gamma\,l_+}{\mu^2}\right) .
\end{equation}

Factorization holds as long as the photon is energetic in the
$B$-meson rest frame, meaning that $E_\gamma$ is of the order of the
$b$-quark mass. In order to prove the factorization formula (\ref{ff})
one needs to show that \cite{Beneke:2000ry}:

\begin{enumerate}
\item 
The decay amplitude can be expanded in powers of transverse momenta
(i.e. in powers of $\Lambda_{\rm QCD}/E_\gamma$) and, at leading order, can be
expressed in terms of a convolution with the $B$-meson LCDA as shown
in (\ref{ff}).
\item
After subtraction of infra-red contributions corresponding to the 
$B$-meson decay constant and LCDA, the leading contributions to the 
amplitude come from hard internal lines, i.e., the hard-scattering kernel 
$T$ is free of infra-red singularities to all orders in perturbation 
theory.
\item
The convolution integral of the hard-scattering kernel with the LCDA is 
convergent.
\item
Non-valence Fock states do not give rise to leading contributions.
\end{enumerate}

\noindent In the next Section we will address all of the above points.

\section{Proof of factorization}

In the full theory, the hadronic part of the decay amplitude for $B
\to \gamma\,l\nu$ is given by the time-ordered product of a weak,
flavor-changing current $(\bar u b)_{\rm V-A}$ and the
electro-magnetic currents $(\bar u\Aslash^{\rm (em)}u)$, $(\bar
b\Aslash^{\rm (em)}b)$. However, attachments of the energetic photon
field to the massive $b$-quark line leads to power suppression
\cite{Hill:2002vw}, and so at leading power in SCET, the time-ordered
product is matched onto trilocal operators of the form
\begin{eqnarray}\label{ops}
   &&\sum_{q=u,b} ie\,Q_q \int d^4x\,T\,\Big\{
    \big[ \bar u\gamma^\mu(1-\gamma_5) b \big](0),
    \big[ \bar q\,\Aslash^{\rm (em)}\,q \big](x) \Big\} \nonumber\\
   &\to& \sum_i \int ds\,dt\,\widetilde C_i(t,s,v\cdot q,m_b,\mu)\,
    \bar\Q_s(tn)\,\calAslash_{c\perp}^{({\rm em})}(s\bar n)\,
    \frac{\nslash}{2}\,\Gamma_i\,\H(0)  \,.
\end{eqnarray}
The Wilson coefficients $\widetilde C_i$ receive contributions only
from Feynman diagrams where the photon is emitted from the spectator
quark line. The fields $\Aslash^{\rm (em)}$ and $\Q_s$ are located at
the appropriate light cones, which reflects the SCET scaling
properties of the photon and spectator-quark
momenta. $\calAslash_{c\perp}^{({\rm em})}$ is the electromagnetic
analog of the gauge-invariant collinear gluon field defined in
(\ref{blocks}). To first order in $e$ we have
\begin{equation}
   \calAslash_{c\perp}^{({\rm em})}(0)
   = \bar n_\alpha\gamma_\mu^\perp
   \int_{-\infty}^0\!dw\,e F^{\alpha\mu}(w\bar n) \,.
\end{equation}
The Feynman rule for this object is simply $e\,\epsslash^*$, where 
$\varepsilon$ is the photon polarization vector. Finally, from the fact 
that the leptonic weak current $\bar\nu\gamma_\mu(1-\gamma_5)\,l$ is 
conserved (in the limit where the lepton mass is neglected) it follows 
that the relevant Dirac structures $\Gamma_i$ in (\ref{ops}) can be taken 
as $\Gamma_1=\gamma^\mu(1-\gamma_5)$ and $\Gamma_2=n^\mu(1+\gamma_5)$.

We can use the trace formalism in HQET and the definition of the
$B$-meson LCDA (\ref{BLCDA}). Performing the relevant traces, we find
that at leading power in $\Lambda_{\rm QCD}/m_b$ the decay amplitude
vanishes if the photon has right-circular polarization, while for a
photon with left-circular polarization it is given by
\begin{eqnarray}\label{ampl}
   {\cal A}(B^-\to\gamma_L\,l^-\bar\nu_l) 
   &=& \frac{iG_F}{\sqrt2}\,V_{ub}\,e\,
    \bar u_l(p_l)\,\epsslash_-^*(1-\gamma_5) v_\nu(p_\nu) \\
   &\times& \sqrt{M_B}\,F(\mu) \int d\omega\,
    C_1(\omega,\bar n\cdot q,v\cdot q,m_b,\mu)\,
    \phi_+^B(\omega,\mu) + \dots \,, \nonumber
\end{eqnarray}
where the dots represent power-suppressed contributions. Only the SCET
operator with Dirac structure $\Gamma_1=\gamma^\mu(1-\gamma_5)$
contributes to the decay amplitude.  $C_1$ denotes the Fourier
transform of $\widetilde C_1$, and we will continue the discussion in
momentum space. The HQET parameter $F(\mu)$ is related to the physical
$B$-meson decay constant through
$f_B\sqrt{M_B}=K_F(m_b,\mu)\,F(\mu)\,[1+O(\Lambda_{\rm QCD}/m_b)]$,
where at next-to-leading order in the $\overline{\rm MS}$ scheme
\cite{Neubert:1991sp}
\begin{equation}
   K_F(m_b,\mu) = 1 + \frac{C_F\,\alpha_s(\mu)}{4\pi}
   \left( 3\ln\frac{m_b}{\mu} - 2 \right) .
\end{equation}
Combining these results, it follows that the terms shown in the second
line of (\ref{ampl}) equal those on the right-hand side of the
factorization formula (\ref{ff}) if we identify the hard-scattering
kernel as
\begin{equation}\label{kernel}
   \frac{Q_u}{l_+}\,T(l_+,E_\gamma,m_b,\mu)
   = K_F^{-1}(m_b,\mu)\,C_1(l_+,2E_\gamma,E_\gamma,m_b,\mu) \,.
\end{equation}
We thus completed the first of the four steps to prove factorization.

We proceed to prove the convergence of the convolution integral in
(\ref{ff}). The key ingredient here is to note that the invariance of
SCET operators under reparameterization of the light-cone basis
vectors $n$ and $\bar n$ can be used to deduce the dependence of
Wilson coefficient functions on the separation $t$ between the
component fields of non-local operators. This has been discussed in
some detail in Section~\ref{sec:RPI}. In our case, invariance of the
operators in (\ref{ops}) under the type~III rescaling transformation
$n^\mu\to n^\mu/\alpha$ and $\bar n^\mu\to\alpha\bar n^\mu$ (with
fixed $v$) implies that
\begin{equation}\label{HiJdef}
   \widetilde C_i(t,\bar n\cdot q,v\cdot q,m_b,\mu)
   = \widetilde C_i(\alpha t,\alpha\bar n\cdot q,v\cdot q,m_b,\mu)
\end{equation}
to all orders in perturbation theory. In other words, the variables
$t$ and $\bar n\cdot q$ can only appear in the combination $\bar
n\cdot q/t$, but not individually. The coefficient functions can be
factorized further by noting that $v\cdot q$ and $m_b$ enter only
through hard or hard-collinear interactions with the heavy quark. The
corresponding modes can be integrated out in a first matching step
onto SCET$_I$ and lead to a functions $\widetilde H_i(v\cdot
q,m_b,\mu)$, which depend on the Dirac structure of the weak current
containing the heavy quark. The non-localities of the component fields
in (\ref{ops}) result from the coupling of hard-collinear fields to
the soft spectator quark in the $B$ meson. These effects live on
scales of order $m_b\Lambda_{\rm QCD}$ and can be integrated out in a
second step by matching onto SCET$_{II}$, leading to the function
$\widetilde J(t,\bar n\cdot q,\mu)$. Therefore
\begin{equation}\label{Ciscal}
   \widetilde C_i(t,\bar n\cdot q,v\cdot q,m_b,\mu)
   = \widetilde H_i\left(\frac{2v\cdot q}{\mu},x_\gamma\right)
   \cdot\widetilde J\left(\frac{\bar n\cdot q}{\mu^2\,t}\right) ,
\end{equation}
where $x_\gamma\equiv 2v\cdot q/m_b=2E_\gamma/m_b$ is a scaling
variable of order 1. The corresponding result for the hard-scattering
kernel obtained after Fourier transformation has the form shown in
(\ref{fact2}) if we identify
\begin{equation}\label{RAids}
\begin{aligned}
   H\left(\frac{2v\cdot q}{\mu},x_\gamma\right)
   &= K_F^{-1}(m_b,\mu)\,
    \widetilde H_1\left(\frac{2v\cdot q}{\mu},x_\gamma\right) , \\
   \frac{Q_u}{l_+}\,J\left(\frac{\bar n\cdot q\,l_+}{\mu^2}\right)
   &= \int dt\,e^{-il_+ t}\,
    \widetilde J\left(\frac{\bar n\cdot q}{\mu^2\,t}\right) ,
\end{aligned}
\end{equation}
where $\bar n\cdot q\,l_+=2E_\gamma\,l_+$. Since the dependence of the
coefficient functions on the renormalization scale is logarithmic, it
follows that to all orders in perturbation theory the Wilson
coefficients in (\ref{Ciscal}) scale like $\widetilde C_i\sim 1$
modulo logarithms. (Correspondingly, the kernel scales like $T\sim 1$
modulo logarithms.) The convergence of the convolution integral in
(\ref{ampl}) in the infra-red region $t\to\infty$, corresponding to
the region $l_+\to 0$ in the factorization formula (\ref{ff}), then
follows to all orders in perturbation theory as long as the integral
converges at tree level. Because the $B$ meson has a spatial size of
order $1/\Lambda_{\rm QCD}$ due to confinement, the bilocal matrix
element must vanish faster than $1/t$ for $t\gg 1/\Lambda_{\rm QCD}$,
and so the integral over $t$ is convergent.
The absence of endpoint divergences in convolution integrals is connected
to the question of whether the infra-red degrees of freedom have
correctly been identified in the effective theory
\cite{Bauer:2002aj}. If we had found endpoint singularities as $l_+
\to 0$, we would have to include other, long-distance modes in the
construction of the effective theory (such as soft-collinear modes),
thus invalidating factorization.

Finally, let us demonstrate that more complicated projections onto the
$B$ meson involving higher Fock states or transverse parton momenta
only enter at subleading order in power counting. This can be seen
from the rules for constructing SCET operators out of gauge-invariant
building blocks (\ref{blocks}), as explained in \cite{Hill:2002vw}.
Projections sensitive to transverse momentum components contain extra
derivatives and so are power suppressed. Projections corresponding to
non-valence Fock states contain insertions of the soft gluon field
$\A_s$. Since $\A_s$ scales like $\Lambda_{\rm QCD}$, such insertions
lead to power suppression unless this field is integrated over a
domain of extension $1/\Lambda_{\rm QCD}$. Reparameterization
invariance dictates the such an insertion must be accompanied by a
factor of $n$ in the numerator, or by a factor of $\bar n$ in the
denominator. The only possibility to insert a factor of $n$ in the
numerator is through the combination $\int du\,\nslash
\calAslash_{s\perp}(un)$, which vanishes since the operator already
contains a factor of $\nslash$, and $n^2=0$. For a factor of $\bar n$
in the denominator, we might insert e.~g.~$\bar n/(u\,\bar n\cdot q)$,
which scales like $\Lambda_{\rm QCD}/m_b$.  One might object that an
insertion of $v\cdot n$ invalidates the above argument. However, such
a factor cannot appear as interactions with the heavy quark can be
integrated out before the final matching onto SCET$_{II}$. This also
ensures that the factor $\nslash$ must appear to the left of
$\Gamma_i$.
We have thus completed the proof of the factorization formula (\ref{ff}) 
to all orders in perturbation theory, and at leading power in 
$\Lambda_{\rm QCD}/m_b$. 

\section{Calculation of the hard-scattering kernel}

We derive the Wilson coefficients $\widetilde C_i$ in (\ref{ops}) by
performing a matching calculation using on-shell external quark
states. To this end, we assign incoming momenta $m_b v$ to the heavy
quark and $l$ with $l^2=0$ to the soft light quark. By construction,
the Wilson coefficients are independent of the nature of the external
states. Soft gluon emission from the external quark lines cancel in
the matching. However, emissions from the internal quark propagator in
the full theory lead to the build-up of the finite-length Wilson line
$S_s(tn)S_s^\dagger(0)$. To one-loop precision, the coefficient
functions can be written in the form \cite{Bosch:2003fc}
\begin{equation}
   C_i = \frac{Q_u}{l_+} \left[ \delta_{i1}
   + \frac{C_F\,\alpha_s(\mu)}{4\pi}\,c_i + \dots \right] .
\end{equation}

\begin{figure}
\epsfxsize=14.0cm
\centerline{\epsffile{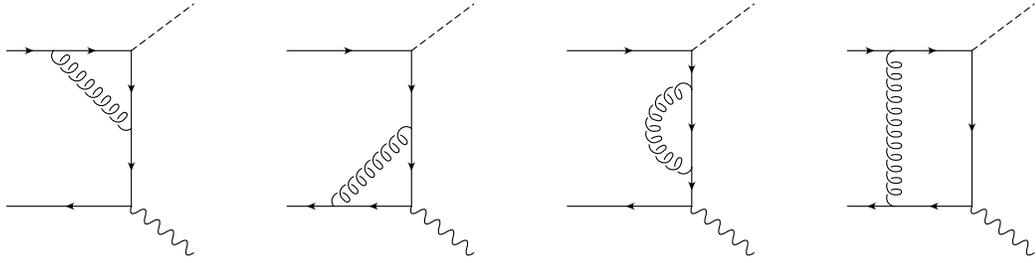}}
\vspace{0.0cm}
\centerline{\parbox{14cm}{\caption[One-loop diagrams in QCD]{\label{fig:QCD}
One-loop diagrams in the full theory contributing at leading power to the 
$B\to\gamma l\nu$ decay amplitude.}}}
\end{figure}

To obtain the NLO corrections $c_i$ we evaluate the one-loop
contributions to the decay amplitude in the full theory and in SCET.
The relevant diagrams in full QCD are shown in
Fig.~\ref{fig:QCD}. In addition there is a contribution from the
wave-function renormalization for the heavy quark. Using
anti-commuting $\gamma_5$ (and, as usual, working in $4-2\epsilon$
dimensions), the diagrams evaluate to the following bare expressions:
\begin{equation} 
\begin{aligned}
   A_1^{\rm QCD}
   &= \left( \frac{m_b}{\mu} \right)^{-2\epsilon}
    \bigg[ - \frac{1}{2\epsilon} - \ln^2\frac{2E_\gamma\,l_+}{m_b^2}
    - 2(1-2\log x_\gamma)\,\ln\frac{2E_\gamma\,l_+}{m _b^2}
    \nonumber
\end{aligned}
\end{equation}
\begin{equation}\label{AiQCD}
\begin{aligned}
   &\hspace{2.6cm}\mbox{}- 4\ln^2 x_\gamma 
    + \frac{2-3x_\gamma}{1-x_\gamma}\,\ln x_\gamma
    - 2L_2(1-x_\gamma) - 2 - \pi^2 \bigg] \\
   &+ \left( \frac{2E_\gamma\,l_+}{\mu^2} \right)^{-\epsilon}\!
    \left( - \frac{2}{\epsilon} - 5 \right) 
    + \left( \frac{-2v\cdot l}{\mu} \right)^{-2\epsilon}\!
    \bigg[ - \frac{1}{\epsilon^2}
    + \ln^2\left(\frac{-2v\cdot l}{l_+} \right) 
    + \frac{7\pi^2}{12} \bigg] \,, \\
   A_2^{\rm QCD}
   &= \left( \frac{m_b}{\mu} \right)^{-2\epsilon}\,
    \frac{x_\gamma\ln x_\gamma}{1-x_\gamma} \,.
\end{aligned}
\end{equation}
The expression $A_1^{\rm QCD}$ receives contributions from the weak
current vertex correction and wave-function renormalization for the
heavy quark (first bracket), the electro-magnetic vertex correction
and self-energy insertion (second term), and the box diagram (last
term, which depends on $v\cdot l$ in addition to $l_+$
\cite{Bosch:2003fc,Korchemsky:1999qb}).

\begin{figure}
\epsfxsize=10.0cm
\centerline{\epsffile{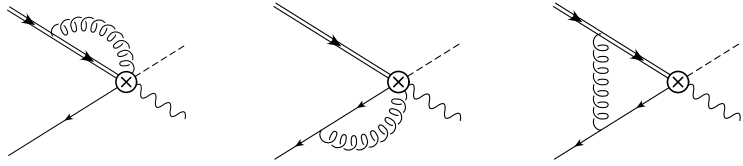}}
\vspace{0.0cm}
\centerline{\parbox{14cm}{\caption[One-loop diagrams in SCET.]{
\label{fig:SCET} One-loop diagrams in the effective theory whose
contribution to the amplitude needs to be subtracted in the
calculation of the Wilson coefficients.}}}
\end{figure}

On the low-energy theory side we evaluate the one-loop order
contributions, which can be written as $C_{\rm 1-loop}\,\otimes
\langle O_{\rm SCET}\rangle_{\rm tree} +C_{\rm tree}\,\otimes \langle
O_{\rm SCET}\rangle_{\rm 1-loop}$. The relevant diagrams are shown in
Fig.~\ref{fig:SCET} and evaluate to
\begin{eqnarray}\label{AiSCET}
   A_1^{\rm SCET}
   &=& \left( \frac{l_+}{\mu} \right)^{-2\epsilon}
    \left( - \frac{1}{\epsilon^2} - \frac{3\pi^2}{4} \right) \\
   &&\mbox{} + \left( \frac{-2v\cdot l}{\mu} \right)^{-2\epsilon}
    \bigg[ - \frac{1}{\epsilon^2}
    + \ln^2\left(\frac{-2v\cdot l}{l_+} \right) 
    + \frac{7\pi^2}{12} \bigg] \,, \nonumber\\
   A_2^{\rm SCET} &=& 0 \,. \nonumber
\end{eqnarray}
The first term in the expression for $A_1^{\rm SCET}$ corresponds to
the first two diagrams in the figure, while the second term is
obtained from the last graph. Note that this is precisely the same
contribution as obtained from the box diagram in the full theory.
The difference of the expressions given in (\ref{AiQCD}) and
(\ref{AiSCET}) determines the NLO contributions to the Wilson
coefficient functions. Subtracting the pole terms in the
$\overline{\rm MS}$ scheme, we find
\begin{equation}\label{ci}
\begin{aligned}
   c_1 &= - 2\ln^2\frac{m_b}{\mu} + (5-4\ln x_\gamma) \ln\frac{m_b}{\mu}
    + \ln^2\frac{2E_\gamma\,l_+}{\mu^2} - 2\ln^2 x_\gamma \\
   &\hspace{0.5cm}\mbox{}+ \frac{2-3x_\gamma}{1-x_\gamma}\,\ln x_\gamma
    - 2L_2(1-x_\gamma) - 7 - \frac{\pi^2}{4} \,, \\
   c_2 &= \frac{x_\gamma\ln x_\gamma}{1-x_\gamma} \,.
\end{aligned}
\end{equation}
These results for the Wilson coefficients agree with the expressions
for the hard-scattering kernel given in \cite{Descotes-Genon:2002mw}
and \cite{Lunghi:2002ju}, apart from the contribution of the HQET
decay constant $F(\mu)$. (The authors of \cite{Descotes-Genon:2002mw}
and \cite{Lunghi:2002ju} assumed an unconventional $m_b$-dependent
definition of the $B$-meson LCDA.) After simplification we find the
hard-scattering kernel at next-to-leading order \cite{Bosch:2003fc}
\begin{eqnarray}\label{RATres}
   T(l_+,E_\gamma,m_b,\mu) &=& 1 + \frac{C_F\,\alpha_s(\mu)}{4\pi}
   \bigg[ - 2\ln^2\frac{2E_\gamma}{\mu} + 2\ln\frac{2E_\gamma}{\mu} +
   \ln^2\frac{2E_\gamma\,l_+}{\mu^2} \nonumber\\
   &&\hspace{1.8cm}\mbox{}- \frac{x_\gamma\ln x_\gamma}{1-x_\gamma} -
   2L_2(1-x_\gamma) - 5 - \frac{\pi^2}{4} \bigg] \,.
\end{eqnarray}
The renormalization-scale dependence is, despite appearance, entirely
given in the form
$2\ln^2(\mu/l_+)-2\ln(\mu/l_+)+\mbox{$\mu$-independent terms}$, which
can therefore be canceled against the $\mu$ dependence of the LCDA
$\phi_+^B(l_+,\mu)$ under the convolution integral in (\ref{ff}). The
remaining large logarithms $\ln(E_\gamma/l_+)$ need to be resummed to
all orders in perturbation theory. This can be done by stating the
result in the above form, factorizing the kernel into a hard function
$H$ (which contains $\ln(2E_\gamma/\mu)$) and a jet function $J$
(which contains $\ln(2E_\gamma\,l_+/\mu^2)$), and solving their RGEs.

\section{Resummation of large logarithms}

The hard-scattering kernel $T$ can be systematically factorized when
performing the matching calculation in two separate steps QCD $\to$
SCET$_I$ $\to$ SCET$_{II}$
\cite{Hill:2002vw,Lunghi:2002ju,Bauer:2002aj}. The hard function
$H(2E_\gamma/\mu, x_\gamma)$ arises when integrating out off-shell
fluctuations of order $m_b$, while fluctuations around the
intermediate scale $m_b\Lambda_{\rm QCD}$ lead to the jet
function. Since the scale $m_b\Lambda_{\rm QCD}$ arises only from a
scalar product of a soft momentum with a collinear momentum, one can
simply set $l=0$ in our previous matching calculation, which ensures
that only the hard fluctuations are integrated out
\cite{Bosch:2003fc}. After the function $H$ has been identified in
this way, the jet function is constructed using that
$J=T/H$. Specifically, for $l=0$ the second and third diagrams in
Fig.~\ref{fig:QCD} involve scaleless integrals that vanish in
dimensional regularization, while the box graph can readily be shown
to vanish at leading power. The remaining contribution from the weak
vertex correction and wave-function renormalization for the heavy
quark yields
\begin{equation}\label{Ail0}
\begin{aligned}
   A_1^{\rm QCD}\Big|_{l=0}
   = \left( \frac{m_b}{\mu} \right)^{-2\epsilon}
   \bigg[ & - \frac{1}{\epsilon^2} - \frac{5}{2\epsilon}
   + \frac{2\ln x_\gamma}{\epsilon} - 2\ln^2 x_\gamma \\
   & + \frac{2-3x_\gamma}{1-x_\gamma}\,\ln x_\gamma
   - 2L_2(1-x_\gamma) - 6 - \frac{\pi^2}{12} \bigg] \,,
   \phantom{\Bigg|}
\end{aligned}
\end{equation}
while the expression for $ A_2^{\rm QCD}\big|_{l=0}$ is the same as
that for $ A_2^{\rm QCD}$ given in (\ref{AiQCD}). After $\overline{\rm
MS}$ subtractions the above result determines the hard function
$\widetilde H_1$ defined in (\ref{HiJdef}). Using the first relation
in (\ref{RAids}), it then follows that
\begin{equation}\label{RAHres}
\begin{aligned}
   H\left(\frac{2E_\gamma}{\mu},x_\gamma\right)
   = 1 + \frac{C_F\,\alpha_s(\mu)}{4\pi}
   \bigg[ & - 2\ln^2\frac{2E_\gamma}{\mu} + 2\ln\frac{2E_\gamma}{\mu}\\
&   - \frac{x_\gamma\ln x_\gamma}{1-x_\gamma} - 2L_2(1-x_\gamma)
   - 4 - \frac{\pi^2}{12} \bigg] \,.
   \phantom{\Bigg|}
\end{aligned}
\end{equation}
The knowledge of the hard-scattering kernel (\ref{RATres}) and the
above result for the hard function yields the jet function, which is
thus given by
\begin{equation}\label{RAJres}
   J\left(\frac{2E_\gamma\,l_+}{\mu^2}\right)
   =  1 + \frac{C_F\,\alpha_s(\mu)}{4\pi}
    \bigg( \ln^2\frac{2E_\gamma\,l_+}{\mu^2} - 1 - \frac{\pi^2}{6}
    \bigg) .
\end{equation}

In order to proceed we need RG equations obeyed by the various
coefficient functions. The fact that the decay amplitude in (\ref{ff})
is scale independent links the scale dependence of the hard-scattering
kernel to the evolution of the LCDA $\phi_+^B(\omega,\mu)$
\cite{Grozin:1996pq}. By analyzing the renormalization properties of this 
function, as we have done in Section~\ref{sec:LCDA:RGEsolution}, it
follows that the hard-scattering kernel satisfies the
integro-differential equation (for
$l_+>0$) \cite{Lange:2003ff}\footnote{For simplicity of notation, we
omit the arguments $E_\gamma$, $m_b$, and $x_\gamma$ for the remainder
of this section. Also, unless otherwise indicated,
$\alpha_s\equiv\alpha_s(\mu)$.}
\begin{equation}\label{me}
   \frac{d}{d\ln\mu}\,T(l_+,\mu)
   = \left[ \Gamma_{\rm cusp}(\alpha_s)\,\ln\frac{\mu}{l_+}
   + \gamma(\alpha_s) \right] T(l_+,\mu)
   + \int_0^\infty\!d\omega\,l_+\,\Gamma(\omega,l_+,\alpha_s)\,
   T(\omega,\mu) ,
\end{equation}
where $\Gamma_{\rm cusp}$ is the universal cusp anomalous dimension
familiar from the theory of the renormalization of Wilson loops
\cite{Korchemsky:wg} and $\Gamma(\omega,\omega^\prime,\alpha_s)$ is
given in equation~(\ref{eq:1-loop}). From the functional forms of the
hard and jet functions given above, it follows that the hard component
and the jet function obey the RG equations
\begin{eqnarray}\label{RAHevol}
   \frac{d}{d\ln\mu}\,H(\mu)
   &=& \left[ - \Gamma_{\rm cusp}(\alpha_s)\,\ln\frac{\mu}{2E_\gamma}
    + \gamma(\alpha_s) - \gamma^\prime(\alpha_s) \right] H(\mu) \,, \\
   \frac{d}{d\ln\mu}\,J(l_+,\mu)
   &=& \left[ \Gamma_{\rm cusp}(\alpha_s)\,
    \ln\frac{\mu^2}{2E_\gamma\,l_+} + \gamma^\prime(\alpha_s) \right]
    J(l_+,\mu) \\
   && \mbox{}\quad\qquad + \int_0^\infty\!d\omega\,l_+\,
    \Gamma(\omega,l_+,\alpha_s)\,J(\omega,\mu) \,. \nonumber
\end{eqnarray}

The anomalous dimensions $\gamma$ and $\gamma^\prime$ do not have a
simple geometric interpretation and must be determined by explicit
calculation, unlike is the case for the cusp anomalous dimension
\cite{Korchemsky:1994jb}. From (\ref{RAHres}) and (\ref{RAJres}) we
find
\begin{equation}
   \gamma(\alpha_s) = -2C_F\,\frac{\alpha_s}{4\pi} + O(\alpha_s^2) \,,
    \qquad
   \gamma^\prime(\alpha_s) = O(\alpha_s^2) \,.
\end{equation}

We now discuss the general solution of the evolution equations
(\ref{me}) and (\ref{RAHevol}). Exact solutions can be written down
analogously to the discussion in Section~\ref{sec:LCDA:RGEsolution}.
We define the dimensionless function 
\begin{equation}
   \F(a,\alpha_s) = \int d\omega\,\omega^\prime\,
   \Gamma(\omega,\omega^\prime,\alpha_s)
   \left( \frac{\omega}{\omega^\prime} \right)^{-a} .
\end{equation}
This is the same function as previously defined in (\ref{eq:F}), and
for convenience we restate the one-loop order result
(\ref{eq:1-loop-F}) as
\begin{equation}
   \F^{(1)}(a,\alpha_s)
   = \Gamma_{\rm cusp}^{(1)}(\alpha_s)\,
   \Big[ \psi(1+a) + \psi(1-a) + 2\gamma_E \Big] \,,
\end{equation}
where $\psi(z)$ is the logarithmic derivative of the Euler
$\Gamma$-function. We start by solving the first equation in
(\ref{RAHevol}) with the initial condition for $H(\mu_h)$ evaluated at
a high scale $\mu_h\sim m_b$, for which it does not contain large
logarithms. We then evolve the function $H(\mu)$ down to an
intermediate scale $\mu_i\sim\sqrt{m_b\Lambda_{\rm QCD}}$ and multiply
it by the result $J(l_+,\mu_i)$ for the jet function, which at the
intermediate scale is free of large logarithms and can be written in
the general form
$J(l_+,\mu_i)\equiv\J[\alpha_s(\mu_i),\ln(2E_\gamma\,l_+/\mu_i^2)]$.
This determines the kernel $T(l_+,\mu_i)$ at the intermediate scale.
Finally, we solve (\ref{me}) and compute the evolution down to a
low-energy scale $\mu\sim\mbox{few}\times\Lambda_{\rm QCD}$. The exact
solution is given by
\begin{equation}\label{magic1}
   T(l_+,\mu) = H(\mu_h)\,\J[\alpha_s(\mu_i),\nabla_\eta]\,
   \exp U(l_+,\mu,\mu_i,\mu_h,\eta) \Big|_{\eta=0} \,,
\end{equation}
where the notation $\J[\alpha_s(\mu_i),\nabla_\eta]$ means that one
must replace each logarithm of the ratio $2E_\gamma\,l_+/\mu_i^2$ by a
derivative with respect to an auxiliary parameter $\eta$. The
evolution function $U$ is given by \cite{Bosch:2003fc}
\begin{eqnarray}\label{magic2}
   U(l_+,\mu,\mu_i,\mu_h,\eta)
   &=& \int\limits_{\alpha_s(\mu_h)}^{\alpha_s(\mu_i)}\!d\alpha\,
    \frac{\Gamma_{\rm cusp}(\alpha)}{\beta(\alpha)}
    \Bigg[ \ln\frac{2E_\gamma}{\mu_h}
     - \int\limits_{\alpha_s(\mu_h)}^\alpha
     \frac{d\alpha^\prime}{\beta(\alpha^\prime)} \Bigg]
    - \int\limits_{\alpha_s(\mu_h)}^{\alpha_s(\mu_i)}\!d\alpha\,
    \frac{\gamma^\prime(\alpha)}{\beta(\alpha)} \nonumber\\
   &&- \int\limits_{\alpha_s(\mu_i)}^{\alpha_s(\mu)}\!d\alpha\,
    \frac{\Gamma_{\rm cusp}(\alpha)}{\beta(\alpha)}
    \Bigg[ \ln\frac{l_+}{\mu}
     + \int\limits_\alpha^{\alpha_s(\mu)}
     \frac{d\alpha^\prime}{\beta(\alpha^\prime)} \Bigg]
    + \int\limits_{\alpha_s(\mu_h)}^{\alpha_s(\mu)}\!d\alpha\,
    \frac{\gamma(\alpha)}{\beta(\alpha)} \nonumber\\
   &&+ \eta\ln\frac{2E_\gamma\,l_+}{\mu_i^2}
    + \int\limits_{\alpha_s(\mu_i)}^{\alpha_s(\mu)}
    \frac{d\alpha}{\beta(\alpha)}\,
    \F\bigg(-\eta+\!\int\limits_{\alpha_s(\mu_i)}^\alpha\!
    d\alpha^\prime\,
    \frac{\Gamma_{\rm cusp}(\alpha^\prime)}{\beta(\alpha^\prime)},
    \alpha\bigg) \,. \nonumber\\
\end{eqnarray}

Given this exact result, it is straightforward to derive approximate
expressions for the kernel at given orders in RG-improved perturbation
theory, by using perturbative expansions of the anomalous dimensions
and $\beta$-function to the required order. Unfortunately, controlling
terms of $O(\alpha_s)$ in the evolution function $U$ would require
knowledge of all anomalous dimensions at two-loop order, which at
present is lacking. (It also requires the cusp anomalous dimension to
three loops, which has been computed only recently \cite{moch:et:al}.)
We will, however, control the dependence on the variables $l_+$ and
$E_\gamma$ to $O(\alpha_s)$. As usual, we write
\begin{equation}
   \beta(\alpha_s) = -2\alpha_s \sum_{n=0}^\infty \beta_n
    \left( \frac{\alpha_s}{4\pi} \right)^{n+1} , \qquad
   \Gamma_{\rm cusp}(\alpha_s) = \sum_{n=0}^\infty \Gamma_n
    \left( \frac{\alpha_s}{4\pi} \right)^{n+1} ,
\end{equation}
and similarly for the anomalous dimensions $\gamma$ and $\gamma^\prime$. The
relevant expansion coefficients are listed in (\ref{cuspExpand}) and
(\ref{betaExpand}), and we also find $\gamma_0=-2C_F$, $\gamma_0^\prime=0$.
Defining the ratios $r_1=\alpha_s(\mu_i)/\alpha_s(\mu_h)$ and
$r_2=\alpha_s(\mu)/\alpha_s(\mu_i)$, we obtain our final result 
\cite{Bosch:2003fc}
\begin{eqnarray}\label{exact1}
   T(&\hspace{-4mm} l_+&\hspace{-5mm},\mu)
   = e^{U_0(\mu,\mu_i,\mu_h)} \left( \frac{l_+}{\mu} \right)^{c\ln r_2}
    \left( \frac{2E_\gamma}{\mu_h} \right)^{-c\ln r_1} \\
   \mbox{}\hspace{0mm}\times\Bigg\{ 1 \hspace{-2mm} &+& 
    \frac{C_F\,\alpha_s(\mu_h)}{4\pi}
    \bigg[ - 2\ln^2\frac{2E_\gamma}{\mu_h} + 2\ln\frac{2E_\gamma}{\mu_h} 
    - \frac{x_\gamma\ln x_\gamma}{1-x_\gamma} - 2L_2(1-x_\gamma)
    - 4 - \frac{\pi^2}{12} \bigg] \nonumber\\
   &+& \frac{C_F\,\alpha_s(\mu_i)}{4\pi}
    \bigg[ \left( \ln\frac{2E_\gamma\,l_+}{\mu_i^2}
    - \psi(1+c\ln r_2) - \psi(1-c\ln r_2) - 2\gamma_E \right)^2
    \nonumber\\
   &&\hspace{3.1cm}\mbox{}- \psi^\prime(1+c\ln r_2) + \psi^\prime(1-c\ln r_2)
    - 1 - \frac{\pi^2}{6} \bigg] \nonumber \\
   &+& \frac{\Gamma_0}{2\beta_0}
    \left( \frac{\Gamma_1}{\Gamma_0} - \frac{\beta_1}{\beta_0} \right)
    \bigg[ \frac{\alpha_s(\mu)-\alpha_s(\mu_i)}{4\pi}\,\ln\frac{l_+}{\mu}
    - \frac{\alpha_s(\mu_i)-\alpha_s(\mu_h)}{4\pi}\,
    \ln\frac{2E_\gamma}{\mu_h} \bigg] \Bigg\} \,, \nonumber
\end{eqnarray}
where $c=\Gamma_0/2\beta_0$, and 
\begin{eqnarray}
   U_0(\mu,\mu_i,\mu_h) 
   &=& \frac{\Gamma_0}{4\beta_0^2}\,\Bigg\{
    (1-\ln r_1)\,\frac{4\pi}{\alpha_s(\mu_h)}
    + (1+\ln r_2)\,\frac{4\pi}{\alpha_s(\mu)}
    - \frac{8\pi}{\alpha_s(\mu_i)} \nonumber \\
   &&\hspace{0mm}\mbox{}+
    \frac{\beta_1}{2\beta_0}\,(\ln^2 r_1+\ln^2 r_2)
    + \left( \frac{\Gamma_1}{\Gamma_0} - \frac{\beta_1}{\beta_0}
    \right) \left( \ln\frac{r_1}{r_2} + 2 - r_1 - \frac{1}{r_2} \right)
    \Bigg\} \nonumber
\end{eqnarray}
\begin{eqnarray}\label{exact2}
   &&\mbox{}- \frac{\gamma_0}{2\beta_0}\,\ln(r_1 r_2) 
    - \ln\frac{\Gamma(1+c\ln r_2)}{\Gamma(1-c\ln r_2)} 
    - 2\gamma_E\,c\ln r_2 + O(\alpha_s)
\end{eqnarray}
corresponds to the function $U$ evaluated with $\eta=0$,
$l_+=\mu$, and $2E_\gamma=\mu_h$. 

\begin{figure}
\epsfxsize=15.0cm
\centerline{\epsffile{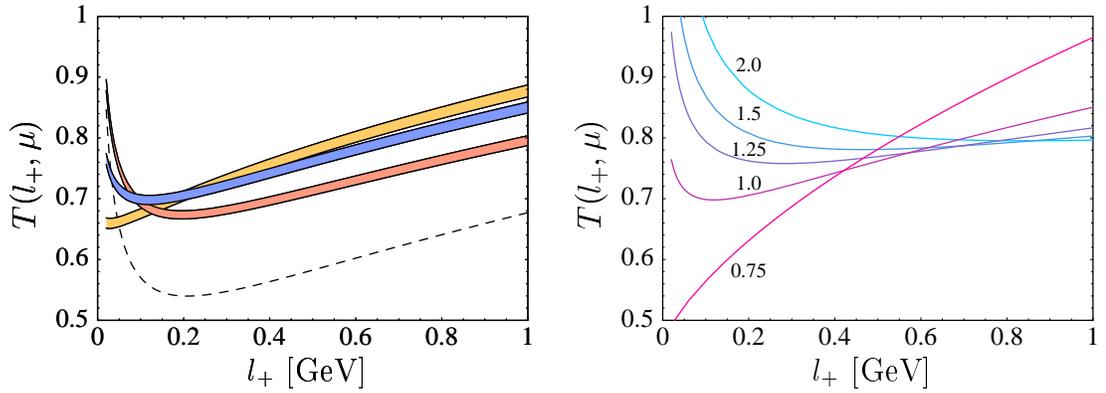}}
\vspace{0.0cm}
\centerline{\parbox{14cm}{\caption[Sudakov resummation for the 
hard-scattering kernel.]{\label{fig:resum} RG-improved predictions for
the hard-scattering kernel at maximum photon energy.  {\em Left:\/}
Results at $\mu=1$\,GeV. The bands refer to different values of the
intermediate matching scale: $\mu_i^2=\Lambda_h m_b$ (center),
$2\Lambda_h m_b$ (top), $0.5\Lambda_h m_b$ (bottom). Their width
reflects the sensitivity to the high-energy matching scale $\mu_h^2$,
varied between $2m_b^2$ and $0.5m_b^2$. The dashed line shows the
result obtained at one-loop order.  {\em Right:\/} Dependence of the
kernel on the renormalization scale $\mu$, varied between 0.75\,GeV
and 2.0\,GeV as indicated on the curves.}}}
\end{figure}

Finally, let us study the effects from Sudakov resummation
numerically. In the left-hand plot in Fig.~\ref{fig:resum} we
compare the resummed hard-scattering kernel (\ref{exact1}) with the
expanded $O(\alpha_s)$ approximation (\ref{RATres}). The effect is
maximized by setting the photon energy to its highest possible value
of $E_\gamma = m_b/2$. (Numerically, we work with the pole mass $m_b =
4.8$ GeV.) The renormalization scale $\mu$ is chosen to 1 GeV, which
is a reasonable scale for the $B$-meson LCDA. The matching scales
$\mu_h$ and $\mu_i$ are varied around their ``natural'' choices
$\mu_h=2E_\gamma$ and $\mu_i=\sqrt{2E_\gamma\Lambda_h}$, where
$\Lambda_h=0.5$\,GeV serves as a typical hadronic scale, are taken as
default values. We use the two-loop running coupling normalized at
$\alpha_s(m_b)=0.22$ and set $n_f=4$ for the number of light quark
flavors. (For simplicity, we do not match onto a three-flavor theory
even for low renormalization scales.) We find that resummation effects
decrease the magnitude of the radiative corrections, i.e., the
resummed kernel is closer to the tree-level value ($T=1$) than the
one-loop result. The fact that after Sudakov resummation the radiative
corrections are moderate in magnitude persists even for asymptotically
large $b$-quark masses. For instance, setting $m_b=50$\,GeV we find
after resummation $T(l_+,\mu)=0.74$ at $l_+=\mu=1$\,GeV. (Fixed-order
perturbation theory breaks down for such large values of the quark
mass.  From (\ref{RATres}) we would obtain $T(l_+,\mu)=0.08$ with
these parameter values.) The figure also exhibits that our results are
stable under variation of the two matching scales. Varying $\mu_i^2$
and $\mu_h^2$ by factors of 2 changes the result for the kernel by
less than 10\%. This suggests that the unknown NNLO corrections to the
function $U_0$ in (\ref{exact2}) are perhaps not very important.

The scale dependence of the resummed expression for the kernel is
illustrated in the right-hand plot in Fig.~\ref{fig:resum}, which
shows the functional dependence of $T(l_+,\mu)$ for maximal photon
energy and several values of $\mu$. The matching scales are set to
their default values $\mu_h=m_b=4.8$\,GeV and $\mu_i=\sqrt{\Lambda_h
m_b}\simeq 1.55$\,GeV. We observe a significant scale dependence of
the kernel, especially as one lowers $\mu$ below the intermediate
scale $\mu_i$. In other words, the second stage of running (for
$\mu<\mu_i$) is numerically significant.

\chapter{Exclusive Semileptonic Decays}\label{chap:formfactor}

\section{Heavy-to-light form factors at large recoil}

Transitions of the $B$ meson into light pseudoscalar or vector mesons
are parameterized in terms of scalar functions, called form factors,
that depend only on the Lorentz invariant quantity $q^2$, where $q =
p_B - p_M$ is the momentum transfer from the $B$-meson onto the light
final meson $M$. They are defined by the following Lorentz
decompositions of bilinear quark current matrix elements:
\begin{equation}
\langle P(p_P)|\bar q \, \gamma^\mu b |\bar{B}(p_B)\rangle =
f_+(q^2)\left[p_B^\mu+p_P^{\mu}-\frac{M_B^2-m_P^2}{q^2}\,q^\mu\right]
+f_0(q^2)\,\frac{M_B^2-m_P^2}{q^2}\,q^\mu,
\label{fvector}
\end{equation}
\begin{equation}
\langle P(p_P)|\bar q \, \sigma^{\mu\nu} q_\nu b|\bar{B}(p_B) \rangle =
\frac{i f_T(q^2)}{M_B+m_P}\left[q^2(p_B^\mu+p_P^{\mu})-
(M_B^2-m_P^2)\,q^\mu\right],
\label{ftensor}
\end{equation}
where $m_P$ the mass of the pseudoscalar meson. The relevant form
factors for $B$ decays into vector mesons with polarization vector
$\varepsilon$ and mass $m_V$ are defined as
\begin{equation}
\langle V(p_V,\varepsilon^\ast)| \bar q \gamma^\mu b | \bar{B}(p_B) \rangle =
 \frac{2iV(q^2)}{M_B+m_V} \,\epsilon^{\mu\nu\rho\sigma}
 \varepsilon^{\ast}_\nu \, p_{V\,\rho} p_{B\,\sigma},
\label{V}
\end{equation}
\begin{eqnarray}
\langle V(p_V,\varepsilon^\ast)| \bar q \gamma^\mu\gamma_5 b | \bar{B}(p_B) 
\rangle &=&
  2m_VA_0(q^2)\,\frac{\varepsilon^\ast\cdot q}{q^2}\,q^\mu + \\
  && (M_B+m_V)\,A_1(q^2)\left[\varepsilon^{\ast\mu}-
  \frac{\varepsilon^\ast\cdot q}{q^2}\,q^\mu\right] \nonumber\\
  && -\,A_2(q^2)\,\frac{\varepsilon^\ast\cdot q}{M_B+m_V}
 \left[p_B^\mu+p_V^{\mu} -\frac{M_B^2-m_V^2}{q^2}\,q^\mu\right],\nonumber
\end{eqnarray}
\begin{equation}
\langle V(p_V,\varepsilon^\ast)| \bar q \sigma^{\mu\nu}q_\nu b | \bar{B}(p_B)
\rangle =
  2\,T_1(q^2)\,\epsilon^{\mu\nu\rho\sigma}\varepsilon^{\ast}_\nu\, 
  p_{B\,\rho} p_{V\,\sigma},
\end{equation}
\begin{eqnarray}
\langle V(p_V,\varepsilon^\ast)| \bar q \sigma^{\mu\nu} \gamma_5 q_\nu b | 
\bar{B}(p_B) \rangle&\hspace{-1mm}=&\hspace{-1mm}
(-i)\,T_2(q^2)\left[(M_B^2-m_V^2)\,\varepsilon^{\ast\mu}-
(\varepsilon^\ast\cdot q)\,(p_B^\mu+p_V^{\mu})\right]
\nonumber\\
 && \hspace*{-2cm}
+\,(-i)\,T_3(q^2)\,(\varepsilon^\ast\cdot q)
\left[q^\mu-\frac{q^2}{M_B^2-m_V^2}(p_B^\mu+p_V^{\mu})\right].
\label{ffdef}
\end{eqnarray}

\begin{figure}
\begin{center}
\epsfig{file=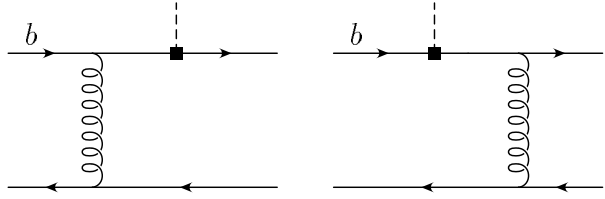,width=8cm} 
\end{center}
\centerline{\parbox{14cm}{\caption[Gluon exchange contributions to
heavy-to-light form factors. ]{\label{fig:twopart} Gluon exchange
contributions to heavy-to-light form factors. The flavor-changing
current is denoted by a dashed line. The lines to the left belong to
the $B$ meson, and those to the right belong to the light meson
$M$.}}}
\end{figure}

In many applications form factors are needed near zero momentum
transfer ($q^2\approx 0$), corresponding to a kinematic situation in
which a flavor-changing weak current turns a heavy $B$ meson at rest
into a highly energetic light meson. Such heavy-to-light transitions
at high recoil are suppressed in QCD by inverse powers of the
heavy-quark mass $m_b$. (This is contrary to the case of
heavy-to-heavy transitions such as $B\to D$, which are unsuppressed in
the heavy-quark limit \cite{Neubert:1993mb}.) The challenge in
understanding the physics of these processes is to describe properly
the transformation of the soft constituents of the $B$ meson into the
fast moving, collinear constituents of the energetic light meson in
the final state. At lowest order in perturbation theory this
transformation can be achieved by the exchange of a gluon, as depicted
in Fig.~\ref{fig:twopart}. Naive power counting suggests that the
nature of the gluons is hard-collinear, as they turn the soft
spectator quark into a collinear constituent of the final meson of
energy
\begin{equation}
   E = \frac{M_B^2+m_M^2-q^2}{2M_B}\gg \Lambda_{\rm QCD}
\end{equation}
in the $B$-meson rest frame. The meson momentum can be written as
$p_M^\mu=E n^\mu+O(m_M^2/4E)$, which is nearly light-like. We assume
that the partons inside the light meson carry a significant fraction
of its total energy, and that the light constituents of the $B$ meson
carry soft momenta of order $\Lambda_{\rm QCD}$.  This justifies to
describe the process using SCET. It would then appear that the form
factor is governed by a sufficiently hard gluon exchange, which can be
dealt with using perturbative methods for hard exclusive QCD processes
\cite{Lepage:1980fj,Efremov:1979qk}. The situation is, as we shall see
below, not quite that simple
\cite{Beneke:2002ph,Hill:2002vw,Beneke:2003pa,Bauer:2002aj,Lange:2003pk,Beneke:2000wa,Beneke:2003xr}. 
Leading contributions to the form factors also arise when the
exchanged gluon is of long-distance nature, leading to the ``soft
overlap'' (also called Feynman mechanism). As an example, a
straightforward calculation of the two diagrams in
Fig.~\ref{fig:twopart} yields
\begin{equation}\label{LO}
   A = -g^2 \left( \Gamma\,
   \frac{m_b(1+\vslash) - x_2 E\nslash}
        {4m_b E^2 x_2^2\,n\cdot l}\,\gamma_\mu
   + \gamma_\mu\,\frac{E\nslash-\lslash}{4E^2 x_2(n\cdot l)^2}\,
   \Gamma \right) t_a * \gamma^\mu t_a + \dots \,,
\end{equation}
where $v^\mu=(1,0,0,0)$ is the $B$-meson velocity, $\Gamma$ denotes
the Dirac structure of the flavor-changing current, the $*$ product
means that the two factors must be sandwiched between quark spinors,
and the dots represent power-suppressed terms. Here we assigned an
incoming momentum $l$ to the light spectator anti-quark, and the
outgoing parton momenta $p_1$ and $p_2$ collinear to the light-meson
momentum are parameterized as
\begin{equation}\label{p1p2}
   p_1^\mu = x_1 E n^\mu + p_\perp^\mu + \dots \,, \qquad
   p_2^\mu = x_2 E n^\mu - p_\perp^\mu + \dots \,,
\end{equation}
with the fractions $x_1$, $x_2$, subject to the constraint
$x_1+x_2=1$. Assuming that $l$ is soft and $x_i$ of order unity one
finds that the amplitude (\ref{LO}) scales like $1/(E^2\Lambda_{\rm
QCD}$. However, assuming linear distributions for $x_2\to 0$ and
$n\cdot l \to 0$ in the endpoints of the meson LCDAs, it is evident
from (\ref{LO}) that the corresponding convolution integrals for a
hadronic amplitude diverge at these endpoints
\cite{Szczepaniak:1990dt,Burdman:hg}. In other words, the above
analysis breaks down for such momentum configurations because the
gluon exchange is no longer of short-distance nature. Naively, form
factors would then be dominated by this soft overlap, since it is not
suppressed by a perturbative coupling constant $\alpha_s$. However, it
has sometimes been argued that the summation of large Sudakov
logarithms associated with the soft gluon exchange mechanism may lead
to a strong suppression of the soft overlap term, essentially
reinstating the perturbative nature of the form factors (see, e.g.,
\cite{Akhoury:uw,Dahm:1995ne,Kurimoto:2001zj}). The assumption of a
short-distance nature of heavy-to-light form factors at large recoil
is the basis of the pQCD approach to exclusive hadronic $B$ decays
\cite{Keum:2000ph}, which is often considered a competitor to QCD
factorization. Leaving aside some conceptual problems associated with
this treatment \cite{Descotes-Genon:2001hm}, the issue of Sudakov
logarithms is an intricate one, because contributions to the form
factors can arise from several different energy scales. Besides the
hard scale $E\sim M_B$ and the hadronic scale $\Lambda_{\rm QCD}$,
interactions between soft and collinear partons involve the
intermediate hard-collinear scale of order $\sqrt{E\Lambda_{\rm
QCD}}$. To settle the question of Sudakov suppression of the soft
overlap contribution is one of the main motivations for our work
\cite{Lange:2003pk}.

\section{Spin-symmetry and factorization}

The discussion of heavy-to-light form factors in the heavy-quark limit
$E\gg\Lambda_{\rm QCD}$ can be summarized by the factorization formula
\cite{Beneke:2000wa}
\begin{equation}\label{ffff}
   f_i^{B\to M}(E) = C_i(E,\mu_{\rm I})\,\zeta_M(\mu_{\rm I},E)
   + T_i(E,\sqrt{E\Lambda_{\rm QCD}},\mu)\otimes\phi_B(\mu)\otimes\phi_M(\mu)
   + \dots ,
\end{equation}
where the dots represent terms that are of subleading order in
$\Lambda_{\rm QCD}/E$. $C_i$ and $T_i$ are calculable short-distance
coefficient functions, $\phi_B$ and $\phi_M$ denote the leading-order
LCDAs of the $B$ meson and the light meson $M$, and the $\otimes$
products imply convolutions over light-cone momentum fractions. The
functions $\zeta_M$ denote universal form factors that only depend on
the nature of the light final-state meson but not on the Lorentz
structure of the currents whose matrix elements define the various
form factors. The first term in the factorization formula therefore
implies spin-symmetry relations between different form factors, which
were first derived in \cite{Charles:1998dr} by considering the
large-energy limit of QCD. The second term in (\ref{ffff}), which arises
from hard-collinear gluon exchange, breaks these symmetry relations
\cite{Beneke:2000wa}.

The arguments $E$ and $\sqrt{E\Lambda_{\rm QCD}}$ in the
short-distance coefficients $C_i$ and $T_i$ are representative for any
of the hard or hard-collinear scales in the problem, respectively. The
form of the factorization formula shown in (\ref{ffff}) assumes that
the factorization scale $\mu_{\rm I}$ in the first term is chosen to
lie between the hard scale $E$ and the hard-collinear scale
$\sqrt{E\Lambda_{\rm QCD}}$, whereas the scale $\mu$ in the second
term is chosen to lie below the hard-collinear scale. The Wilson
coefficients $C_i$ receive contributions at tree level, whereas the
hard-scattering kernels $T_i$ start at first order in
$\alpha_s(\sqrt{E\Lambda_{\rm QCD}})$. Naively, one would conclude
that the spin-symmetry preserving term provides the leading
contribution to the form factors in the heavy-quark limit. Only then
the notion of an approximate spin symmetry would be
justified. However, the situation is more complicated because as
written in (\ref{ffff}) the universal functions $\zeta_M(\mu_{\rm
I},E)$ still depend on the short-distance scale $\sqrt{E\Lambda_{\rm
QCD}}$; in fact, their $E$-dependence remains unspecified. We will see
how this dependence on the large energy enters the amplitudes when our
technology of scale separation is applied. The matching of a QCD form
factor onto matrix elements in SCET is done in two steps:
QCD\,$\to$\,SCET$_I$\,$\to$\,SCET$_{II}$.  In close analogy to the
previous applications discussed in this thesis, hard fluctuations with
virtualities on the scale $E\sim M_B$ are integrated out in a first
step, while in the second step hard-collinear modes with virtualities
of order $\sqrt{E\Lambda_{\rm QCD}}$ are removed. Both steps of this
matching are understood for the hard-scattering term in the
factorization formula, for which the kernels $T_i$ can be factorized
as \cite{Bauer:2002aj}
\begin{equation}\label{Tifact}
   T_i(E,\sqrt{E\Lambda_{\rm QCD}},\mu) = \sum_j H_{ij}(E,\mu)
   \otimes J_j(\sqrt{E\Lambda_{\rm QCD}},\mu) \,. \vspace{-0.3cm}
\end{equation}
Below we will focus on the universal functions $\zeta_M(E,\mu_{\rm
I})$ entering the first term in (\ref{ffff}). (The notation $\mu_{\rm
I}$ serves as a reminder that the factorization scale for this term is
defined in the intermediate effective theory SCET$_I$.) In
\cite{Bauer:2002aj,Beneke:2003xr} these functions are defined in terms
of matrix elements in the intermediate effective theory SCET$_I$,
which leaves open the possibility that they could be dominated by
short-distance physics. Here we show \cite{Lange:2003pk} that the
functions $\zeta_M$ renormalized at a hard-collinear scale
$\mu_{hc}\sim\sqrt{E\Lambda_{\rm QCD}}$ can be factorized further
according to
\begin{equation}\label{scet2ff}
   \zeta_M(\mu_{hc},E) = \sum_k D_k^{(M)}(\sqrt{E\Lambda_{\rm
   QCD}},\mu_{hc},\mu) \otimes\xi_k^{(M)}(\mu,E)\;, 
\end{equation}
where the functions $\xi_k^{(M)}$ are defined in terms of hadronic
matrix elements of SCET$_{II}$ operators. By solving the RG equation
for these operators we show that the universal form factors $\zeta_M$
do not receive a significant perturbative suppression, neither by a
power of $\alpha_s(\sqrt{E\Lambda_{\rm QCD}})$ nor by resummed Sudakov
logarithms. This statement holds true even in the limit where $M_B$ is
taken to be much larger than the physical $B$-meson mass. Perhaps
somewhat surprisingly, we find that there exists a leading
contribution to the soft functions $\zeta_M$, for which the
hard-collinear scale $\sqrt{E\Lambda_{\rm QCD}}$ is without physical
significance. Switching from SCET$_I$ to SCET$_{II}$, one describes
the same physics using a different set of degrees of freedom. This
observation leads us to the most important conclusion of this work,
namely that {\em the long-distance, soft overlap contribution to
heavy-to-light form factors exists}. However, we point out that even
at low hadronic scales $\mu\sim\Lambda_{\rm QCD}$ the functions
$\xi_k^{(M)}$ contain a dependence on the large recoil energy $E$
which is of long-distance nature and cannot be factorized using RG
techniques. As a result, it is impossible to determine the asymptotic
behavior of the QCD form factors $f_i^{B\to M}$ as $E\to\infty$ using
short-distance methods.

In order to prove the factorization formula (\ref{ffff}) one needs to show
that all contributions to the form factors that do not obey spin-symmetry 
relations can be written in terms of convolution integrals involving the 
leading-order LCDAs of the $B$ meson and the light meson. Specifically, 
this means showing that (i) no higher Fock states (or higher-twist 
two-particle distribution amplitudes) contribute at leading power, and 
(ii) the convolution integrals are convergent to all orders in 
perturbation theory. Point (i) can be dealt with using the power-counting 
rules of SCET along with reparameterization invariance 
\cite{Bauer:2002aj,Beneke:2003xr}. Here we use our formalism to complete 
step (ii) of the factorization proof. (The convergence of the convolution 
integral over the $B$-meson LCDA can also be shown using arguments along 
the lines of the discussion around equation~(\ref{HiJdef}).)

\section{Matching calculations}
\label{sec:matching}

In the matching of the intermediate effective theory SCET$_I$ onto the
final low-energy effective theory SCET$_{II}$, hard-collinear modes
with virtuality of order $\sqrt{E\Lambda_{\rm QCD}}$ are integrated
out, and their effects are included in short-distance coefficient
functions. Our primary goal in this section is to construct a basis of
operators relevant to the matching of the universal functions
$\zeta_M$ in (\ref{ffff}) onto the low-energy theory (at leading power
in $\lambda = \Lambda_{\rm QCD}/E$), and to calculate their Wilson
coefficients at lowest order in perturbation theory. As always,
matching calculations can be done using on-shell external quark and
gluon states rather than the physical meson states, whose matrix
elements define the form factors. The results for the Wilson
coefficients are insensitive to infra-red physics. 

\subsection{Spin-symmetric contributions}

Heavy-to-light form factors are defined in terms of $B\to M$ matrix
elements of flavor-changing currents $\bar q\,\Gamma\,b$. The
spin-symmetric contributions to the form factors are characterized by
the fact that in the intermediate effective theory (i.e., after hard
fluctuations with virtuality $\mu\sim E\sim M_B$ are integrated out)
they contain the Dirac structure $\Gamma$ sandwiched between the two
projection operators $\frac14\nbslash\nslash$ and
$\frac12(1+\vslash)$. This implies that $\Gamma$ can be decomposed
into a linear combination of only three independent Dirac structures,
which leads to symmetry relations between various form factors
\cite{Charles:1998dr}. As was shown in the equations
(\ref{HtCcurrents}) and (\ref{DiracReduction}), the relevant operators
in SCET$_I$ can be written as time-ordered products of the effective
Lagrangian with effective current operators
$J_M^{(0)}(x)=[\bar\xi_{hc} W_{hc}](x)\,\Gamma_M\,h(x_-)$ defined by
the matching relation
\begin{equation}\label{ffcurrent}
   \bar q\,\Gamma\,b = \sum_M K_M^\Gamma(m_b,E,\mu)\,J_M^{(0)}
   + \dots \,, \vspace{-0.3cm}
\end{equation}
where the dots denote power-suppressed terms. Here
$\Gamma_M=1,\gamma_5,\gamma_{\perp\nu}$ is one of the three Dirac
basis matrices that remain after the projections onto the
two-component spinors $\bar\xi_{hc}$ and $h$. The variable $E$
entering the coefficients $K_M^\Gamma$ is the total energy carried by
collinear particles (more precisely, $2E=\bar n\cdot p_c^{\rm tot}$),
which in our case coincides with the energy of the final-state meson
$M$. Since the Lagrangian is a Lorentz scalar, it follows that for
$B\to M$ transitions each of the three possibilities corresponds to a
particular choice of the final-state meson (hence the label ``$M$'' on
$\Gamma_M$), namely $\Gamma_M=1$ for $M=P$ a pseudoscalar meson,
$\Gamma_M=\gamma_5$ for $M=V_\parallel$ a longitudinally polarized
vector meson, and $\Gamma_M=\gamma_{\perp\nu}$ for $M=V_\perp$ a
transversely polarized vector meson. The coefficients $K_M^\Gamma$ for
the various currents that are relevant in the discussion of
heavy-to-light form factors are summarized in Table~\ref{tab:coefs},
where we use the definitions (\ref{transtensors}) for the transverse
tensors $g_\perp^{\mu \nu}$ and $\epsilon_\perp^{\mu\nu}$. Once a
definition for the heavy-to-light form factors is adopted, it is an
easy exercise to read off from the table which of the coefficient
functions $C_i$ contribute to a given form factor. This determines the
functions $C_i(E,\mu_{\rm I})$ in (\ref{ffff}). (In general, these
functions are linear combinations of the $C_i$ in
Table~\ref{tab:coefs}.)

\begin{table}
\centerline{\parbox{14cm}{\caption[Matching of flavor-changing
currents from QCD onto SCET.]{\label{tab:coefs} Coefficients
$K_M^\Gamma$ arising in the leading-order matching of flavor-changing
currents from QCD onto SCET$_I$. The coefficients $C_i$ are defined in
equation (\ref{DiracReduction}) and given beyond tree-level in
\cite{Bauer:2000yr}. We denote $\hat q=q/M_B$ and $\hat E=E/M_B$,
where $q=p_B-p_M$.}}}
\vspace{0.1cm}
\begin{center}\small
\begin{tabular}{|c|ccc|}
\hline\hline
Current & $M=P$ & $M=V_\parallel$ & $M=V_\perp$ \\[-0.1cm]
$\bar q\,\Gamma\,b$ & ($\Gamma_M=1$) & ($\Gamma_M=\gamma_5$)
 & ($\Gamma_M=\gamma_{\perp\nu}$) \\
\hline\hline
$\bar q\gamma^\mu b$ & $(n^\mu C_4 + v^\mu C_5)$ & ---
 & $g_\perp^{\mu\nu} C_3$ \\
$\bar q\gamma^\mu\gamma_5 b$ & --- & $-(n^\mu C_7 + v^\mu C_8)$
 & $-i\epsilon_\perp^{\mu\nu} C_6$ \\
$\bar q\,i\sigma^{\mu\nu}\hat q_\nu b$
 & $[v^\mu - (1-\hat E) n^\mu]\,C_{11}$ & ---
 & $-g_\perp^{\mu\nu} [C_9 + (1-\hat E) C_{12}]$ \\
$\bar q\,i\sigma^{\mu\nu}\gamma_5\hat q_\nu b$
 & --- & $[v^\mu - (1-\hat E) n^\mu]\,C_{10}$
 & $-i\epsilon_\perp^{\mu\nu} (C_9 + \hat E C_{12})$ \\
\hline\hline
\end{tabular}
\end{center}
\vspace{-0.2cm}
\end{table}

Next, we match the time-ordered product $i\int d^4x\,\mbox{T}\,\{
J_M^{(0)}(0),{\cal L}_{\rm SCET_I}(x)\}$ onto operators in
SCET$_{II}$, focusing first on operators which include soft and
collinear fields only. The insertion of a subleading interaction from
the SCET$_I$ Lagrangian is required to transform the soft $B$-meson
spectator anti-quark into a hard-collinear final-state parton. In the
resulting SCET$_{II}$ interaction terms the soft and collinear fields
must be multipole expanded, since their momenta have different scaling
properties with the expansion parameter $\lambda$
\cite{Beneke:2002ph}, see Section~\ref{sec:scetPowCou}. Soft fields
are expanded about $x_+=0$ while collinear ones are expanded about
$x_-=0$. In general, collinear fields can live on different $x_+$
positions while soft fields can live on different $x_-$ positions. The
relevant operators can be written as matrix elements of color
singlet-singlet four-quark operators multiplied by position-dependent
Wilson coefficients, i.e.  \cite{Hill:2002vw,Becher:2003kh}
\begin{eqnarray}
   \int ds\,dt\,\widetilde D_k(s,t,E,\mu)
   &&\!\!\! [(\bar\xi\,W_c)(x_+ +x_\perp)\dots
    (W_c^\dagger\,\xi)(x_+ +x_\perp+s\bar n)] \nonumber\\[-0.2cm]
   \times &&\!\!\! [(\bar q_s\,S_s)(x_- +x_\perp+tn)\dots
    (S_s^\dagger\,h)(x_- +x_\perp)] \,,
\end{eqnarray}
where the dots represent different Dirac structures. There is no need to 
include color octet-octet operators, since they have vanishing 
projections onto physical hadron states and do not mix into the color 
singlet-singlet operators under renormalization. The Fourier transforms 
of the coefficient functions defined as
\begin{equation} \label{eq:FTofDk}
   D_k(\omega,u_2,E,\mu) = \int ds\,e^{2iE u_2 s}
   \int dt\,e^{-i\omega t}\,\widetilde D_k(s,t,E,\mu) 
\end{equation}
coincide with the momentum-space Wilson coefficient functions, which
we will calculate below. (In the matching calculation, $\omega$ is
identified with the component $n\cdot l$ of the incoming spectator
momentum, and $u_2$ is identified with the longitudinal momentum
fraction $x_2$ carried by the collinear anti-quark in the final-state
meson.) When we rewrite the operators above in terms of the
gauge-invariant building blocks defined in (\ref{blocks}), all factors
of the soft-collinear Wilson lines $S_{sc}(0)$ and $W_{sc}(0)$ cancel
out, as explained after equation~(\ref{singlet}), since the collinear
fields are evaluated at $x_-=0$ and the soft fields at $x_+=0$. Hence,
we can rewrite the operators in the form (setting $x=0$ for
simplicity)
\begin{equation}
   \int ds\,dt\,\widetilde D_k(s,t,E,\mu)\,
   [\bar\X(0)\dots\X(s\bar n)]\,[\bar\Q_s(tn)\dots\H(0)] \,.
\end{equation}

At leading order in $\lambda$, operators can also contain insertions
of transverse derivatives or gauge fields between the collinear or
soft quark fields. Since $\partial_\perp^\mu\sim\lambda$ and
$\A_{c\perp}^\mu\sim\A_{s\perp}^\mu\sim\lambda$, such transverse
insertions must be accompanied by a factor of
$\nslash/in\cdot\partial_s\sim\lambda^{-1}$. The inverse derivative
operator acts on a light soft field and implies an integration over
the position of that field on the $n$ light-cone, the effect of which
can be absorbed into the Wilson coefficient functions. The appearance
of $\nslash$ in the numerator is enforced by reparameterization
invariance, as derived in equation~(\ref{eq:RPIgivesNslash}). It then
follows that in our case only single insertions of transverse objects
are allowed \cite{Hill:2002vw,Bosch:2003fc}. Finally, the multipole
expansion of the soft fields ensures that the component $\bar n\cdot
p_s$ of soft momenta does not enter Feynman diagrams at leading
power. Likewise, at leading power there are no operators that contain
$in\cdot\partial_c\sim\lambda^2$ (acting on collinear fields) or
$n\cdot\A_c\sim\lambda^2$, since these are always suppressed with
respect to the corresponding transverse quantities. The new structures
containing gluon fields are of the form
\begin{equation}
\begin{aligned}
   &\int dr\,ds\,dt\,\widetilde D_k(r,s,t,E,\mu)\,
    [\bar\X(0)\dots\A_{c\perp}^\mu(r\bar n)\dots\X(s\bar n)]\,
    [\bar\Q_s(tn)\dots\H(0)] \,, \\
   &\int ds\,dt\,du\,\widetilde D_k(s,t,u,E,\mu)\,
    [\bar\X(0)\dots\X(s\bar n)]\,
    [\bar\Q_s(tn)\dots\A_{s\perp}^\mu(un)\dots\H(0)] \,,
\end{aligned}
\end{equation}
and we define the corresponding Fourier-transformed coefficient
functions in analogy to (\ref{eq:FTofDk}) as
$D_k(u_2,u_3,\omega,E,\mu)$ and $D_k(\omega_1,\omega_2,u_2,E,\mu)$.
In matching calculations, $u_3$ is identified with the longitudinal 
momentum fraction $x_3$ carried by a final-state collinear gluon, while
$\omega_1$ and $\omega_2$ are associated with the components $n\cdot l_q$ 
and $n\cdot l_g$ of the incoming soft anti-quark and gluon momenta. 

We are now in a position to construct a basis of four-quark operators 
relevant to the discussion of the universal functions $\zeta_M$ in 
(\ref{ffff}) \cite{Lange:2003pk}. 
These operators must 
transform like the current $J_M^{(0)}$ under Lorentz transformations. 
Also, the soft and collinear parts of the four-quark operators must have 
non-zero projections onto the $B$ meson and the final-state meson $M$. We 
set the transverse momenta of the mesons to zero, in which case there is 
no need to include operators with transverse derivatives acting on the 
products of all soft or collinear fields.

\newpage
\paragraph{Case \boldmath$M=P$\unboldmath:\/}
The resulting operators must transform as a scalar ($\Gamma_M=1$). A 
basis of such operators is
\begin{equation}\label{Pbasis}
\begin{aligned}
   O_1^{(P)}
   &= g^2\,[\bar\X(0)\,\frac{\nbslash}{2}\,\gamma_5\,\X(s\bar n)]\,
    [\bar\Q_s(tn)\,\frac{\nbslash\nslash}{4}\,\gamma_5\,\H(0)] \,, \\
   O_2^{(P)}
   &= g^2\,[\bar\X(0)\,\frac{\nbslash}{2}\,\gamma_5\,
    i\delslash_\perp\X(s\bar n)]\,
    [\bar\Q_s(tn)\,\frac{\nslash}{2}\,\gamma_5\,\H(0)] \,, \\
   O_3^{(P)}
   &= g^2\,[\bar\X(0)\,\frac{\nbslash}{2}\,\gamma_5\,
    \calAslash_{c\perp}(r\bar n)\,\X(s\bar n)]\,
    [\bar\Q_s(tn)\,\frac{\nslash}{2}\,\gamma_5\,\H(0)] \,, \\
   O_4^{(P)}
   &= g^2\,[\bar\X(0)\,\frac{\nbslash}{2}\,\gamma_5\,\X(s\bar n)]\,
    [\bar\Q_s(tn)\,\calAslash_{s\perp}(un)\,\frac{\nslash}{2}\,\gamma_5\,
    \H(0)] \,.
\end{aligned}
\end{equation}
The soft and collinear currents both transform like a pseudoscalar.

\vspace{-0.2cm}
\paragraph{Case \boldmath$M=V_\parallel$:\unboldmath}
The resulting operators must transform as a pseudo-scalar 
($\Gamma_M=\gamma_5$). A basis of such operators is obtained by omitting 
the $\gamma_5$ between the two collinear spinor fields in (\ref{Pbasis}), 
so that the collinear currents transform like a scalar. The Wilson 
coefficients for the cases $M=P$ and $M=V_\parallel$ coincide up to an 
overall sign due to parity invariance.

\vspace{-0.2cm}
\paragraph{Case \boldmath$M=V_\perp$:\unboldmath}
The resulting operators must transform as a transverse vector 
($\Gamma_M=\gamma_{\perp\nu}$). A basis of such operators is
\begin{equation}
\begin{aligned}
   O_1^{(V_\perp)}
   &= g^2\,[\bar\X(0)\,\frac{\nbslash}{2}\,
    \gamma_{\perp\nu}\gamma_5\,\X(s\bar n)]\,
    [\bar\Q_s(tn)\,\frac{\nbslash\nslash}{4}\,\gamma_5\,\H(0)] \,, \\
   O_2^{(V_\perp)}
   &= g^2\,[\bar\X(0)\,\frac{\nbslash}{2}\,
    (i\epsilon_{\nu\alpha}^\perp - g_{\nu\alpha}^\perp\gamma_5)\,
    i\partial_\perp^\alpha\,\X(s\bar n)]\,
    [\bar\Q_s(tn)\,\frac{\nslash}{2}\,\gamma_5\,\H(0)] \,, \\
   O_3^{(V_\perp)}
   &= g^2\,[\bar\X(0)\,\frac{\nbslash}{2}\,
    (i\epsilon_{\nu\alpha}^\perp - g_{\nu\alpha}^\perp\gamma_5)\,
    \A_{c\perp}^\alpha(r\bar n)\,\X(s\bar n)]\,
    [\bar\Q_s(tn)\,\frac{\nslash}{2}\,\gamma_5\,\H(0)] \,, \\
   O_4^{(V_\perp)}
   &= g^2\,[\bar\X(0)\,\frac{\nbslash}{2}\,
    \gamma_{\perp\nu}\gamma_5\,\X(s\bar n)]\,
    [\bar\Q_s(tn)\,\calAslash_{s\perp}(un)\,\frac{\nslash}{2}\,
    \gamma_5\,\H(0)] \,.
\end{aligned}
\end{equation}
The soft currents transform like a pseudoscalar, while the collinear
currents transform like a transverse axial-vector. (The relative sign
between the two terms in $O_{2,3}^{(V_\perp)}$ can be determined by
considering left and right-handed spinors $\bar\X$ and using relation
(\ref{wonderful}) below.)

It is not necessary to include operators with a derivative on the soft 
spectator anti-quark. All such operators would contain 
$\bar Q_s\,(-i\overleftarrow{\delslash}_{\!\!\perp})\,\nslash$, which can 
be related to a linear combination of the operators $O_1^{(M)}$ and 
$O_4^{(M)}$ using the equation of motion for the light-quark field.

\subsubsection{Wilson coefficients of four-quark operators}

We now calculate the momentum-space Wilson coefficients of the
operators $O_i^{(M)}$ at tree level \cite{Lange:2003pk}. The relevant
graphs must contain a hard-collinear gluon exchange, which turns the
soft $B$-meson spectator anti-quark into a collinear parton that can
be absorbed by the final-state hadron. These hard-collinear gluons are
integrated out when SCET$_I$ is matched onto SCET$_{II}$. However, at
$O(\alpha_s)$ we can obtain the Wilson coefficients in SCET$_{II}$ by
directly matching QCD amplitudes onto the low-energy theory, without
going through an intermediate effective theory.

As we have seen, at leading power the universal form factors $\zeta_M$
receive contributions from ordinary four-quark operators as well as
from four-quark operators containing an additional collinear or soft
gluon field. We start with a discussion of the matching calculation
for the operators $O_{1,2}^{(M)}$, whose coefficients can be obtained
by analyzing the diagrams shown in Fig.~\ref{fig:twopart} in the
kinematic region where the outgoing quarks are collinear and the
incoming quarks are soft. The resulting expression for the amplitude
in the full theory has already been given in (\ref{LO}). Note that, by
assumption, $n\cdot l\sim\Lambda_{\rm QCD}$ and $x_2\sim 1$, so that
the result is well defined and it is consistent to neglect subleading
terms in the gluon propagators.

We wish to express the result for the QCD amplitude as a combination of 
SCET$_{II}$ matrix elements multiplied by coefficient functions. To 
this end, we must first eliminate the superficially leading term in 
(\ref{LO}) using the equations of motion. Assigning momenta to the two 
outgoing collinear lines as shown in (\ref{p1p2}), it follows that
\begin{equation}\label{eom}
   \gamma_\mu E\nslash\,\Gamma * \gamma^\mu
   = \frac{\pslash_\perp}{x_1}\,\gamma_\mu\Gamma * \gamma^\mu
   - 2\Gamma * \frac{\pslash_\perp}{x_2} + \dots \,,
\end{equation}
where the dots denote power-suppressed terms. In the next step we use a 
Fierz transformation to recast the amplitude into a form that is 
convenient for our analysis. Taking into account that between collinear 
spinors the Dirac basis contains only three independent matrices, we 
find for general matrices $M$ and $N$
\begin{eqnarray}\label{Fierz}
  2 (\bar u_\xi M u_h)\,(\bar v_q N v_\xi) 
  &=& (\bar u_\xi \frac{\nbslash}{2} v_\xi)\,
   (\bar v_q N \frac{\nslash}{2} M u_h) 
   + (\bar u_\xi \frac{\nbslash}{2}\gamma_5 v_\xi)\,
   (\bar v_q N \gamma_5 \frac{\nslash}{2} M u_h) \nonumber \\
  &&\mbox{}+ (\bar u_\xi \frac{\nbslash}{2}\gamma_{\perp\alpha} v_\xi)\,
   (\bar v_q N \gamma_\perp^\alpha \frac{\nslash}{2} M u_h) \,.
\end{eqnarray}
To simplify the Dirac algebra we use the identities
\begin{equation}\label{wonderful}
   \gamma_\perp^\mu\gamma_5\,\nslash
    = i\epsilon_\perp^{\mu\nu}\gamma_{\perp\nu}\,\nslash \,, \qquad
   \gamma_\perp^\mu\gamma_\perp^\nu\,\nslash
    = (g_\perp^{\mu\nu} - i\epsilon_\perp^{\mu\nu}\gamma_5)\,\nslash \,.
\end{equation}
Finally, we include a minus sign from fermion exchange under the Fierz
transformation and project the collinear and soft ``currents'' in the 
resulting expressions onto color-singlet states.

Once in this form, the different contributions to the amplitude can be
readily identified with matrix elements of the operators $O_{1,2}^{(M)}$.
For $M=P,V_\parallel$ we obtain
\begin{equation}
   D_1^{(P)} = - D_1^{(V_\parallel)}
    = - \frac{C_F}{N}\,\frac{1+u_2}{4E^2 u_2^2\,\omega} \,, \qquad
   D_2^{(P)} = - D_2^{(V_\parallel)}
    = - \frac{C_F}{N}\,\frac{1}{4E^2 u_1 u_2^2\,\omega^2} \,,
\end{equation}
while for $M=V_\perp$ we find
\begin{equation}
   D_1^{(V_\perp)}
    = \frac{C_F}{N}\,\frac{1}{4E^2 u_2^2\,\omega} \,, \qquad
   D_2^{(V_\perp)} = \frac{C_F}{N}\,\frac{1}{4E^2 u_2^2\,\omega^2} \,.
\end{equation}

\subsubsection{Wilson coefficients of four-quark operators with an extra 
gluon}

\begin{figure}
\begin{center}
\epsfig{file=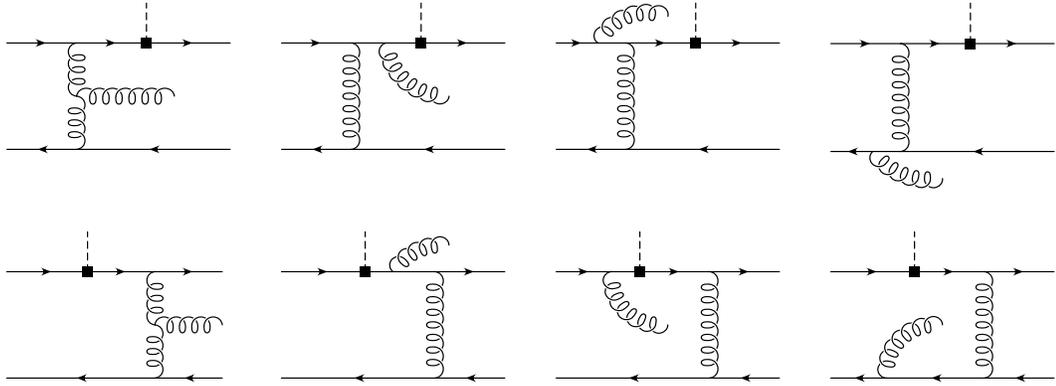,width=14cm}
\end{center}
\centerline{\parbox{14cm}{\caption[Diagrams with an extra gluon.]{\label{fig:1}
Diagrams relevant to the matching calculation for operators containing an 
extra collinear gluon.}}}
\end{figure}

The matching calculation for the operators $O_{3,4}^{(M)}$ proceeds along 
the same lines. However, in this case it is necessary to study diagrams 
with four external quarks and an external gluon, which we treat as a 
background field. Let us first discuss the case with three collinear 
particles in the final state. The relevant QCD graphs are shown in 
Fig.~\ref{fig:1}. Physically, they correspond to non-valence Fock 
states of the light final-state meson, but this interpretation is 
irrelevant for the matching calculation, which can be done with free 
quark and gluon states. 

By assumption, each of the three collinear particles carries large 
momentum components along the $n$ direction. We denote the corresponding 
longitudinal momentum fractions by $x_1$ (quark), $x_2$ (anti-quark) and
$x_3$ (gluon), where $x_1+x_2+x_3=1$. The calculation of the diagrams in 
Fig.~\ref{fig:1} exhibits that at leading power the amplitude depends 
only on the light-cone components $n\cdot l$ and $\bar n\cdot p_i=2E x_i$ 
of the external momenta, and that only transverse components of the 
external gluons fields must be kept. These observations are in accordance 
with the structure of the operator $O_3^{(M)}$. One must add to the 
diagrams shown in the figure a contribution arising from the application 
of the equation of motion for the collinear quark fields that led to 
(\ref{eom}). One way of obtaining it is to include diagrams with gluon 
emission from the external collinear quark lines in the matching 
calculation. Performing the Fierz transformation and projecting onto the 
color singlet-singlet operators $O_3^{(M)}$, we obtain the coefficient 
functions
\begin{equation}
\begin{aligned}
   D_3^{(P)} = - D_3^{(V_\parallel)}
   &= \frac{1}{8E^2(u_2+u_3)^2\omega^2} \left[
    \frac{u_3}{u_2} - 1 + \frac{2}{N^2} - \frac{2C_F}{N}
    \frac{(u_2+u_3)^2}{u_2(u_1+u_3)} \right] , \\
   D_3^{(V_\perp)}
   &= - \frac{1}{8E^2(u_2+u_3)^2\omega^2} \left[
    \frac{u_3}{u_2} - 1 + \frac{2}{N^2}
    + \frac{1}{N^2}\,\frac{(u_2+u_3)^2}{u_2(u_1+u_2)} \right] .
\end{aligned}
\end{equation}

Next, we calculate the contributions from diagrams with three external 
soft particles, which physically correspond to three-particle Fock states
of the $B$ meson. Again there are eight diagrams, analogous to those 
in Fig.~\ref{fig:1}. The initial soft gluon is attached to either an 
off-shell intermediate line or a collinear line. The diagrams with 
external gluons must again be supplemented by a contribution resulting 
from the application of the equation of motion $\bar v_q\,\lslash=O(g)$ 
used in the analysis of the four-quark amplitude in (\ref{LO}). After 
Fierz transformation and projection onto the color singlet-singlet 
operators $O_4^{(M)}$, we find the coefficient functions
\begin{equation}
\begin{aligned}
   D_4^{(P)} = - D_4^{(V_\parallel)}
   &= \frac{1}{8E^2 u_2^2(\omega_1+\omega_2)^2} \left[
    \left( 1 - \frac{2C_F}{N}\,u_2 \right) 
    \frac{\omega_2}{\omega_1} + \frac{1}{N^2} \right] , \\
   D_4^{(V_\perp)} 
   &= - \frac{1}{8E^2 u_2^2(\omega_1+\omega_2)^2} \left[
    \left( 1 + \frac{1}{N^2}\,\frac{u_2}{u_1} \right)
    \frac{\omega_2}{\omega_1}
    + \frac{1}{N^2}\,\frac{1}{u_1} \right] .
\end{aligned}
\end{equation}
The variables $\omega_1$ and $\omega_2$ correspond to the light-cone
components $n\cdot l_q$ and $n\cdot l_g$ of the incoming momenta of the 
soft anti-quark and gluon, respectively.

\subsubsection{Endpoint singularities}
\label{subsec:divs}

The Wilson coefficients of the SCET$_{II}$ four-quark operators 
become singular in the limit where some of the momentum components of the 
external particles tend to zero. When the $B\to M$ matrix elements of the 
operators $O_k^{(M)}$ are evaluated, these singularities give rise to
endpoint divergences of the resulting convolution integrals with LCDAs.
For instance, the matrix elements of the operators $O_1^{(M)}$ and 
$O_4^{(M)}$ involve the leading twist-2 projection onto the light meson
$M$. The corresponding LCDAs $\phi_M$ are expected to vanish linearly as 
$u_2\to 0$, whereas the corresponding coefficients contain terms that 
grow like $1/u_2^2$, giving rise to logarithmically divergent convolution 
integrals. Similarly, the operators $O_2^{(M)}$ and $O_3^{(M)}$ involve 
the leading-order projection onto the $B$-meson LCDA $\phi_B^{(+)}$, 
which is expected to vanish linearly as $\omega\to 0$ 
\cite{Lange:2003ff}. Once again, logarithmic singularities arise because 
the corresponding coefficients contain terms that grow like $1/\omega^2$. 

While the logarithmic divergences in the convolution integrals could be 
avoided by introducing some infra-red regulators, they indicate that 
leading-order contributions to the amplitudes arise from momentum regions 
that cannot be described correctly in terms of collinear or soft fields. 
In Section~\ref{sec:toy}, we will explain that in SCET$_{II}$ these 
configurations are accounted for by matrix elements of operators 
containing the soft-collinear messenger fields.

\subsection{Spin-symmetry breaking contributions}

The two-particle amplitude in (\ref{LO}) also includes contributions
for which the Dirac structure is different from
$\nslash\,\Gamma\,h$. These give rise to symmetry-breaking
contributions to the form factors. We shall not derive a complete
basis of all possible symmetry-breaking operators (there are many) but
rather list the ones that enter at first order in $\alpha_s$. Since
the spin-symmetry violating terms can be factorized in the form of the
second term in (\ref{ffff}), they are associated with a short-distance
coupling constant $\alpha_s(\sqrt{E\Lambda_{\rm QCD}})$. It is
therefore appropriate in this case to include the coupling constant in
the Wilson coefficient functions.

Let $\Gamma$ denote the Dirac structure of the flavor-changing currents
$\bar q\,\Gamma\,b$, whose matrix elements define the form factors. To 
first order in $\alpha_s$, the spin-symmetry breaking terms can then be 
obtained from the matrix elements of two operators given by
\begin{equation}\label{Q1Q2}
\begin{aligned}
   Q_1 &= [\bar\X(0)\,\frac{\nbslash}{2} \left(
    \begin{array}{c} 1 \\ \gamma_5 \\ \gamma_{\perp\alpha} \end{array}
    \right) \X(s\bar n)]\,
    [\bar\Q_s(tn)\,\frac{\nslash}{2}\,\gamma_\mu\! \left( 
    \begin{array}{c} 1 \\ -\gamma_5 \\ -\gamma_\perp^\alpha \end{array}
    \right) \Gamma\,\gamma^\mu\nslash\,\H(0)] \,, \\
   Q_2 &= [\bar\X(0)\,\frac{\nbslash}{2} \left(
    \begin{array}{c} 1 \\ \gamma_5 \\ 0 \end{array} \right) \X(s\bar n)]\,
    [\bar\Q_s(tn)\,\frac{\nslash\nbslash}{4}
    \left( \begin{array}{c} 1 \\ -\gamma_5 \\ 0 \end{array} \right)
    \Gamma\,\H(0)] \,,
\end{aligned}
\end{equation}
where each line contributes for a different final-state meson $M$. The 
corresponding Wilson coefficients are
\begin{equation}
   \hat T_1 = - \frac{C_F}{N}\,\frac{\pi\alpha_s}{2M_B E u_2\,\omega} \,,
    \qquad
   \hat T_2 = \frac{C_F}{N}\,\frac{\pi\alpha_s}{E^2 u_2\,\omega} \,.
\end{equation}
Linear combinations of $\hat T_1$ and $\hat T_2$ determine (up to 
prefactors) the hard-scattering kernels $T_i$ in (\ref{ffff}). The matrix 
elements of the operators $Q_{1,2}$ can be expressed in terms of the 
leading-order LCDAs of the $B$ meson and the light meson $M$. Only the 
$B$-meson LCDA called $\phi_B^{(+)}$ contributes \cite{Grozin:1996pq} 
because of the factor $\nslash$ next to $\bar Q_s$. This property holds
true to all orders in perturbation theory \cite{Bauer:2002aj}. The 
resulting convolution integrals are convergent. Evaluating these matrix 
elements we reproduce the spin-symmetry breaking terms obtained in 
\cite{Beneke:2000wa}.

\section{Physics of endpoint singularities -- a toy model}
\label{sec:toy}

Our strategy in the previous sections has been to perform matching 
calculations by expanding QCD amplitudes in powers of $\Lambda_{\rm QCD}/E$, 
assuming that collinear and soft external momenta have the scaling 
assigned to them in SCET. We have then matched the results onto 
SCET$_{II}$ operators containing soft and collinear fields and read 
off the corresponding Wilson coefficient functions in momentum space. A 
problematic aspect of this procedure has been the observation that, if 
the matrix elements of the effective-theory operators are expressed in
terms of meson LCDAs, then the resulting convolution integrals do not 
converge. Endpoint singularities arise, which correspond to exceptional 
momentum configurations in which some of the partons inside the external
hadrons carry very small momentum. The question naturally arises how one 
should interpret these singularities, and whether the results we found 
for the short-distance coefficient functions are in fact correct.

\begin{figure}
\begin{center}
\epsfig{file=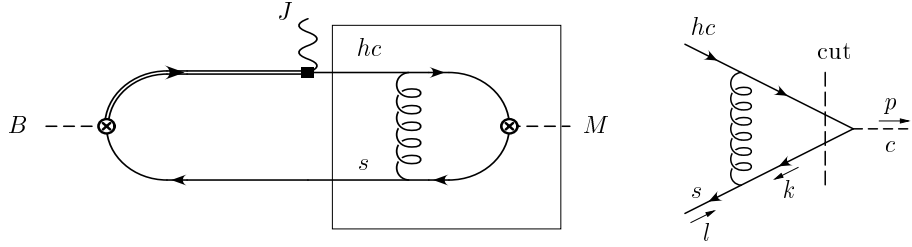,width=12cm} 
\end{center}
\centerline{\parbox{14cm}{\caption[Subgraph study of endpoint
singularities.]{\label{fig:toy} Triangle subgraph whose spectral
function can be used to study endpoint singularities on the collinear
side.}}}
\end{figure}

The fact that endpoint configurations are not kinematically suppressed 
points to the relevance of new momentum modes. In the limit $x_2\to 0$, 
the scaling associated with the collinear anti-quark in the final state
of Fig.~\ref{fig:twopart} changes from $(\lambda^2,1,\lambda)$ to
$(\lambda^2,\lambda,\dots)$. Likewise, in the limit $n\cdot l\to 0$, the 
scaling associated with the soft spectator anti-quark in the initial 
state changes from $(\lambda,\lambda,\lambda)$ to 
$(\lambda^2,\lambda,\dots)$. The soft-collinear messenger fields in the
SCET$_{II}$ Lagrangian have precisely the scaling properties 
corresponding to these exceptional configurations. (Since the modes in an 
effective theory are always on-shell, the transverse components of 
soft-collinear momenta scale like $\lambda^{3/2}$. We will see below that 
this is not really relevant.)

The purpose of this section is to analyze the interplay of the various
modes present in SCET$_{II}$ and to see how soft-collinear messengers
are connected with the phenomenon of endpoint singularities. We will
do this with the help of a toy example. Consider the triangle subgraph
enclosed by the box in the diagram shown on the left-hand side in
Fig.~\ref{fig:toy}, which is one of the two gluon-exchange graphs
relevant to the $B\to M$ form factors (see Fig.~\ref{fig:twopart}).
To understand the physics of endpoint singularities we focus on the
side of the light meson $M$ (an analogous discussion could be given
for the $B$-meson side). We study the discontinuity of the triangle
diagram in the external collinear momentum $p^2$. The resulting
spectral density $\varrho(p^2)$ models the continuum of light
final-state hadrons that can be produced on the collinear side. We
will not bother to project out a particular light meson from this
spectral density. For simplicity, we will also ignore any numerator
structure in the triangle subgraph and instead study the corresponding
scalar triangle, as discussed in
\cite{Becher:2003qh,Lange:2003pk}. 

Let us define the discontinuity
\begin{equation}
   \frac{1}{2\pi i}\,\Big[ I(p^2+i0) - I(p^2-i0) \Big]
   \equiv D\cdot\theta(p^2)
\end{equation}
of the scalar triangle integral
\begin{equation}
   I = i\pi^{-d/2} \mu^{4-d} \int d^dk\,
   \frac{2l_+\cdot p_-}{(k^2+i0)\,[(k+l)^2+i0]\,[(k+p)^2+i0]}
\end{equation}
in $d=4-2\epsilon$ space-time dimensions, where $p$ is the outgoing
collinear momentum and $l$ the incoming soft momentum. It will be
convenient to define the invariants $L^2\equiv-l^2-i0$ and
$Q^2\equiv-(l-p)^2=2l_+\cdot p_-+\dots$, which scale like
$L^2\sim\lambda^2$ and $Q^2\sim\lambda$. (In physical units,
$L^2\sim\Lambda_{\rm QCD}^2$ and $Q^2\sim E\Lambda_{\rm QCD}$ with
$E\gg\Lambda_{\rm QCD}$.) We will assume that these quantities are
non-zero. From \cite{Becher:2003qh}, we can obtain explicit results
for the discontinuity $D$ and for the various regions of loop momentum
$k$ that give a non-vanishing contribution. We find that
\begin{equation}
   D = \ln\frac{Q^2}{L^2} + O(\epsilon,\lambda) \,,
\end{equation}
and that the momentum configurations that contribute to this result are 
those where the loop momentum is either collinear, meaning that
$k\sim(\lambda^2,1,\lambda)$, or soft-collinear, meaning that
$k\sim(\lambda^2,\lambda,\lambda^{3/2})$. The contributions from these 
two regions are
\begin{equation}\label{DcDsc}
\begin{aligned}
   D_{\rm C} &= \frac{\Gamma(-\epsilon)}{\Gamma(1-2\epsilon)}
    \left( \frac{\mu^2}{p^2} \right)^\epsilon
    = - \frac{1}{\epsilon} + \gamma_E - \ln\frac{\mu^2}{p^2}
    + O(\epsilon) \,, \\
   D_{\rm SC} &= \Gamma(\epsilon)
    \left( \frac{\mu^2 Q^2}{p^2 L^2} \right)^\epsilon
    = \frac{1}{\epsilon} - \gamma_E + \ln\frac{\mu^2 Q^2}{p^2 L^2}
    + O(\epsilon) \,,
\end{aligned}
\end{equation}
which add up to the correct answer. 

It is instructive to rewrite these results in a more transparent form. To 
this end we use Cutkosky rules to evaluate the discontinuities of the 
diagrams directly and perform all phase-space integrations except the 
integral over the light-cone component $\bar n\cdot k$ of the loop 
momentum, which we parameterize as 
$\bar n\cdot k\equiv-x_2\bar n\cdot p$. As in previous sections, $x_2$ 
denotes the fraction of longitudinal momentum carried by the anti-quark 
in the final-state. The exact result for the discontinuity of the scalar
triangle is ($\bar x_2\equiv 1-x_2$)
\begin{eqnarray}\label{Dfull}
   D &=& \frac{1}{\Gamma(1-\epsilon)}
    \left( \frac{\mu^2}{p^2} \right)^\epsilon
    \int_0^1\!dx_2\,\frac{(x_2\bar x_2)^{-\epsilon}}{x_2+L^2/Q^2}
    + O(\lambda) \nonumber\\
   &=& \int_0^1\!dx_2\,\frac{1}{x_2+L^2/Q^2} + O(\epsilon,\lambda)
    = \int_{L^2/Q^2}^1 \frac{dx_2}{x_2} + O(\epsilon,\lambda) \,.
\end{eqnarray}
The collinear and soft-collinear contributions separately are divergent 
even though they correspond to tree diagrams (after the two propagators 
have been cut). We obtain
\begin{equation}\label{DcDsc2}
\begin{aligned}
   D_{\rm C} &= \frac{1}{\Gamma(1-\epsilon)}
    \left( \frac{\mu^2}{p^2} \right)^\epsilon
    \int_0^1 \frac{dx_2}{x_2}\,(x_2\bar x_2)^{-\epsilon} \,, \\
   D_{\rm SC} &= \frac{1}{\Gamma(1-\epsilon)}
    \left( \frac{\mu^2}{p^2} \right)^\epsilon
    \int_0^\infty\!dx_2\,\frac{x_2^{-\epsilon}}{x_2+L^2/Q^2}
    \,.
\end{aligned}
\end{equation}
The collinear contribution is infra-red singular for $x_2\to 0$, which is 
an example of an endpoint singularity. In the present case, the 
singularity is regularized dimensionally by keeping $\epsilon$ non-zero. 
The soft-collinear contribution is infra-red finite but ultra-violet 
divergent, since $x_2$ runs up to $\infty$. Again, this divergence is 
regularized dimensionally. Evaluating the remaining integrals one 
recovers the exact results in (\ref{DcDsc}). 

\begin{figure}
\begin{center}
\epsfig{file=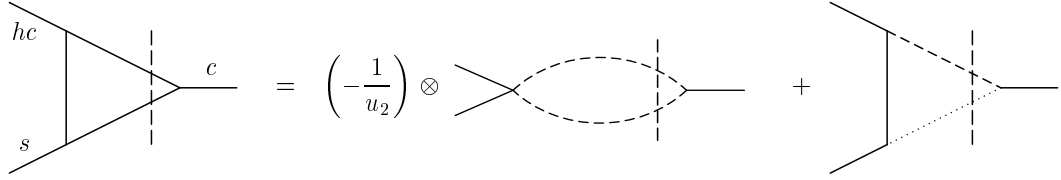,width=14cm} 
\end{center}
\centerline{\parbox{14cm}{\caption[Endpoint singularities in a toy
model.]{\label{fig:toy2} Collinear and soft-collinear contributions to
the scalar triangle graph in SCET$_{II}$. Dashed (dotted) lines
represent collinear (soft-collinear) propagators.}}}
\end{figure}

Several comments are in order:

i)\;
The result for the spectral density in the full theory is finite. The
endpoint singularity is regularized by keeping subleading terms in the
hard-collinear propagator $1/[-(k+l)^2]\simeq 1/(x_2 Q^2+L^2)$. When the 
subleading terms are dropped based on naive power counting (as we did in 
the analysis of the previous sections) the full-theory result reduces to 
the contribution obtained from the collinear region. An endpoint 
singularity arises in this case, which however can be regularized 
dimensionally. In order for this to happen in the realistic case of
external meson states, it would be necessary to perform the projections 
onto the meson LCDAs in $d\ne 4$ dimensions. The factor 
$(x_2\bar x_2)^{-\epsilon}$ in (\ref{DcDsc2}) would then correspond to a
modification of the LCDAs, which renders the convolution integrals 
finite.

ii)\;
The collinear approximation fails for values $x_2=O(\lambda)$. For such 
small momentum fractions the exact full-theory result is reproduced by 
the contribution obtained from the soft-collinear region. In fact, the 
collinear and soft-collinear contributions in (\ref{DcDsc2}) coincide 
with the first terms in the Taylor expansion of the full-theory result in 
(\ref{Dfull}) in the limits where $x_2=O(1)\gg L^2/Q^2$ and 
$x_2=O(\lambda)\ll 1$. The fact that $x_2$ runs up to $\infty$ in the 
soft-collinear contribution is not a problem. In dimensional 
regularization the integral receives significant contributions only from 
the region where $x_2\sim L^2/Q^2=O(\lambda)$. To see this, one can 
introduce a cutoff $\delta$ to separate the collinear and soft-collinear 
contributions, chosen such that $1\gg\delta\gg\lambda$. This cutoff is 
introduced as an infra-red regulator in the collinear integral 
($\int_0^1\to\int_\delta^1$) and as an ultra-violet regulator in the 
soft-collinear integral ($\int_0^\infty\to\int_0^\delta$). The integrals 
can then be evaluated setting $\epsilon\to 0$. This yields 
$D_{\rm C}=-\ln\delta$ and 
$D_{\rm SC}=\ln\delta+\ln\frac{Q^2}{L^2}+O(\lambda/\delta)$. The sum of 
the two contributions once again reproduces the exact result.

iii)\; 
Next, note that the soft-collinear contribution {\em precisely\/}
reproduces the endpoint behavior of the full-theory amplitude. It is
irrelevant in this context that the soft-collinear virtuality
$k_{sc}^2\sim E^2\lambda^3$ is parametrically smaller than the QCD
scale $\Lambda_{\rm QCD}^2$. What matters is that the plus and minus
components of the soft-collinear momentum,
$k_{sc}\sim(\lambda^2,\lambda,\dots)$, are of the same order as the
corresponding components of a collinear momentum in the endpoint
region, where $\bar n\cdot p\sim\lambda$ rather than being $O(1)$.

iv)\;
Finally, we see that the endpoint divergences we encountered in the 
Section~\ref{subsec:divs} were not regularized because we dropped the 
dimensional regulator when performing the projections onto meson LCDAs. 
This, however, does not affect the results for the Wilson coefficients of 
the SCET$_{II}$ operators, which remain valid. In the toy model, the 
collinear contribution is represented in the effective theory as the 
discontinuity of the integral (corresponding to the first diagram on the 
right-hand side in Fig.~\ref{fig:toy2})
\begin{eqnarray}
   I_{\rm C}
   &=& i\pi^{-d/2} \mu^{4-d} \int d^dk\,
    \frac{2l_+\cdot p_-}{(k^2+i0)\,(2k_-\cdot l_+)\,[(k+p)^2+i0]}
    \nonumber\\
   &=& i\pi^{-d/2} \mu^{4-d} \int d^dk\,\frac{(-1)}{x_2}\,
    \frac{1}{(k^2+i0)\,[(k+p)^2+i0]} \,,
\end{eqnarray}
which shows that the Wilson coefficient resulting from integrating out 
the hard-collinear propagator is simply $-1/u_2$.

To summarize this discussion, we stress that to reproduce the behavior of 
the full amplitude it is necessary to include in the effective theory 
contributions involving collinear fields and those involving 
soft-collinear messengers, as indicated in Fig.~\ref{fig:toy2}. The 
latter ones represent the exact behavior of the amplitude in the endpoint 
region. Whether or not endpoint configurations contribute at leading 
order in power counting is equivalent to the question of whether or not 
operators involving soft-collinear fields can arise at leading power. 
This shows the power of our formalism. Operators containing 
soft-collinear fields can be used to parameterize in a systematic way the 
long-distance physics associated with endpoint configurations in the 
external mesons states. This will be discussed in more detail in the next 
section. It also follows that the scaling properties of operators 
containing soft-collinear fields can be used to make model-independent 
statements about the convergence of convolution integrals that arise in 
QCD factorization theorems such as (\ref{ffff}). We will exploit this 
connection in Section~\ref{sec:fffact}.

\section{Soft-collinear messengers and the soft overlap}
\label{sec:softoverlap}

In Section~\ref{sec:matching} we have derived four-quark operators built 
out of collinear and soft fields, which contribute at leading order to 
the universal soft functions $\zeta_M$ in (\ref{ffff}). Power counting 
shows that the products of these operators with their Wilson coefficients 
scale like $\lambda^4$. (Here one uses that $dr\sim ds\sim 1$ and 
$dt\sim du\sim\lambda^{-1}$.) Taking into account the counting of the 
external hadron states, $|B\rangle\sim\lambda^{-3/2}$ and 
$\langle M|\sim\lambda^{-1}$, it follows that the corresponding 
contributions to the universal form factors scale like $\lambda^{3/2}$. 
In other words, heavy-to-light form factors are quantities that vanish at 
leading order in the large-energy expansion. The fact that the resulting 
convolution integrals were infra-red singular suggested that there should 
be other contributions of the same order in power counting, which cannot 
be described in terms of collinear or soft fields.

An important observation made in the previous section was that the 
endpoint behavior of QCD amplitudes is described in the low-energy theory 
in terms of diagrams involving soft-collinear messenger fields. For 
instance, in the last graph in Fig.~\ref{fig:toy2} the soft spectator 
anti-quark inside the $B$ meson turns into a soft-collinear anti-quark by 
the emission of a soft gluon. The soft-collinear anti-quark is then 
absorbed by the final-state meson. In a similar way, endpoint 
singularities on the $B$-meson side would correspond to a situation where 
the initial state contains a soft-collinear spectator anti-quark, which 
absorbs a collinear gluon and turns into a collinear anti-quark. The 
SCET$_{II}$ Lagrangian contains the corresponding interaction terms 
only at subleading order in $\lambda$. However, because the universal 
form factors $\zeta_M$ themselves are power-suppressed quantities, these 
subleading interactions will nevertheless give rise to leading-power 
contributions.

We will need the first two non-vanishing orders in the interactions that 
couple a soft-collinear quark to a soft or collinear quark. The results 
are most transparent when expressed in terms of the gauge-invariant 
building blocks introduced in (\ref{blocks}). They are 
\cite{Becher:2003qh} 
\pagebreak
\begin{equation}
   {\cal L}_{\bar q\theta}^{(1/2)}
   = \bar\Q_s\,\calAslash_{s\perp} W_{sc}^\dagger\,\theta \,, \qquad 
   {\cal L}_{\bar\theta\xi}^{(1/2)}
   = \bar\sigma\,S_{sc}\,\calAslash_{c\perp}\,\X \,, \vspace{-0.2cm}
\end{equation}
and
\begin{equation}
\begin{aligned}
   {\cal L}_{\bar q\theta}^{(1)}
   &= \bar\Q_s\,\calAslash_{s\perp} W_{sc}^\dagger\,
    (x_\perp\cdot D_{sc}\,\theta + \sigma) 
    + \bar\Q_s\,\frac{\nslash}{2}\,\bar n\cdot\A_s\,
    W_{sc}^\dagger\,\sigma \,, \\
   {\cal L}_{\bar\theta\xi}^{(1)}
   &= \bar\sigma\,x_\perp\cdot\overleftarrow{D}_{sc}\,S_{sc}\,
    \calAslash_{c\perp}\,\X
    - \bar\theta\, S_{sc}\,\calAslash_{c\perp}\,\frac{\nbslash}{2}\,
    \frac{1}{i\bar n\cdot\partial}\,
    (i\delslash_\perp + \calAslash_{c\perp})\,\X
    + \bar\theta\,S_{sc}\,\frac{\nbslash}{2}\,n\cdot\A_c\,\X \,,
\end{aligned}
\end{equation}
where 
\begin{equation}\label{sigmasol}
   \sigma = - \frac{\nbslash}{2}\,\frac{1}{i\bar n\cdot D_{sc}}\,
   i\Dslash_{sc\perp}\,\theta
\end{equation}
contains the ``small components'' of the soft-collinear quark field, 
which are integrated out in the construction of the effective Lagrangian.
The longitudinal components of the calligraphic gluon fields are defined 
in \cite{Becher:2003qh}. (Note also that $n\cdot\A_s=0$ and 
$\bar n\cdot\A_c=0$.) The soft-collinear fields enter these results in 
combinations such as $W_{sc}^\dagger\,\theta$ or 
$S_{sc}^\dagger\,\theta$, which are gauge invariant. Soft and collinear 
fields live at position $x$, while soft-collinear fields are evaluated at 
position $x_+$ for ${\cal L}_{\bar q\theta}$ and $x_-$ for 
${\cal L}_{\bar\theta\xi}$. The measure $d^4x$ associated with these 
interactions scales like $\lambda^{-4}$. The superscript on the 
Lagrangians indicates at which order in power counting ($\lambda^{1/2}$ 
or $\lambda$) the corresponding terms contribute to the action.

Next, we need the representation of the flavor-changing SCET$_I$ 
current $J_M^{(0)}$ in (\ref{ffcurrent}) in terms of operators in the 
low-energy theory SCET$_{II}$. As shown in
\cite{Becher:2003kh}, at leading power, and at the matching scale 
$\mu=\mu_{hc}$, the relation reads
\begin{equation}\label{Jmatch}
   J_M^{(0)}(x) 
   \to {\cal J}_M^{(0)}(x) = \bar\X(x_+ + x_\perp)\,\Gamma_M\,
   (S_{sc}^\dagger\,W_{sc})(0)\,\H(x_- + x_\perp)
\end{equation}
with a Wilson coefficient equal to unity. The anomalous dimensions of the 
currents are the same in the two theories. The product 
$(S_{sc}^\dagger\,W_{sc})(0)$ arises since soft-collinear messenger 
fields cannot be decoupled from the current operator ${\cal J}_M^{(0)}$ 
in SCET$_{II}$, contrary to the case of the color singlet-singlet 
four-quark operators discussed in Section~\ref{sec:matching}.

\begin{figure}
\begin{center}
\epsfig{file=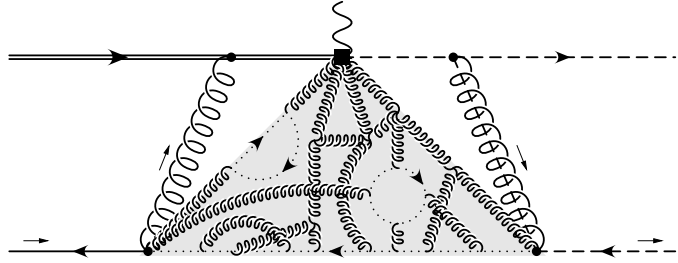,width=9cm} 
\end{center}
\centerline{\parbox{14cm}{\caption[An artist's view of the
soft-collinear messenger contribution to the form
factors. ]{\label{fig:SCblob} An artist's view of the soft-collinear
messenger contribution to the form factors. The shaded region contains
soft-collinear interactions. The arrows indicate the flow of the
components $n\cdot p_s$ (left) and $\bar n\cdot p_c$ (right) of soft
and collinear momenta, which do not enter the soft-collinear block.}}}
\end{figure}

Using these results, we can write down a tri-local operator whose matrix 
element provides a long-distance contribution to the universal form 
factors. It is
\begin{equation}\label{3T}
   O_5^{(M)} = i^2\!\int d^4x\,d^4y\,\mbox{T} \left\{
    {\cal L}_{\bar q\theta}^{(1/2)}(x)\,
    {\cal L}_{\bar\theta\xi}^{(1)}(y)\,{\cal J}_M^{(0)}(0)
    + {\cal L}_{\bar q\theta}^{(1)}(x)\,
    {\cal L}_{\bar\theta\xi}^{(1/2)}(y)\,{\cal J}_M^{(0)}(0) \right\} .
\end{equation}
Even after the decoupling transformation (\ref{blocks}) this operator 
contains arbitrarily complicated soft-collinear exchanges, as illustrated 
in the cartoon in Fig.~\ref{fig:SCblob}.

Note that the superficially leading term in the time-ordered product 
cancels \cite{Becher:2003qh}. This can be understood as follows: 

After the decoupling transformation the strong-interaction part of the
SCET$_{II}$ Lagrangian no longer contains unsuppressed interactions
between soft-collinear messengers and soft or collinear fields. In
order to preserve a transparent power counting it is then convenient
to define hadron states in the effective theory as eigenstates of one
of the two leading-order Lagrangians ${\cal L}_s$ and ${\cal
L}_c$. For instance, we define a ``SCET pion'' to be a bound state of
only collinear fields, and a ``SCET $B$ meson'' to be a bound state of
only soft fields. The SCET pion state coincides with the true pion,
because the collinear Lagrangian is equivalent to the QCD Lagrangian
\cite{Beneke:2002ph}.\footnote{The endpoint region of the pion wave
function is not described in terms of a Fock component containing a
soft-collinear parton, but rather in terms of a time-ordered product
of the SCET pion state with an insertion of the Lagrangian ${\cal
L}_{\bar\theta\xi}$. This insertion is non-zero only in processes
where also soft partons are involved.}  The SCET $B$ meson coincides
with the asymptotic heavy-meson state as defined in heavy-quark
effective theory. It is important in this context that time-ordered
products of soft-collinear fields with only collinear or only soft
fields vanish to all orders as a consequence of analyticity
\cite{Becher:2003qh}. Hence, soft-collinear modes do not affect the
spectrum of hadronic eigenstates of the collinear or soft Lagrangians.

Once the SCET$_{II}$ hadron states are defined in this way, each term
in the time-ordered product (\ref{3T}) can be factorized into a part
containing all soft and collinear fields times a vacuum correlation
function of the soft-collinear fields. These vacuum correlators must
be invariant under rotations in the transverse plane. It follows that
the correlator arising from the superficially leading term in the
time-ordered product vanishes, since
\begin{equation}\label{vacuum}
   \langle\Omega|\,\mbox{T}\,\{ (\bar\sigma\,S_{sc})_i(y_-)\,
   (S_{sc}^\dagger\,W_{sc})_{jk}(0)\,(W_{sc}^\dagger\,\theta)_l(x_+)
   \}\,|\Omega\rangle
\end{equation}
contains a single transverse derivative (see (\ref{sigmasol})).

The leading terms in the time-ordered product in (\ref{3T}) scale like
$\lambda^4$, since ${\cal J}_M^{(0)}\sim\lambda^{5/2}$. They thus 
contribute at the same order in power counting to the universal functions 
$\zeta_M$ as the four-quark operators discussed in 
Section~\ref{sec:matching}. When $O_5^{(M)}$ is added to the list of 
four-quark operators a complete description of the soft-overlap 
contribution is obtained. The Wilson coefficient of this new operator 
follows from the fact that there is no non-trivial matching coefficient 
in (\ref{Jmatch}), that the current operators have the same anomalous 
dimensions in SCET$_I$ and SCET$_{II}$, and that the 
SCET$_{II}$ Lagrangian is not renormalized. Hence, to all orders in 
perturbation theory
\begin{equation}\label{D5}
   D_5^{(M)}(\mu_{hc},\mu) = \frac{C_i(\mu)}{C_i(\mu_{hc})}
   \equiv D_5(\mu_{hc},\mu) \,,
\end{equation}
which is in fact a universal function, independent of the labels ``$i$''
and ``$M$''. As before, $\mu_{hc}$ denotes the hard-collinear matching 
scale, at which the transition SCET$_I$\,$\to$\,SCET$_{II}$ is
made. The appearance of $C_i(\mu_{hc})$ in the denominator of this 
relation is due to the fact that this coefficient was factored out in the 
definition of $D_k^{(M)}$, see~(\ref{ffff}) and (\ref{scet2ff}).

If we define by $\xi_k^{(M)}$ the $B\to M$ hadronic matrix elements of
the operators $O_k^{(M)}$, then the sum $\zeta_M=\sum_k
D_k^{(M)}\,\xi_k^{(M)}$ describes the entire soft overlap
contribution. Each term in this sum gives a contribution of order
$\lambda^{3/2}$ to the form factors, in accordance with the scaling
law obtained a long time ago in the context of QCD sum rules
\cite{Chernyak:ag}. We have thus completed the derivation of the new
factorization formula (\ref{scet2ff}). Whereas the Wilson coefficients
$D_k^{(M)}$ depend on the renormalization scale $\mu$ as well as on
the hard-collinear scale $\mu_{hc}\sim\sqrt{E\Lambda_{\rm QCD}}$, the
characteristic scale of the hadronic matrix elements $\xi_k^{(M)}$ is
the QCD scale $\Lambda_{\rm QCD}$, not the hard-collinear scale. While
this is obvious for the matrix elements of the operator $O_5^{(M)}$,
it also holds true for the remaining matrix elements, for which the
sensitivity to long-distance physics is signaled by the presence of
endpoint singularities. It remains to discuss how our results are
affected by single and double logarithmic corrections arising in
higher orders of perturbation theory.

\section{Operator mixing and Sudakov logarithms}
\label{sec:Sudakov}

Because the operators $O_k^{(M)}$ share the same global quantum numbers
they can mix under renormalization. This mixing is governed by a 
$5\times 5$ matrix of anomalous dimension kernels, which are in general
complicated functions of the light-cone variables $u_i$ and $\omega_i$. 
The anomalous dimension matrix governing this mixing has the structure
\begin{equation}\label{gamma}
   \bm{\gamma} = \left( \begin{array}{cccc|c}
    \gamma_{11} & 0 & 0 & \gamma_{14} & 0 \\
    0 & \gamma_{22} & \gamma_{23} & 0 & 0 \\
    0 & \gamma_{32} & \gamma_{33} & 0 & 0 \\
    \gamma_{41} & 0 & 0 & \gamma_{44} & 0 \\
    \hline
    \gamma_{51} & \gamma_{52} & \gamma_{53} & \gamma_{54} & \gamma_{55}
   \end{array} \right) .
\end{equation}
Because the operators $O_{1\dots 4}^{(M)}$ consist of products of soft 
and collinear currents (after decoupling of soft-collinear messengers), 
the entries in the upper left $4\times 4$ sub-matrix can all be written 
as sums of soft and collinear anomalous dimensions for the corresponding 
current operators. We have taken into account that the operators 
$O_2^{(M)}$ and $O_3^{(M)}$ mix under renormalization (this mixing is 
determined by the mixing of twist-3 two-particle and three-particle LCDAs 
for light mesons), as do the operators $O_1^{(M)}$ and $O_4^{(M)}$ (this 
mixing has not yet been worked out but is allowed on general grounds). 
The operator $O_5^{(M)}$ consisting of a triple time-ordered product 
requires the ``local'' operators as counter-terms; however, those 
operators do not mix back into $O_5^{(M)}$.

Because of the structure of the anomalous dimension matrix it follows
that the coefficient $\gamma_{55}$ is one of the eigenvalues, which
governs the scale dependence of the Wilson coefficient $D_5$ in
(\ref{D5}). This coefficient coincides with the well-known anomalous
dimension of the current operators $J_M^{(0)}$ and ${\cal J}_M^{(0)}$
computed in Section~\ref{sec:currents}. The corresponding ``operator
eigenvector'' can be written as a linear combination (in the
convolution sense)
\begin{equation}\label{calO5}
   {\cal O}_5^{(M)} = O_5^{(M)} + \sum_{k=1}^4 d_k^{(M)}
   \otimes O_k^{(M)} \,, \vspace{-0.2cm}
\end{equation}
where the coefficients $d_k^{(M)}$ are independent of the high-energy 
matching scale $\mu_{hc}$. The other four eigenvectors are linear 
combinations of the operators $O_k^{(M)}$ with $k\ne 5$. This observation 
has an important consequence: 
The decomposition of the form factors in terms of matrix elements of
these eigenvectors implies that there exists a contribution from 
${\cal O}_5^{(M)}$. The remaining contributions can be written as
endpoint-finite convolution integrals that involve both two- and
three-particle LCDAs of the initial and final mesons. Leaving such
factorizable terms aside, it follows that the combination of operators
contributing to the universal form factors must, up to an overall
factor, coincide with the eigenvector ${\cal O}_5^{(M)}$, and hence
\begin{equation}
   d_k^{(M)} = \frac{D_k^{(M)}}{D_5}
\end{equation}
to all orders in perturbation theory. 
With a slight abuse of notation, let us now denote by $\zeta_M$ the 
$B\to M$ hadronic matrix element of the eigenvector ${\cal O}_5^{(M)}$ in
SCET$_{II}$. Combining (\ref{ffff}) and (\ref{D5}), we then find that 
the spin-symmetric universal form-factor term can be rewritten as
\begin{equation}\label{ffwow}
   C_i(E,\mu_{\rm I})\,\zeta_M(\mu_{\rm I},E) \big|_{\rm SCET_I}
   =  C_i(E,\mu)\,\zeta_M(\mu,E) \big|_{\rm SCET_{II}} \,.
\end{equation}
This relation is not as dull as it seems; rather, it contains the
remarkable message that for the soft overlap contribution to
heavy-to-light form factors the intermediate hard-collinear scale is
without {\em any\/} physical significance. Switching from SCET$_I$ to
SCET$_{II}$ we merely describe the same physics using a different set
of degrees of freedom. In other words, there is no use of going
through an intermediate effective theory. The RG evolution of the soft
functions $\zeta_M$ remains the same all the way from the high-energy
scale $E\sim M_B$ down to hadronic scales $\mu\sim\Lambda_{\rm
QCD}$. The physics of the soft overlap term is thus rather different
from the physics of the spin-symmetry breaking corrections in the
factorization formula (\ref{ffff}), for which the hard-collinear scale
is of physical significance. Any spin-symmetry breaking contribution
involves at least one hard-collinear gluon exchange, and the amplitude
factorizes below the scale $\mu_{hc}$.

The result (\ref{ffwow}) allows us to systematically resum the 
short-distance logarithms arising in the evolution from high energies 
down to hadronic scales. The RG equation obeyed by the Wilson coefficient 
functions is \cite{Bauer:2000yr,Becher:2003kh}
\begin{equation}\label{RGE}
   \frac{d}{d\ln\mu}\,C_i(E,\mu)
   = \left( \Gamma_{\rm cusp}[\alpha_s(\mu)]\,\ln\frac{2E}{\mu}
   + \gamma[\alpha_s(\mu)] \right) C_i(E,\mu) \,,
\end{equation}
where the coefficient of the logarithmic term is determined by the cusp 
anomalous dimension \cite{Korchemsky:wg}. Its solution is
\begin{equation}\label{Cisol}
   C_i(E,\mu) = C_i(E,\mu_h)\,\exp U(\mu_h,\mu,E) \,,
\end{equation}
where $\mu_h\sim 2E$ is the high-energy matching scale for the transition 
from QCD to SCET, at which the values of the Wilson coefficients can be 
reliably computed using fixed-order perturbation theory. The RG evolution 
function can be written as
\begin{equation}\label{Uevol}
   U(\mu_h,\mu,E)
   = \int\limits_{\alpha_s(\mu_h)}^{\alpha_s(\mu)}\!d\alpha\,
    \frac{\Gamma_{\rm cusp}(\alpha)}{\beta(\alpha)}
    \Bigg[ \ln\frac{2E_\gamma}{\mu_h}
    - \int\limits_{\alpha_s(\mu_h)}^\alpha
    \frac{d\alpha^\prime}{\beta(\alpha^\prime)} \Bigg]
   + \int\limits_{\alpha_s(\mu_h)}^{\alpha_s(\mu)}\!d\alpha\,
    \frac{\gamma(\alpha)}{\beta(\alpha)} \,,
\end{equation}
where $\beta(\alpha_s)=d\alpha_s/d\ln\mu$. The dependence on the 
high-energy matching scale $\mu_h$ cancels against that of the Wilson 
coefficients $C_i(E,\mu_h)$ in (\ref{Cisol}). Note that after 
exponentiation the evolution function contains an energy and 
scale-dependent factor $\exp U(\mu_h,\mu,E)\propto E^{a(\mu)}$, where
\begin{equation}
   a(\mu) = \int^{\alpha_s(\mu)}\!d\alpha\,
   \frac{\Gamma_{\rm cusp}(\alpha)}{\beta(\alpha)} \,.
\end{equation} 
In order for this scale dependence to be canceled, the low-energy 
hadronic  matrix element must carry an energy dependence of the form 
$(\Lambda_{\rm QCD}/E)^{a(\mu)}$.

Evaluating the hadronic matrix elements at a low scale means that all 
{\em short-distance\/} dependence on the large energy $E$ is extracted 
and resummed in the coefficient functions. However, in the present case 
the matrix elements still contain a long-distance dependence on the large 
scale $E$, which cannot be factorized \cite{Becher:2003kh}. The reason is 
that the large energy enters the effective theory as an external variable 
imprinted by the particular kinematics of soft-to-collinear transitions. 
Because of the large Lorentz boost $\gamma=E/m_M$ connecting the rest 
frames of the $B$ meson and the light meson $M$, the low-energy effective 
theory knows about the large scale $E$ even though hard quantum 
fluctuations have been integrated out. This is similar to applications of 
heavy-quark effective theory to $b\to c$ transitions, where the fields 
depend on the external velocities of the hadrons containing the heavy 
quarks, and $\gamma=v_b\cdot v_c$ is an external parameter that appears 
in the matrix elements of velocity-changing current operators 
\cite{Falk:1990yz,Neubert:1993mb}. In the present case, it is perhaps 
reasonable to assume that the primordial energy dependence of the 
hadronic matrix elements at some low hadronic scale might be moderate, so
that the dominant $E$ dependence is of short-distance nature and can be 
extracted into the Wilson coefficient functions. However, there will 
always be some energy dependence left in the matrix elements; even if we 
assume that it is absent for some value of $\mu$, it will unavoidably be 
reintroduced when we change the scale. As a result, it is impossible to 
determine the asymptotic behavior of the QCD form factors 
$f_i^{B\to M}(q^2)$ using short-distance methods.\footnote{The same 
phenomenon is known to occur in the case of the Sudakov form factor, for 
which the coefficient of the double logarithm is sensitive to infra-red 
physics \cite{Korchemsky:1988hd,Kuhn:1999nn}.}

Let us now proceed to study the numerical importance of short-distance 
Sudakov logarithms. Given the exact results in (\ref{Cisol}) and 
(\ref{Uevol}), it is straightforward to derive approximate expressions 
for the resummed Wilson coefficients at a given order in RG-improved 
perturbation theory by using perturbative expansions of the anomalous 
dimensions and $\beta$-function. Unfortunately, controlling terms of 
$O(\alpha_s)$ in the evolution function $U$ would require knowledge of 
the cusp anomalous dimension to three-loop order (and knowledge of 
$\gamma$ to two-loop order), which at present is lacking. We can, 
however, control the dependence on the recoil energy $E$ to 
$O(\alpha_s)$. Following \cite{Bosch:2003fc}, we define the ratio 
$r=\alpha_s(\mu)/\alpha_s(\mu_h)$ and obtain
\begin{equation}
   e^{U(\mu_h,\mu,E)} = e^{U_0(\mu_h,\mu)}
   \left( \frac{2E}{\mu_h} \right)^{-\frac{\Gamma_0}{2\beta_0} \ln r}
   \left[ 1 - \frac{\alpha_s(\mu_h)}{4\pi}\,\frac{\Gamma_0}{2\beta_0}
   \left( \frac{\Gamma_1}{\Gamma_0} - \frac{\beta_1}{\beta_0} \right)
   (r-1)\,\ln\frac{2E}{\mu_h} \right] ,
\end{equation}
where
\begin{eqnarray}
   U_0(\mu_h,\mu) 
   &=& \frac{\Gamma_0}{4\beta_0^2} \bigg[
    \frac{4\pi}{\alpha_s(\mu_h)} \left( 1 - \frac{1}{r} - \ln r \right)
    + \frac{\beta_1}{2\beta_0}\,\ln^2 r \\
   &&\mbox{}\hspace{8mm} - \left( \frac{\Gamma_1}{\Gamma_0} 
    - \frac{\beta_1}{\beta_0} \right) (r-1-\ln r) \bigg] 
   - \frac{\gamma_0}{2\beta_0}\,\ln r + O(\alpha_s) \,.
   \nonumber
\end{eqnarray}
The only piece missing for a complete resummation at 
next-to-next-to-leading order is the $O(\alpha_s)$ contribution to $U_0$, 
which is independent of $E$. The relevant expansion coefficients are 
$\Gamma_0=\frac{16}{3}$, 
$\Gamma_1=\frac{1072}{9}-\frac{16}{3}\pi^2-\frac{160}{27}n_f$, 
$\gamma_0=-\frac{20}{3}$, and $\beta_0=11-\frac23\,n_f$, 
$\beta_1=102-\frac{38}{3}n_f$. We set $n_f=4$ in our numerical work.

\begin{figure}
\begin{center}
\epsfig{file=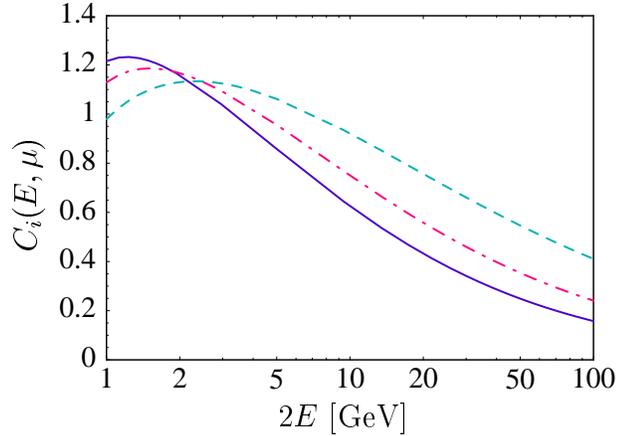,width=8cm} 
\end{center}
\centerline{\parbox{14cm}{\caption[Energy dependence of the Wilson
coefficients $C_i(E,\mu)$ at next-to-leading order in RG-improved
perturbation theory. ]{\label{fig:Sudakov} Energy dependence of the
Wilson coefficients $C_i(E,\mu)$ at next-to-leading order in
RG-improved perturbation theory. The three curves correspond to
$\alpha_s(\mu)=1$ (solid), $\alpha_s(\mu)=0.75$ (dashed-dotted), and
$\alpha_s(\mu)=0.5$ (dashed), which are representative of typical
hadronic scales.}}}
\end{figure}

In Fig.~\ref{fig:Sudakov} we show the dependence of the Wilson 
coefficients $C_i(E,\mu)$ on the large energy $E$. We choose $\mu_h=2E$ 
for the high-energy matching scale and use the tree-level initial
conditions $C_i(E,\mu_h)=1$, which is consistent at next-to-leading 
order. We fix $\mu$ at a low hadronic scale in order to maximize the 
effect of Sudakov logarithms. The maximum recoil energy in $B\to\pi$ 
transitions is such that $2E_{\rm max}\simeq 5.3$\,GeV. Obviously, for 
such values the perturbative resummation effects are very moderate. In 
the energy range $1\,\mbox{GeV}<2E<M_B$ the Wilson coefficients differ 
from unity by no more than about 20\%. The extrapolation to larger energy 
values shows that Sudakov suppression remains a moderate effect even for 
very large recoil energy.

\section{Factorization of spin-symmetry breaking effects}
\label{sec:fffact}

Using the close connection between messenger exchange and endpoint 
singularities discussed in Section~\ref{sec:toy}, the formalism of 
soft-collinear fields can be used to demonstrate the convergence of 
convolution integrals in QCD factorization theorems. If messenger 
exchange is unsuppressed, the convolution integrals diverge at the 
endpoints, spoiling factorization. By the same reasoning, convolution 
integrals are finite if soft-collinear messenger contributions are absent 
at leading power. 

Let us apply this method to show, to all orders in perturbation theory,
that the convolution integrals entering the spin-symmetry breaking term
in the factorization formula (\ref{ffff}) are free of endpoint 
singularities. As mentioned in the Introduction, this is an essential
ingredient still missing from the proof of this formula. With our 
technology the proof is rather simple and consists of only the following 
two steps:

1.\;
Soft-collinear messenger fields can be decoupled from the four-quark
operators mediating spin-symmetry breaking effects, which are of the type 
shown in (\ref{Q1Q2}). The reason is that in the color singlet-singlet 
case the operators are invariant under the field redefinition 
(\ref{blocks}) \cite{Becher:2003kh}. It follows that their matrix 
elements factorize into separate matrix elements of soft and collinear 
currents, which can be written in terms of the leading-order LCDAs of the 
external mesons. If the messengers did not decouple, they would introduce 
unsuppressed interactions between the soft and collinear parts of the 
four-quark operators, thereby spoiling factorization.

2.\;
Time-ordered products containing interactions of soft-collinear 
messengers with soft or collinear fields do not contribute to the 
spin-symmetry violating term in (\ref{ffff}). This follows from SCET power 
counting. The insertions of the Lagrangians ${\cal L}_{\bar q\theta}$ and 
${\cal L}_{\bar\theta\xi}$ in (\ref{3T}) cost a factor $\lambda^{3/2}$,
meaning that they can only come together with a leading-order current
operator of the type $\bar\X\dots\H\sim\lambda^{5/2}$. (Note that there 
are no $O(\lambda^{1/2})$ corrections to the current that could make the 
vacuum correlator (\ref{vacuum}) involving 
${\cal L}_{\bar q\theta}^{(1/2)}$ and 
${\cal L}_{\bar\theta\xi}^{(1/2)}$ non-zero \cite{Becher:2003kh}.) 
However, such a current will always be of the form (\ref{Jmatch}) because 
of the projection properties of the heavy-quark and collinear-quark 
spinors. It thus respects the spin-symmetry relations. 

These two observations imply that at leading order in $\Lambda_{\rm QCD}/E$ 
soft-collinear exchanges do not contribute to the spin-symmetry breaking 
term in the factorization formula, and hence the corresponding 
convolution integrals are convergent.

\chapter{Conclusion}\label{chap:conclusion}

QCD-Factorization theorems are statements about leading power
properties of amplitudes {\em to all orders in the strong coupling
constant}. The use of effective field theories enables one to prove
such theorems rigorously, and thus take a significant step further
from past fixed-order proofs. In this thesis we demonstrated this
technique on some decay modes that are particularly clean in the sense
of simplicity of QCD-factorization formulae. The first part of this
work was devoted to a detailed introduction to Soft-Collinear
Effective Theory (SCET), which describes the infra-red physics of
strong interactions between light, but highly energetic partons with
soft heavy or light degrees of freedom. The crucial advantage of this
theory is that it allows for a systematic expansion in inverse powers
of the heavy-quark mass already at the Lagrangian level, therefore
allowing for all-order proofs. Furthermore, SCET enables us to perform
perturbative calculations for systems in which local operator
expansions break down due to long-distance effects. This is a
significant improvement of our ability to compute the
strong-interaction effects of weak decays of heavy mesons.
Furthermore we investigated two non-perturbative structure functions:
the $B$-meson light-cone distribution function, which enters the
calculation of exclusive decay amplitudes in the high-recoil region,
and the shape function, which encodes the Fermi-motion of the heavy
quark inside the $B$-meson for the calculation of inclusive decay
modes. In particular, we have derived important relations that link the
(finite-interval) moments of the shape function to the values of local
HQET operators.

To gain full control over the separation of physics of different
energy scales, it is necessary to perform a resummation of large
logarithms. We have demonstrated this technique for inclusive $B\to
X_u\,l^-\bar\nu$ and exclusive $B^-\to\gamma\,l^-\bar\nu$ and $\bar
B\to M\,l^-\bar\nu$ (where $M$ is a light pseudoscalar or vector
meson), using a sophisticated multi-step matching procedure and
solving renormalization-group equations (RGEs). In the inclusive decay
mode we matched QCD $\to$ SCET$_I$ $\to$ HQET. The intermediate theory
SCET$_I$ contains soft and hard-collinear degrees of freedom, the
latter being momentum modes with large energy $E$ and moderate
invariant momentum square of order $E\Lambda_{\rm QCD}$. In a second
step, these modes are removed, resulting in a description within
heavy-quark effective theory (HQET). The triple differential decay
rate was then expressed at leading power in $\Lambda_{\rm QCD}/E$ and
at next-to-leading order in renormalization-group improved
perturbation theory in terms of Wilson coefficient functions $H$ and
$J$, convoluted with the matrix element of a single HQET operator, the
shape function $S$. This formulation is valid in the shape-function
region of phase space, where the $X_u$ system is energetic and light,
as anticipated. Model-independent results have been derived for
various event fractions and spectra, including a cut on the
charged-lepton energy, the hadronic variable $P_+ = E_H-|\vec P_H|$,
and the hadronic invariant mass. Finally we have computed numerical
predictions using models for the shape function that are consistent
with all known constraints.  As a result, event fractions needed for
the extraction of the CKM matrix element $|V_{ub}|$ are found to be
much larger than previously anticipated. This in turn implies that the
numerical value of $|V_{ub}|$ determinations from inclusive $B$ decays
might become smaller. To make a conclusive statement it will be
necessary to include subleading power corrections (and possibly
higher-order corrections) to the predictions derived here. For future
determinations we proposed to use the $P_+$ variable to discriminate
against the charm background. A detailed study of the current
theoretical uncertainties and the nature of the charm background was
also performed.

QCD-Factorization was proved to all orders in the coupling constant
and to leading power for the exclusive $B^-\to\gamma\,l^-\bar\nu$
decay amplitude. This mode provides a clean and realistic environment
in which to study the aforementioned methodology. We performed a
detailed matching calculation QCD $\to$ SCET$_{II}$. Large logarithms
were resummed by first factorizing the hard-scattering kernel into a
hard and a jet function. This is accomplished by matching in two steps
QCD $\to$ SCET$_I$ $\to$ SCET$_{II}$. While the first matching step
gives rise to the hard function, the jet function arises in the second
step. Sudakov resummation was then achieved by solving the RGEs of the
hard function and the combined hard-scattering kernel. We found that
the resummation effects enhance the amplitude (as opposed to a
``Sudakov suppression'').

Finally we turned our attention to amplitudes that do not
factorize. Form factors encode the exclusive decay amplitudes for $B$
$\to$ $P\,l^-\bar\nu$ and $B$ $\to$ $V\,l^-\bar\nu$ \linebreak ($P=$
light pseudoscalar meson, $V=$ light vector meson). There are three
different form factors for $B\to\pi$ and seven for $B\to\rho$
transitions. In the large recoil limit they obey approximate
spin-symmetry relations (corrections of which factorize), leaving only
three independent universal form factors. In our formalism, the proof
of this statement was rather simple; however, our analysis revealed
also that there exists a soft contribution to the form factors which
cannot be factorized. In SCET$_{\rm II}$ this can be understood as the
soft-collinear sector does not decouple. As a result, there exists a
long-distance operator $O_5$, which mixes into the short-distance
operators under renormalization. Since the short-distance operators do
not mix back into $O_5$, the anomalous dimension of $O_5$, which is
identical to the one of heavy-to-collinear current operators, is also
an Eigenvalue of the entire anomalous dimension matrix. The matrix
element of the corresponding ``Eigenvector'', a linear combination of
all operators, is endpoint-finite and contributes to the form factor
without any suppression. In fact, the knowledge of the anomalous
dimension Eigenvalue allows to control the short-distance dependence
on the large recoil energy. We found that there is no sizable Sudakov
suppression to the universal form factors. In summary, the
spin-symmetric soft overlap exists, and dominates over the
factorizable hard-scattering contributions that break spin-symmetry.

In light of these few examples of $B$ decay processes, the power of
the QCD-Factorization approach to strong-interaction effects with more
than one relevant physical energy scale is apparent. While we are
unable to solve QCD analytically, an approximation to leading power in
the expansion parameter $\Lambda_{\rm QCD}/m_b$ can be of great help
in our ability to predict Standard-Model processes. In some cases
(e.g.~inclusive decays) we are even provided with a framework in which
corrections to the leading-power factorization formulae can be
calculated systematically. In the future, we will be able to use such
frameworks to improve the precision of theoretical input quantities
necessary for the extraction of Standard Model parameters and for
searches of New Physics effects. We shall mention here the important
examples of flavor-changing neutral decay modes such as $B \to
X_s\,\gamma$ and $B \to X_s\,l^+l^-$, which are particularly sensitive
to possible New Physics, but suffer from sizable QCD-uncertainties
which can spoil valuable information. It will certainly be worthwhile
to apply the general methodology presented in this Thesis and in the
corresponding references to these processes, as well as to improve the
accuracy of our results in both the power expansion and the coupling
expansion to higher orders. On the exclusive side we have dealt with a
few exemplary decay channels. It will be exciting to study how the
lessons we learned are applicable to decay modes of bigger
phenomenological interest, such as the radiative $B\to K^*\gamma$, or
the non-leptonic $B\to \pi \pi$ and $B\to K \pi$. As far as searches
for New Physics effects are concerned, precision in Flavor Physics can
provide us with valuable information. In both the experimental and
theoretical community this fact is reflected in the ongoing discussion
on the potential of a ``Super $B$-factory'', that might complement the
direct searches of New Physics at next-generation particle colliders.

\appendix
\chapter{List of Abbreviations and Acronyms}

\begin{tabular}{l@{ - }p{0.6\linewidth}}
c & collinear \\
CKM & Cabibbo, Kobayashi, Maskawa  \\
CLCG & Collinear Light-Cone Gauge \\
h & hard \\
hc & hard-collinear \\
HQET & Heavy-Quark Effective Theory \\
NLO & Next-to-Leading Order \\
NNLO & Next-to-Next-to-Leading Order \\
pQCD & perturbative QCD \\
QCD & Quantum Chromo Dynamics \\
RG & Renormalization Group \\
RGE & Renormalization-Group Equation \\
s & soft \\ 
sc & soft-collinear \\
SCET & Soft-Collinear Effective Theory \\
SLCG & Soft Light-Cone Gauge \\
us & ultrasoft 
\end{tabular}

\end{document}